\documentclass[trackchanges, twocolumn]{aastex701}

\usepackage{amsmath}

\usepackage{subcaption}

\usepackage{graphicx}

\usepackage{eso-pic}
\usepackage{xcolor}
\usepackage{graphicx}   
\usepackage{transparent}

\begin{document}

\definecolor{darkorange}{rgb}{1.0, 0.55, 0.0}
\newcommand{\ed}[1]{\textcolor{darkorange}{[#1]}}

\title{JWST Edge-on Disk Ice (JEDIce): An Analysis of PAH Signatures in Edge-on Class II T Tauri Disks}

\author[orcid=0000-0002-8023-2834,sname='North America']{Charles Mentzer}
\affiliation{Institute of Astronomy, Department of Physics, National Tsing Hua University, Hsinchu, Taiwan}
\email[show]{cmentzer@gapp.nthu.edu.tw}  

\author[0000-0001-6307-4195]{Daniel Harsono}
\affiliation{Institute of Astronomy, Department of Physics, National Tsing Hua University, Hsinchu, Taiwan}
\email{dharsono@gapp.nthu.edu.tw}

\author[0000-0002-8716-0482]{Jennifer B.~Bergner}
\affiliation{University of California, Berkeley, Berkeley CA 94720, USA}
\email{jbergner@berkeley.edu}

\author[0000-0003-1197-7143]{Emmanuel Dartois}
\affiliation{Institut des Sciences Mol\'{e}culaires d’Orsay, CNRS, Univ. Paris-Saclay, 91405 Orsay, France}
\email{emmanuel.dartois@universite-paris-saclay.fr}

\author[0000-0003-4985-8254]{Jennifer A. Noble}
\affil{Physique des Interactions Ioniques et Mol\'{e}culaires, CNRS, Aix-Marseille Univ., 13397 Marseille, France}
\email{jennifer.noble@univ-amu.fr}

\author[0000-0002-7402-6487]{Zhi-Yun Li}
\affiliation{Department of Astronomy, University of Virginia, Charlottesville, VA 22903, USA}
\affiliation{Virginia Institute of Theoretical Astronomy, University of Virginia, Charlottesville, VA 22903, USA}
\email{zl4h@virginia.edu}

\author[0000-0002-2131-4346]{Korash Assani}
\affiliation{Department of Astronomy, University of Virginia, Charlottesville, VA 22903, USA}
\affiliation{Virginia Institute of Theoretical Astronomy, University of Virginia, Charlottesville, VA 22903, USA}
\email{ka8km@virginia.edu}

\author[0000-0001-7479-4948]{Maria Drozdovskaya}
\affiliation{Department of Chemistry, Biochemistry and Pharmaceutical Sciences (DCBP), Universit{\"a}t Bern
Freiestrasse 3, 3012 Bern, Switzerland}
\email{}

\author[0000-0002-0554-1151]{Mayank Narang}
\affiliation{Jet Propulsion Laboratory, California Institute of Technology, 4800 Oak Grove Drive, Pasadena, CA 91109, USA}
\email{mayank.narang@jpl.nasa.gov}

\author[0000-0001-8227-2816]{Yao-Lun Yang}
\affiliation{Star and Planet Formation Laboratory, RIKEN Pioneering Research Institute, Wako-shi, Saitama, 351-0106, Japan 351-0198}
\email{yaolunyang.astro@gmail.com}

\author[0000-0003-1878-327X]{Melissa McClure}
\affiliation{Leiden Observatory, Leiden University, PO Box 9513, 2300 RA Leiden, The Netherlands}
\email{mcclure@strw.leidenuniv.nl}

\author[0009-0008-8810-6577]{Lukas Welzel}
\affiliation{Leiden Observatory, Leiden University, PO Box 9513, 2300 RA Leiden, The Netherlands}
\email{welzel@strw.leidenuniv.nl}

\author[0000-0002-0786-7307]{Will E. Thompson}
\affiliation{University of California, Berkeley, Berkeley CA 94720, USA}
\email{willet@berkeley.edu}

\author[0000-0002-3401-5660]{Julia C.~Santos}
\affiliation{Center for Astrophysics, Harvard \& Smithsonian, 60 Garden St., Cambridge, MA 02138, USA}
\email{julia.santos@cfa.harvard.edu}

\begin{abstract}

 Polycyclic aromatic hydrocarbons (PAHs) are ubiquitous carbonaceous species in the interstellar medium (ISM) that have been elusive in low-mass protoplanetary disks. In higher-mass protoplanetary disks, they are widely used as diagnostics of disk surface chemistry, UV processing, and ionization structure. Recent \emph{James Webb Space Telescope (JWST)} observations have begun detecting PAHs in edge-on protoplanetary disks around T Tauri stars. Using NIRSpec and MIRI IFU observations from the JEDIce survey, we analyze near- and mid-infrared emission features commonly attributed to PAHs. We investigate the spatial distribution of PAHs in these edge-on disks utilizing band ratios (i.e., 3.3/11.3~$\mu$m, 11.3/7.7~$\mu$m, and 3.4/3.3~$\mu$m) to constrain PAH representative sizes, ionization, and the aliphatic fraction ($\eta_{\rm ali}$). PAHs are detected toward 3 out of 6 disks in our sample, whose band ratios have constrained their sizes to $\approx 30–100~N_\mathrm{C}$, with ionization fractions typically lower than 50\% and $\eta_{\rm ali}$ above typical ISM processing levels of 4--6\%. Towards two systems, Oph~163131 and ESO H$\alpha$ 574, the central position seems devoid of PAH emission in contrast to further radii, likely due to shielding. Including these targets, we have roughly doubled the detections of PAHs around T Tauri stars and have shown through the combination of edge-on disks and JWST constraints on gas excitation and ice composition, PAHs emerge as a powerful diagnostic of radiation-driven evolution, ionization balance, and the processing of aromatic carbon during planet formation supporting the inheritance of interstellar carbonaceous material as seen in the $\rho$ Oph disks.

\end{abstract}


\section{Introduction} \label{sec:intro}

Understanding the evolution of carbonaceous material from the interstellar medium (ISM) to planet-forming environments is a central problem in star and planet formation \citep{vanDishoeck2014,Oberg_2021, Bergin2026}. Protoplanetary disks (PPDs) represent a transitional stage between diffuse interstellar clouds and planetary systems \citep{henning2013}. In PPDs, carbon is distributed among volatile species (e.g., CO), refractory carbonaceous grains, and complex organics formed by energetic processing of ice and dust \citep{henning2013, Oberg_2021, vanDishoeck2014}. Polycyclic aromatic hydrocarbons (PAHs) are a distinct, spectroscopically accessible part of the small-grain carbon budget, linking solid-state carbonaceous material to molecular-scale aromatic structures. PAHs are chemically important as potential precursors to more carbon-rich complex organic material \citep{Bernstein1999, Kwok2016}. They are relatively common in the interstellar medium \citep{leger1984,All1989}.  Thus far, space- and ground-based observations have had difficulties detecting PAHs toward protoplanetary disks around T Tauri stars \citep[e.g.,][]{Geers2006,Tielents2008}.  Recent \emph{James Webb Space Telescope (JWST)} observations have detected the presence of PAHs toward edge-on disks around HH~48~NE \citep{Sturm_2024} and Tau~042021 \citep[2MASS~J04202144+2813491,][]{Arulanantham2024} via their mid-IR spectral features.  Due to the lack of characterization of PAHs we still do not fully understand how PAHs are distributed in disks around T-Tauris.

PAH vibrational modes produce characteristic infrared emission features near 3.3, 6.2, 7.7, 8.6, and 11.3~$\mu$m, corresponding to aromatic C–H stretching (3.3~$\mu$m), C–C stretching (6.2 and 7.7~$\mu$m), in-plane C–H bending (8.6 $\mu$m), and out-of-plane C–H bending (11.3~$\mu$m) modes. Ionized PAHs enhance the 6.2, 7.7, and 8.6 $\mu$m bands, while neutral PAHs dominate the 3.3 and 11.3~$\mu$m features, making band ratios sensitive probes of the radiation field, electron density, equilibrium temperature, and molecular properties \citep{All1989,Li2001,Draine2007}. PAHs should be understood as an ensemble population rather than a single, well-defined species, as they exhibit variation in size, structure, and charge state.  As a result, commonly observed PAH tracers are effective diagnostics of the underlying population’s charge balance, composition, and size distribution rather than indicators of a single molecular carrier.  A key open question is whether the PAH population observed in the ISM is inherited intact into disks, chemically processed during collapse, or substantially reconstituted within the disk environment \citep{Tielents2008, Visser2009, henning2013}. 

To trace the level of chemical processing within the system, we use the 3.4/3.3~$\mu$m band ratio to estimate the aliphatic fraction ($\eta_{\rm ali}$) \citep{Tielents2008}. Following \citet{yang2023}, we define $\eta_{\rm ali}$ as the fraction of carbon atoms in aliphatic chains relative to those in aromatic rings (e.g., in PAHs). The 3.3/11.3~$\mu$m ratio traces PAH size. Because both features mainly arise from neutral PAHs, their ratio traces the characteristic PAH size (in number of carbon atoms $N_\mathrm{C}$) via excitation temperature differences \citep{Draine2007}. In this work, we refer to PAHs with $N_\mathrm{C}\sim20-40$ as small, those with $N_\mathrm{C}\sim50-90$ as intermediate-sized, and those with $N_\mathrm{C}\gtrsim100$ as large. The 11.3/7.7~$\mu$m ratio traces PAH charge. The 7.7~$\mu$m emission is strongly enhanced in ionized PAHs relative to the 11.3~$\mu$m emission, making the 11.3/7.7~$\mu$m ratio a sensitive diagnostic of the PAH ionization state \citep{Szczepanski1993, Li2001, Draine2007, 2020AL}.

Observational studies of PAHs in disks have traditionally focused on systems with relatively face-on orientations ($<10^{\circ}$), where the disk surface layers are directly visible and emission features are strong against a bright infrared continuum \citep{Geers2007, van_der_Marel_2023,vanderMarel2026}. Early detections of PAH emission in protoplanetary disks were enabled by the \textit{Infrared Space Observatory} (ISO), which revealed prominent aromatic bands in Herbig Ae/Be systems and established PAHs as an important tracer of irradiated disk surfaces \citep{Meeus2001,Acke2004}. Subsequent observations with the \textit{Spitzer Space Telescope} expanded these surveys, enabling statistical studies of PAH detection rates, feature strengths, and correlations with stellar properties and disk structure \citep{Geers2006,Geers2007,Oliveira2010,Seok2017}. Ground-based observations with the \textit{Very Large Telescope} (VLT) further provided spatially resolved constraints on the radial extent of PAH, demonstrating that the emission often originates from the warm, UV-irradiated surface layers of the disk within tens of AU from the star \citep{Lagage2006, Habart2006,Geers2007}. These early studies showed that PAH emission is generally stronger and more frequently detected around Herbig Ae/Be stars than around the low mass T Tauri stars, consistent with the expected dependence of PAH excitation on ultraviolet radiation fields \citep{Geers2006,Geers2007}. 

These spatially integrated spectra we can reveal global trends in PAH luminosity and band ratios as a function of stellar type, disk mass, or evolutionary stage \citep{Seok2017}. However, disentangling the vertical distribution of the emission is challenging in face-on systems, since line-of-sight integration blends contributions from a range of physical regions \citep{Dullemond2004,Woitke2016}. Moreover, the bright inner disk can dominate the continuum, complicating efforts to isolate faint extended emission from the outer regions \citep{Robitaille2006}. 

For a few T Tauri stars with PAH detections so far, the PAH emission is consistently weaker than those features found toward high-mass counterparts with spectra often characterized by relatively strong 11.3~$\mu$m emission and high 11.3/7.7~$\mu$m band ratios indicative of predominantly neutral PAH populations \citep{Geers2006, Arulanantham2024, Sturm_2024}. The inferred PAH abundances are typically lower than ISM values, likely due to PAH sequestration via clustering and adsorption onto larger grains followed by settling, destruction by stellar X-ray irradiation, and/or the weaker photospheric FUV fields of T Tauri stars \citep{Geers2006,Lange2025}. 

Several recent detections in highly inclined and edge-on systems suggest that these viewing geometries are particularly favorable for detecting PAH emission in T Tauri disks \citep{Arulanantham2024,Sturm_2024}. Edge-on PPDs provide a unique laboratory for characterizing PAH. The optically thick mid-plane obscures the bright central star and inner disk, suppressing continuum emission that can otherwise overwhelm PAH features \citep{Pontoppidan2007,Dullemond2010}. The observed infrared emission instead traces scattered light and thermal emission from the disk surface layers, enhancing the contrast of PAHs emission originating in the disk atmosphere \citep{Acke2004, Watson2007}. Because the line of sight probes a large vertical column of material, edge-on disks also enable constraints on how PAH properties vary with disk height. However, PAH excitation depends strongly on the local UV radiation field, which is shaped by disk geometry, shadowing, and stellar irradiation. As a result, spatial variations in PAH band ratios may reflect both intrinsic molecular properties and radiative transfer effects \citep{Draine2007, Woitke2016, Sturm_2024}.

The `JWST Edge-on Disk Ice' (JEDIce) program observes six edge-on disks, about half of the known edge-on disks around low-mass stars, with  JWST NIRSpec \citep{NIRSpec} and MIRI instruments \citep{MIRI_2,MIRI}. 

In this work, we present a systematic empirical investigation of PAH emission in the sample of edge-on PPDs observed by the JEDIce collaboration initially reported in \cite{bergner2026}. Our objectives are threefold: First, we characterize the observed PAH band profiles and ratios to infer size distributions, ionization states, and processing across the disks. This is done via the 3.3/11.3~$\mu$m, 11.3/7.7~$\mu$m band ratios, and $\eta_{\rm ali}$. Second, we map the vertical and radial distribution of PAH emission, quantifying its scale height and spatial extent relative to the dust continuum. Third, we use these observational constraints to assess PAH abundances and their role in disk energetics, placing PAHs in the broader context of disk structure and carbon chemistry. In doing so, we aim to clarify how aromatic carbon evolves from the interstellar medium to planet-forming environments and to assess its significance for the composition of emerging planetary systems, without yet invoking a detailed radiative transfer analysis. This paper is organized as follows: Section 2 presents the data preparation and the methodology used to fit the observed PAH emission features.  Our results can be found in Section 3 for detections, their fits using PAHFIT, and spatial analysis when available. We discuss the variation of the observed PAH features across the different disks and their properties in Section 4. Finally, we summarize our results and conclusions in Section 5. 


\section{Methodology} \label{sec:methods}

\subsection{Data Preparation} \label{sec:red}

All data presented in this work were processed following the steps described in \cite{bergner2026}, with several modifications such as rigorous outlier line removal and NIRSpec to MIRI scaling tailored to PAH emission to address the specific scientific goals of this study. Given the large number of gas emission lines present in the data, a detailed outlier analysis on the uncalibrated data was conducted on a per-dither basis: 12 dithers for all NIRSpec observations and 4 for MIRI. This approach enabled robust discrimination between genuine spectral features and transient artifacts.  Finally, when scaling between instruments (i.e., NIRSpec to MIRI) was required, the scaling was anchored at the blue end of of MIRI Channel 2 Long to ensure proper scaling of the 11.3~$\mu$m PAH feature. The resulting scaling factors were generally small, ranging from 1 to 2, indicating good overall agreement between the instruments (see Appendix~\ref{Shifting}).

\subsection{Spatial Isolation and Analysis of PAH Emission} \label{sec:origin}

 Among the targets, HV~TAU~C and LkH$\alpha$~263C are members of multiple stellar systems \citep{Jayawardhana2002,Stapelfeldt2003,Villenave2020}. Meanwhile, Oph~163131, the Flying~Saucer, and OphE~MM3 are all located within the broader $\rho$~Oph star forming region \citep{Brandner2000,Grosso2003,Guilloteau2016,Villenave2020,Wolff2021}. To differentiate the disks' PAH emission from the background or neighboring contamination (as in the cases of HV~TAU~C and LkH$\alpha$~263C) we constructed 2, 5, and 10$\sigma_{RMS}$ contours of the 3.3~$\mu$m PAH emission to identify the different components: background emission, extended disk emission, and the brightest peak emission. These levels were chosen to visually match these distinct emission components present in the data. We calculate 10 radial slices that support these choices, with the 2$\sigma_{\rm RMS}$ level consistently lying above the baseline flux, providing a lower cutoff for background emission, while 10$\sigma_{\rm RMS}$ isolates the central peak. The 5$\sigma_{\rm RMS}$ level therefore provides a reasonable intermediate threshold for defining the extended emission. The 2 to 10$\sigma$ region is subtracted from the data cube and residuals are inspected to assess the quality of the subtraction (see Appendix \ref{ConAnn}). Any features that appeared over-subtracted were interpreted as likely belonging to the extended disk structure and were therefore retained for subsequent disk analysis. 

Within the MIRI data, longer mid-infrared wavelengths are less optimal for identifying subtle spatial structures because the larger diffraction-limited PSF and coarser spatial sampling reduce angular resolution. Therefore, we adopt the spatial masks used for NIRSpec, after aligning the WCS of both instruments with that of the archival ALMA data to account for pointing differences between the instruments (see Appendix~\ref{Shifting}).

PAH emission features were spectrally decomposed using PAHFIT \citep{PAHFIT} for the NIRSpec and MIRI data separately. The spectra were extracted using the 5$\sigma_{RMS}$ emission contour to define the aperture for photometry, with each pixel subdivided into a $5\times5$ grid of sub-pixels to account for fractional pixel coverage at the aperture boundaries, following the approach implemented in \texttt{Photutils} \citep{photoutils}.

Because PAHFIT is primarily designed for mid-infrared spectra, we adapted the fitting procedure for these edge-on disks. The observed spectra are modeled as a combination of thermal dust, stellar continua, PAH/aliphatic emission features, and extinction along the line of sight. The continuum is represented by multiple modified blackbodies at different temperatures to approximate the thermal dust emission, together with a simple blackbody stellar component. This continuum treatment is preferred because it provides a more physically motivated representation of the underlying emission. PAH and aliphatic bands are modeled using Drude profiles, following the functional form commonly used for PAH emission features. Drude profiles are well suited to these broad vibrational bands because they reproduce their extended wings and broad spectral profiles more naturally than Gaussian functions. Each Drude component is characterized by its central wavelength, width, and amplitude, with its integrated flux providing the corresponding feature strength.

We adopt a foreground-screen extinction geometry in which the continuum and emission components are attenuated by intervening disk material. For the MIRI spectra, we adopt the KVT extinction curve based on the silicate profile of \cite{Kemper2004}. For NIRSpec, we additionally include a simplified ice-extinction component to account for the strong near-infrared ice absorption observed toward several targets. Details of the ice-extinction model are provided in Appendix~\ref{ice_abs}.

The NIRSpec and MIRI regions are fitted independently, allowing their thermal dust and extinction parameters to vary, while the stellar continuum identified in the MIRI spectrum is retained in the NIRSpec fit. The principal features considered are the 3.3, 6.2, 7.7, and 11.3~$\mu$m aromatic PAH bands and the 3.4~$\mu$m aliphatic band. Multiple Drude components are used where necessary to reproduce blended or non-canonical profiles. The 3.3/11.3 and 11.3/7.7~$\mu$m band ratios are calculated from the integrated fluxes of the corresponding fitted components, with the 3.3/11.3~$\mu$m ratio therefore combining feature measurements from the independently optimized NIRSpec and MIRI fits.

The continuum, feature, and extinction parameters are optimized simultaneously using Levenberg-Marquardt least-squares minimization within the PAHFIT fitting framework.  This simultaneous optimization allows the continuum, extinction, and discrete emission features to be fit self-consistently rather than determining the continuum and features sequentially.

Given the large spatial extent of the emission contours (see Figures~\ref{fig:coll_PAH} and \ref{fig:coll_null}), we also are able to identify PAH emission originating in the disk based on an analysis of the radial and vertical behavior. Indeed, variations of PAHs within the disk can arise from disk emission, whereas the background would be expected to be nearly uniform. Therefore, we define rectangular aperture slices via aperture photometry with \texttt{Photutils} \citep{photoutils} to examine the PAH and aliphatic emission. These apertures enable the behavior of the 11.3/7.7, 3.3/11.3, and 3.4/3.3~$\mu$m band ratios to be spatially examined while mitigating and accounting for PSF smearing.


\section{Results} \label{sec:results}

\begin{figure*}[htb!]
\centering

\includegraphics[width=\textwidth]{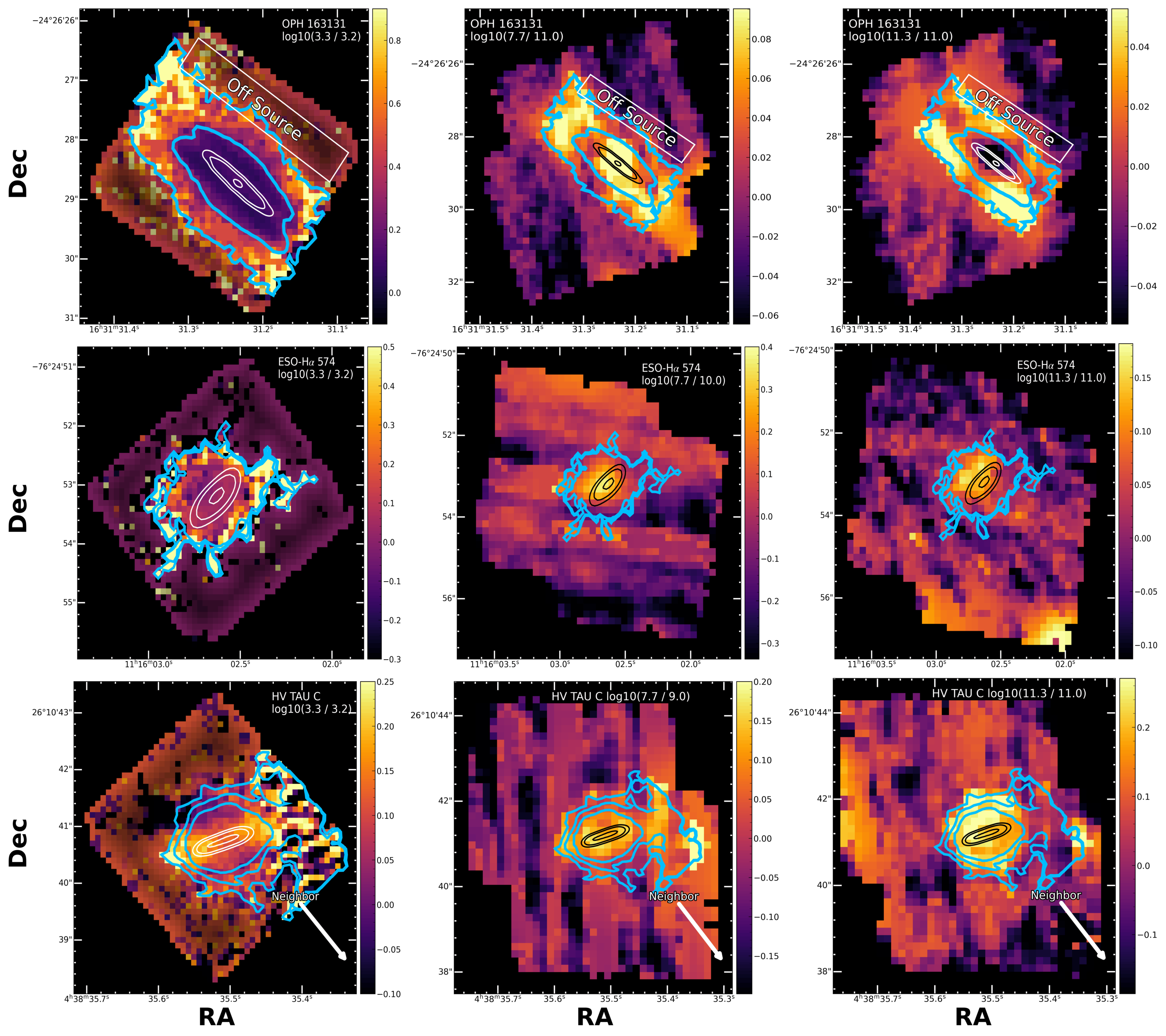}

 \caption{PAH emission at 3.3, 7.7, and 11.3~$\mu$m seen in logarithmic intensity maps normalized to neighboring PAH-free wavelength channels. Blue contours correspond to PAH emission at 3.3~$\mu$m for 2, 5, and 10 $\sigma_{RMS}$ while the inner black/white contours show the 30\% 70\% and 90\% of the peak intensity of the ALMA millimeter continuum observations of ESO-H$\alpha$574: 225 GHz, HV~TAU~C: 284 GHz, and Oph~163131: 224 GHz. For Oph~163131 an Off-Source aperture is placed to investigate any possible background PAH emission that is examined in \S~\ref{sec:Off_Source}.} 

\label{fig:coll_PAH}
\end{figure*}

We first identify PAH emission associated with the disk components using normalized logarithmic intensity maps at 3.3, 7.7, and 11.3~$\mu$m (Figures~\ref{fig:coll_PAH} and \ref{fig:coll_null}). These integrated PAH excess maps are normalized to continuum reference points at 3.2, $\sim$10, and 11.0~$\mu$m, respectively, with the middle reference point selected for each source to minimize noise, see \cite{Dartois2026}. We then examine the PAH band strengths and ratios using PAHFIT, followed by a spatial analysis of $\eta_{\rm ali}$ and ratios involving the 11.3~$\mu$m emission for sources with detected disk PAHs, where the signal-to-noise ratio permits. Finally, we compare these signatures with off-source extractions to investigate the PAH emission and conditions surrounding the disks.

\subsection{PAH Detections}\label{sec:detection}

From these normalized maps, several sources exhibit clear PAH concentration around the disk. In particular, ESO-H$\alpha$ 574, HV~TAU~C, and Oph~163131 show strong emission consistent with typical disk morphology across the canonical PAH bands, supporting their classification as robust detections. The 3.3~$\mu$m PAH feature is covered by NIRSpec, while the 7.7 and 11.3~$\mu$m fall within the MIRI band. 

\subsubsection*{OPH~163131}

The NIR normalized image shows a clear PAH excess ($>$5$\sigma$) that surrounds the millimeter disk as shown in Figure~\ref{fig:coll_PAH}. In addition, the the over-subtraction at 3.3 and 3.4 $\mu$m further supports it is not a background PAH feature (see Appendix~\ref{ConAnn}).  

To further investigate this structure, we examine the MIRI data by the integrated 11.3/11.0 and 7.7/11.0 $\mu$m normalized images seen in Figure~\ref{fig:coll_PAH}. Both the 7.7/11.0 and 11.3/11.0 $\mu$m images show a similar extended emission component as is seen in the 3.3/3.2 $\mu$m image. In addition, the 7.7/11.0 $\mu$m image shows a centrally peaked on-source component, while the 11.3/11.0 $\mu$m image shows a pronounced dip at the center. This could be the stronger ionizing radiation field present within the 10$\sigma$, strengthening the 7.7~$\mu$m feature \citep{All1989}. Because of this we focus on two distinct regions for subsequent analysis: the central 2\arcsec x 3\arcsec ~and the surrounding larger 4\arcsec x 5\arcsec ~regions. This division allows us to compare the PAH properties of the inner disk with those of the more extended cloud surrounding the system.

\subsubsection*{ESO-H$\alpha$ 574}

From the normalized 3.3~$\mu$m image, PAH emission is detected in the upper and lower lobes of the disk that sharply falls with distance (see Figure~\ref{fig:coll_PAH}), indicating pronounced PAH emission in this target. The lack of apparent background contamination for the prominent 3.3~$\mu$m feature as observed with NIRSpec strongly indicates that PAH emission originates from the disk. 

The MIRI images show that the upper region of the disk appears brighter, consistent with an absolute flux enhancement in the upper portion of the disk, given this source's less edge-on nature compared to the rest of the sample. Within the region enclosed by the 10$\sigma$ contour, we detect 11.3$\mu$m PAH emission originating from the disk. This aperture also includes the 7.7~$\mu$m PAH feature, which is strongly localized to the disk as well.

\subsubsection*{HV~Tau~C}

For the region around HV~TAU~C, an extended cloud is located between HV~TAU~C and HV Tau AB SW of the disk as seen in the 3.3~$\mu$m normalized image. This cloud does not appear to exhibit significant PAH emission; instead, its emission is dominated by a blackbody continuum. After subtraction, the over-subtraction of the absolute flux levels suggests that any PAH signatures found within the 5$\sigma$ are related to the disk. 

The MIRI images show that the 7.7 and 11.3~$\mu$m emission is strongly concentrated in the disk and is not significantly attributed to the extended cloud SW of the disk. The consistent morphology and its confinement to the central bright contours further support that the PAH emission is localized to the disk around HV~TAU~C.

\subsection{PAH Non-Detections}\label{sec:null_detection}
\label{subsec:nondet}

\begin{figure*}[htb!]
\centering

\includegraphics[width=\textwidth]{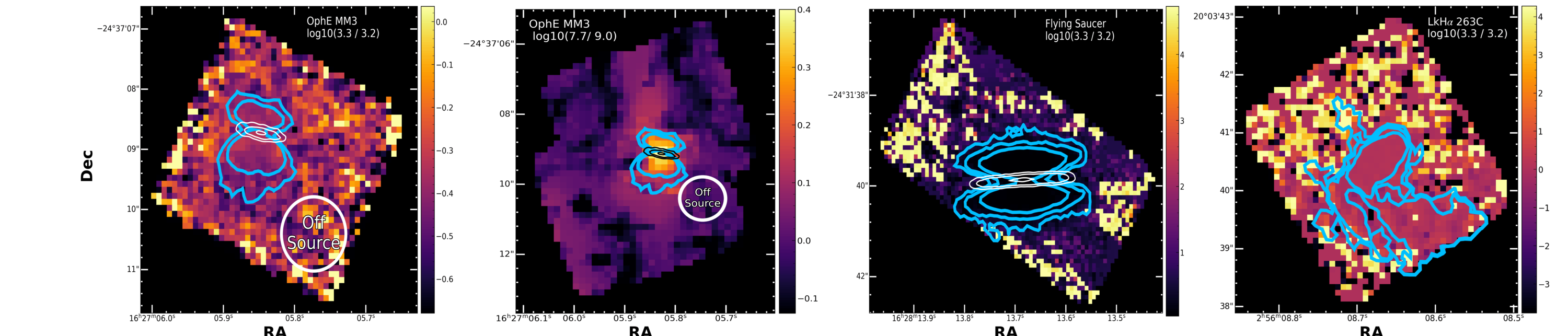}

\caption{Logarithmic normalized intensity maps for targets without conclusive PAH detection. Contour definitions remain consistent with Figure~\ref{fig:coll_PAH} using ALMA millimeter continuum observations of Flying~Saucer: 347 GHz and OphE~MM3: 671 GHz. ALMA data for LkH$\alpha$ 263C were not available. For OphE~MM3 an Off-Source aperture is placed to investigate the foreground cloud not in the direction of the disk that is examined in \S~\ref{sec:Off_Source}.}

\label{fig:coll_null}
\end{figure*}

In this sample, three sources show non- or ambiguous signal localization (see Figure~\ref{fig:coll_null}). OphE~MM3 may show localized PAH emission in MIRI at 7.7~$\mu$m, but NIRSpec indicates very large extinction \citep{vanKempen2009}, suggesting the source is embedded; we therefore cannot separate the foreground cloud from disk emission and classify it as a non-detection. The Flying~Saucer, lying in the foreground of the $\rho$ Ophiuchus molecular cloud \citep{Brandner2000,Grosso2003,Guilloteau2016,Villenave2020,Wolff2021}, suffers from strong background contamination and shows no localized PAH emission features. Toward LkH$\alpha$ 263C, PAH emission arises in a nearby cloud rather than the disk. Thus, all three are treated as non-detections of localized PAH emission.

\subsubsection*{OphE~MM3}

When examining the images as seen in Figure~\ref{fig:coll_null}, the 3.3~$\mu$m emission is spread throughout the image and appears not localized to the disk. In contrast, the 7.7~$\mu$m emission appears concentrated on the source. Although the 11.3~$\mu$m feature appears relatively clean in the spectra, the maps are affected by pronounced standing waves at $>$12 $\mu$m impacting the analysis of the relatively weak feature so it is not considered. Because both the 3.3 and 7.7~$\mu$m features are required to robustly confirm a detection, the absence of a secure 3.3~$\mu$m feature introduces ambiguity. Consequently, since the emission cannot be unequivocally attributed to the disk, we classify this case as a non-detection.

\subsubsection*{Flying~Saucer}

The Flying~Saucer disk, like ESO-H$\alpha$ 574 and OphE~MM3, is in the foreground of the $\rho$ Ophiuchus molecular cloud which contributes to background PAH emission and must be subtracted \citep{Brandner2000,Grosso2003,Guilloteau2016,Villenave2020,Wolff2021}. The 3.3~$\mu$m emission is diffuse around the source without any evidence of localized emission (see Figure \ref{fig:coll_null}). To further quantify this non-detection, no features are seen after the removal of diffuse PAH emission (see Appendix \ref{ConAnn}) in agreement with Figure~\ref{fig:coll_null} and the observed 3.3~$\mu$m emission is background.

\subsubsection*{LkH$\alpha$ 263C}

LkH$\alpha$ 263C, similar to HV~TAU~C, resides in a multiple-star system \citep{Jayawardhana2002,Stapelfeldt2003,Villenave2020}. The normalized 3.3~$\mu$m images reveals the presence of an intervening cloud structure located between the target and one of the nearby stellar components that exhibits extended PAH emission (see Figure~\ref{fig:coll_null}). The intervening cloud displays a pronounced excess at 3.3~$\mu$m.  However, localized enhancement towards LkH$\alpha$ 263C is missing. An over-subtraction feature is seen after the removal of background emission associated with the cloud (see Appendix~\ref{ConAnn}). Rather than obscuring a potential disk signal, this over-subtraction reinforces the interpretation that the detected PAH emission is spatially localized within the cloud structure, not in the central object.

To isolate any potential disk's contribution, we restricted the analysis to the central $5\sigma$ region around LkH$\alpha$ 263C. From this the apparent 3.3~$\mu$m signal vanishes entirely. Therefore LkH$\alpha$ 263C does not exhibit PAH emission, despite the strong PAH signal present in the surrounding cloud. The 11.3~$\mu$m PAH feature is absent with a weak possible 7.7~$\mu$m but no other PAH bands. Furthermore, the normalized log image (see Figure \ref{fig:coll_null}) shows the cloud separated from the source with pronounced background contamination, showing a lack of localization to LkH$\alpha$ 263C.

\subsection{PAH Analysis Using PAHFIT}\label{sec:PAHFIT}

\begin{figure*}[htb!]
\centering

\includegraphics[width=0.45\textwidth]{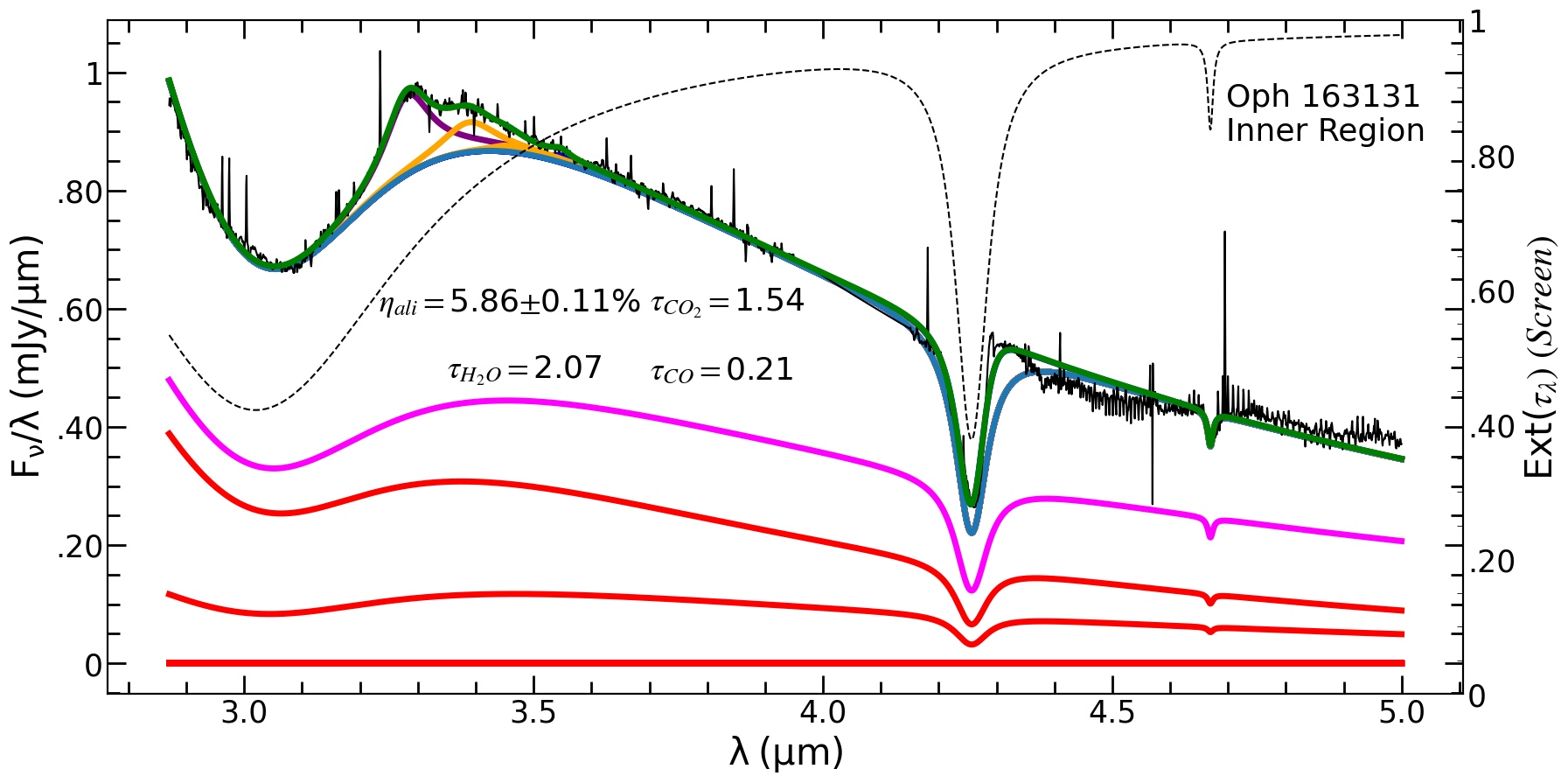}
\includegraphics[width=0.45\textwidth]{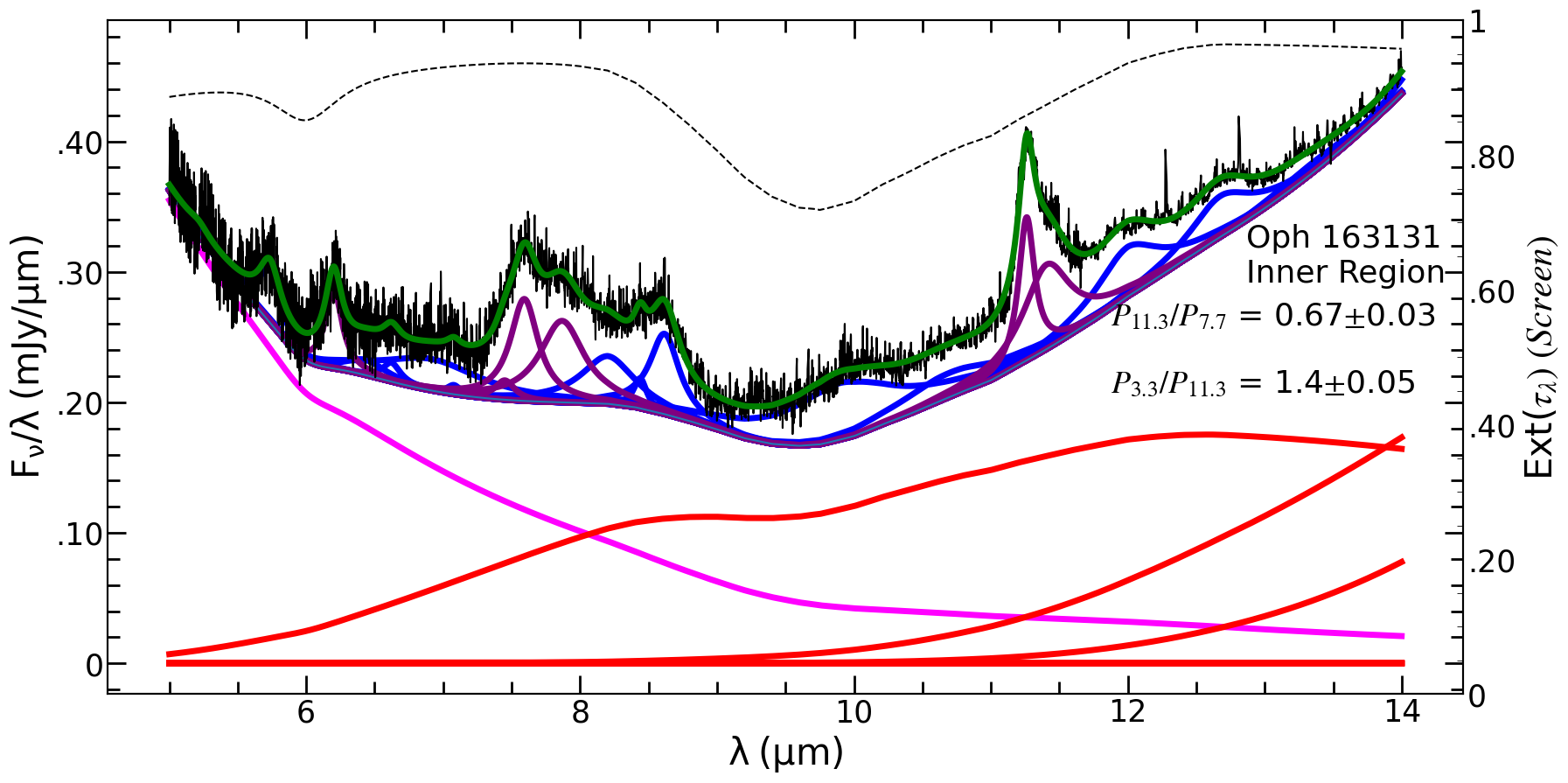}

\includegraphics[width=0.45\textwidth]{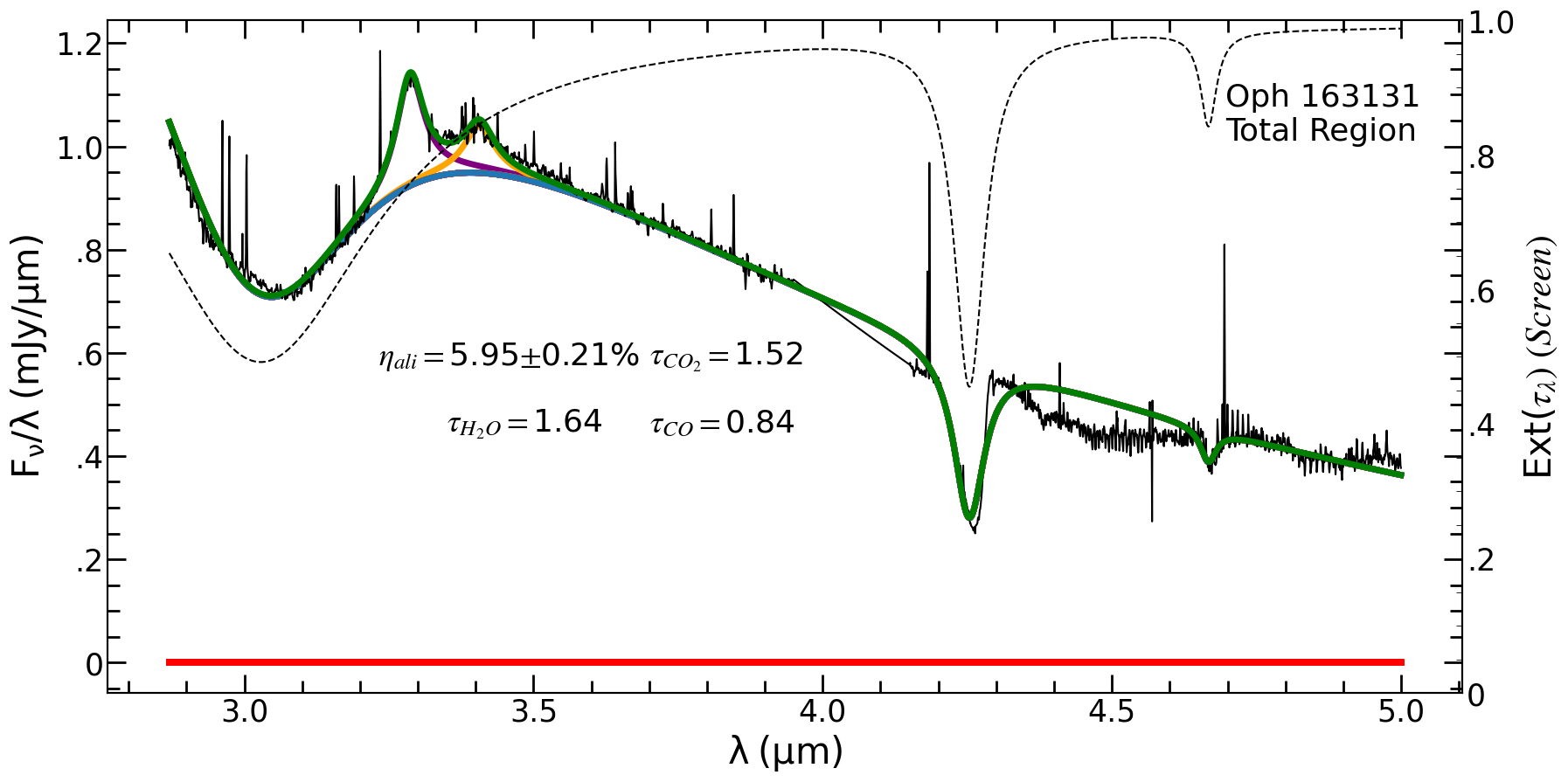}
\includegraphics[width=0.45\textwidth]{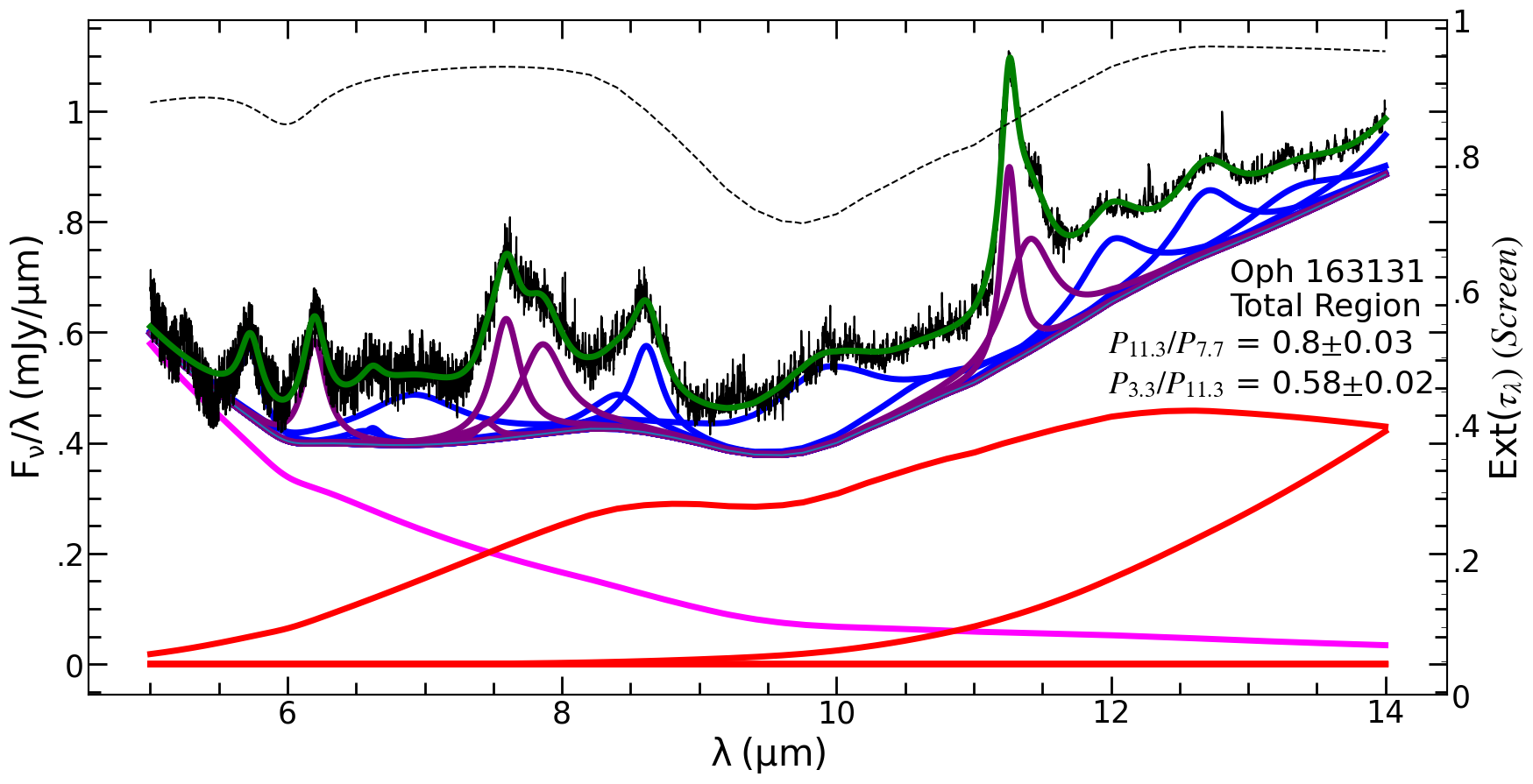}

\includegraphics[width=0.45\textwidth]{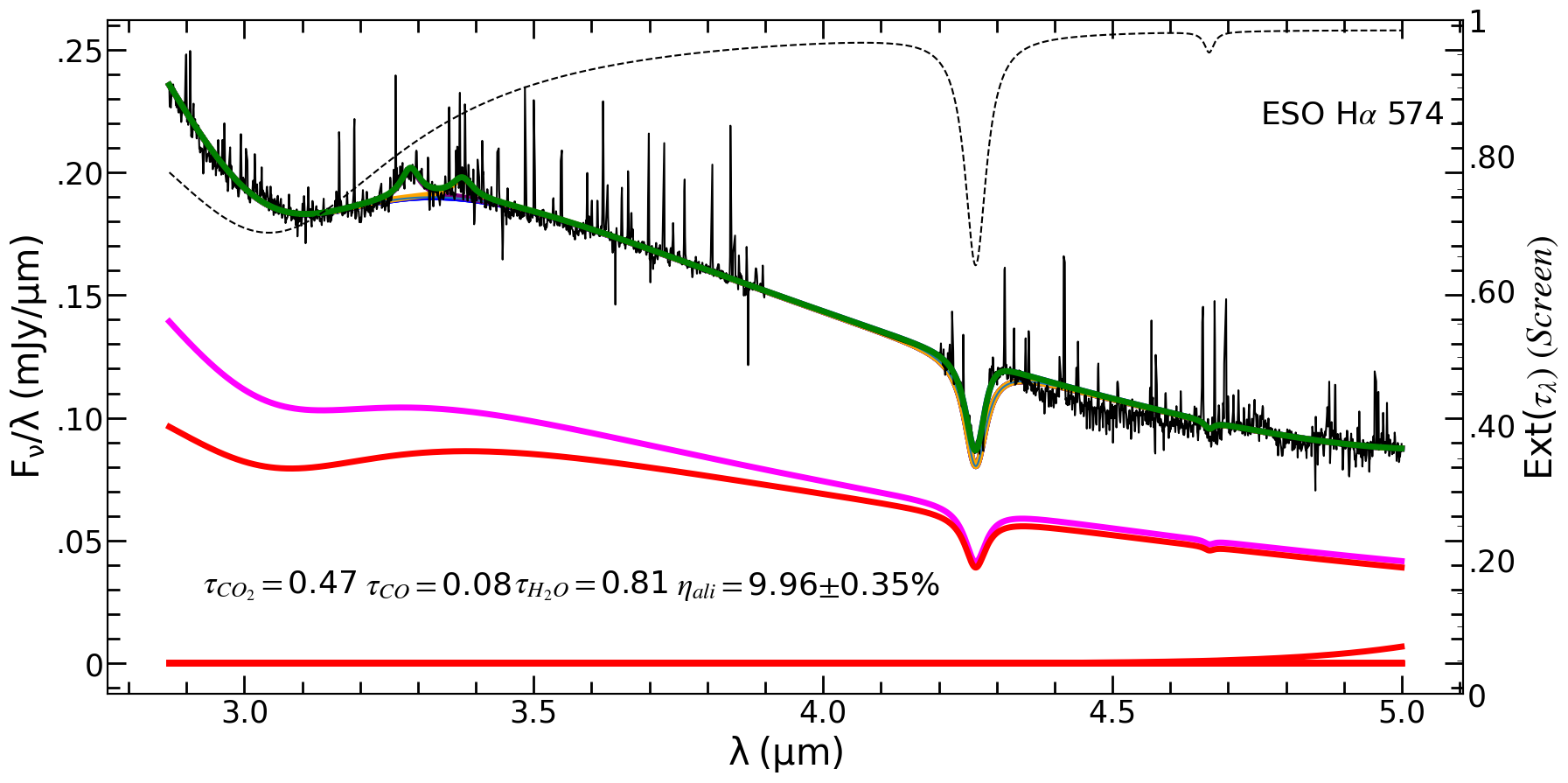}
\includegraphics[width=0.45\textwidth]{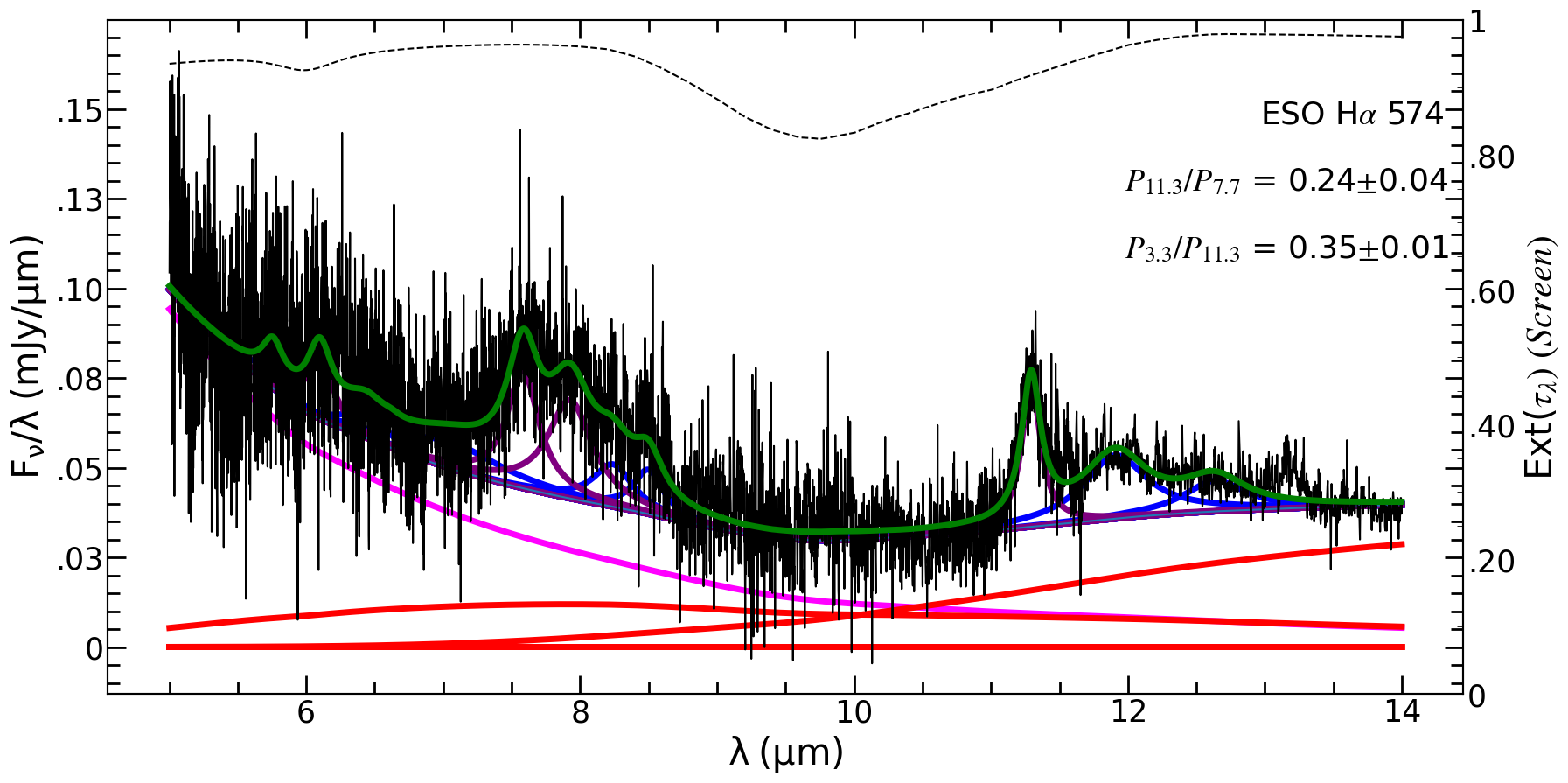}

\includegraphics[width=0.45\textwidth]{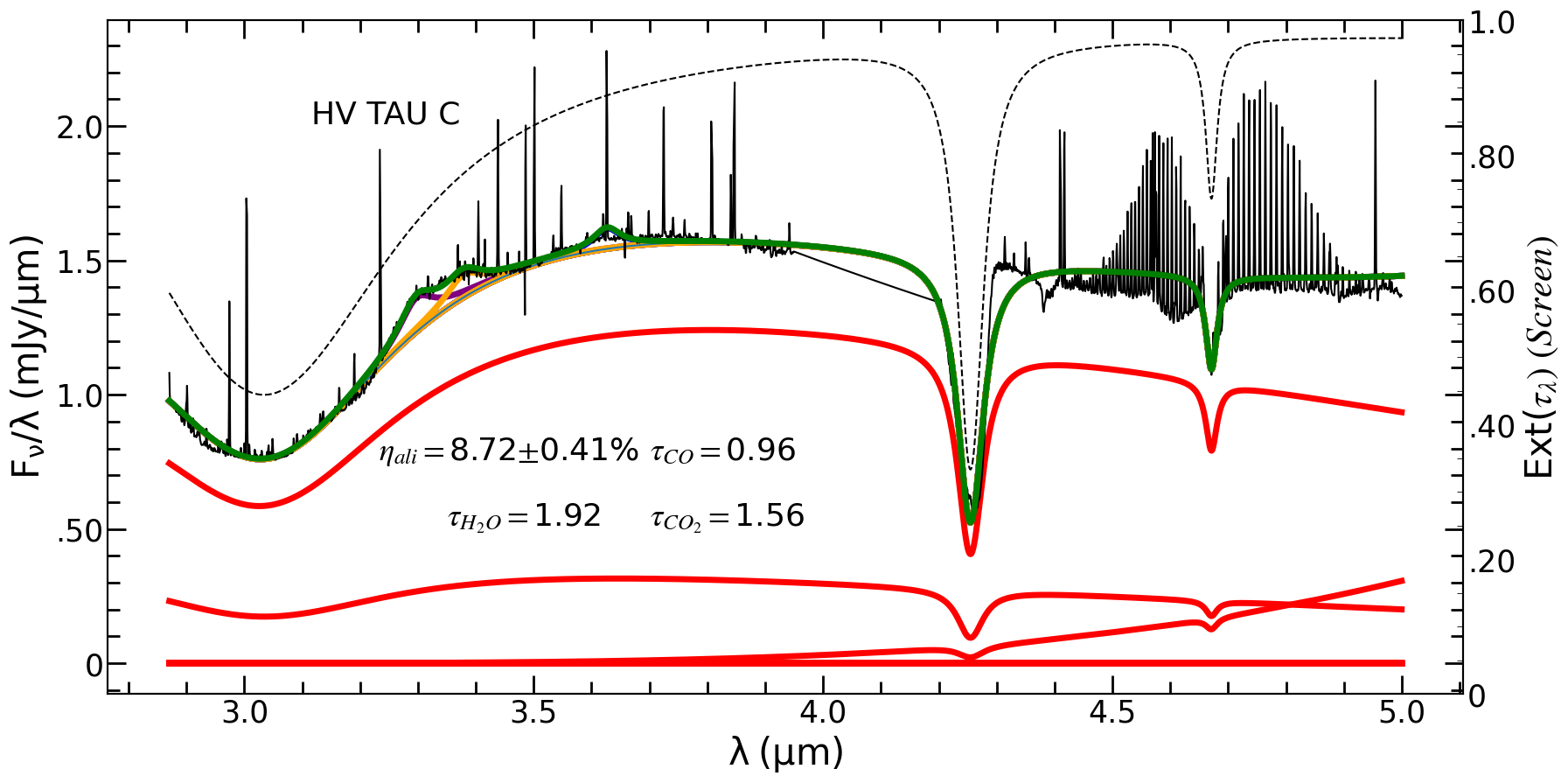}
\includegraphics[width=0.45\textwidth]{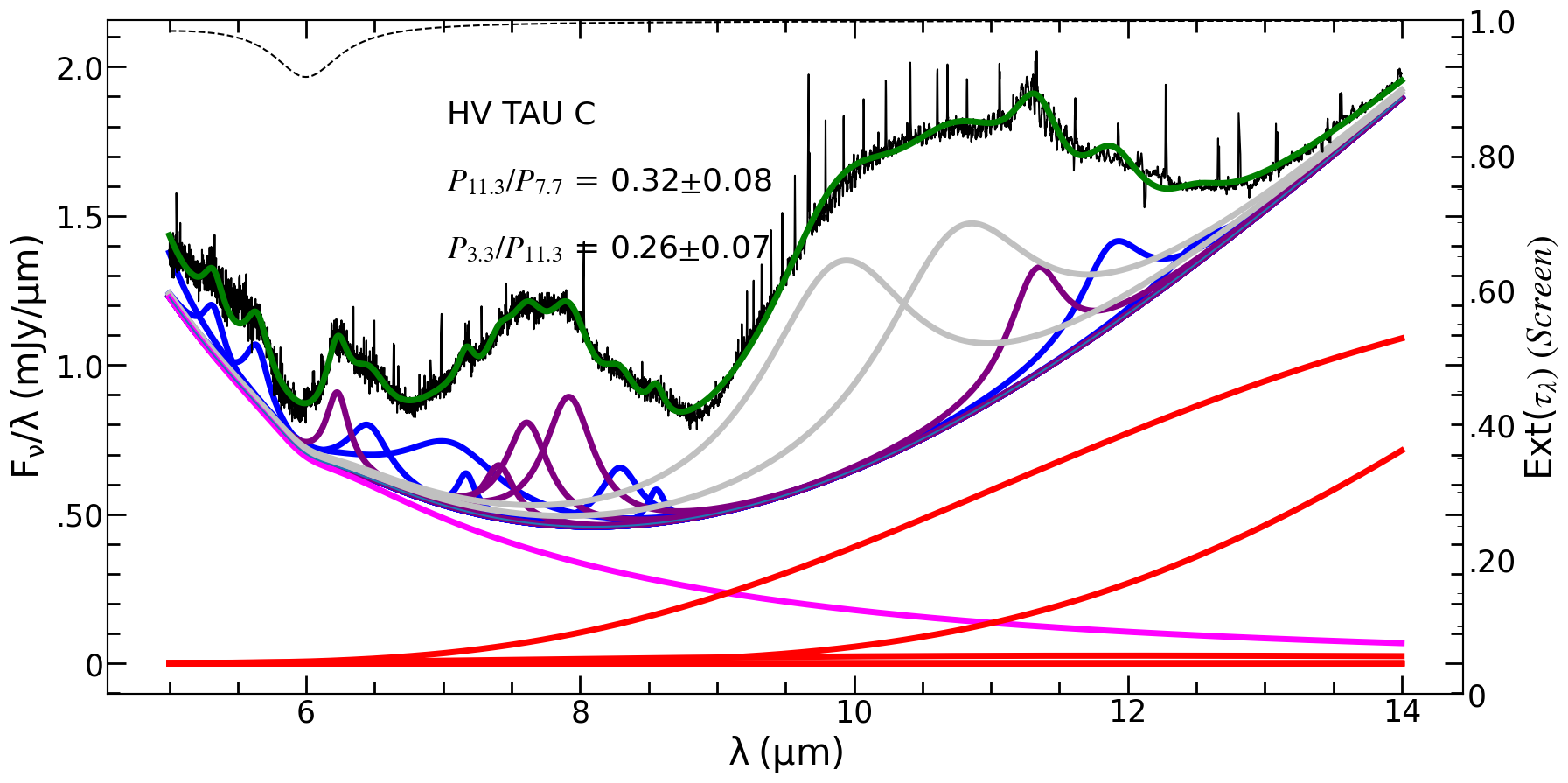}

\caption{PAHFIT results for PAH emission localized to the disks. The spectra show the individual components used in the spectral decomposition: red, modified blackbody components representing thermal dust emission; magenta, stellar emission; blue, Drude profiles representing general molecular emission, with PAH aromatic and aliphatic emission features shown in purple and yellow, respectively, to distinguish these components from other molecular emission; silver, amorphous silicate emission; dashed black, the extinction profile; and green, the total model including all fitted components.}
\label{fig:PAHFIT}
\end{figure*}

\begin{figure*}[htb!]
\centering

\includegraphics[width=0.49\textwidth]{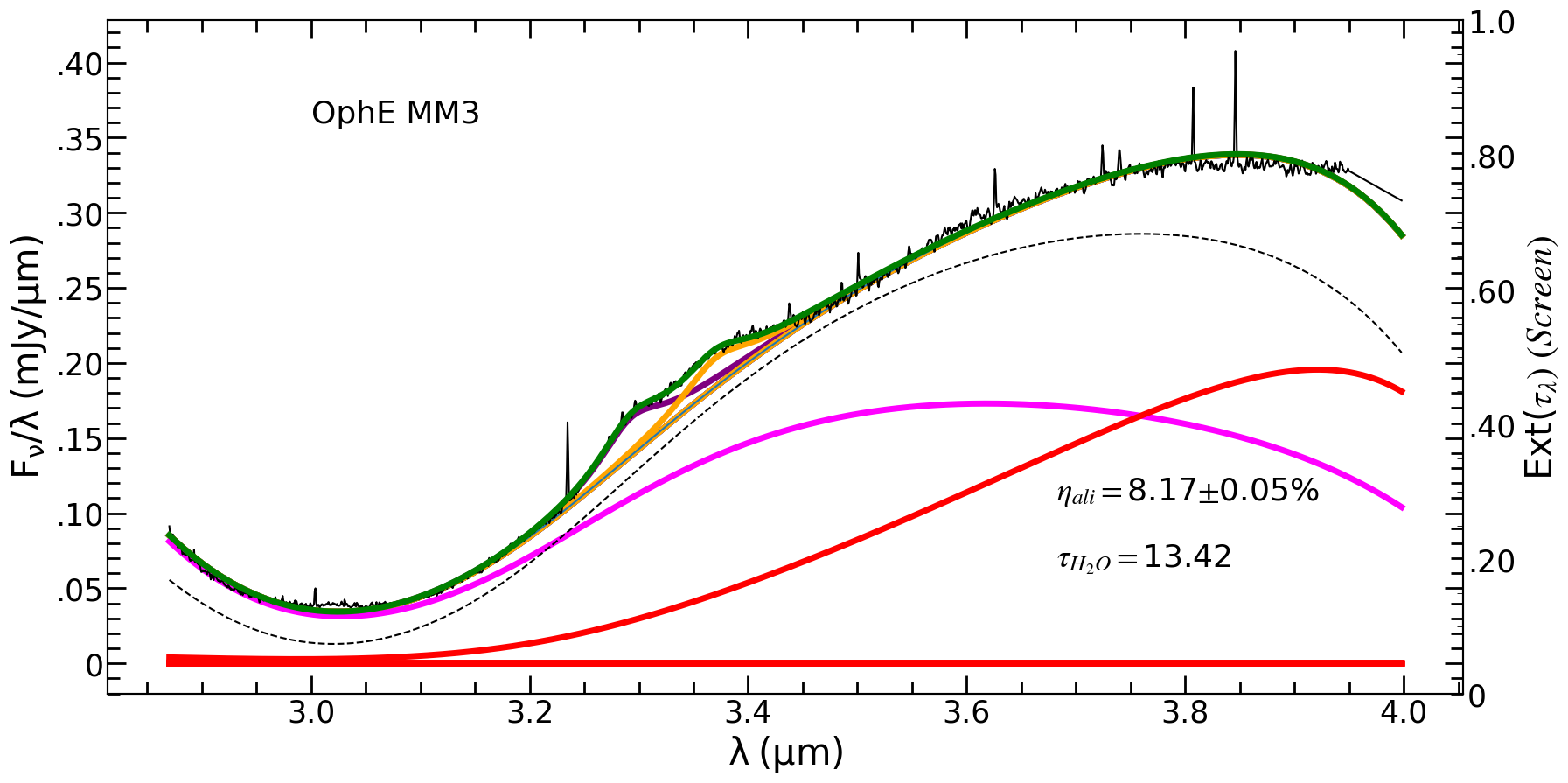}
\includegraphics[width=0.49\textwidth]{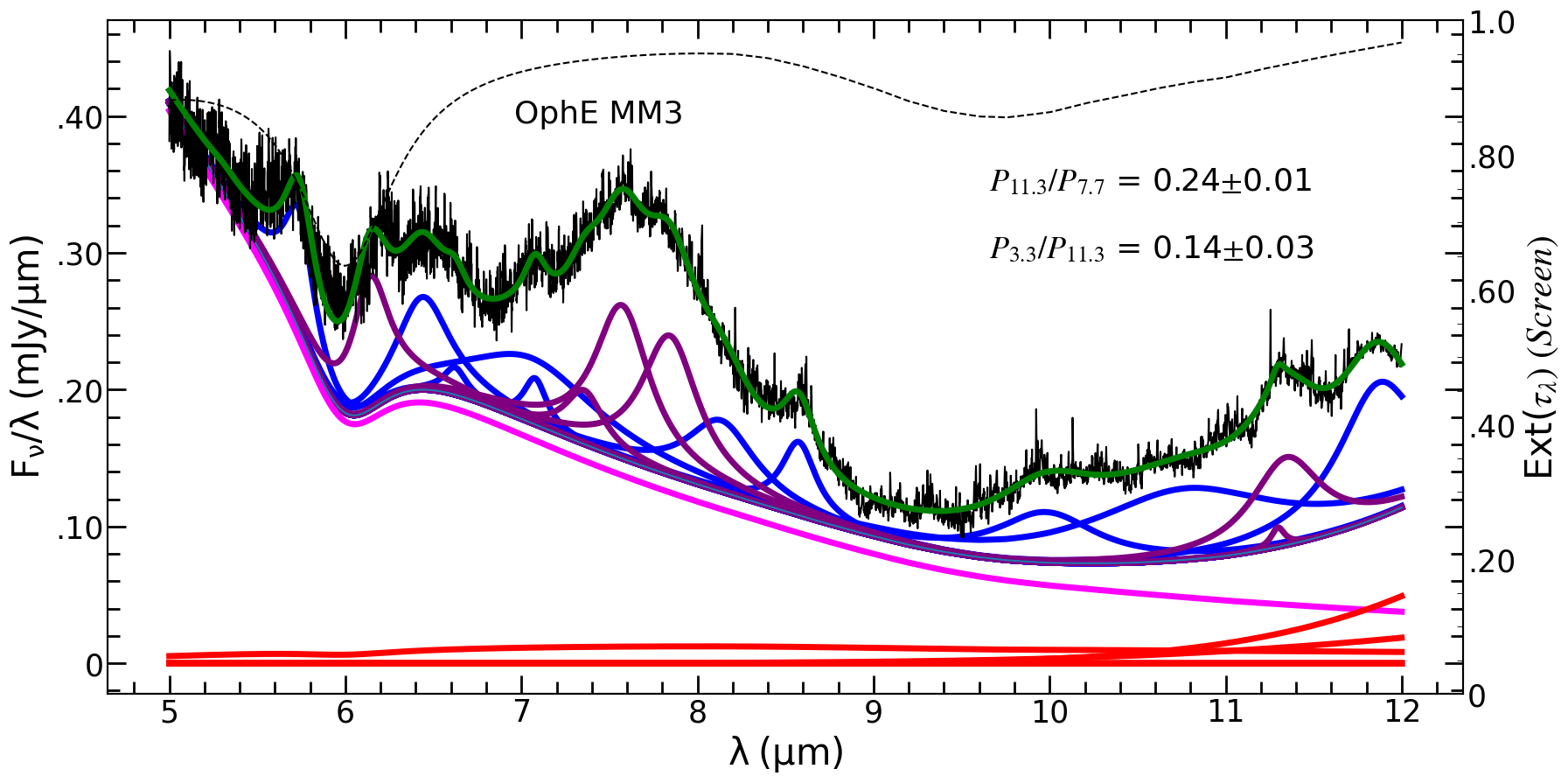}

\includegraphics[width=0.49\textwidth]{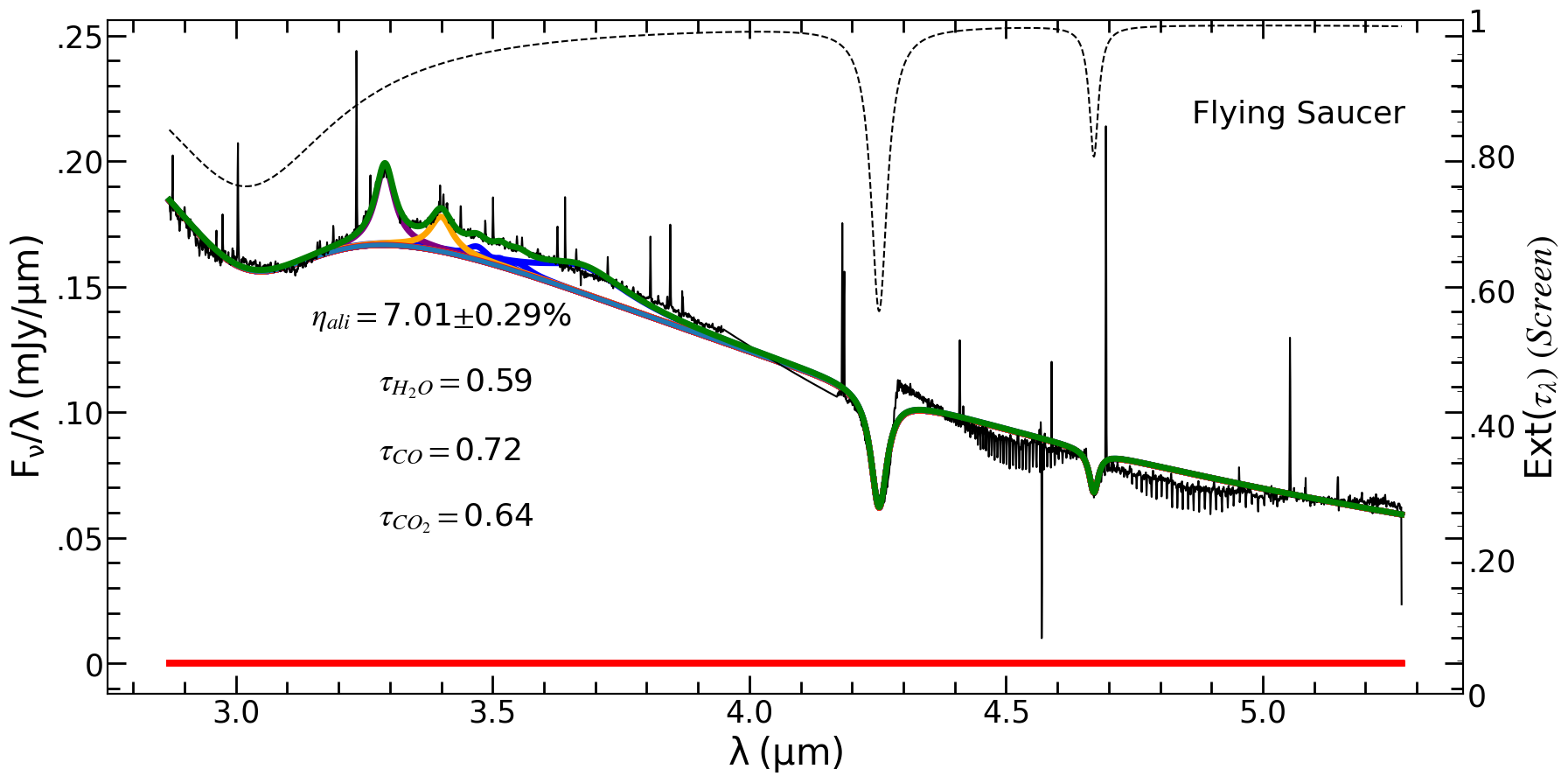}
\includegraphics[width=0.49\textwidth]{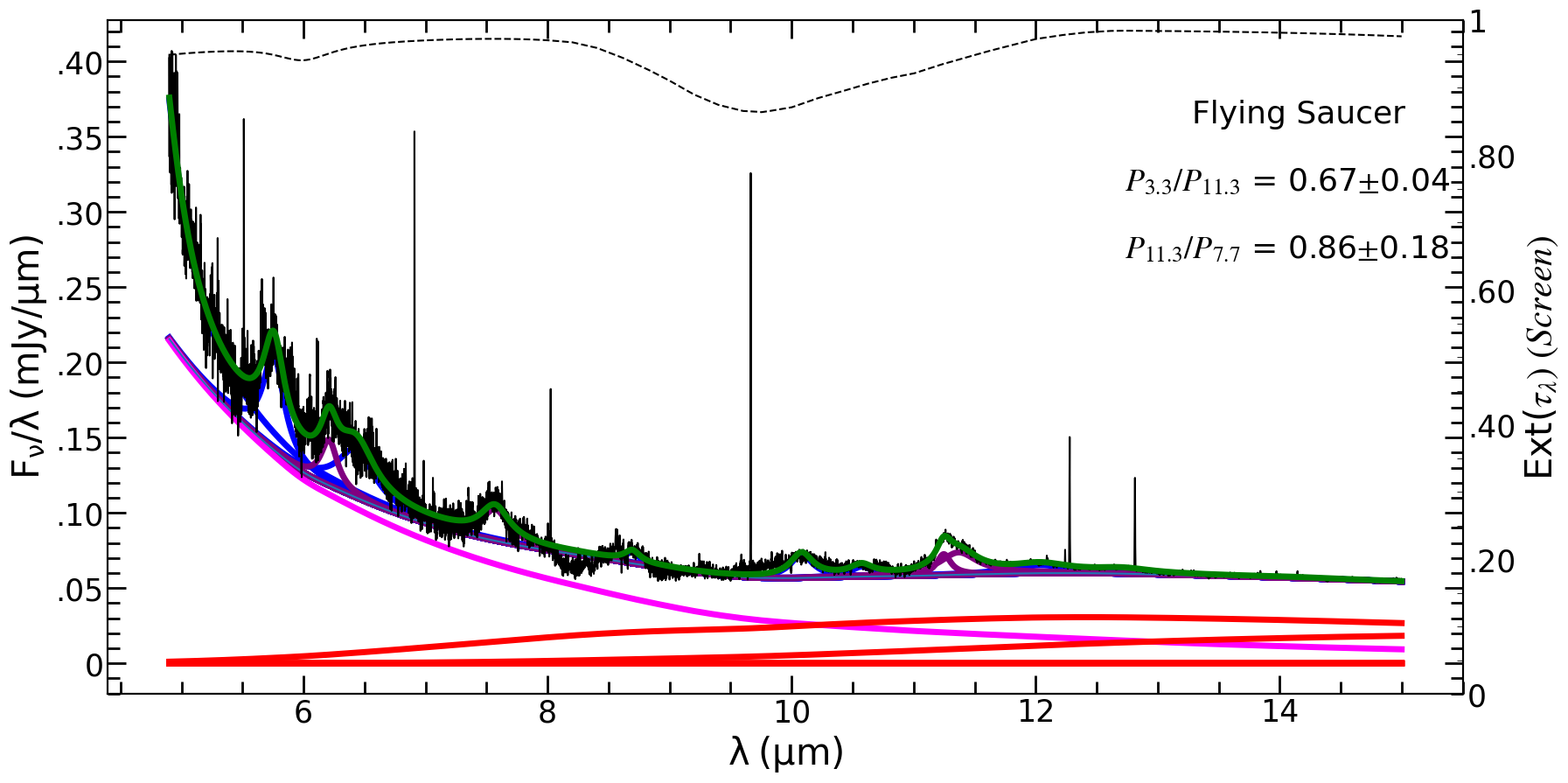}

\includegraphics[width=0.49\textwidth]{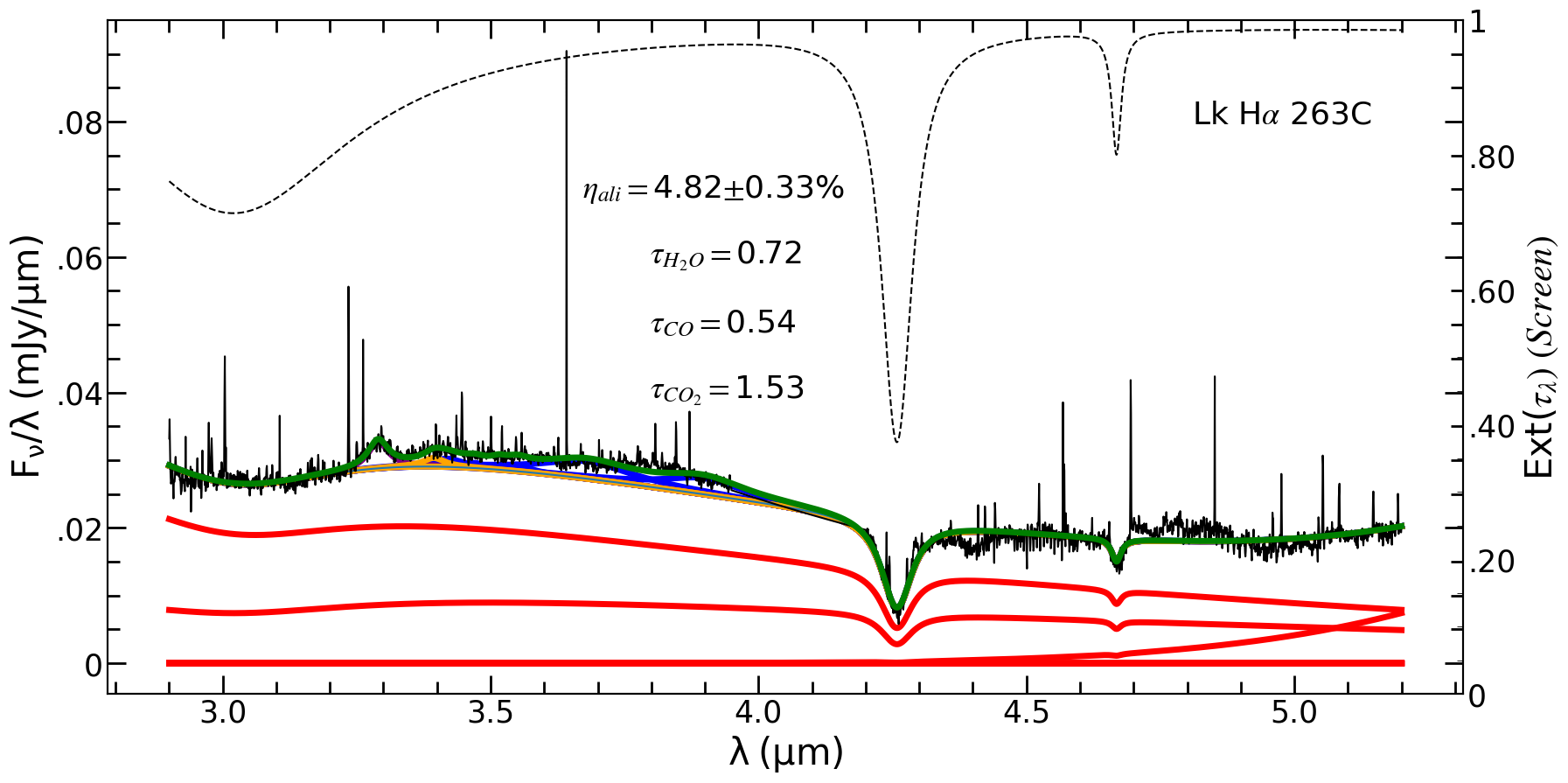}
\includegraphics[width=0.49\textwidth]{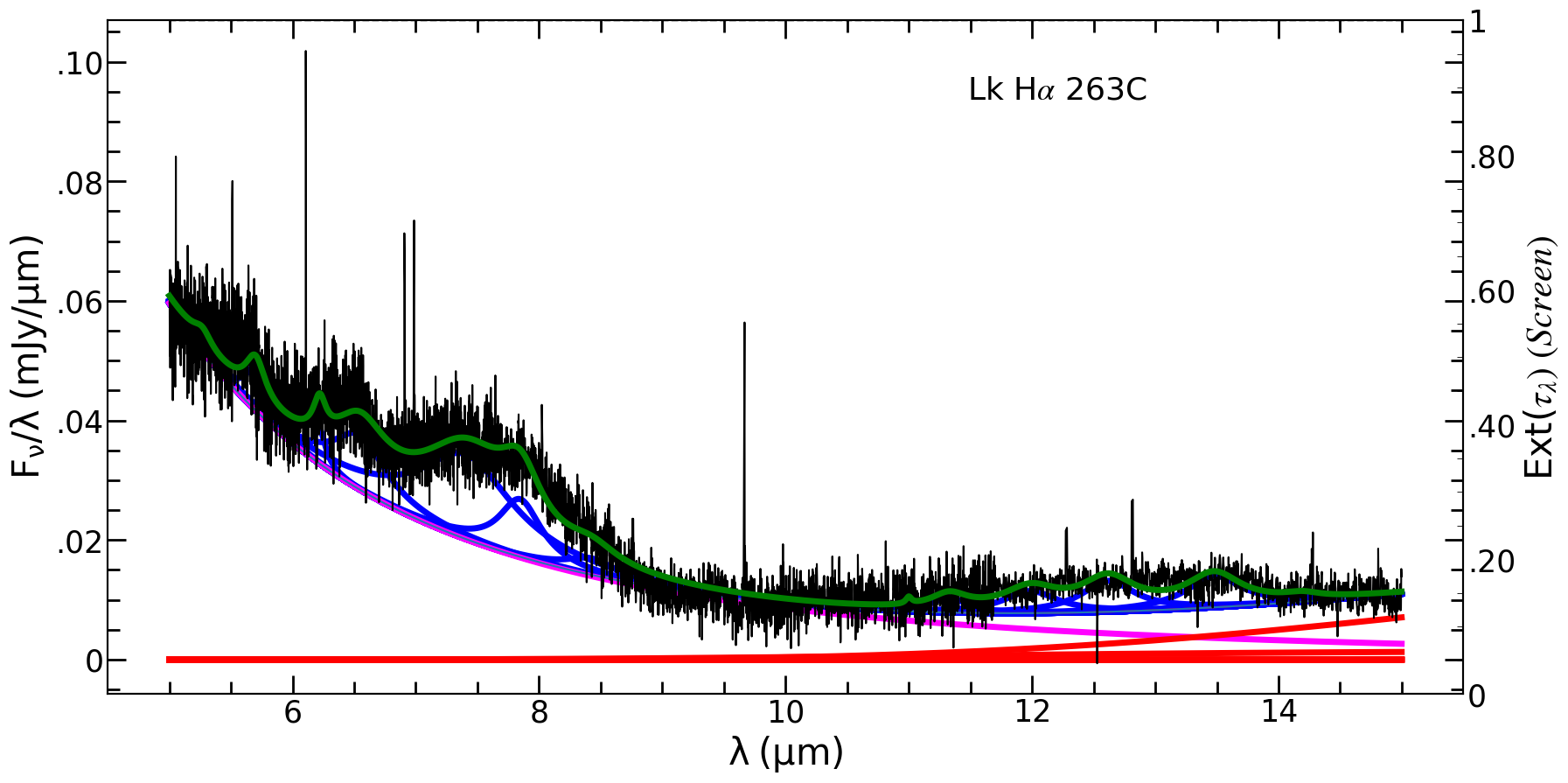}

\caption{PAHFIT results for disks where non-localized PAHs were found. Line definitions remain consistent with Figure~\ref{fig:coll_PAH}.}
\label{fig:PAHFIT_null}
\end{figure*}

\begin{figure}[htb!]
\centering

\includegraphics[width=0.4\textwidth]{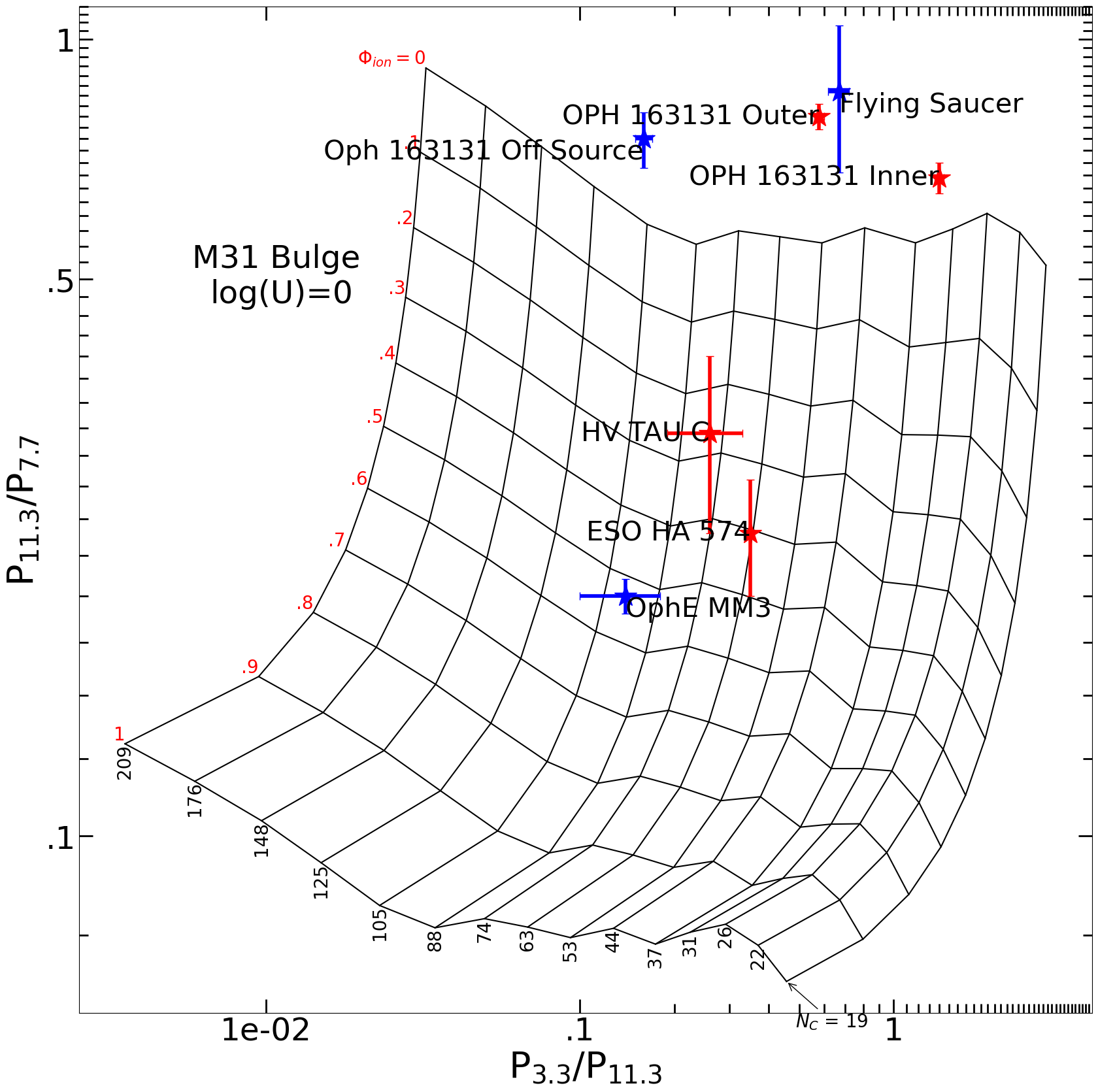}

\caption{Grid constructed for this work from the M31 model of \cite{DVN/LPUHIQ_2021}, selecting the lowest-energy model available in $N_\mathrm{C}$--$\phi_{\rm ion}$ space. The horizontal axis spans $N_\mathrm{C}=19$--209, and the vertical axis spans $\phi_{\rm ion}=0$ (fully neutral) to $\phi_{\rm ion}=1$ (fully ionized). Band ratios associated with the disk are indicated in red, whereas those corresponding to background-dominated or otherwise inconclusive regions are indicated in blue.}
\label{fig:grid}
\end{figure}

Figure~\ref{fig:PAHFIT} presents the PAHFIT decomposition of the NIRSpec and MIRI spectra for each disk with localized PAH, where the observed spectra are separated into PAH emission features, thermal dust continuum components, and ice/silicate extinction contributions fit individually for extinction accuracy as described in \S~\ref{sec:origin}. The fits reproduce the overall near- and mid-IR spectral shape and the major aromatic and aliphatic features well. For disks where no disk-localized PAH was detected, we can still characterize the composition of PAHs in the ambient cloud, shown in Figure~\ref{fig:PAHFIT_null}. We interpret the observed PAH band ratios using the model framework developed for the quiescent bulge of M31 by \cite{DVN/LPUHIQ_2021}. We adopt the $U=1$ model, which represents the lowest radiation-field intensity considered and provides the best match to the relatively low-energy radiation fields and photon hardness expected for our T Tauri systems. The model predicts PAH band ratios for different PAH sizes, parameterized by the number of carbon atoms ($N_\mathrm{C}$), and ionization fractions ($\phi_\mathrm{ion}$). We calculate the corresponding model band ratios and construct a grid in $N_\mathrm{C}$--$\phi_\mathrm{ion}$ space (see Figure~\ref{fig:grid}). The observed band ratios are then compared with this grid to identify the range of PAH sizes and ionization fractions that can reproduce the measured values within the observational band ratio errors.

For Oph~163131, within the central region, there appears to be a stronger ionizing radiation field compared to the extended cloud, as indicated by the 11.3/7.7 $\mu$m band ratio of 0.67, compared to the surrounding cloud's 0.80. However both these values indicate very neutral PAHs overall, possibly entirely. The $\eta_{\rm ali}$ is consistent with 5.86\% within the center region and 5.95\% in the surroundings; both within typical, albeit at the higher end of, ISM values \citep{yang2023}. The 3.3/11.3 $\mu$m band ratio is very large at the central region of 1.4, then drops to 0.58 in the surrounding region, suggesting a smaller number of $N_\mathrm{C}$ of $\approx30$ internally and $\approx50$ in the extended cloud.

Towards ESO-H$\alpha$ 574, the $\eta_{\rm ali}$ was found to be 9.96\%, indicating a relatively enhanced aliphatic component in the emitting PAH population. If the aliphatic-to-aromatic composition is primarily governed by UV photo-processing, this higher aliphatic fraction may instead indicate that the PAH population has experienced less extensive aromatization during the disk formation phase. The 3.3/11.3 $\mu$m band ratio of 0.35 and 11.3/7.7 $\mu$m band ratio of 0.24 are consistent with medium-sized PAHs containing approximately 50--70 $N_\mathrm{C}$ and reveal a low ionized PAH component within the disk most likely around 30-50\%.

Towards HV~TAU~C, we obtain an $\eta_{\rm ali}$ of 8.72\%, which is again moderately elevated. This enhancement is consistent with a somewhat less photo-processed PAH population and is not unexpected given the possible presence of less compact PAH molecules as suggested by the increased 3.3~$\mu$m FWHM, already suggesting a reduced influence of hard photons within the system \citep{All1989, 2020AL}. The 3.3/11.3~$\mu$m band ratio of 0.26 and 11.3/7.7~$\mu$m band ratio of 0.32 are indicative of medium-to-large PAHs containing on the order of $N_\mathrm{C} \approx 75-90$ and the PAH species present at $\approx 25-50\%$ ionization with a larger uncertainty due to the overlap of the PAH emission  with the silicate feature.

The analysis of the foreground and background cloud emission provides a useful environmental baseline against which the embedded disks can be interpreted. For OphE~MM3, the measured 11.3/7.7~$\mu$m band ratio of 0.20 and 3.3/11.3~$\mu$m band ratio of 0.19 indicate a PAH population dominated by relatively large ($N_\mathrm{C} \approx 90$--100), moderately ionized molecules with an ionization fraction of approximately 50--60\%. The inferred $\eta_{\rm ali}$ of 8.17\% is modestly elevated compared to typical diffuse ISM values \citep{yang2023}. Similarly, the background PAH emission toward Flying~Saucer serves as a useful local ISM reference baseline for the $\rho$ Ophiuchus molecular cloud and has previously been used to probe the disk in silhouette at 3.3~$\mu$m \citep{Dartois2026}. Its derived $\eta_{\rm ali}$ of 7.01\%, together with an 11.3/7.7~$\mu$m band ratio of 0.86 and a 3.3/11.3 ratio of 0.67, suggests a population of predominantly neutral, somewhat smaller PAHs that appear slightly less photo-processed. In contrast, the intermediate cloud toward LkH$\alpha$~263C exhibits a typical $\eta_{\rm ali}$ of 4.82\%, but the absence of detectable 11.3~$\mu$m emission precludes the use of the 11.3/7.7 and 3.3/11.3 $\mu$m band ratio diagnostics. The suppressed 11.3~$\mu$m feature is nevertheless consistent with a highly ionized PAH population, as expected for typical ISM conditions \citep{Tielents2008}. 

These findings are summarized for all disks in Figure~\ref{fig:grid}. In general, we see a clustering of PAH size around $\sim$20-60 carbon atoms, and ionization fractions ranging from 0 to 0.5.  There are no obvious systematic differences between on- vs. off-source PAH populations, nor between the $\rho$ Ophiuchus disks versus other star-forming environments.

\subsection{Spatial Information}\label{sec:spatial}

Spatial variations in band ratios are measured using apertures placed along four orthogonal directions: parallel and perpendicular to the disk mid-plane with position angles matched to the apparent disk orientation (see Appendix~\ref{apps}). The rapid S/N decline limits smaller sources to only 2–3 usable apertures beyond the central brightest extraction. Owing to Oph~163131’s brightness and extent, we can dissect the disk using 10 vertical and 13 radial apertures. For ESO-H$\alpha$ 574 and HV~Tau~C, PSF limitations restrict the spatial analysis to $\eta_{\mathrm{ali}}$, whereas for Oph~163131 all three ratios are analyzed, as the extended cloud justifies additional PSF-limited apertures. 

\begin{figure}[htb!]
\centering

\includegraphics[width=0.45\textwidth]{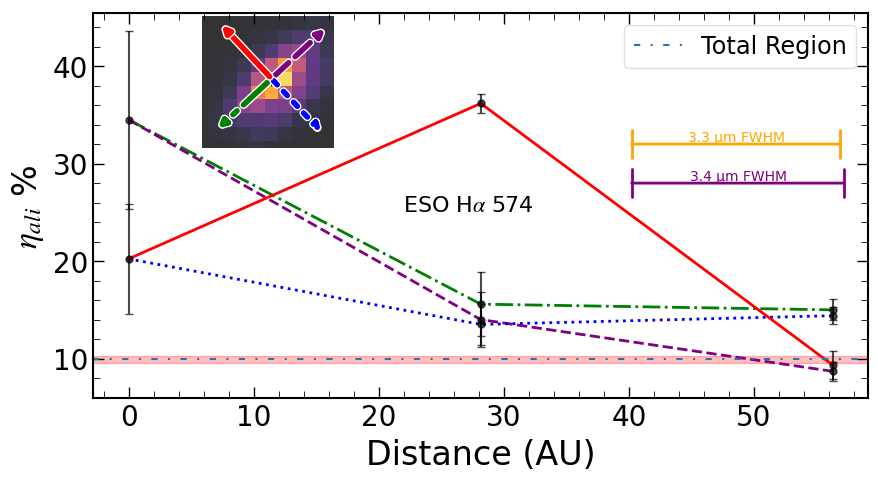}
\includegraphics[width=0.45\textwidth]{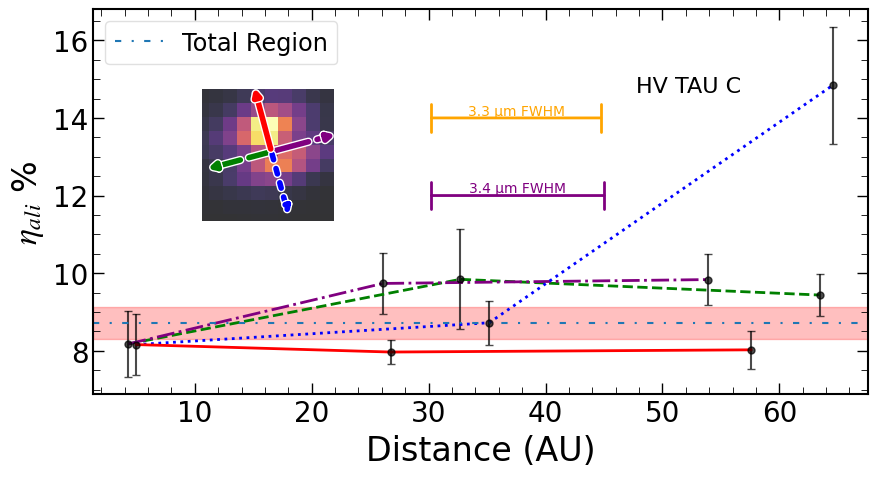}
\includegraphics[width=0.45\textwidth]{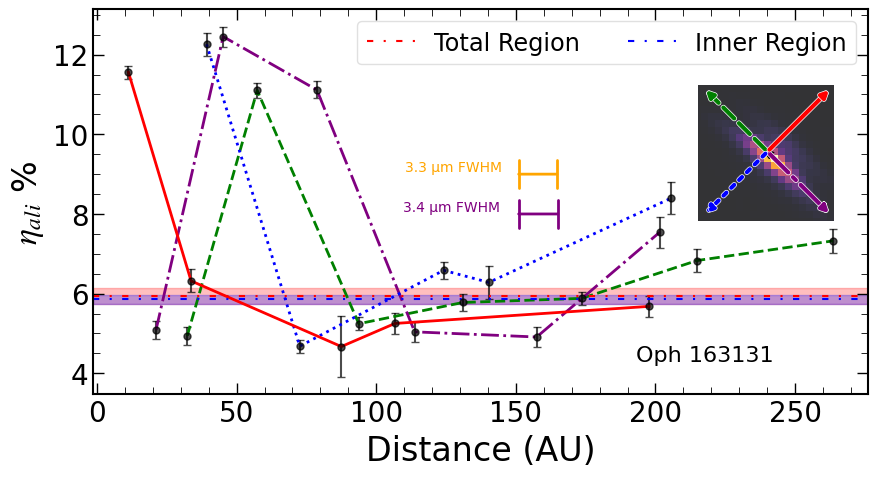}

\caption{$\eta_{\rm ali}$ in four orthogonal directions from the central point (red solid: upper surface, blue dotted: lower surface, green dashed: eastward, purple dash-dotted: westward) for ESO-H$\alpha$ 574, HV~TAU~C, and Oph~163131. Each point represents the $\eta_{\rm ali}$ derived from the PAHFIT spectral decomposition at the position and distance for its aperture (see Appendix~\ref{apps} \& \ref{PAHFIT}). Varying uncertainties arise from not only the differing S/N but also the quality of the spectral decomposition via PAHFIT.}
\label{fig:spatial}
\end{figure}

\begin{figure}[htb!]
\centering

\includegraphics[width=0.45\textwidth]{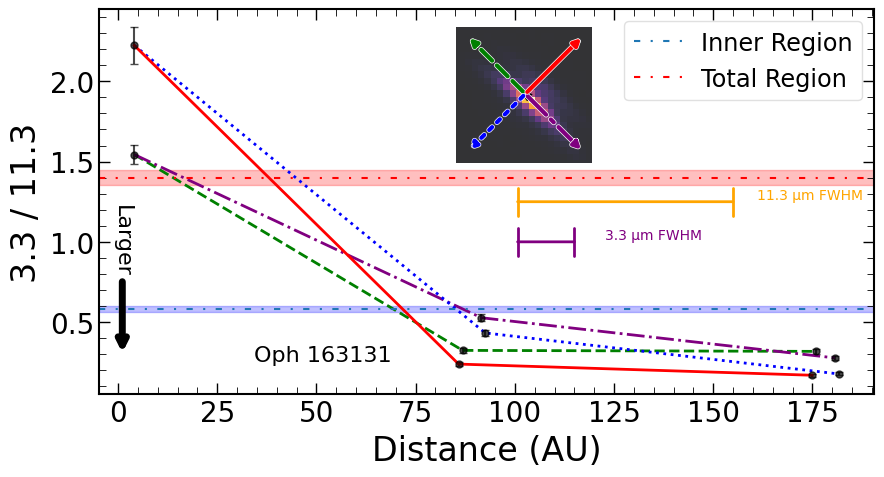}
\includegraphics[width=0.45\textwidth]{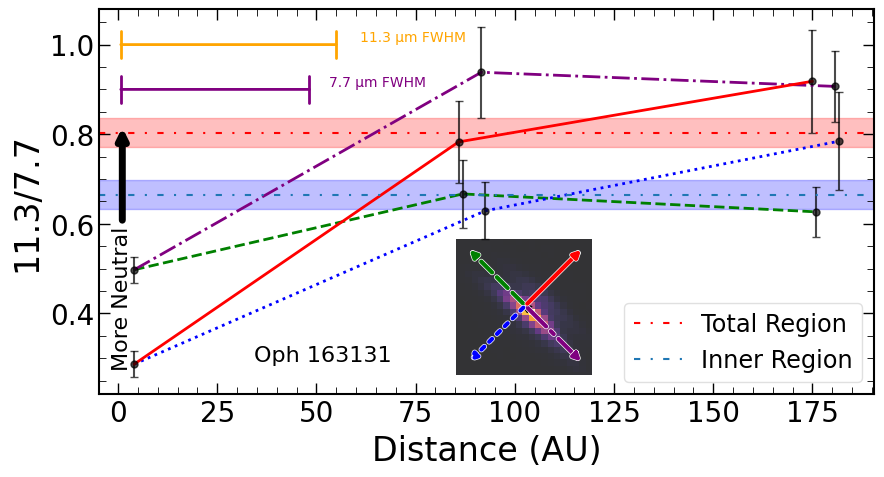}
 
\caption{Similar to Figure~\ref{fig:spatial} with identical arrow definitions, showing the 3.3/11.3~$\mu$m and 11.3/7.7~$\mu$m band ratios of Oph~163131. Each point represents the ratio measured within the larger PSF justified  apertures at the indicated position and distance from the star (see Appendix~\ref{apps} \& \ref{PAHFIT}).}
\label{fig:psf_just}
\end{figure}

\paragraph{ESO-H$\alpha$ 574}

The $\eta_{\rm ali}$ appears to start at 25-45\%, which decreases radially down to 8-9\% at $\approx$ 80~AU. Sampling the height, it appears to start lower at \mbox{25-35\%} with the upper surface region exhibiting a substantial increase to $\approx$ 20\%, a behavior not seen in the lower surface region. This asymmetry may indicate non-uniform physical and chemical processing in the upper and lower surface lobes of the disk, although variations in excitation conditions and line-of-sight effects may also contribute to the observed trend. The large difference in $\eta_{\rm ali}$ between the central apertures, although they are within the error bars of one another, suggests spatially variable chemistry around the disk, with small differences in aperture placement sampling potentially distinct regions near the energetic center.

\paragraph{HV~Tau~C}

The $\eta_{\rm ali}$ remains roughly constant at around 10\% in the upper and lower surface lobes; in the lower layer of the disk, it is constant at $\sim$8\% before it spikes to $\sim$14\% at $\sim$65~AU (see Figure~\ref{fig:spatial}) towards HV TAU AB. These values are slightly above the typical galactic level, which could potentially indicate a decrease in typical ISM photo-processing within this system, especially towards HV Tau AB. 

\paragraph{OPH~163131}

For $\eta_{\rm ali}$, within $\pm$100~AU of the central star we find values higher than expected if photo-processing by the stellar radiation field preferentially reduces the aliphatic component, before $\eta_{\rm ali}$ decreases to 4--8\% at larger distances. In the radial direction, the fraction starts near the 5--8\% level close to the star, but at around 50~AU it rises sharply to 10--12\%, roughly matching the values found within the innermost aperture of the upper and lower lobes of the disk. This suggests a roughly symmetric, elevated $\eta_{\rm ali}$ within the inner 50~AU of the star, which may reflect the formation or reprocessing of carbonaceous material within the dense inner disk, competing with the effects of photo-processing. A similar, but likewise symmetric, behavior is observed in the 11.3/7.7~$\mu$m band ratio, which increases uniformly both with height and radius. This is consistent with the expected behavior of PAHs in a radiation field where the ionization balance is important: the increasing 11.3/7.7~$\mu$m band ratio indicates a higher fraction of neutral PAHs, which then stabilizes beyond $\approx$100~AU. Notably, this is roughly the same radius at which the $\eta_{\rm ali}$ also stabilizes in all directions, pointing to a physical boundary at $\sim$100~AU in this system. A final indication that this boundary at this radius exists comes from the 3.3/11.3~$\mu$m band ratio, which shows a pronounced decline with radius. At the center, the high value (see Figure~\ref{fig:psf_just}) suggests a population dominated by small PAHs that increases in size steadily until it levels off at $\sim$80--100~AU. The symmetry of this radial trend, which is also reflected in the $\eta_{\rm ali}$ and ionization patterns, suggests that the observed increase in PAH size with radius may be related to the same radiation-driven processing that shapes these other PAH properties.

\subsection{Off Source}\label{sec:Off_Source}

\begin{figure*}[htb!]
\centering

\includegraphics[width=0.32\textwidth]{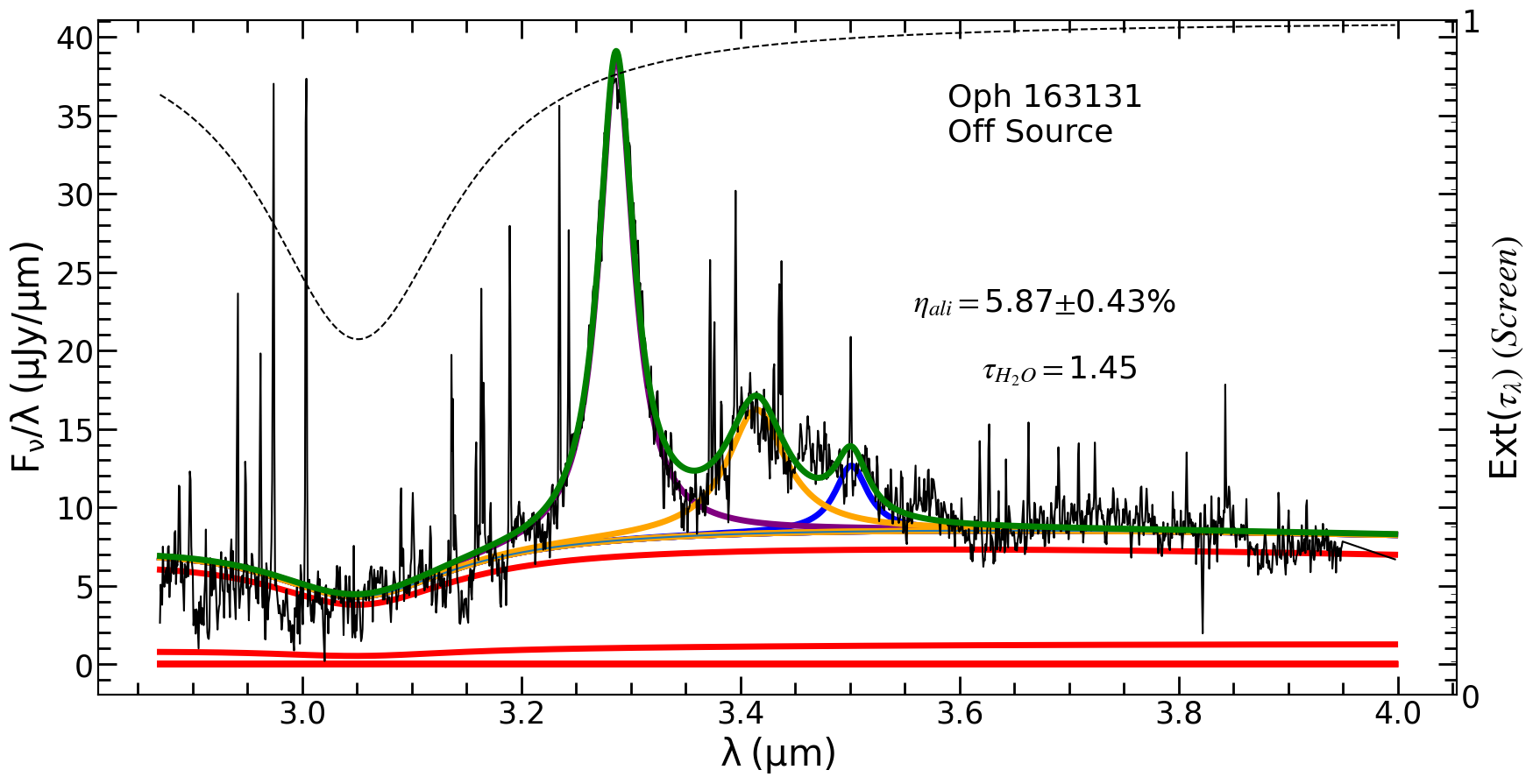}
\includegraphics[width=0.32\textwidth]{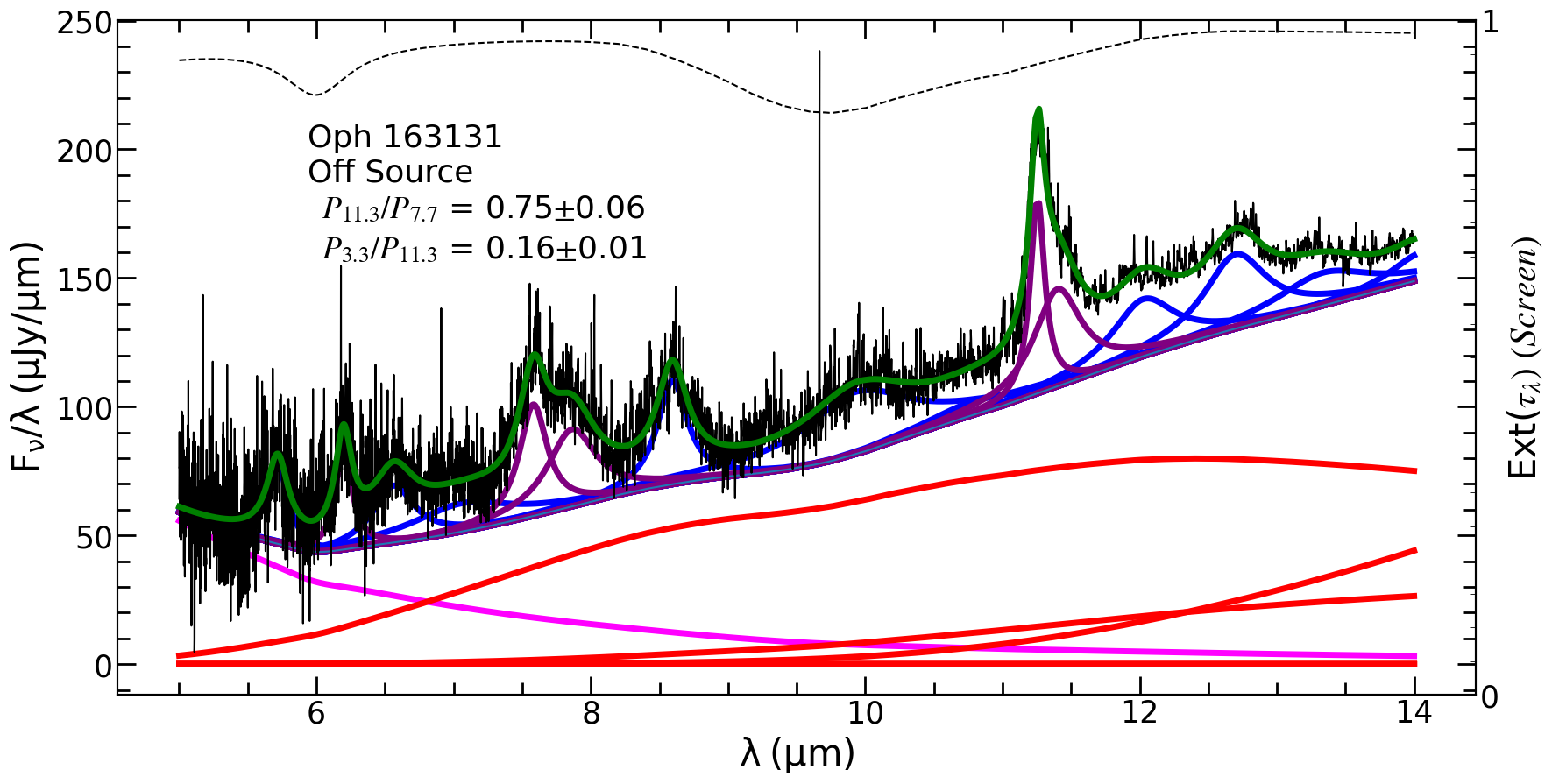}
\includegraphics[width=0.32\textwidth]{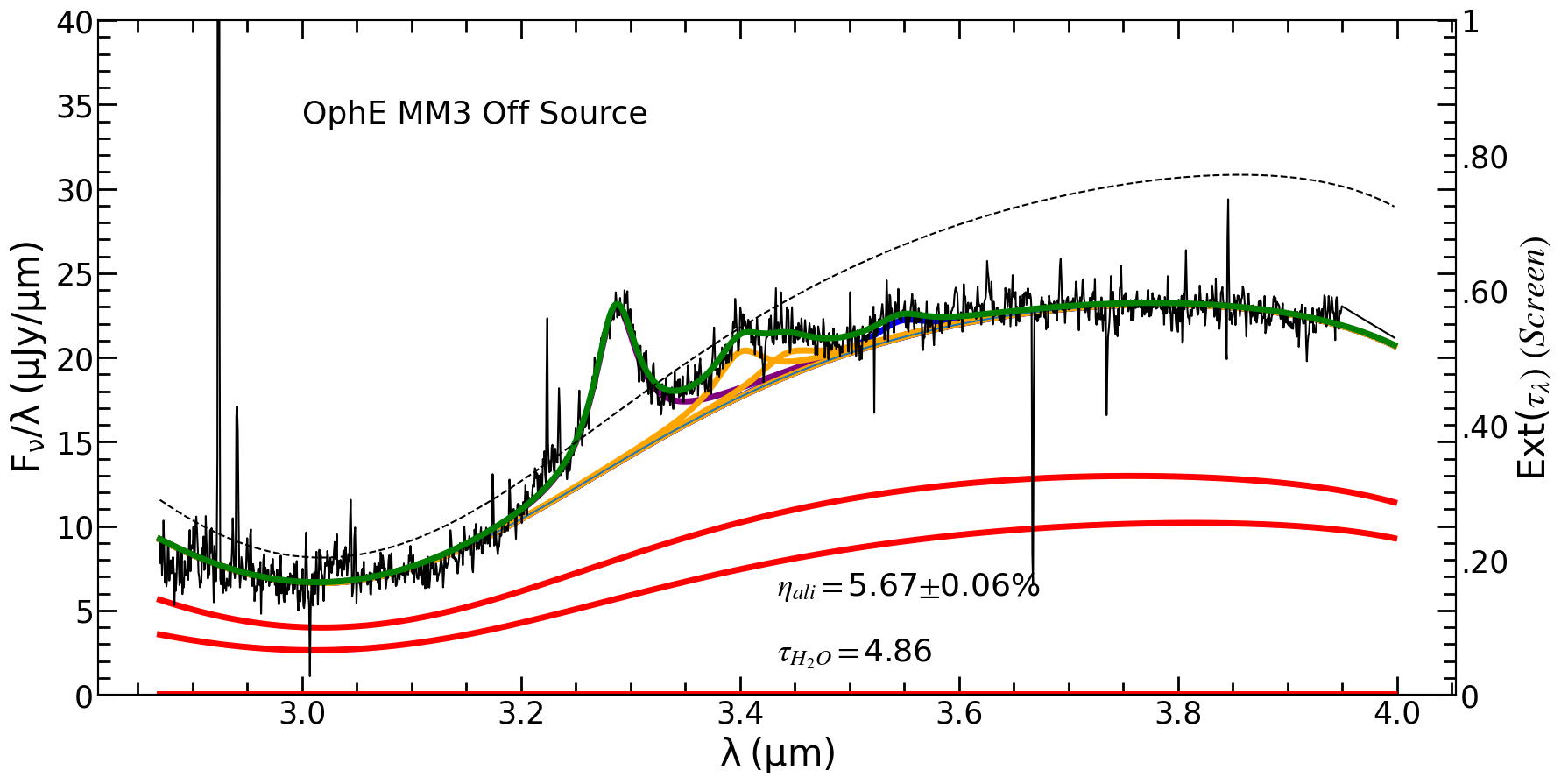}

\caption{PAHFIT fits to spectra extracted from external apertures in Figures~\ref{fig:coll_PAH} and~\ref{fig:coll_null} for Oph~163131 and OphE~MM3, where off-source PAH emission could be extracted. Line definitions remain consistent with Figure~\ref{fig:coll_PAH}.}
\label{fig:off_source}
\end{figure*}

For two targets, Oph~163131 and OphE~MM3, we additionally evaluate the off-source PAH emission (see Figures~\ref{fig:coll_PAH} and~\ref{fig:coll_null}) and band ratios extracted (see Figure~\ref{fig:off_source}). Although OphE~MM3 yields an inconclusive determination of the spatial localization of the PAH features, we define an aperture oriented away from the apparent disk's location in order to probe any potential influence on the foreground cloud. For Oph~163131, we position an aperture well outside the surrounding cloud to investigate PAH emission originating from the background. Owing to standing-wave artifacts in the OphE~MM3 data, only the near-IR extraction is reliable, whereas for Oph~163131, both the mid- and near-IR extractions are usable. 

Off-source extractions can provide valuable constraints on the ambient conditions of the regions hosting these objects. For Oph~163131, the off-source spectra exhibit changes in the 11.3/7.7 and 3.3/11.3~$\mu$m band ratios of 0.75 and 0.16, respectively. These differences imply a modestly larger and more ionized PAH population than the other extractions (see Figure~\ref{fig:grid}), consistent with a shift toward typical diffuse ISM behavior \citep{Tielents2008}. In the near-IR data, the $\eta_{\rm ali}$ is roughly 5.87\%, consistent with values derived from the other extractions suggesting the aliphatic compounds within Oph~163131 are inherited and remain unprocessed.

For OphE~MM3 two principal differences emerge from "off-source" compared to "on-source". First, the 3.3~$\mu$m feature is substantially broader in the direction of the disk, closely resembling the emission observed from the inner region of Oph~163131 and from HV~TAU~C. Second, there is a pronounced change in the $\eta_{\rm ali}$: while the "on-source" value is elevated, the off-source $\eta_{\rm ali}$ is lower, at 5.67\%, consistent with typical diffuse ISM levels (see Figures~\ref{fig:PAHFIT_null} \& \ref{fig:off_source}) suggesting an increase of aliphaticity from OphE~MM3 behind the cloud.  

\section{Discussion} \label{sec:disc}

The analysis presented here demonstrates that PAH emission is detected towards three sources that are consistent across NIRSpec and MIRI data. The detection of multiple PAH bands with consistent spatial morphology across two independent instruments provides strong evidence that the emission is astrophysical rather than an observational artifact or background contamination. Although the present data does not uniquely determine the vertical location of the emitting PAHs, the observations are consistent with the expected origin of PAH emission in the UV-irradiated surface layers of protoplanetary disks, where these molecules are most readily excited \citep{Acke2004,Geers2006, Tielents2008, Visser2009}.

\subsection{Source-specific PAH Carrier and Chemical Diagnostics}

\begin{figure}[htb!]
\centering

\includegraphics[width=0.45\textwidth]{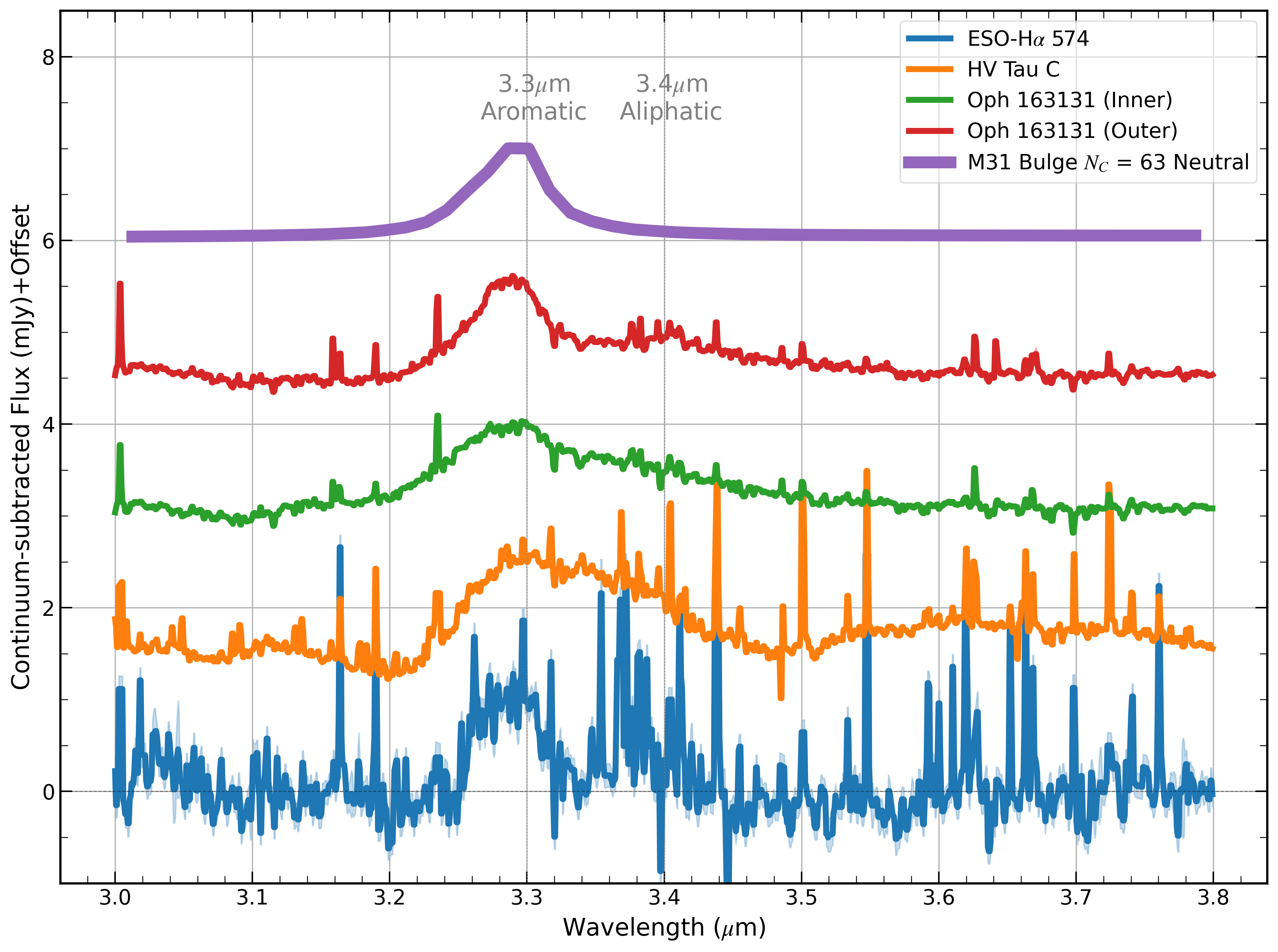}

\caption{Normalized near-IR spectra for all PAH-confirmed targets, comparing the aromatic 3.3~$\mu$m and aliphatic 3.4~$\mu$m emission features. Spectra are vertically offset for clarity. A representative model determined from Figure~\ref{fig:grid} is shown for comparison of the typical ISM FWHM.}
\label{fig:NIR_compare}
\end{figure}

A broad feature is present at 3.3–3.4~$\mu$m in the central region of Oph~163131 and in HV~TAU~C (see Figure~\ref{fig:NIR_compare}). Through a detailed PAHFIT analysis, we find that this broad feature is consistent with enhanced blending of aromatic and aliphatic emission components (see Figure~\ref{fig:PAHFIT}). The increased FWHM compared to the compact, canonical PAH models used for M31 Bulge considered here may indicate that the dominant PAH species present are not well represented by the PAH molecules typically assumed in models of galactic PAH emission \citep{All1989, 2020AL}. The prominent amorphous silicate emission near 10 $\mu$m in HV~TAU~C occurs alongside the unusual 3.3 and 3.4 $\mu$m emission. Both may reflect local disk conditions in which increased dust shielding reduces UV processing, allowing less processed or more aliphatic-rich hydrocarbons to survive.

\begin{figure}[htb!]
\centering

\includegraphics[width=0.45\textwidth]{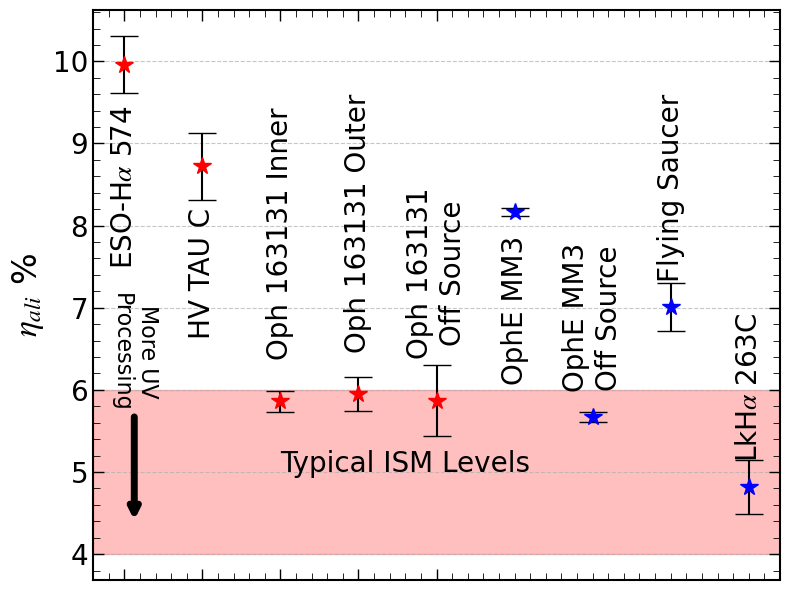}

\caption{Comparison of the derived $\eta_{\rm ali}$ values for all six disks. As in Figure~\ref{fig:grid}, red denotes disk-bound PAHs, while blue denotes PAHs associated with background regions. The derived values are compared with the typical interstellar range of $\eta_{\rm ali}=$ 4--6\%. \citep{yang2023}.}
\label{fig:ali}
\end{figure}

To place the derived aliphatic fractions in context, Figure~\ref{fig:ali} compares the inferred $\eta_{\rm ali}$ values for the JEDIce disks with the typical ISM values reported from \cite{yang2023}. This provides a benchmark for assessing whether the carbonaceous material in edge-on T Tauri disks exhibits an aliphatic fraction similar to that observed in UV-processed ISM, or instead shows evidence for a relatively higher aliphatic fraction that may indicate less extensive UV processing or chemical reprocessing within the disk.

For the case of Oph~163131, $\eta_{\rm ali}$ provides an interesting comparison when contrasted with the extended cloud, where at larger radii the 3.3~$\mu$m exhibits the more canonical shape while both regions remain at nearly identical $\eta_{\rm ali}$ (see Figure~\ref{fig:PAHFIT}). This suggests that, although the physical properties of the PAH carriers may vary, perhaps with a less compact PAH population in the center, the overall PAH emission characteristics remain broadly similar between the two regions. This interpretation is supported by the MIRI data, where the 6.2, 7.7, and 11.3~$\mu$m emission features show broadly similar structures between the two regions (see Figure~\ref{fig:PAHFIT}).

Within the background of The Flying~Saucer, the derived $\eta_{\rm ali}$ is slightly higher than typical ISM values (see Figure~\ref{fig:ali}), indicating weaker-than-expected photo-processing in this background source. The band ratios, particularly the high 11.3/7.7\,$\mu$m value, point to a predominantly neutral PAH population. This is notable because PAHs in the diffuse ISM are usually ionized due to strong UV fields and low electron recombination rates (e.g., \citealt{Tielents2008}). A shift toward neutral PAHs implies a lower ionization fraction, likely caused by UV attenuation or increased local electron density. In a molecular cloud, this is most naturally explained by partial shielding, which suppresses photoionization and favors recombination. The background emission characterized here is physically significant in its own right. It also provides an environmental benchmark for interpreting the PAH emission from nearby disk sources embedded in the same cloud, including Oph~163131 and its unusually high 11.3/7.7~$\mu$m band ratio. 

In contrast, for the potential foreground cloud of OphE~MM3 the derived large and more ionized PAHs (See Figure~\ref{fig:grid}) align well with expected typical ISM PAHs \citep{Draine2007}. This interpretation is physically consistent with the inferred ionization conditions: a harder or less attenuated radiation field preferentially destroys smaller, more fragile PAHs through photodissociation, leaving behind a more resilient, larger ionized population.

These results motivated the additional off-source extractions presented in Figure~\ref{fig:off_source}. For Oph~163131, the similar $\eta_{\rm ali}$ measured on- and off-source (see Figure~\ref{fig:ali}) suggest that the observed level of aliphatic processing is representative of the local environment, with little evidence for additional processing within the disk itself. In contrast, the off-source extraction for OphE~MM3 returns a more typical ISM $\eta_{\rm ali}$. This may indicate that the foreground cloud retains typical diffuse ISM properties, while the elevated $\eta_{\rm ali}$ measured toward the source results from an additional contribution associated with the embedded disk.

From Figure~\ref{fig:grid}, it can be seen that some ratios do not fall within the bounds of this model, underlying not only the lack of a perfect match for the model's parameters, but also the limitations of applying it to these systems. For Oph~163131 and the Flying~Saucer, both within the $\rho$ Ophiuchus molecular cloud, the high likelihood of very neutral species being present may be explained by reducing the photon hardness even more than in the M31 bulge, shifting the solution space upwards and bringing it into better agreement with the calculated band ratios.

\subsection{Disk-to-Disk Variations and Commonality in PAH Properties}

The diversity of environments probed by the JEDIce sample enables a first-order assessment of how local conditions modulate PAH charge states, size distributions, and chemical compositions in edge-on disks. Within the $\rho$~Oph sub-sample, Oph~163131 is dominated by largely neutral PAH populations, consistent with the background PAH found around Flying~Saucer, whereas "on-source" OphE~MM3 exhibits clear signatures of ionized PAH (see Figure~\ref{fig:grid}). Although these sources reside in the same star-forming region and therefore experience similar large-scale environmental conditions, they exhibit different PAH charge states. This suggests that disk-specific properties, such as optical depth, disk geometry, or the local UV radiation field, play a larger role in determining PAH ionization than the shared ambient environment.

Despite variations in ionization, the PAH size distribution across these sources appears remarkably constant, with all three disks dominated by small- to intermediate-sized PAHs ($N_\mathrm{C}\approx20-40$). This similarity may indicate that the PAH population is largely inherited from the molecular cloud material from which these disks formed, consistent with the survival of interstellar carbonaceous material throughout the star formation process \citep{Tielents2008, Oberg_2021}. This inheritance scenario is plausible because, although PAHs in exposed disk surface layers can undergo rapid ionization and photochemical processing, the bulk PAH population can survive in shielded regions on timescales comparable to or exceeding the Class II disk lifetime, allowing molecular-cloud PAHs to remain an important reservoir of carbonaceous material \citep{Tielents2008,Oberg_2021}. The comparable $\eta_{\rm ali}$ ($\sim5-6\%$) further suggest limited chemical processing of the carbonaceous component, although the persistence of aliphatic bonds can depend on local irradiation and thermal processing conditions (\citealt{Kwok2013,yang2023}). Thus, while the PAH ionization state responds rapidly to local radiation fields and disk geometry, the overall molecular structure may retain a memory of earlier cloud and disk formation conditions.

Although the present observations do not directly measure absolute PAH abundances, the detection of spatially extended aromatic emission demonstrates that carbonaceous nanoparticles remain an important component of the disk surface layers in several Class II T Tauri disks. Together with the broadly uniform PAH size distributions, these results support a scenario in which a substantial fraction of the small-grain carbon reservoir survives from the parent molecular cloud while undergoing local modification by the disk radiation field.

\subsection{Comparison with Previous Studies}

Previous observational studies of PAHs in protoplanetary disks, particularly those carried out with \textit{Spitzer} and ground-based facilities such as the VLT, established the presence of PAH emission in a subset of disks. PAH emission was detected in $\sim$50--70\% of Herbig Ae/Be disks \citep{Acke2004,Acke2010}, but in only $\sim$8\% of T Tauri disks \citep{Geers2006,Geers2007}, highlighting the relative rarity of detectable PAHs in low-mass systems. These studies further demonstrated that PAH emission depends on both the incident ultraviolet radiation field and disk geometry, with flared disks generally exhibiting stronger emission because their surface layers are more effectively irradiated.

The results presented here extend earlier findings by leveraging JWST's spatial and spectral capabilities. The detection of multiple PAH bands indicates that PAH emission is not limited to only the most strongly irradiated systems. This is consistent with recent JWST PAH detections in edge-on systems such as Tau 042021 and HH~48~NE \citep{Arulanantham2024,Sturm_2024, Dartois2025}, where emission is identified in low-mass environments when the viewing geometry reduces contamination from the inner disk continuum. Compared to earlier \textit{Spitzer} observations, the JWST data reveal weaker and more spatially extended emission components that were previously unresolved.

The ability to characterize PAH band ratios further provides constraints on the emitting population. As discussed in recent JWST studies, these ratios are consistent with predominantly neutral PAHs with sizes comparable to those inferred in earlier work, while still exhibiting variations between sources. In addition, PAH emission can be spatially resolved and may be associated with distinct emitting regions rather than appearing as a single unresolved component.

The observations presented here, particularly those obtained with \textit{NIRSpec}, provide complementary constraints by accessing shorter wavelengths, including the 3.3~$\mu$m feature and the associated aliphatic emission. This enables a more complete characterization of the carbonaceous material by directly probing both aromatic and aliphatic components (see Figure \ref{fig:NIR_compare}). By incorporating constraints on the $\eta_{\rm ali}$ alongside established PAH diagnostics, this work refines the interpretation of PAH emission in these systems and provides additional context for understanding how these features arise under the physical conditions present in T Tauri disks.

Taken together, these results build upon both legacy \textit{Spitzer} studies and recent JWST observations by demonstrating that PAH emission in T Tauri disks can be detected under a broader range of conditions than previously established, while remaining broadly consistent with earlier constraints on PAH size and charge state.

\subsection{Effects of inclination on PAH detection in T Tauri Disks}

A key question arising from this work is the role of disk inclination in the detectability of PAH emission. PAH features are commonly detected in Herbig Ae/Be disks across a wide range of viewing geometries, likely because their stronger ultraviolet radiation fields efficiently excite PAH molecules. In contrast, PAH emission remains comparatively rare in T Tauri systems \citep{Habart2006, Geers2006, Geers2007, Seok_2016,Yoffe_2023}. Notably, when PAH emission is detected in T Tauri disks, it appears to be preferentially associated with highly inclined or edge-on geometries such as in Tau 042021 and HH~48~NE \citep{Arulanantham2024,Sturm_2024} and this work. This suggests that inclination may introduce a significant observational bias, such that PAH emission is not necessarily absent in face-on systems but instead overwhelmed by more intense internal sources of radiation.

One plausible explanation is that inclination alters the balance between PAH features and the underlying dust continuum emission. In T Tauri disks, the intrinsically weak UV field results in comparatively faint PAH emission \citep{Li2001,Geers2006,Geers2007}, making these features difficult to distinguish from the underlying continuum. In edge-on systems, the optically thick mid-plane obscures direct sightlines to the bright inner disk and may also reduce the scattered continuum observed from the outer disk, thereby enhancing the feature-to-continuum ratio of PAH bands produced locally in the UV-irradiated surface layers \citep{Acke2004,Geers2006, Tielents2008, Visser2009}. Prior to JWST, the detection of such weak PAH emission was further limited by the sensitivity and spectral resolution of previous infrared facilities, which often lacked the ability to separate faint PAH bands from the strong dust continuum and silicate emission features in T Tauri disks. JWST's improved sensitivity and spectroscopic capabilities therefore enable the detection of PAH emission that may have been present but remained below previous observational limits.

Wavelength-dependent extinction through the disk may also affect the observed PAH band ratios and therefore the inferred PAH properties. Shorter-wavelength PAH bands are more strongly attenuated than longer-wavelength bands, so in edge-on systems with large line-of-sight column densities, the 3.3~$\mu$m feature can be suppressed relative to the 11.3~$\mu$m feature. Differential extinction could therefore lower the observed 3.3/11.3~$\mu$m ratio and lead to an overestimate of the characteristic PAH size if extinction is not accounted for. Similarly, extinction can modify the 11.3/7.7~$\mu$m band ratio and thus affect the inferred PAH ionization state. In this work, we account for extinction when fitting the PAH spectra with PAHFIT, allowing the band strengths used to calculate these ratios to be determined from the extinction-corrected model components rather than directly from the observed spectra. The resulting ratios should therefore be less sensitive to the wavelength-dependent extinction along the line of sight, although uncertainties in the extinction correction remain an important source of uncertainty, particularly for the shorter-wavelength features. In contrast, the 3.3 and 3.4~$\mu$m features are affected by similar extinction, so their ratio, and consequently $\eta_{\rm ali}$, should be comparatively insensitive to this effect.

The excitation and visibility of PAHs are also strongly dependent on stellar UV luminosity, disk structure (e.g., flaring and gaps), and the intrinsic PAH abundance. As such, inclination should be interpreted primarily as a mechanism that enhances the detectability of weak emission rather than as evidence that PAHs are preferentially present in edge-on disks. Previous radiative-transfer models of PAH emission in protoplanetary disks have shown that PAHs are primarily excited in UV-irradiated surface layers, with the observed feature strengths controlled by disk flaring, vertical optical depth, and viewing geometry \citep{Dullemond2004, Visser2007}. Next generation radiative transfer models that simultaneously treat PAH abundance, disk geometry, and continuum opacity will be essential to disentangle these effects and quantify the extent to which inclination biases current observational samples similar to the model done in \cite{Sturm_2024}. With the increasing number of spatially and spectrally resolved PAH detections from JWST, more detailed PAH emission models that couple disk structure, dust radiative transfer, and PAH chemistry are now needed to interpret these observations and establish a more complete understanding of PAH evolution in protoplanetary disks.

\subsection{Comparison with H$_2$ lines}

Within this sample, we also detect numerous near-IR H$_2$ emission lines. Building on the analysis of \cite{Assani2026} and focusing on the O- and S-branch H$_2$ transitions at 2.627, 2.803, and 1.958~$\mu$m, we find widespread H$_2$ emission in both Oph~163131 and HV~Tau~C, despite their very different ionization and $\eta_{\rm ali}$ (see Figures~\ref{fig:grid} and \ref{fig:spatial}). This tentatively indicates no link between H$_2$ emission and PAH ionization, implying differing excitation pathways for these species.

\begin{figure}[htb!]
\centering

\includegraphics[width=.45\textwidth]{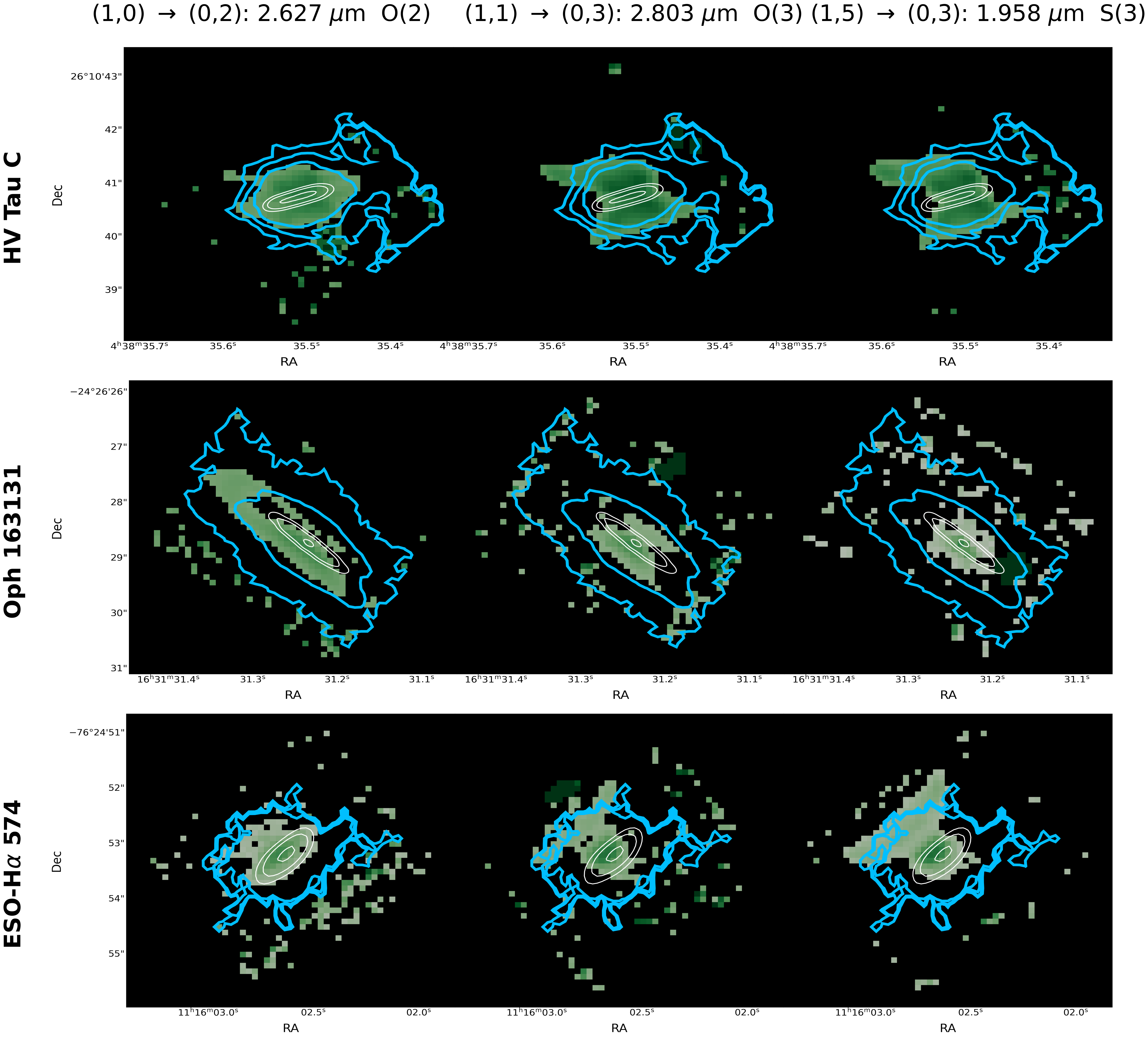}

\caption{Spatial distribution of the O- and S-branch H$_2$ emission (green) for ESO-H$\alpha$~574, HV~Tau~C, and Oph~163131, overlaid with the PAH emission and ALMA continuum contours from Figure~\ref{fig:coll_PAH}. The H$_2$ emission traces the O(2), O(3), and S(3) transitions at 2.627, 2.803, and 1.958~$\mu$m, respectively.}
\label{fig:H2plots}
\end{figure}

A notable observation appears in ESO-H$\alpha$~574, where a pronounced cone of H$_2$ in the surface layer of the disk coincides with the top of the extended PAH emission (see Figure~\ref{fig:H2plots}). In Figure~\ref{fig:spatial}, this region also shows a strong spike in the $\eta_{\rm ali}$, suggesting enhanced aliphaticity associated with the outflow. Future work is needed to fully investigate this.

In addition, there does not appear to be any clear spatial correlation between the H$_2$ and PAH emission. In the cases of Oph~163131 and ESO-H$\alpha$~574, the PAH emission is significantly more spatially extended. For HV~TAU~C, although the H$_2$ emission is also more extended than in the other two sources, the regions of strongest PAH emission coincide with an apparent deficit of O(3) and S(3) H$_2$ emission.

\subsection{Neon emission lines}

\begin{figure}[htb!]
\centering

\includegraphics[width=0.45\textwidth]{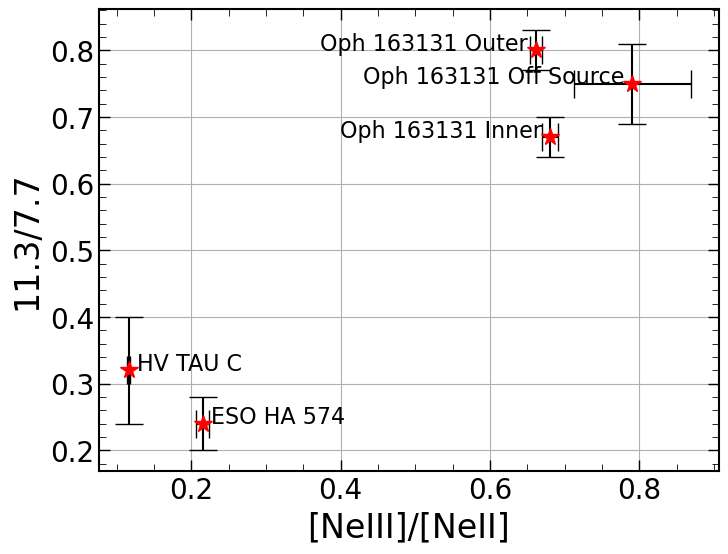}

\caption{Comparison of the 11.3/7.7~$\mu$m PAH band ratio with the [NeIII]/[NeII] line ratio. The 11.3/7.7~$\mu$m ratio traces the PAH ionization state, while [NeIII]/[NeII] provides a complementary tracer of the hardness of the local radiation field. [NeIII] emission is detected in only the brightest regions of the sample because production requires photons with energies $\gtrsim41$~eV \citep{Glassgold_2007,Gudel2010}.}
\label{fig:Neon}
\end{figure}

Within the brightest regions of the sample, [Ne III] emission at 15.6 $\mu$m is detected, providing a useful tracer of hard photons and a natural comparison with the 11.3/7.7~$\mu$m PAH band ratio as a proxy for PAH ionization state when compared to [NeII] (see Figure~\ref{fig:Neon}). For ESO-H$\alpha$ 574 and HV Tau, the [NeIII]/[NeII] ratio is 0.22 and 0.12 respectively, which remains consistent with T Tauri disk environments dominated by relatively soft stellar radiation fields and limited production of photons with energies $\gtrsim$ 41 eV required to generate [Ne III] emission \citep{Glassgold_2007,Gudel2010}. In contrast, Oph~163131 exhibits a higher ratio, ranging from 0.66–0.79.

Oph~163131 is also the region exhibiting the highest 11.3/7.7~$\mu$m band ratio in the sample. This coincidence may suggest a coupling between local ionization conditions traced by both neon fine-structure emission and PAH band ratios, consistent with the established sensitivity of PAH ionization balance to electron density, radiation field hardness, and grain charging processes \citep{Bakes1994, Tielents2008}.  

Notably, Oph~163131 also shows a PAH $\eta_{\rm ali}$ closer to typical diffuse ISM values (see Figure~\ref{fig:ali}), in contrast to the more processed spectra seen in the other two PAH disks. This may indicate comparatively less carbonaceous restructuring or UV-driven aromatization in this system, potentially linking PAH chemical evolution to the observed ionization diagnostics \citep{All1989,2020AL,yang2023}. 

However, given the degeneracy of PAH band ratios with size distribution, local extinction, and processing history, this correspondence should not be interpreted as a one-to-one mapping between [Ne III] abundance, PAH ionization fraction, and $\eta_{\rm ali}$ as the  chemical interplay is non-trivial and likely reflects a combination of local UV processing, grain growth, and reformation pathways. A detailed Neon analysis will be explored in Assani et al. (2026, in prep).

\section{Conclusions}

This paper presents the near- to mid-IR spectra of six edge-on PPDs observed as part of the JWST Edge-on Disk Ice (JEDIce) program. The focus of this work has been to empirically analyze the PAH features found within these sources and where PSF and S/N permit the examination of the spatial variations within the sample. Our main findings include:

\begin{itemize}

\item We expand the sample of PAHs detected in edge-on disks for JWST spectroscopic  observations with 3 detections in our observations. Following those in Tau 042021 and HH~48~NE \citep{Arulanantham2024,Sturm_2024} this brings the total number of detections to 5 supporting that PAHs are present around these low-mass systems.

\item Our analysis advances our understanding of PAH behavior in low-UV T Tauri environments, providing empirical constraints on their presence, survival, and excitation within disk structures. This work informs how PAHs interact with local radiation fields and disk geometry, contributing to the broader picture of organic chemical reservoirs during planet formation.  

\item We propose that the apparent observational bias in PAH detections towards edge-on disks arises from geometric effects, favorable scattering surfaces, and improved contrast against the stellar continuum. Consequently, PAH non-detections in face-on systems may reflect low contrast rather than true chemical depletion.  

\item Globally, the observed PAH population is predominantly small to intermediate neutral PAHs, with ionization fractions below 50\% and $N_\mathrm{C} \approx$ 30–100. 

\item Disks exhibiting PAH emission indicate an enhanced $\eta_{\rm ali}$, with 2 displaying values significantly exceeding typical ISM levels and 1 on the higher end of typical ISM values. This result suggests that T Tauri environments are characterized by intensified aliphaticity during the disk formation phase. 

\item The PAH emission profiles, especially at 3.3~$\mu$m, indicate that  some disks (Oph~163131 inner region and HV~TAU~C) seem to host less compact, more open and fragile PAHs shaped by weak local UV fields, distinct from the more canonical PAH population seen in other regions, possibly from active chemistry occurring adding moieties to the recently exposed PAHs. The strongest deviations from canonical profiles occurs at shorter wavelengths, implying that small, open PAHs dominate the near-IR while larger, compact PAHs dominate the mid-IR, showing potential spatially dependent chemical evolution within the disks.

\item The spatial variation of the PAH ionization fraction emphasizes the role of PAHs as electron sinks in protoplanetary disks. For the singular source capable of 11.3~$\mu$m spatially resolved emission, Oph~163131, we find PAHs become increasingly neutral with distance from the central star, reflecting the sensitivity of the ionization fraction to local conditions such as the radiation field, gas temperature, and electron density.  

\item Oph~163131 additionally exhibits a clear radial gradient in characteristic PAH size, with the 3.3/11.3~$\mu$m band ratio indicating an increase from $\sim$30 carbon atoms in the inner disk to $\sim$100 carbon atoms at larger radii and higher altitudes. This trend suggests spatially varying PAH processing within the disk.

\item Oph~163131 finally exhibits a nearly constant PAH ionization fraction at larger radii, despite strong PAH emission from its surrounding cloud. The flat 11.3/7.7~$\mu$m band ratio suggests that photoionization and recombination reach an approximate equilibrium over the extended emitting region.

\end{itemize}

Expanding on the empirical results presented here, future modeling efforts will be crucial for placing quantitative constraints on PAH abundances, charge states, and spatial distributions within these PPDs, as well as their interaction with the surrounding gas and radiation field. These models will also help refine our understanding of the role of PAHs in gas heating, their contribution to mid-infrared opacity, and their evolution across different UV irradiation regimes. In this framework, JWST observations of edge-on disks offer a distinct advantage by providing heightened sensitivity to vertical structure and line-of-sight stratification. Collectively, these findings emphasize the value of PAHs as diagnostics of disk chemistry and energetics, and reinforce their significance in connecting chemical processing to the physical environments that shape planet formation.

\begin{acknowledgments}

This work is based on observations made with the NASA/ESA/CSA James Webb Space Telescope. The data were obtained from the Mikulski Archive for Space Telescopes at the Space Telescope Science Institute, which is operated by the Association of Universities for Research in Astronomy, Inc., under NASA contracts NAS 5–26555 and NAS 5-03127.  The specific observations analyzed can be accessed via \dataset[doi:10.17909/q1z5-vz46]{https://doi.org/10.17909/q1z5-vz46}.

This paper makes use of the following ALMA data: 2023.1.00634.S, 2016.1.00771.S, 2023.1.00907.S, 2022.1.01302.S, and 2024.1.00928.S. ALMA is a partnership of ESO (representing its member states), NSF (USA) and NINS (Japan), together with NRC (Canada), NSTC and ASIAA (Taiwan), and KASI (Republic of Korea), in cooperation with the Republic of Chile. The Joint ALMA Observatory is operated by ESO, AUI/NRAO and NAOJ.

Support for Program number 5299 was provided by NASA through a grant from the Space Telescope Science Institute, which is operated by the Association of Universities for Research in Astronomy, Inc., under NASA contract NAS 5-03127.

We thank the anonymous reviewer for their constructive comments and valuable suggestions 

C.M.  acknowledges support from National Tsing Hua University Research Talent Resource Center.

D.H. is supported by the Ministry of Education of Taiwan (Center for Informatics and Computation in Astronomy grant and grant number 110J0353I9) and the National Science and Technology Council, Taiwan (Grant NSTC111-2112-M-007-014-MY3, NSTC113-2639-M-A49-002-ASP, and NSTC113-2112-M-007-027). 

JAN and ED acknowledge support from the Thematic Action `Physique et Chimie du Milieu Interstellaire' (PCMI) of INSU Programme National `Astro', with contributions from CNRS Physique \& CNRS Chimie, CEA, and CNES.

MND's work is funded by the European Union. Views and opinions expressed are however those of the author(s) only and do not necessarily reflect those of the European Union or the European Research Council Executive Agency. Neither the European Union nor the granting authority can be held responsible for them. This work is supported by ERC grant PSII (DOI: 10.3030/101230593).

ZYL is supported in part by NSF AST-2307199.  

Part of this research was carried out at the Jet Propulsion Laboratory, California Institute of Technology, under a contract with the National Aeronautics and Space Administration (80NM0018D0004).

\end{acknowledgments}

\software{Astropy \citep{2013A&A...558A..33A,2018AJ....156..123A,2022ApJ...935..167A}, Matplotlib \citep{Hunter2007}, NumPy \citep{Numpy2020}, Scipy \citep{Scipy2020}}

\appendix

\section{Alignment with ALMA and between MIRI and NIRSpec} \label{Shifting}

Table~\ref{tab:target_shifts} compares the pointing differences for NIRSpec and MIRI when compared to the available ALMA archival data. This was necessary to accurately transfer the NIRSpec contours onto the MIRI observations, thereby ensuring that similar physical regions were being examined in all datasets.

\begin{deluxetable}{l c c c c c}
\tablecaption{Target Shifts and Scaling Factors for targets with available ALMA data to correct for pointing and flux discrepancies between the instruments.\label{tab:target_shifts}}
\tablewidth{0pt}
\tablehead{
\colhead{Target} & 
\colhead{NIRSpec $\Delta$RA (\arcsec)} & 
\colhead{NIRSpec $\Delta$Dec (\arcsec)} & 
\colhead{MIRI $\Delta$RA (\arcsec)} & 
\colhead{MIRI $\Delta$Dec (\arcsec)} & 
\colhead{NIRSpec to MIRI Scale}
}
\startdata
ESO H$\alpha$ 574    & -0.385 &  0.107 & -0.158 &  0.011 & 1.26 \\
HV~TAU~C      & -0.043 & -0.389 &  0.257 & -0.01  & 0.99 \\
Oph 163131    & -0.123 & -0.313 &  0.257 & -0.145 & 0.99, 1.60 \\
Flying~Saucer &  0.351 & -0.408 &  0.352 & -0.07  & 1.25 \\
OphE~MM3      & -0.008 &  0.44  &  0.08  &  0.068 & 1.29 \\
\enddata
\end{deluxetable}

\onecolumngrid

\section{Constructed Annuli for Targets}\label{ConAnn}

\begin{figure*}[htb!]
\centering
\begin{subfigure}{0.30\textwidth}
    \includegraphics[width=\linewidth]{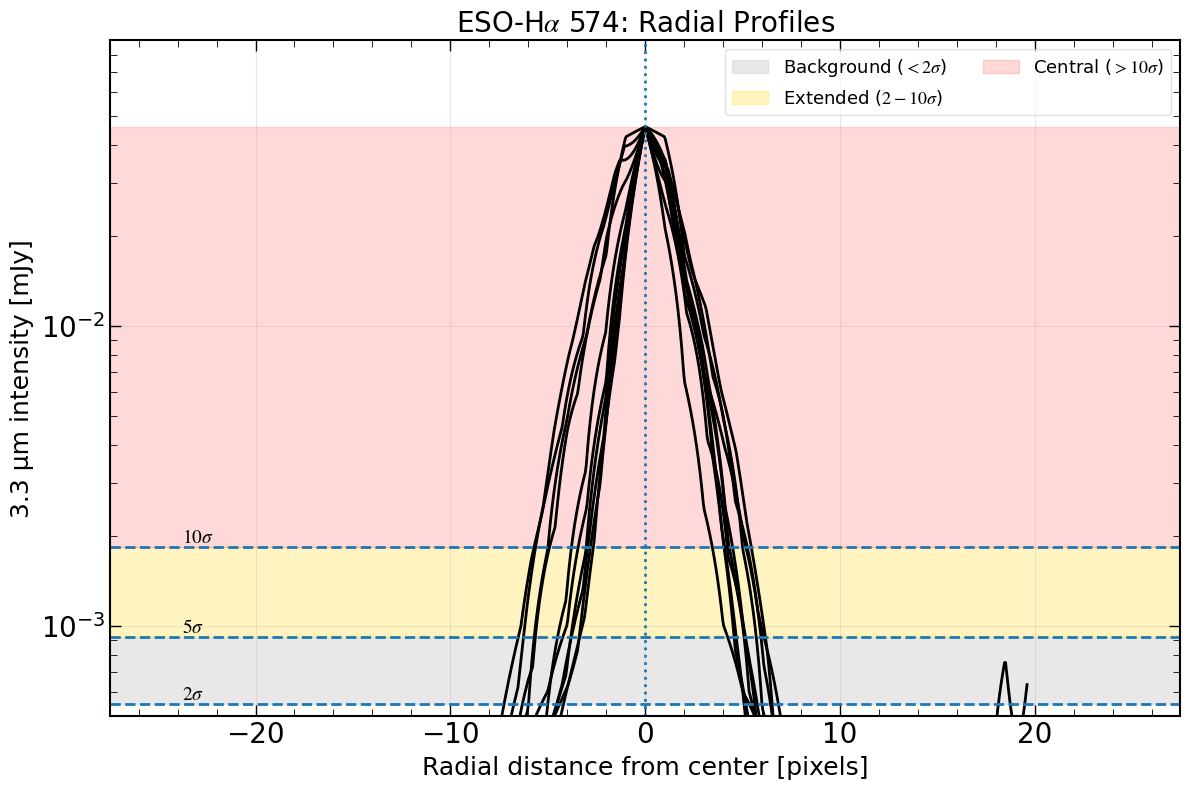}
    \caption{ESO-H$\alpha$ 574}
\end{subfigure}\hfill
\begin{subfigure}{0.30\textwidth}
    \includegraphics[width=\linewidth]{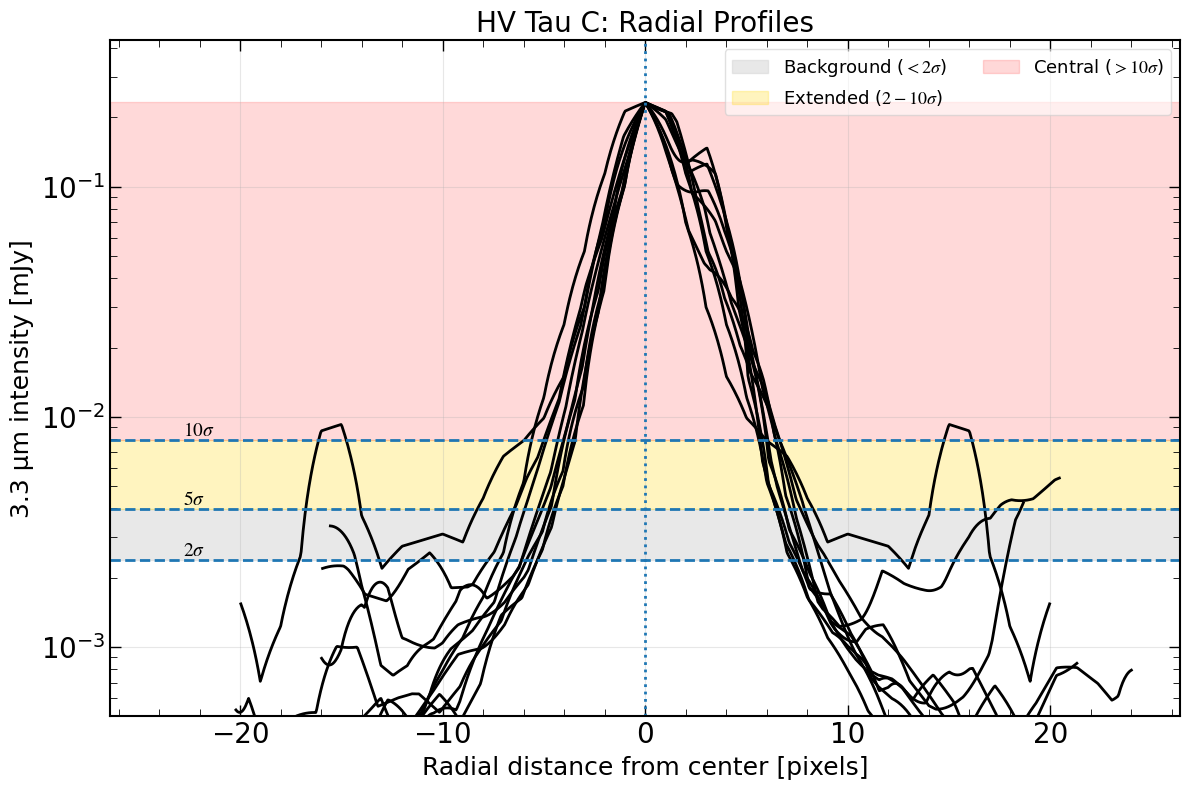}
    \caption{HV~TAU~C}
\end{subfigure}\hfill
\begin{subfigure}{0.30\textwidth}
    \includegraphics[width=\linewidth]{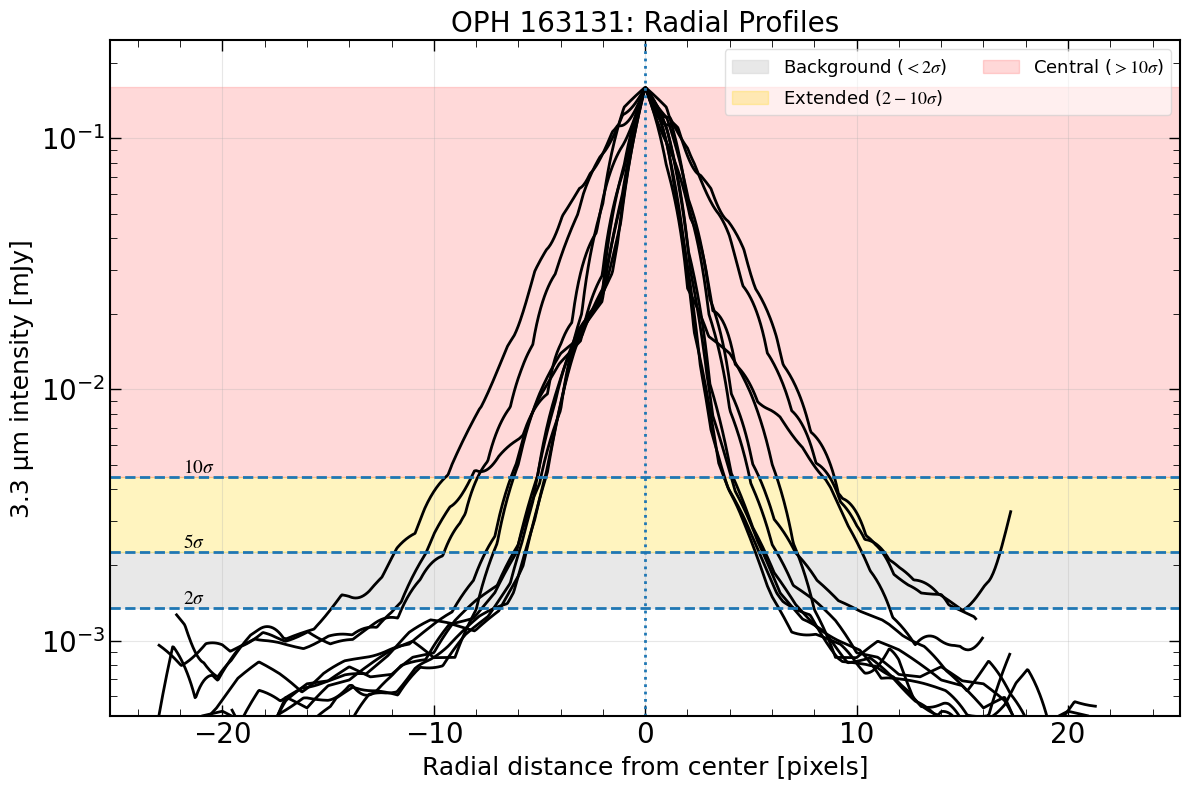}
    \caption{Oph 163131}
\end{subfigure}\\[1ex]
\begin{subfigure}{0.30\textwidth}
    \includegraphics[width=\linewidth]{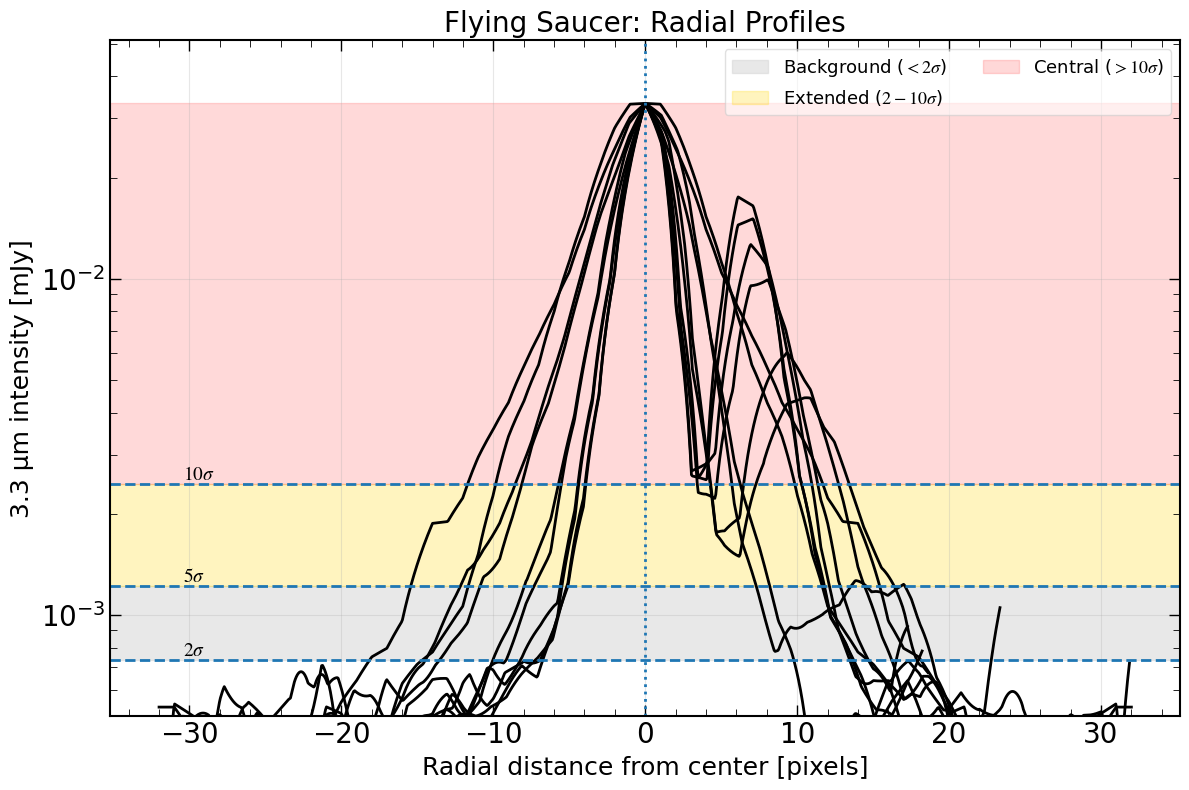}
    \caption{Flying~Saucer}
\end{subfigure}\hfill
\begin{subfigure}{0.30\textwidth}
    \includegraphics[width=\linewidth]{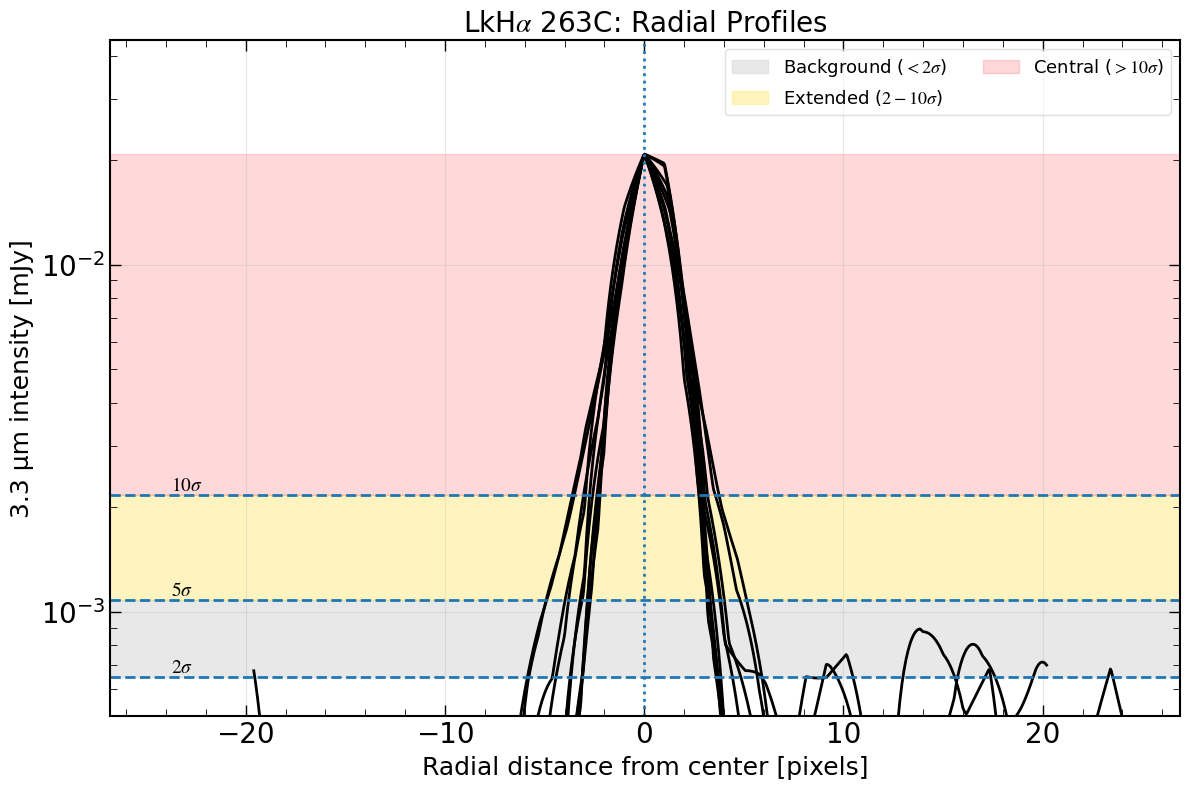}
    \caption{LkH$\alpha$ 263C}
\end{subfigure}\hfill
\begin{subfigure}{0.30\textwidth}
    \includegraphics[width=\linewidth]{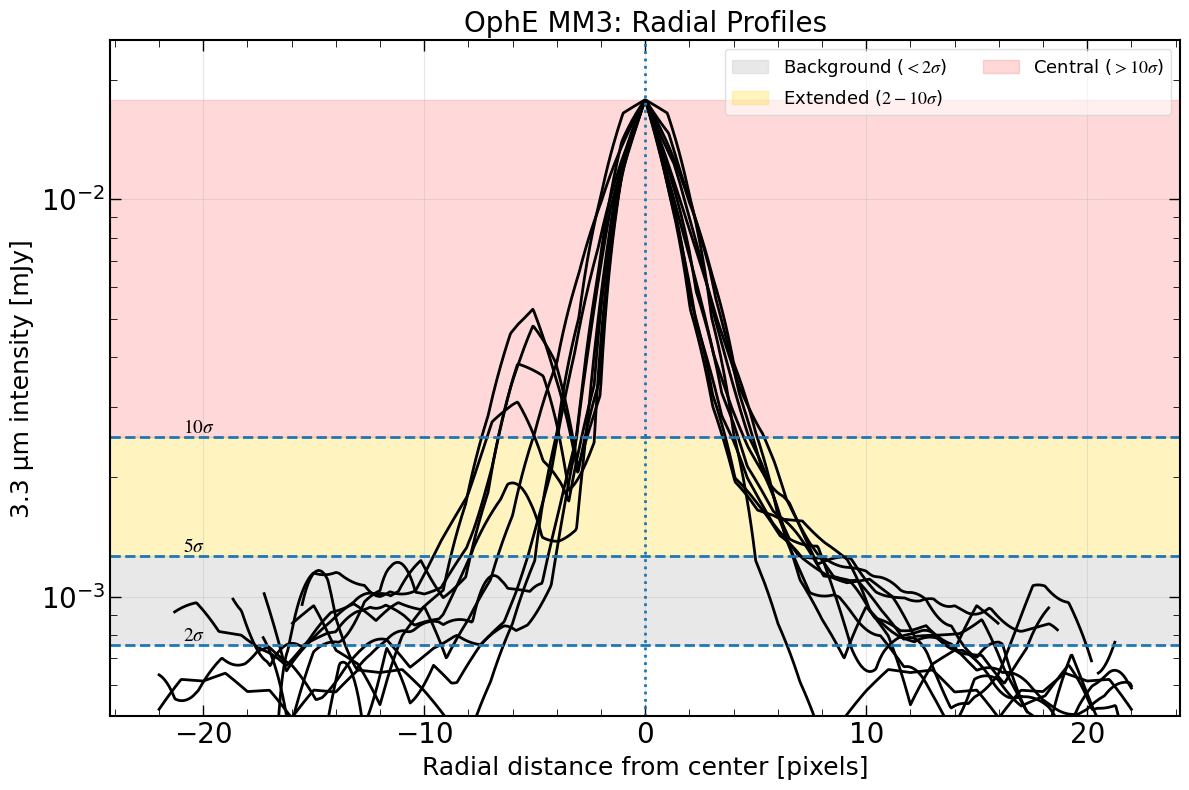}
    \caption{OphE~MM3}
\end{subfigure}\\[1ex]
\caption{Radial flux plots generated to provide a quantitative justification for the selected 2,5 and 10$\sigma_{RMS}$ levels.}
\label{fig:radial_check}
\end{figure*}

\begin{figure*}[htb!]
\centering
\begin{subfigure}{0.30\textwidth}
    \includegraphics[width=\linewidth]{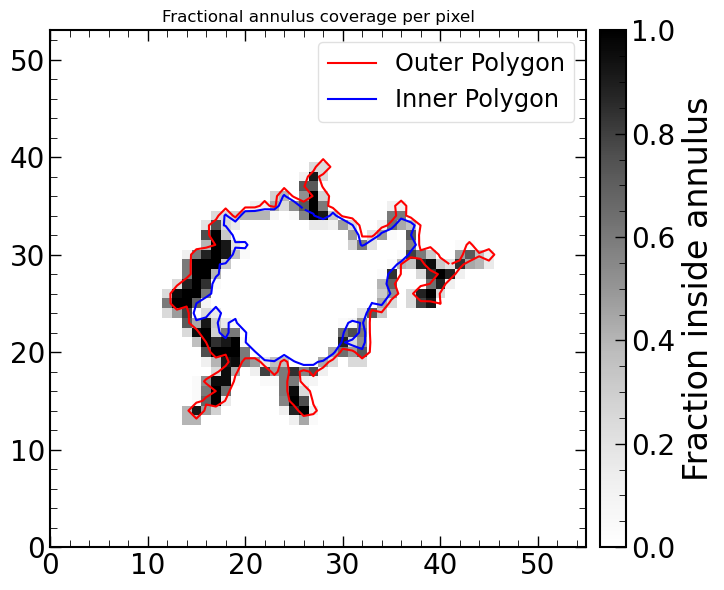}
    \caption{ESO-H$\alpha$ 574 annulus}
\end{subfigure}\hfill
\begin{subfigure}{0.30\textwidth}
    \includegraphics[width=\linewidth]{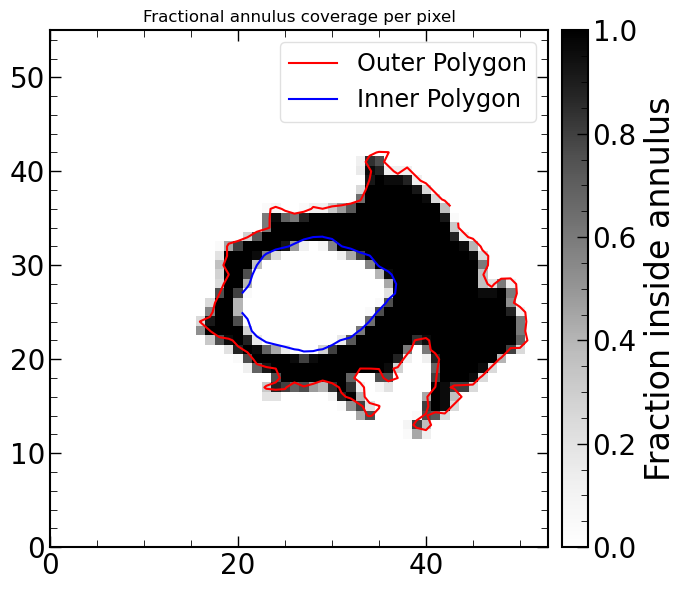}
    \caption{HV~TAU~C annulus}
\end{subfigure}\hfill
\begin{subfigure}{0.30\textwidth}
    \includegraphics[width=\linewidth]{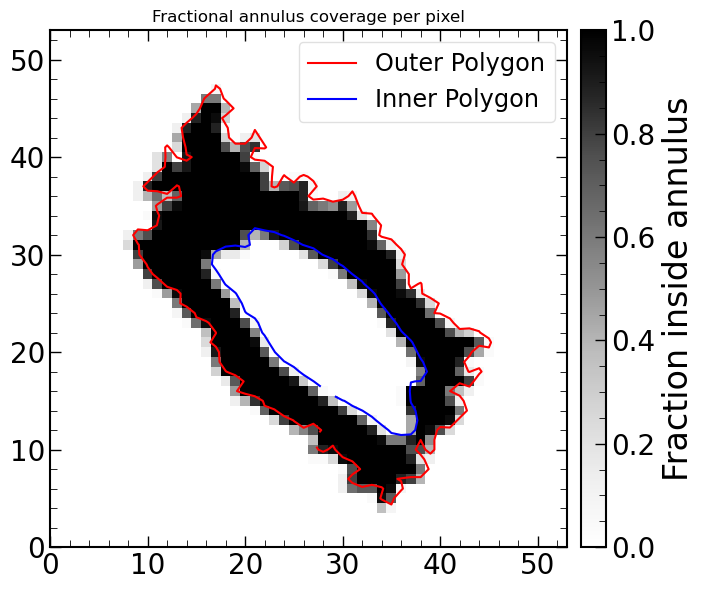}
    \caption{Oph 163131 annulus}
\end{subfigure}\\[1ex]
\begin{subfigure}{0.30\textwidth}
    \includegraphics[width=\linewidth]{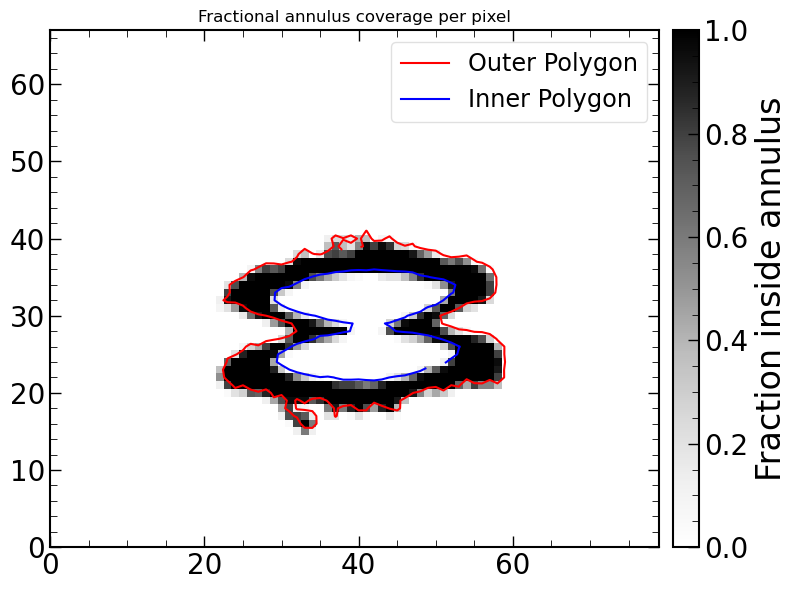}
    \caption{Flying~Saucer annulus}
\end{subfigure}\hfill
\begin{subfigure}{0.30\textwidth}
    \includegraphics[width=\linewidth]{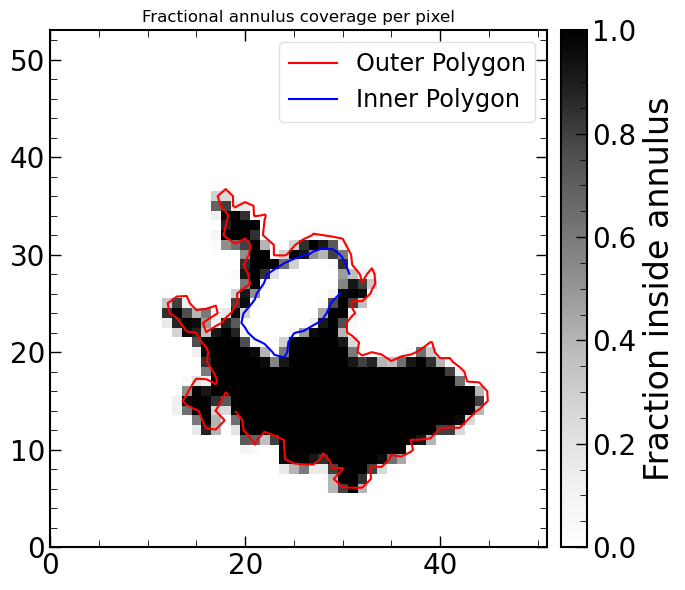}
    \caption{LkH$\alpha$ 263C annulus}
\end{subfigure}\hfill
\begin{subfigure}{0.30\textwidth}
    \includegraphics[width=\linewidth]{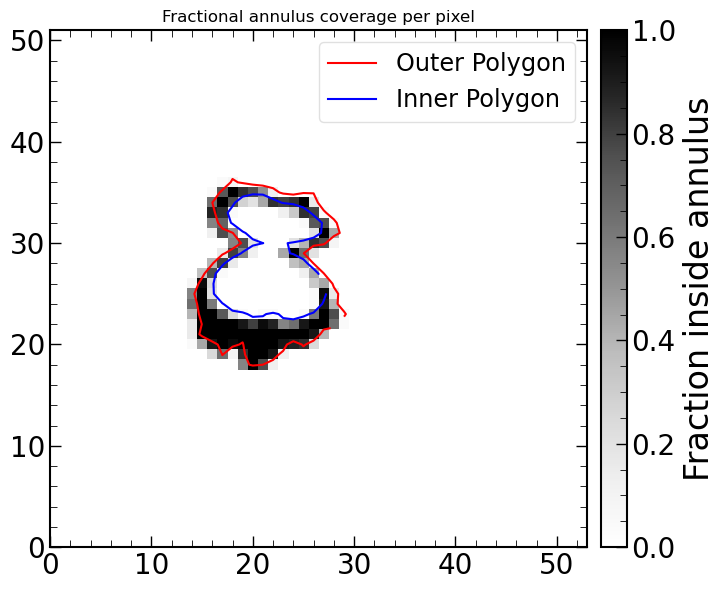}
    \caption{OphE~MM3 annulus}
\end{subfigure}\\[1ex]
\begin{subfigure}{0.30\textwidth}
    \includegraphics[width=\linewidth]{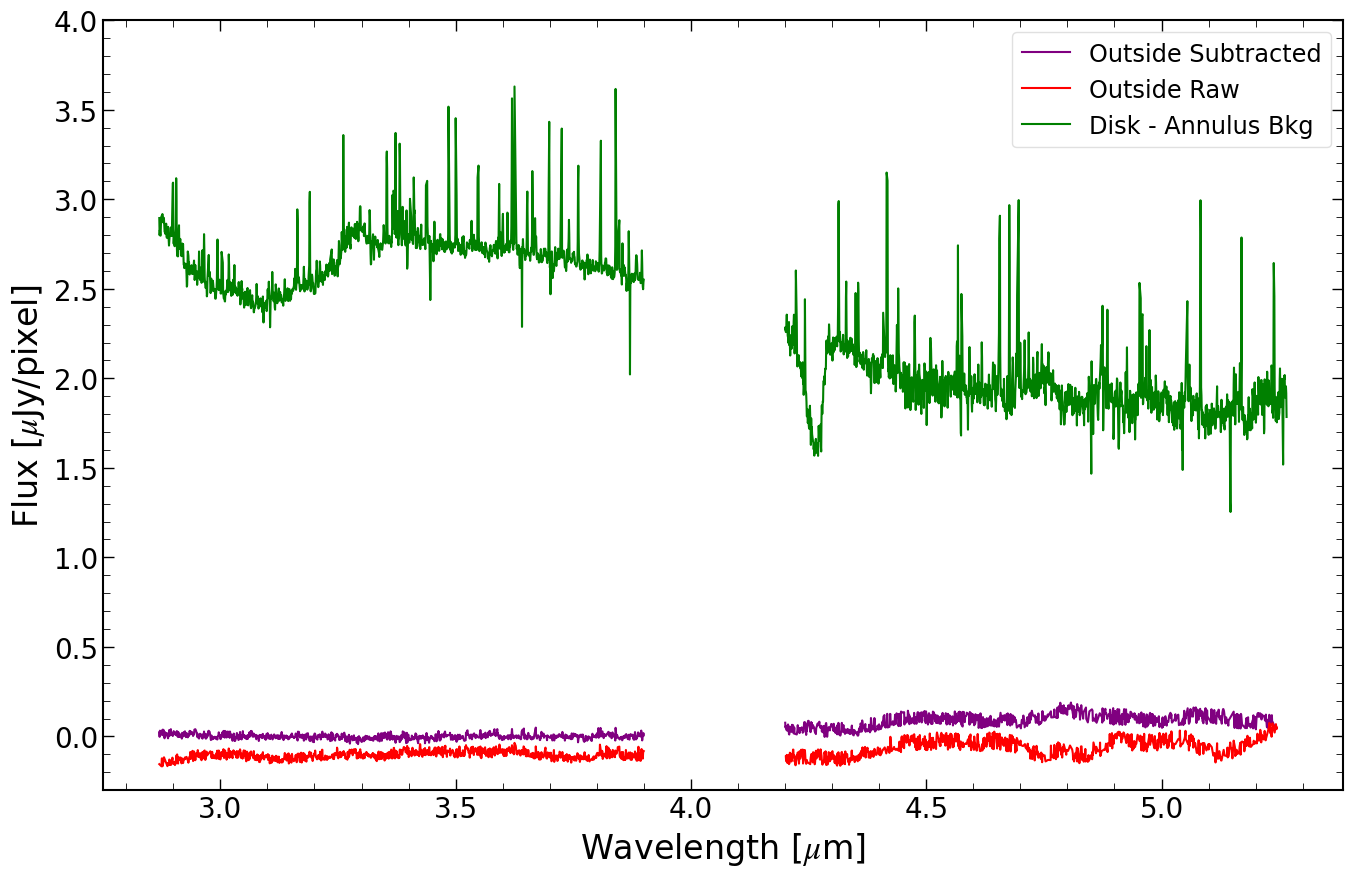}
    \caption{ESO-H$\alpha$ 574 subtraction results}
\end{subfigure}\hfill
\begin{subfigure}{0.30\textwidth}
    \includegraphics[width=\linewidth]{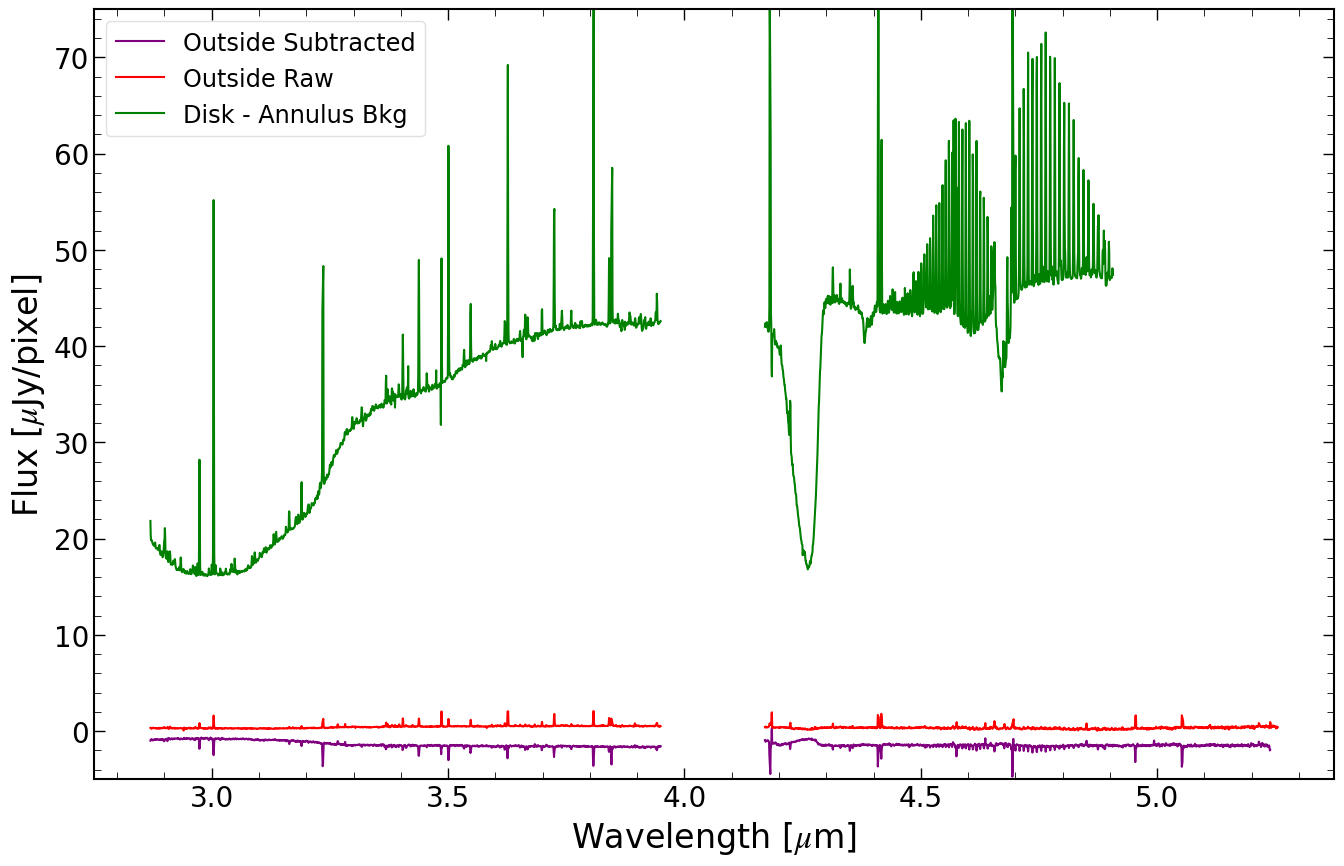}
    \caption{HV~TAU~C subtraction results}
\end{subfigure}\hfill
\begin{subfigure}{0.30\textwidth}
    \includegraphics[width=\linewidth]{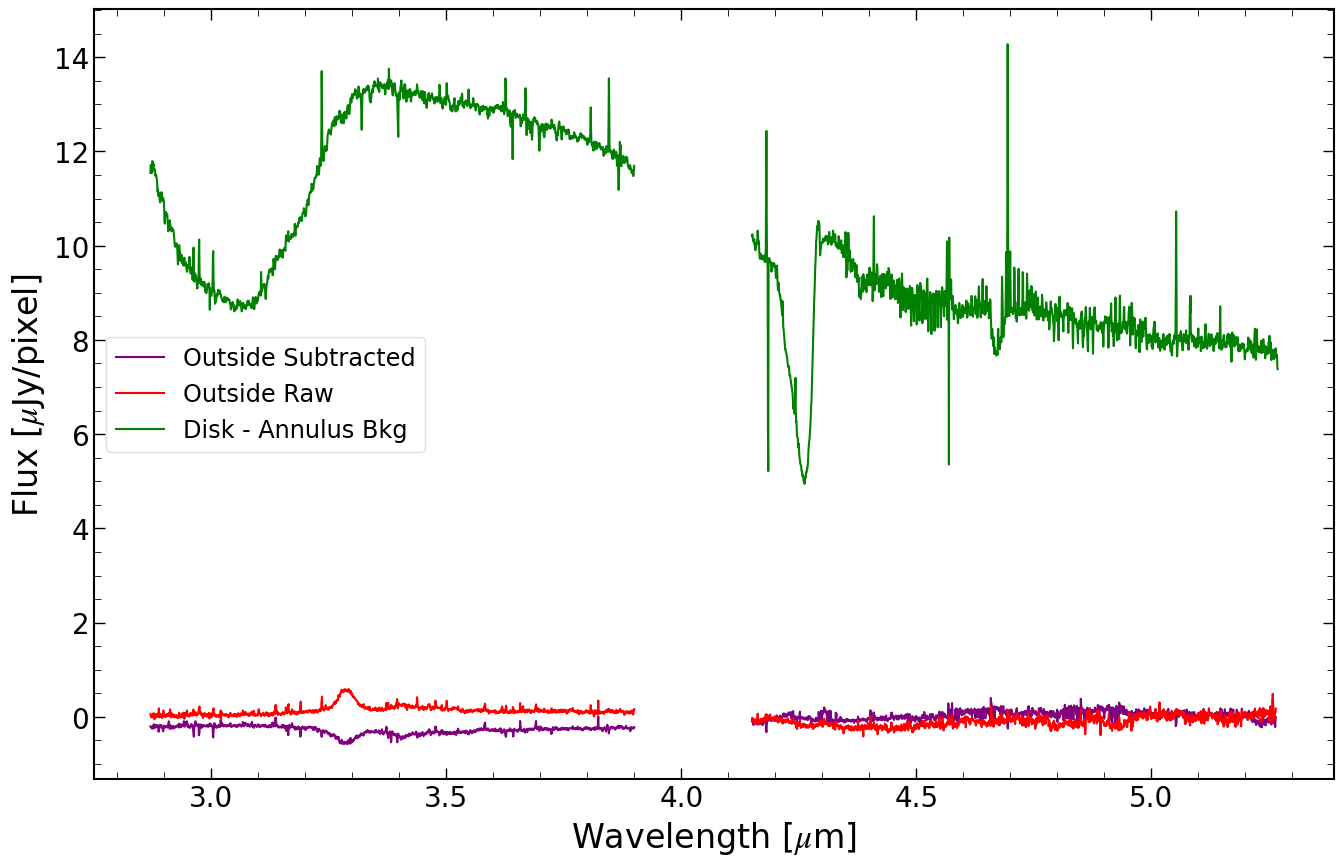}
    \caption{Oph 163131 subtraction results}
\end{subfigure}\\[1ex]
\begin{subfigure}{0.30\textwidth}
    \includegraphics[width=\linewidth]{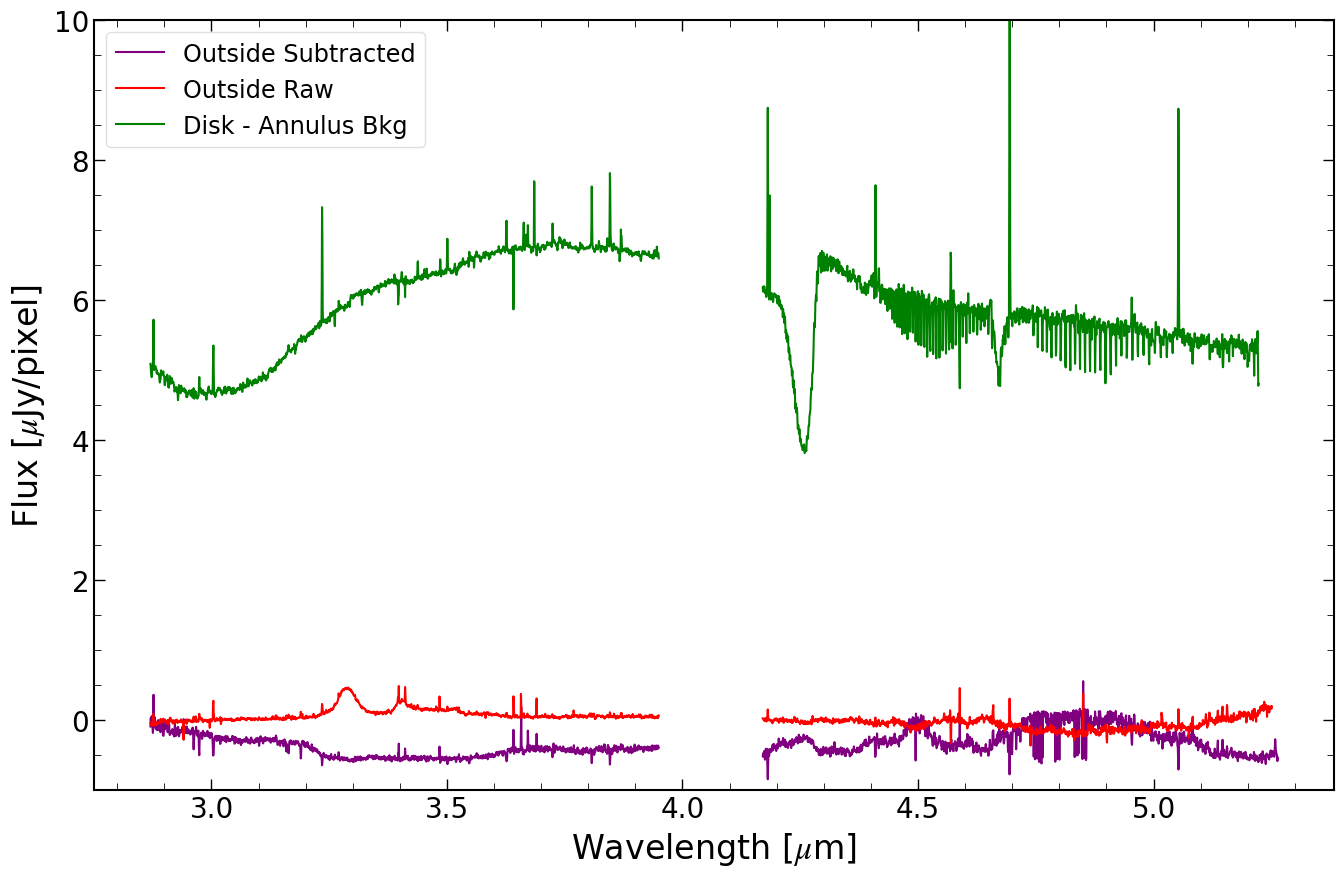}
    \caption{Flying~Saucer subtraction results}
\end{subfigure}\hfill
\begin{subfigure}{0.30\textwidth}
    \includegraphics[width=\linewidth]{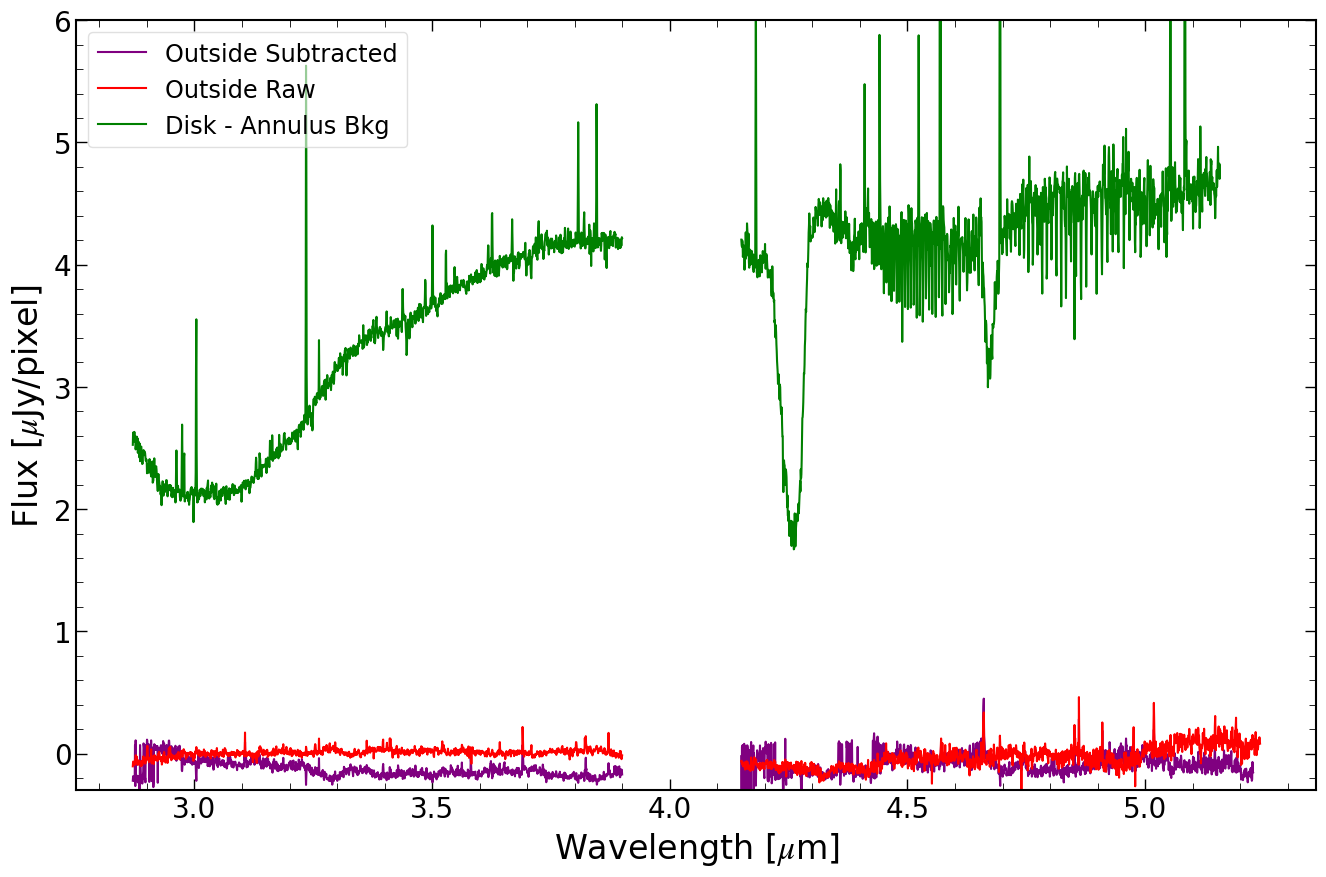}
    \caption{LkH$\alpha$ 263C subtraction results}
\end{subfigure}\hfill
\begin{subfigure}{0.30\textwidth}
    \includegraphics[width=\linewidth]{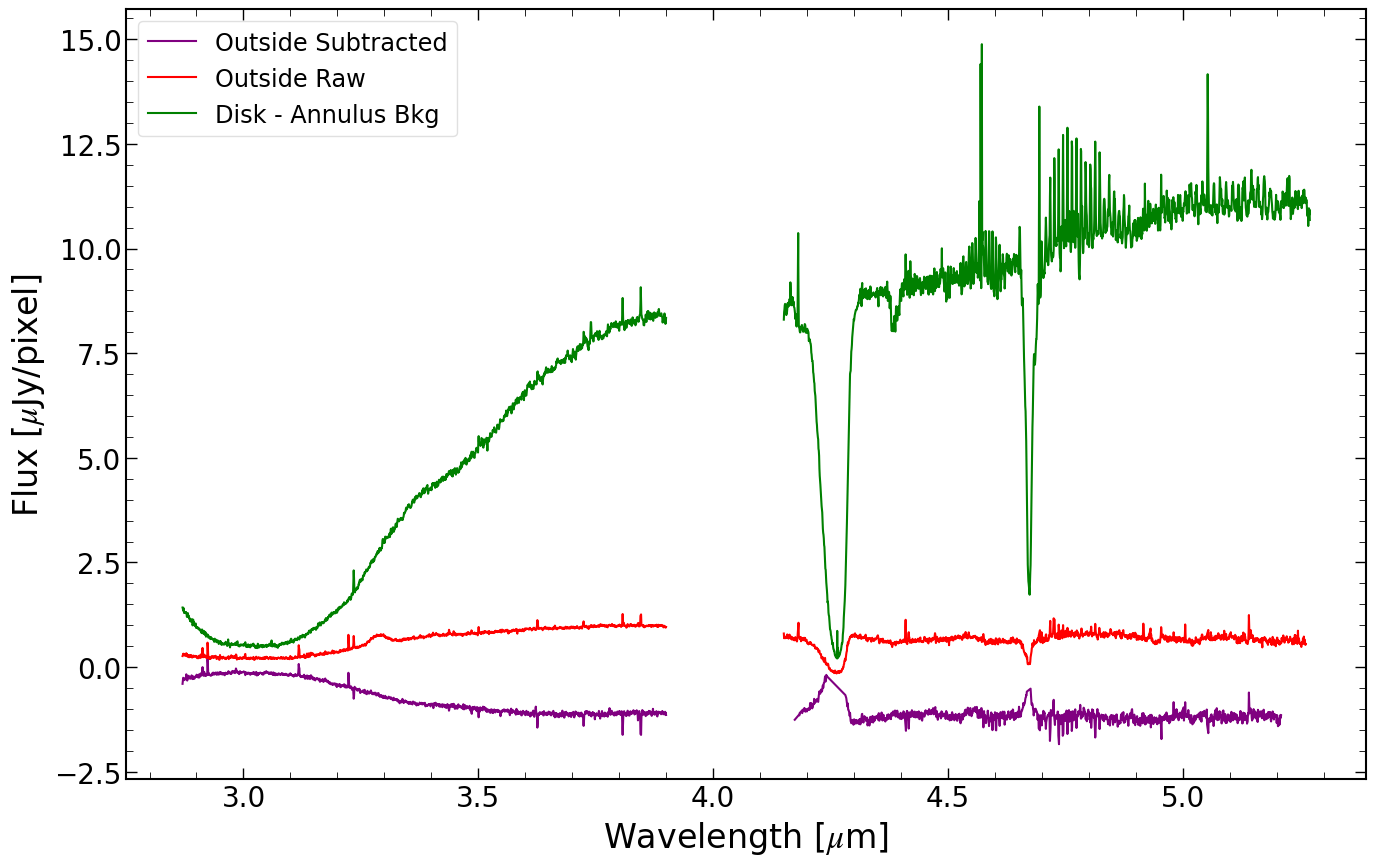}
    \caption{OphE~MM3 subtraction results}
\end{subfigure}
\caption{Annuli constructed and the results of the background subtraction to the data as well as the regions outside the contours to determine validity of the background. The contours are also applied to the integrated 11.3/11.0 and 7.7/$\sim$10 MIRI image to determine the validity of these contours in the MIR.}
\label{fig:sancheck}
\end{figure*}

We generated 2, 5, and 10$\sigma_{rms}$ contours of the 3.3~$\mu$m PAH emission with these levels justified in Figure~\ref{fig:radial_check}. We subtract the 2–10$\sigma_{rms}$ region from the data cube and examined the residuals to evaluate the quality of the subtraction in Figure~\ref{fig:sancheck}. 

\section{Ice Absorption}\label{ice_abs}

To model ice absorption features, we define the optical depth $\tau_\text{ice}$ as a sum over distinct absorption bands:

\begin{equation}
\tau_\text{ice} = N_{\text{H}_2\text{O}} \sum_{j=1}^{n_\text{bands}} \frac{A_j E_j \text{(X}_j/X_{H_20}) \left(\frac{2 \gamma_j}{\pi}\right)}{(\lambda - \lambda_j^2 / \lambda)^2 + \gamma_j^2},
\end{equation}

where:
\begin{itemize}
    \item $N_{\text{H}_2\text{O}} = \frac{\tau_{3.1}\pi \cdot \gamma_{3.1}}{2 A_{3.1}}$ is the column density of H$_2$O given that $\tau_{3.1}$ is the optical depth at 3.1 $\mu$m. 
    \item $\lambda_j$ are the central wavelengths of ice absorption bands \citep{Gerakines1995},
    \item $\gamma_j$ are the full-widths at half-maximum in wavenumbers (cm$^{-1}$) \citep{Li1997},
    \item $A_j$ are the band strengths (cm/molecule) \citep{Gerakines1995},
    \item $\text{X}_j$ represents the abundance ratio relative to H$_2$O \citep{Gibb2004, Boogert2015},
    \item $E_j$ is the enhancement factor, which can be adjusted based on the ice species' abundance.
\end{itemize}

The wavelength input, $\lambda$, is first converted to centimeters to ensure consistency with the absorption parameters, which are defined in cgs units. The resulting optical-depth spectrum provides a quantitative measure of ice absorption across multiple bands, enabling detailed analysis of ice compositions in astrophysical environments. A key component is the factor $E_j$, which accounts for variations in ice abundances. While the baseline abundances are set to typical Milky Way values, astrophysical sources can exhibit substantial variation. The $E_j$ term therefore captures local deviations in the abundances of species such as CO and CO$_2$. In practice, $E_j$ can be treated as a free parameter in PAHFIT, allowing one to simultaneously determine both the optical depth and the relative abundances of these species with respect to H$_2$O ice.

\section{Placement of Apertures}\label{apps}

\begin{figure*}[htb!]
\centering

\includegraphics[width=0.24\textwidth]{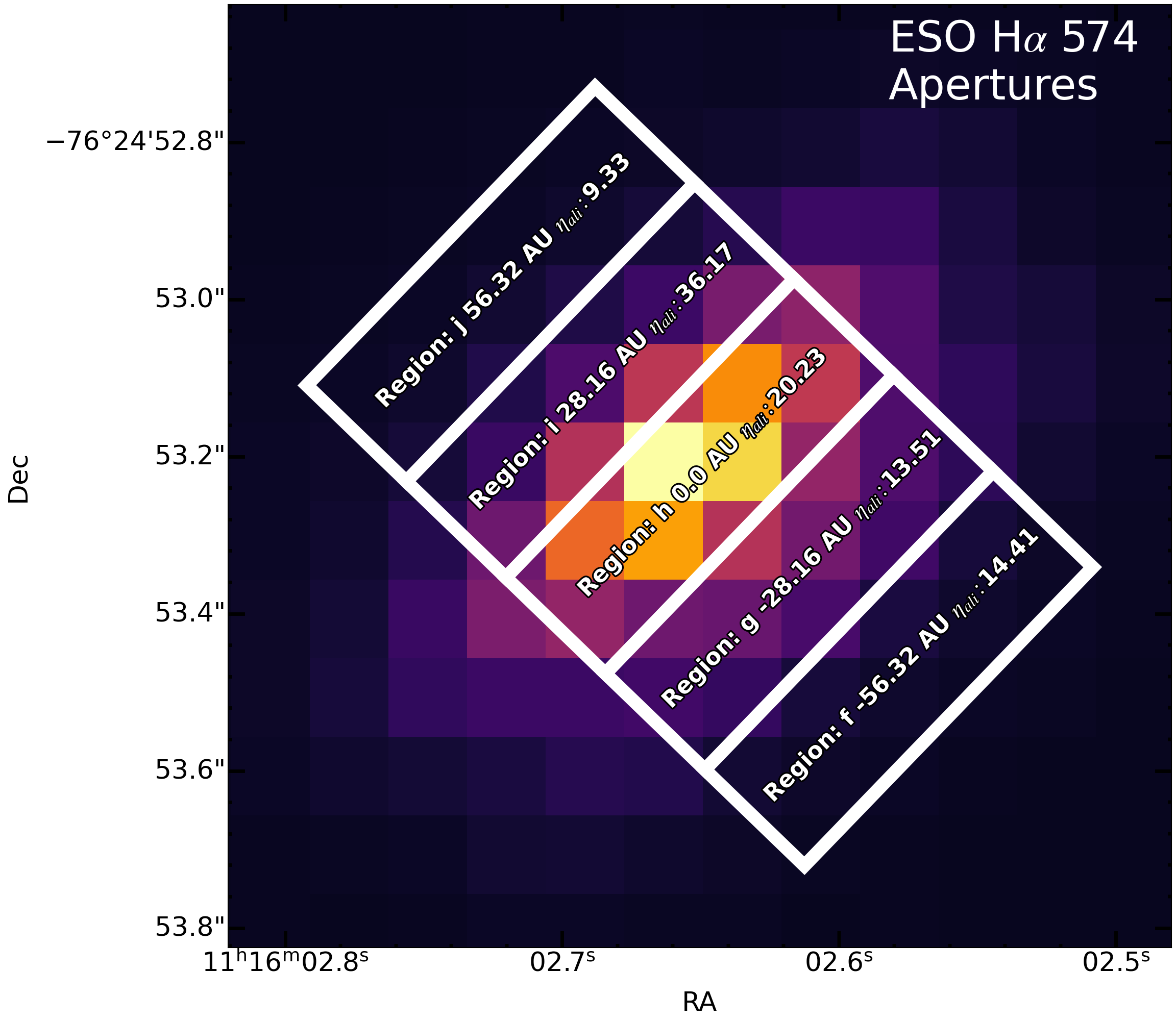}
\includegraphics[width=0.24\textwidth]{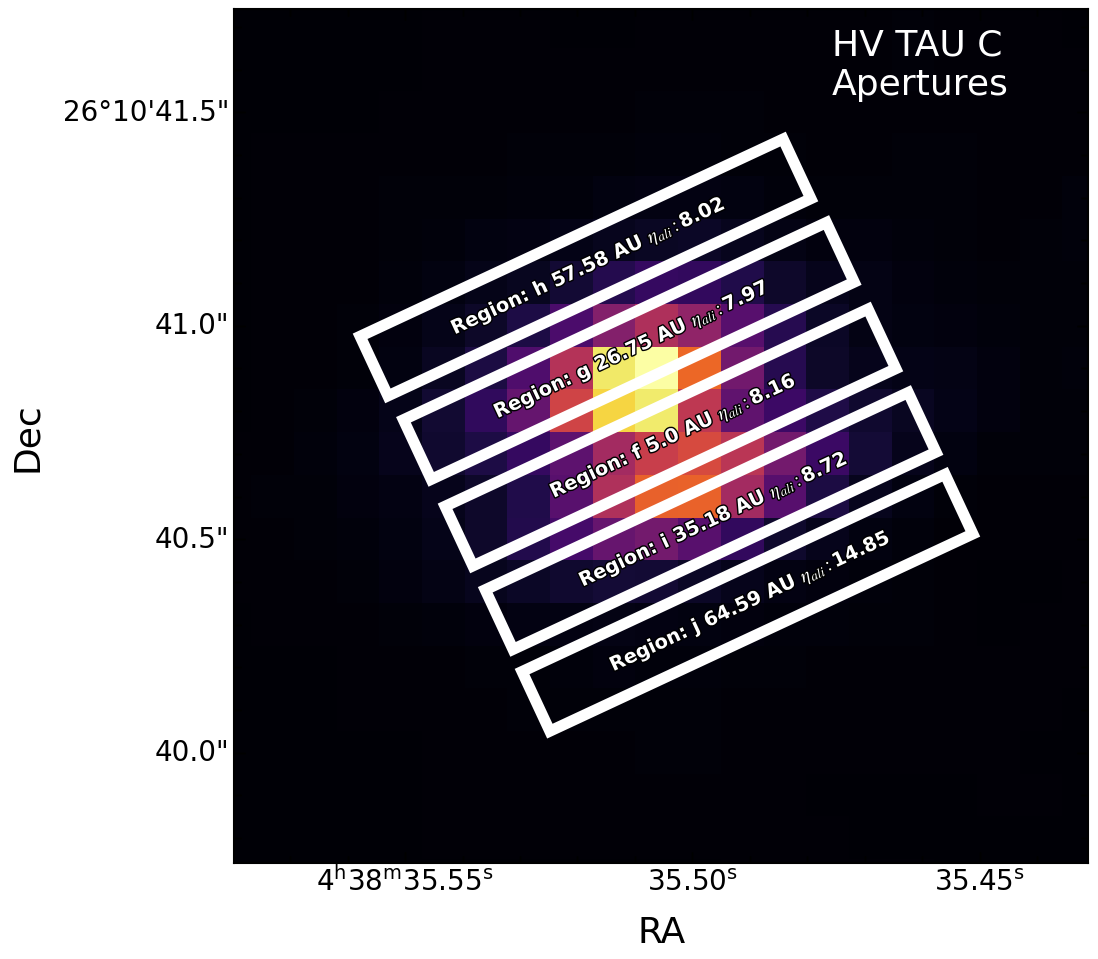}
\includegraphics[width=0.24\textwidth]{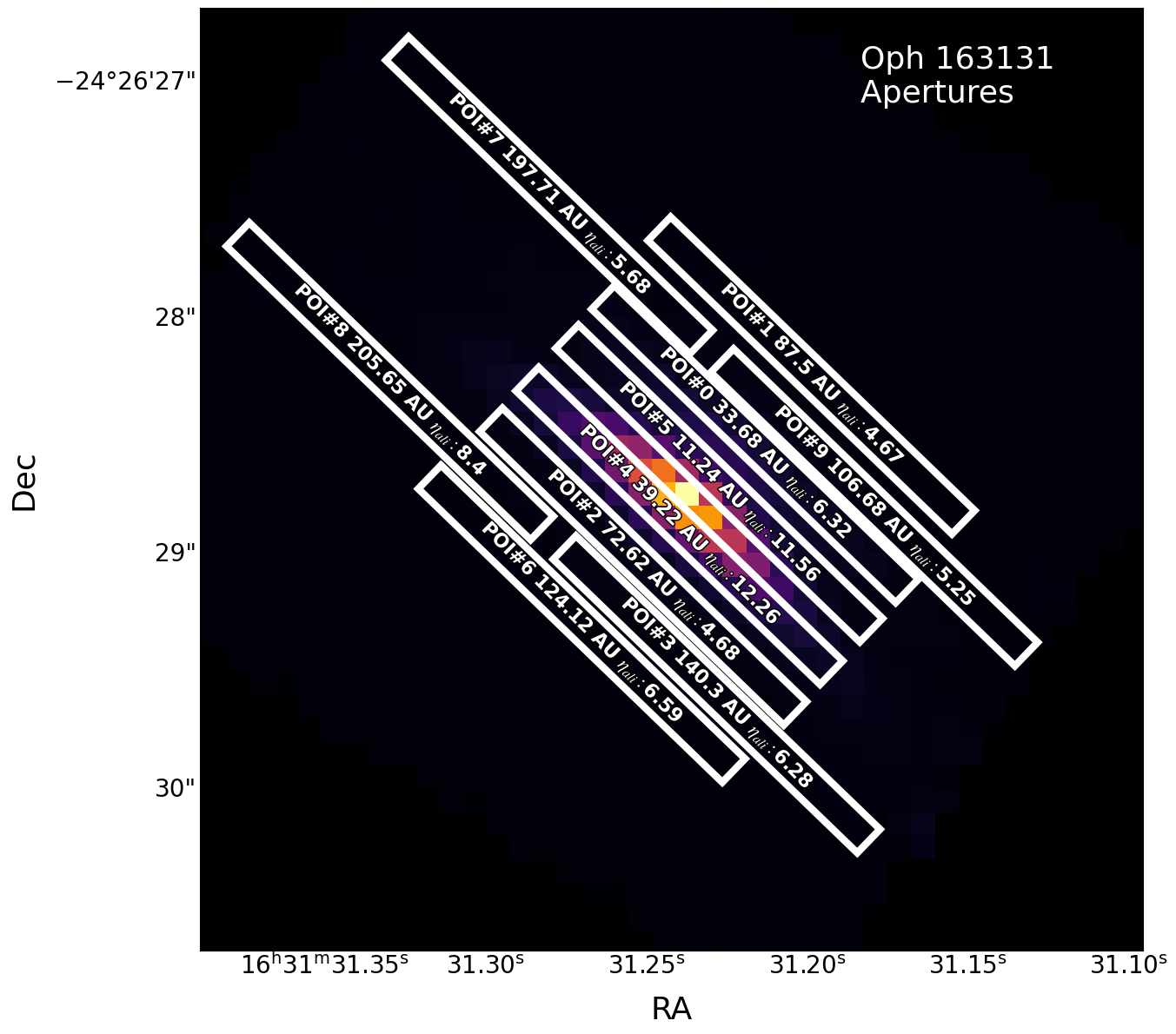}
\includegraphics[width=0.24\textwidth]{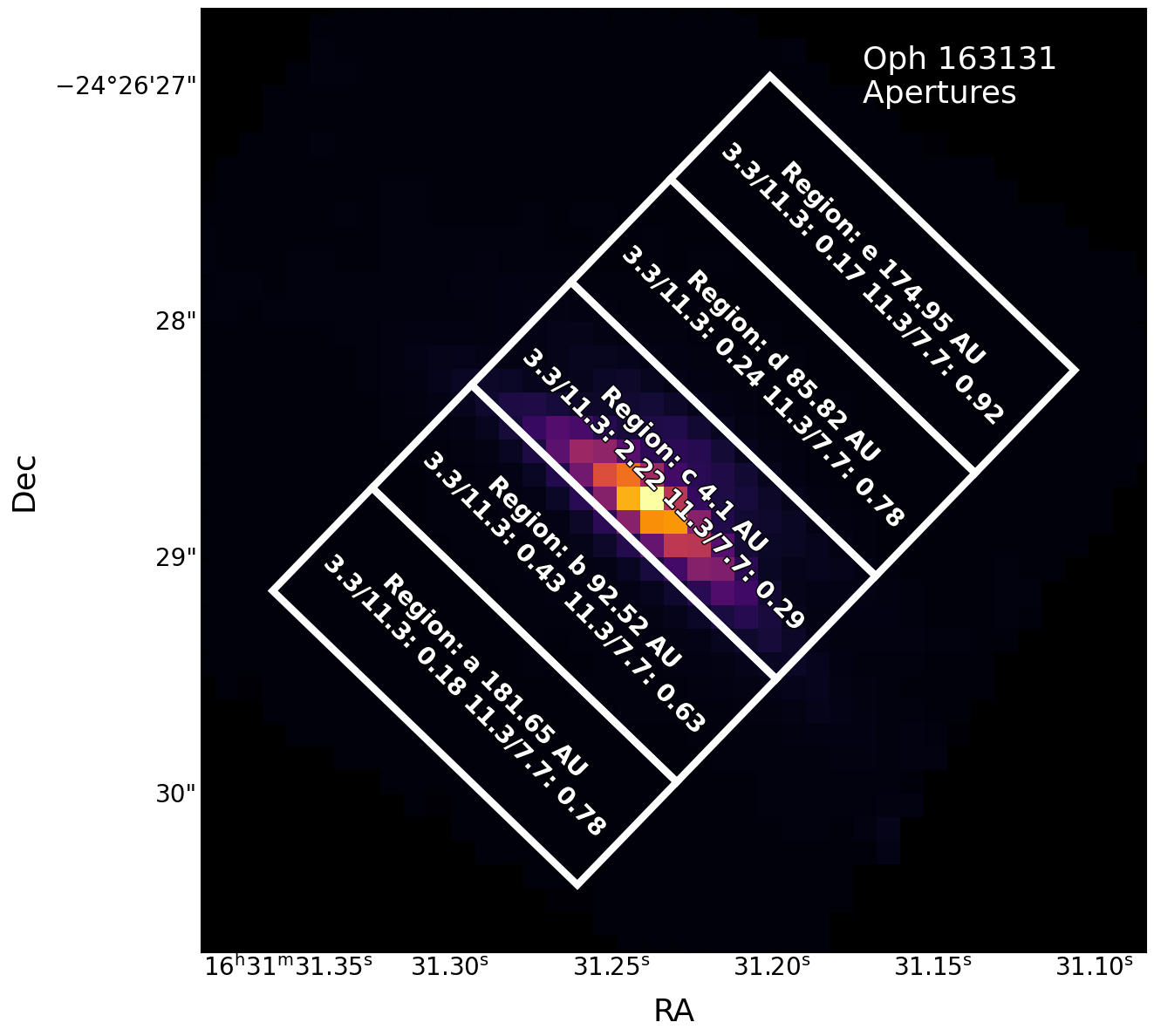}
\includegraphics[width=0.24\textwidth]{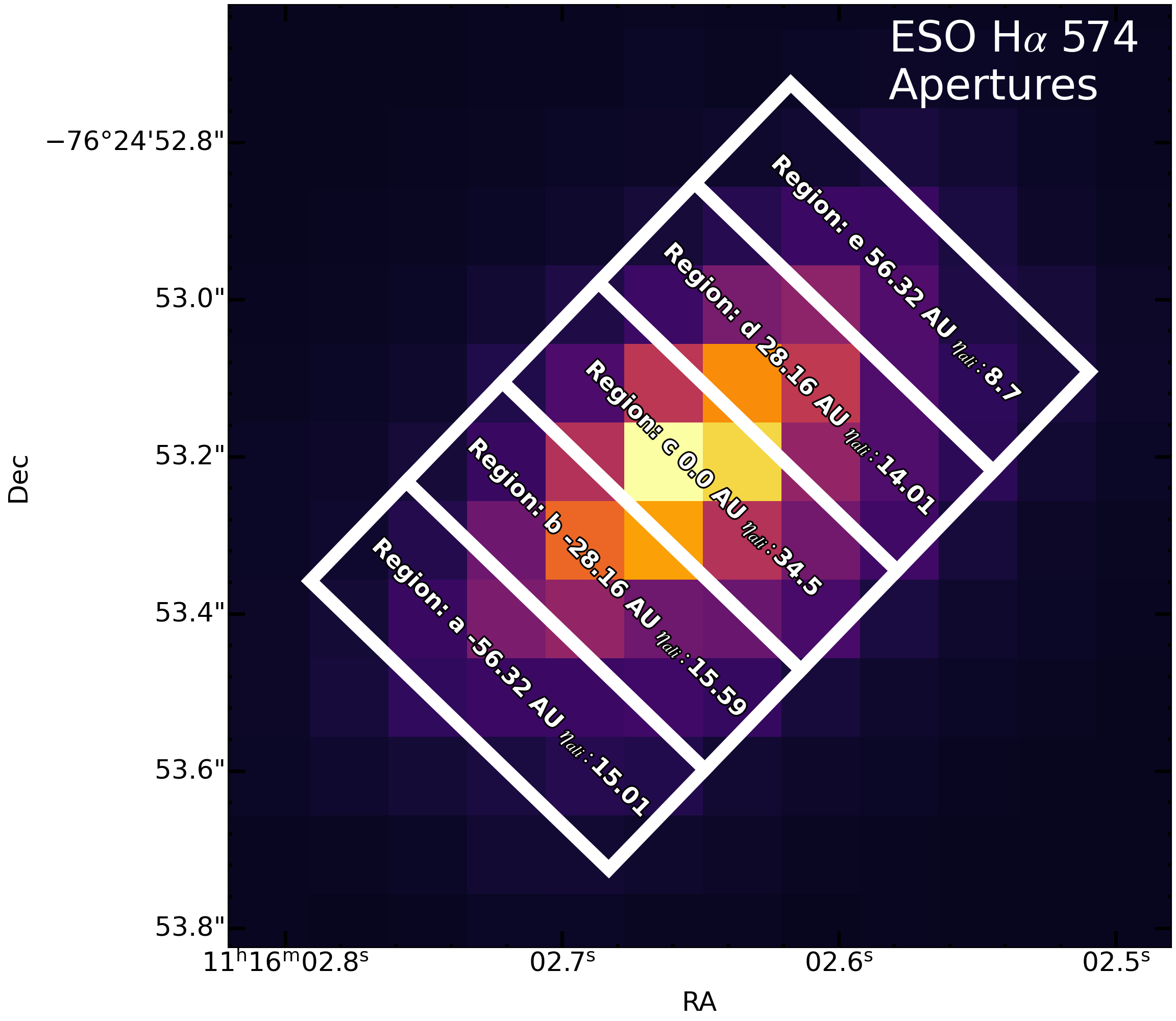}
\includegraphics[width=0.24\textwidth]{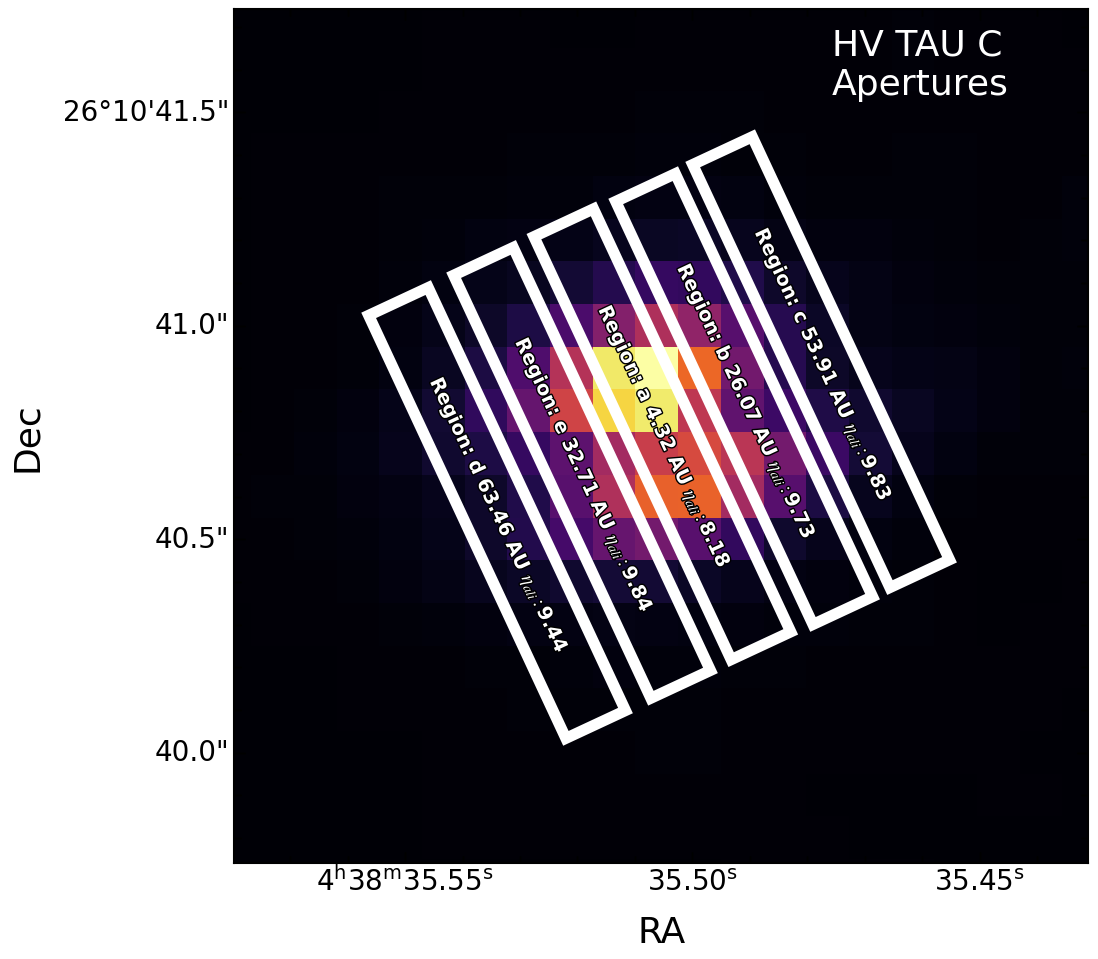}
\includegraphics[width=0.24\textwidth]{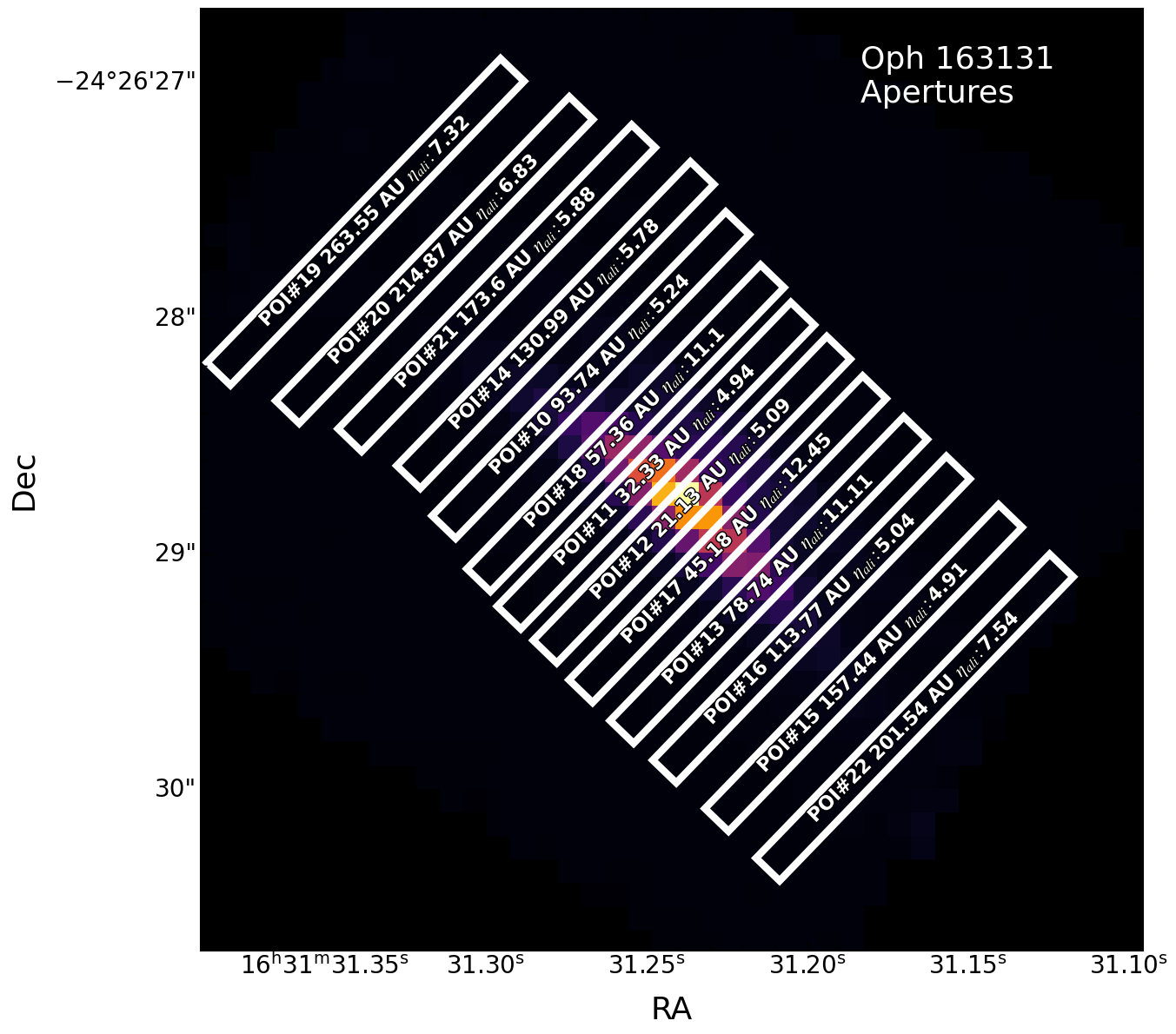}
\includegraphics[width=0.24\textwidth]{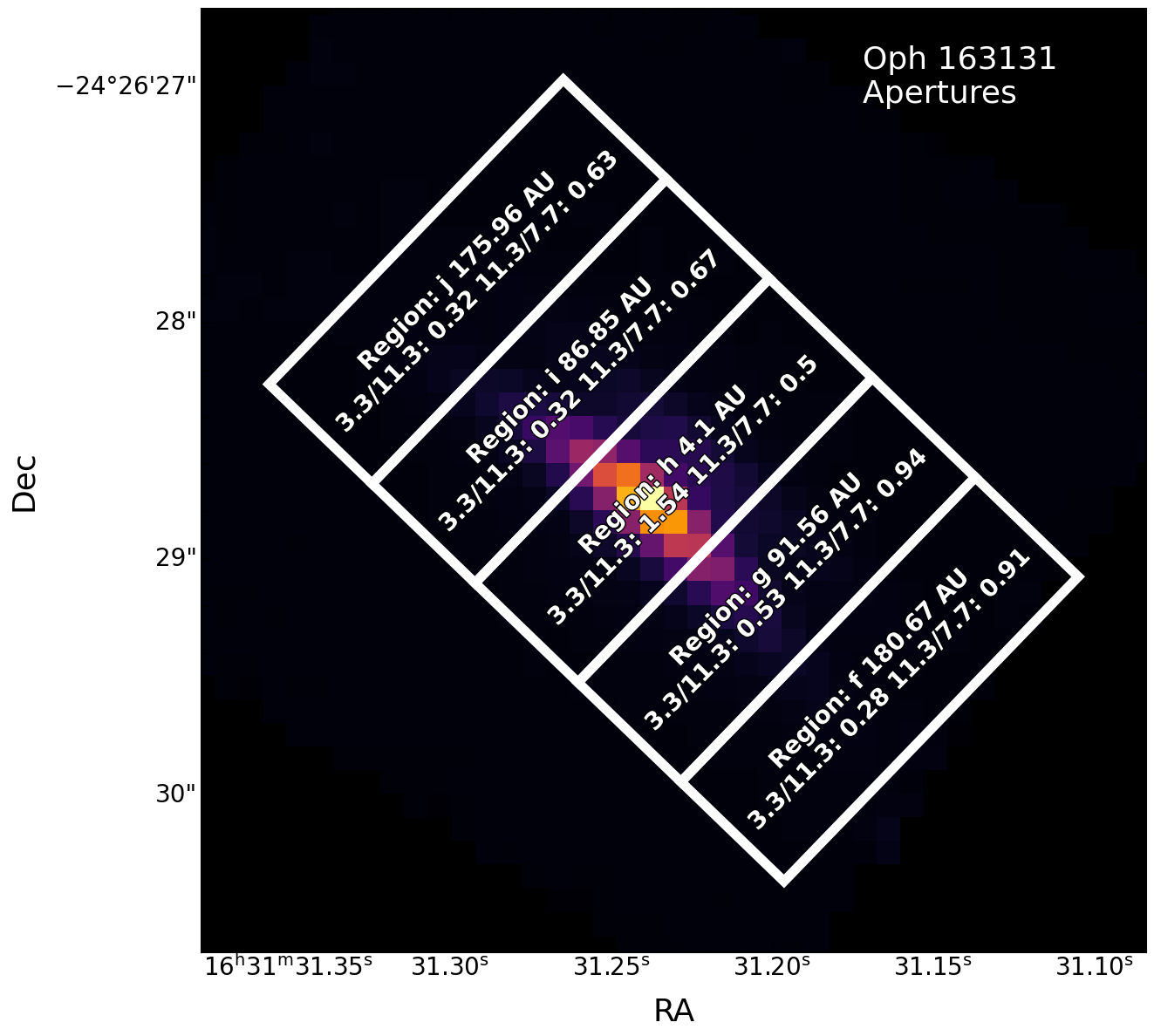}

\caption{Apertures used for all sources determined to have localized PAH to the disks. Smaller apertures (Left three) are for the smaller FWHM of the PSF at 3.4 $\mu$m while the larger right-most apertures used for Oph~163131 are set to minimize the PSF smearing at 11.3~$\mu$m.}
\label{fig:Apertures_total}
\end{figure*}

We have used several rectangular apertures of 0.176\arcsec~in width tailored to the 3.4 $\mu$m JWST PSF of 0.110\arcsec~\citep{STPSF} placed along four orthogonal directions: parallel and perpendicular to the disk mid-plane with position angles matched to the apparent disk orientation. Figure \ref{fig:Apertures_total} showcases the smaller apertures to analyze the $\eta_{\rm ali}$. For the special case of Oph~163131 the larger S/N allows apertures of 0.60\arcsec~, large enough for the 11.3~$\mu$m PSF of 0.379\arcsec~\citep{STPSF}, to be accounted for (see right most figure of Figure~\ref{fig:Apertures_total}).

\section{Gallery of Radial PAHFIT Results}\label{PAHFIT}

\begin{figure*}[htb!]
\centering

\includegraphics[width=0.19\textwidth]{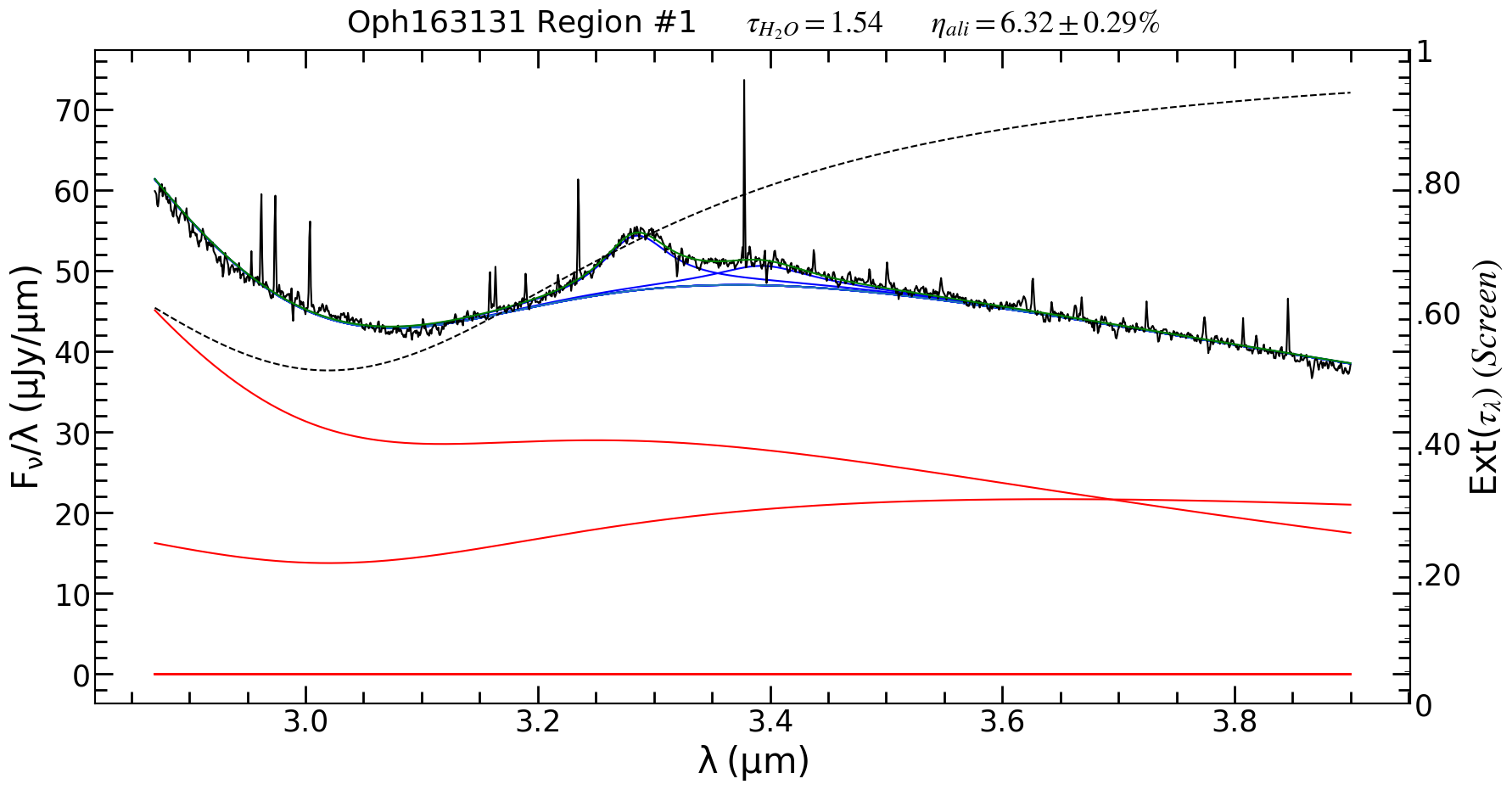}
\includegraphics[width=0.19\textwidth]{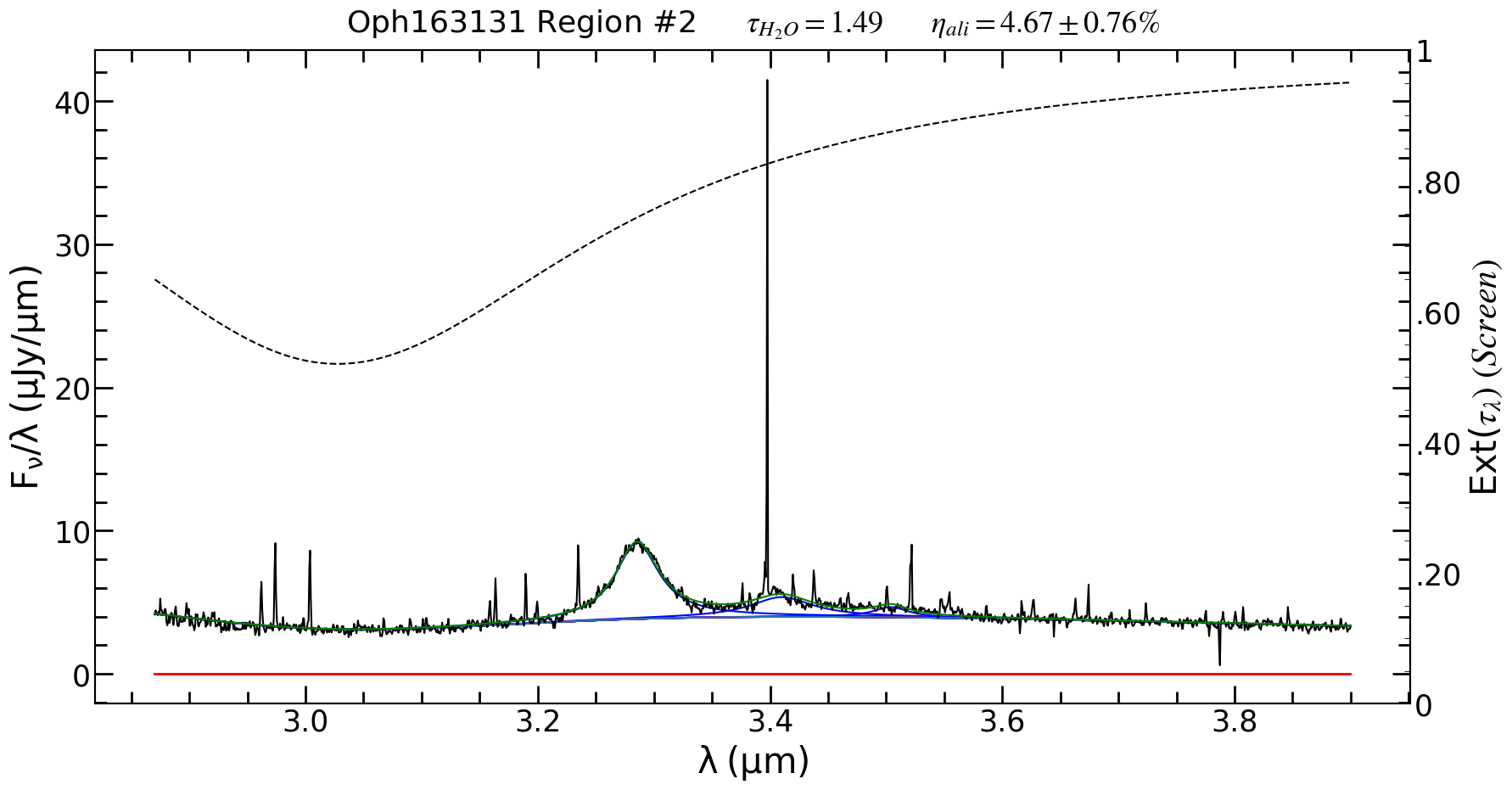}
\includegraphics[width=0.19\textwidth]{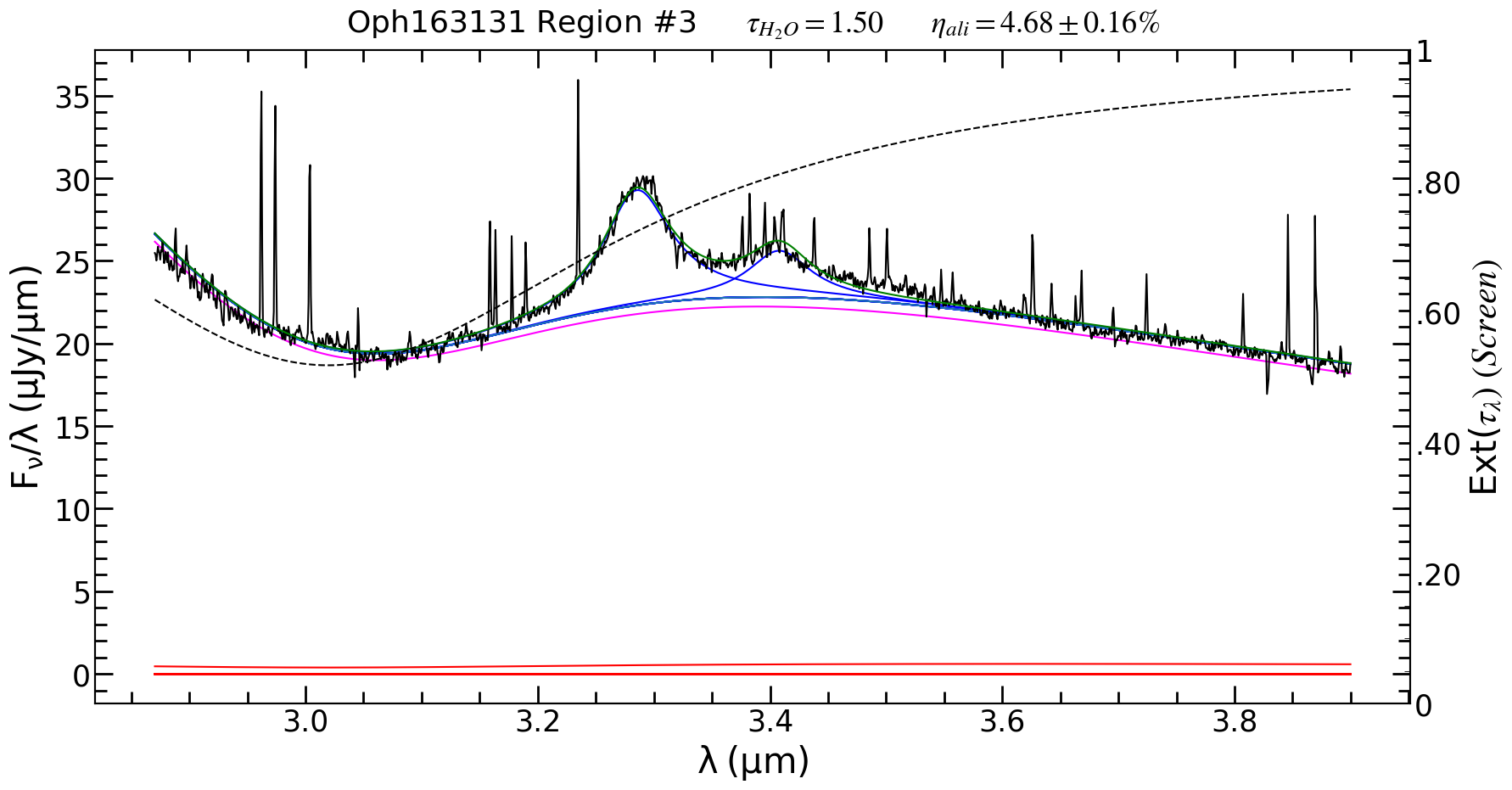}
\includegraphics[width=0.19\textwidth]{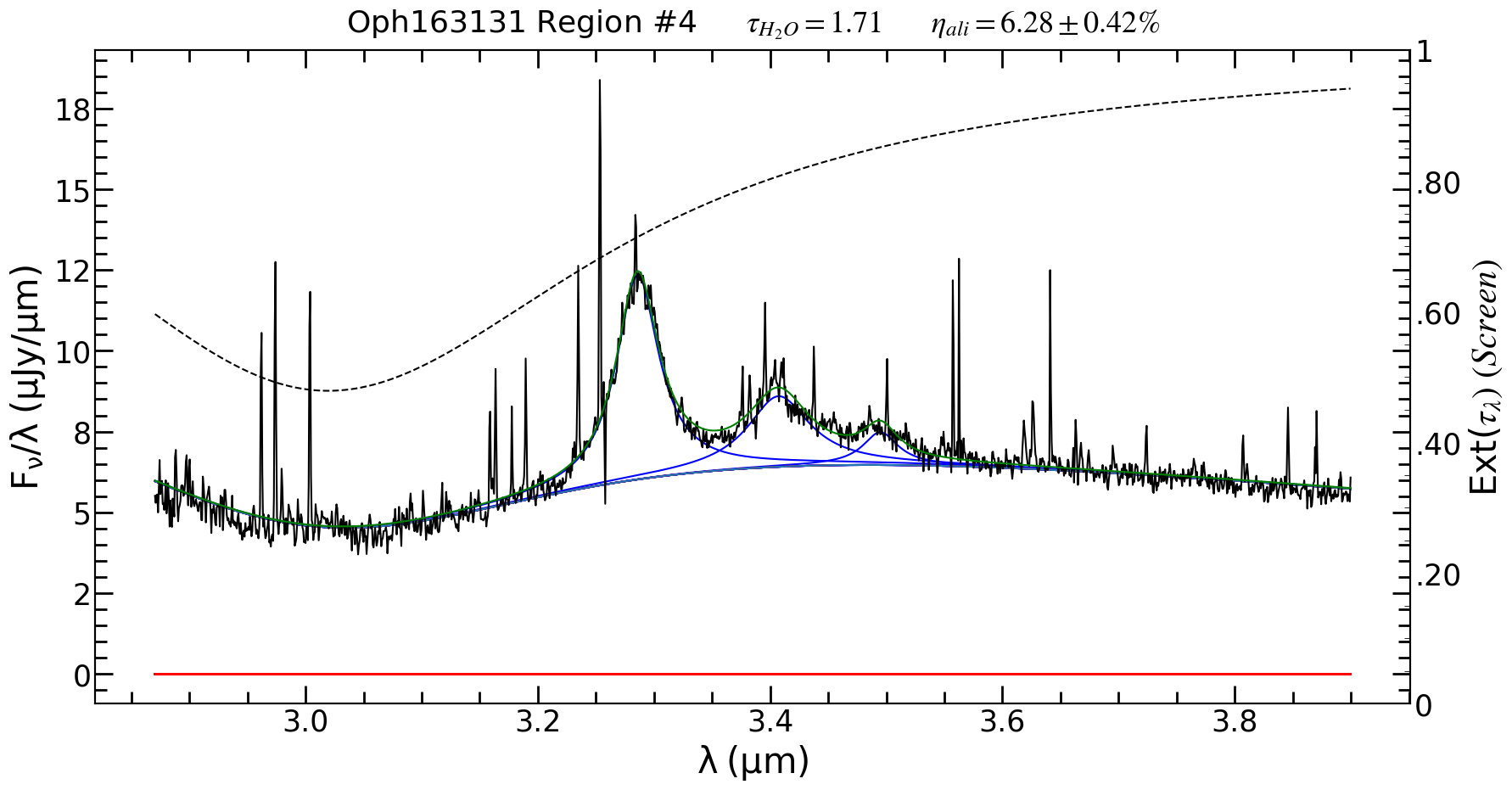}
\includegraphics[width=0.19\textwidth]{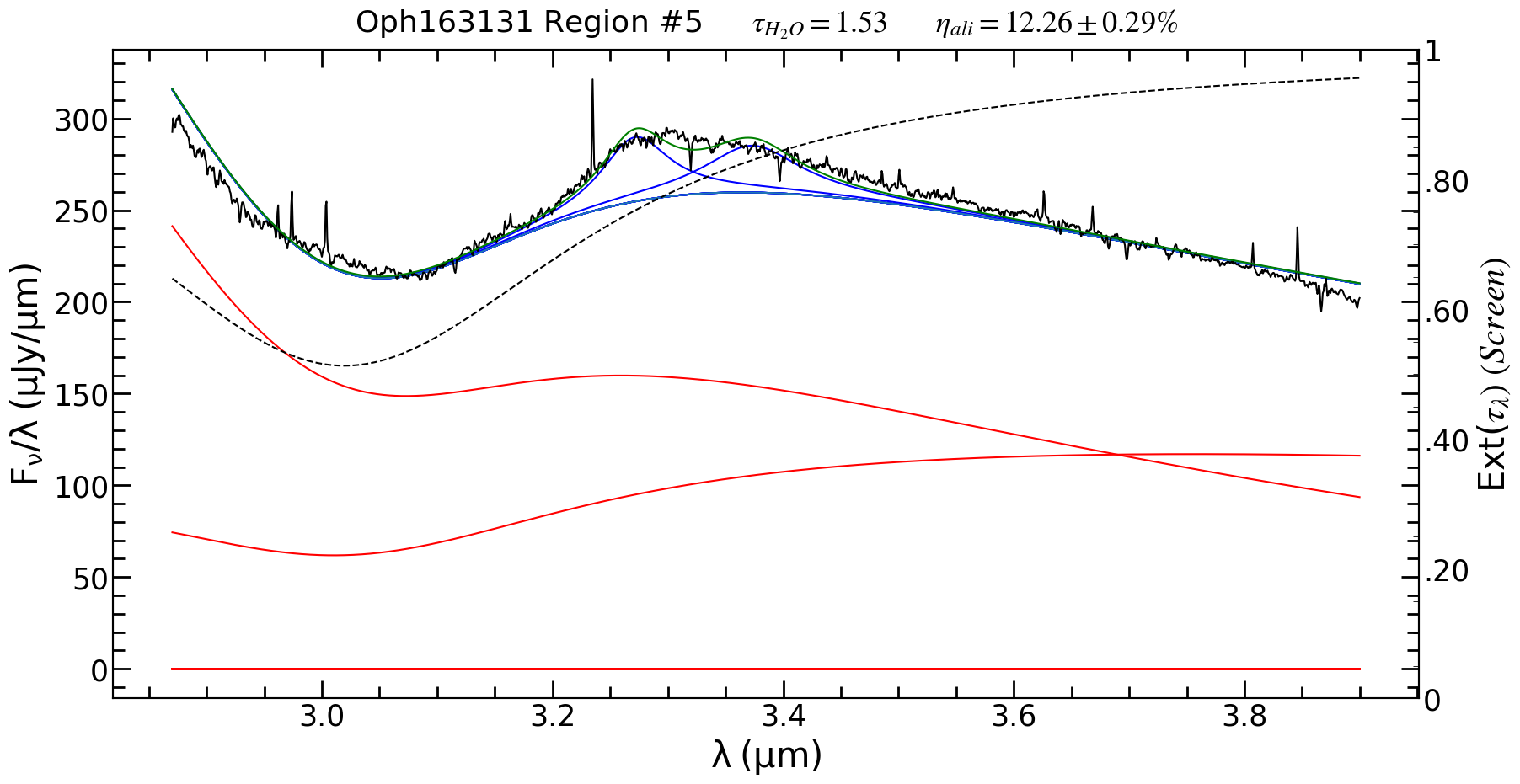}
\includegraphics[width=0.19\textwidth]{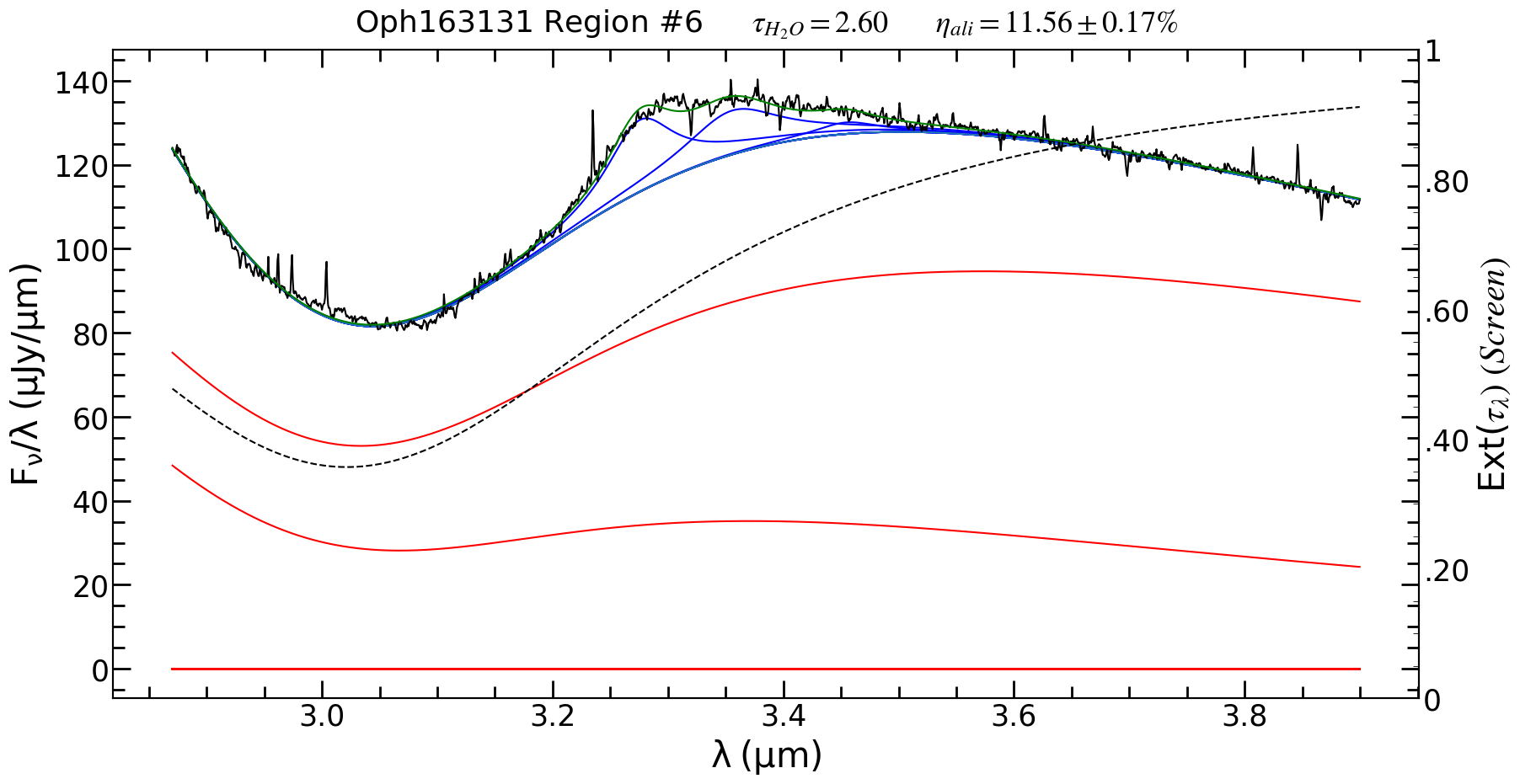}
\includegraphics[width=0.19\textwidth]{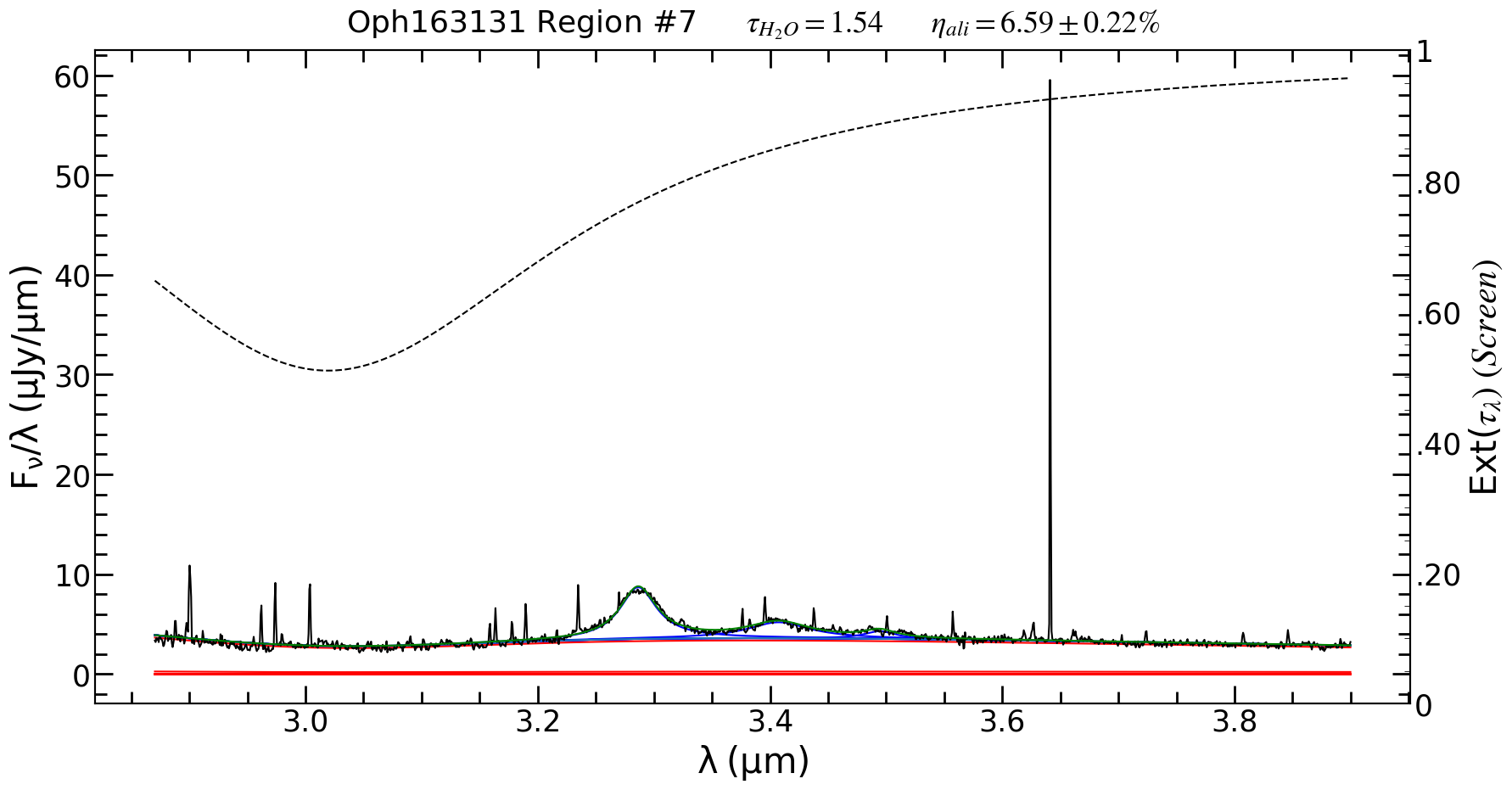}
\includegraphics[width=0.19\textwidth]{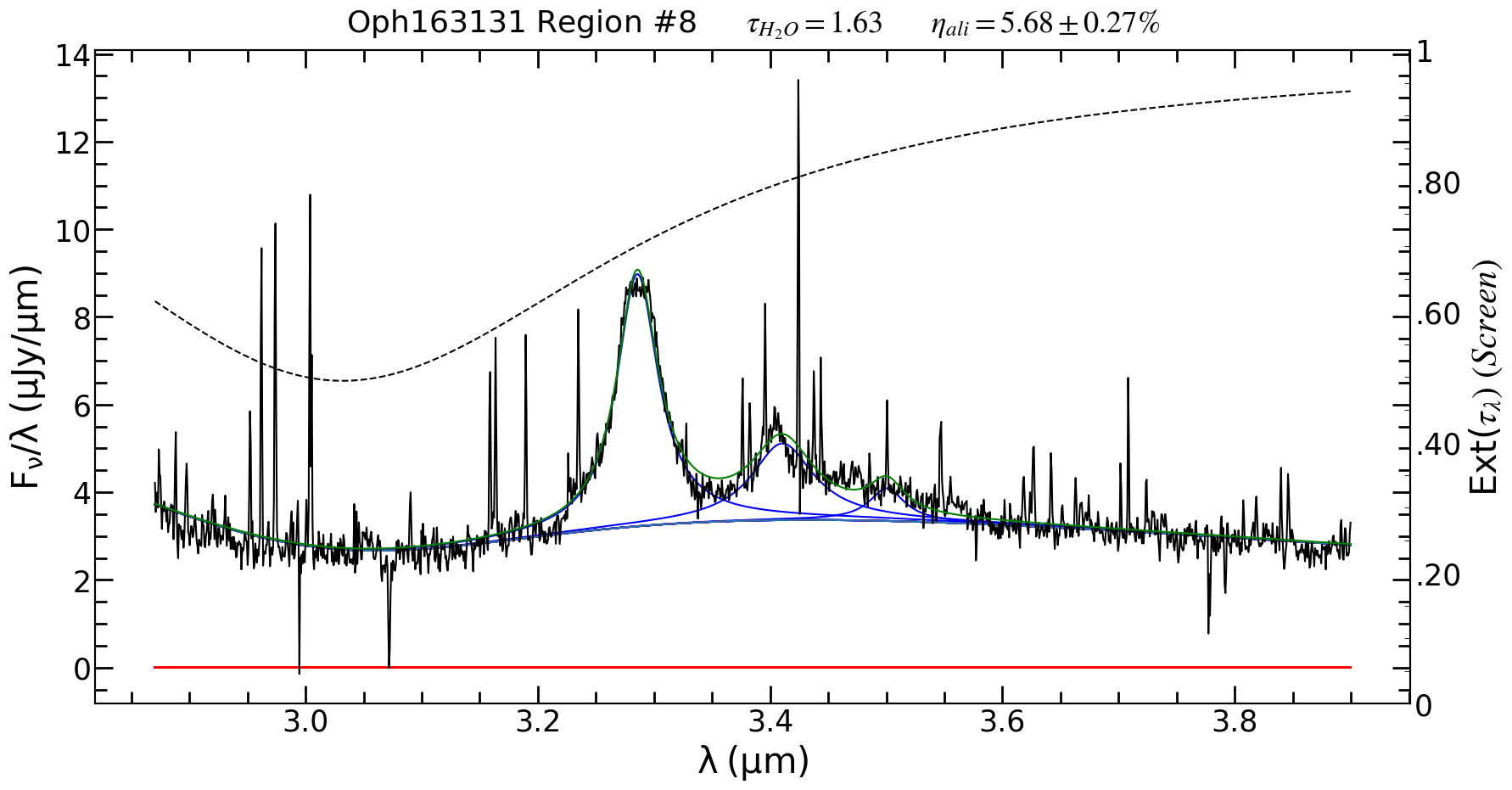}
\includegraphics[width=0.19\textwidth]{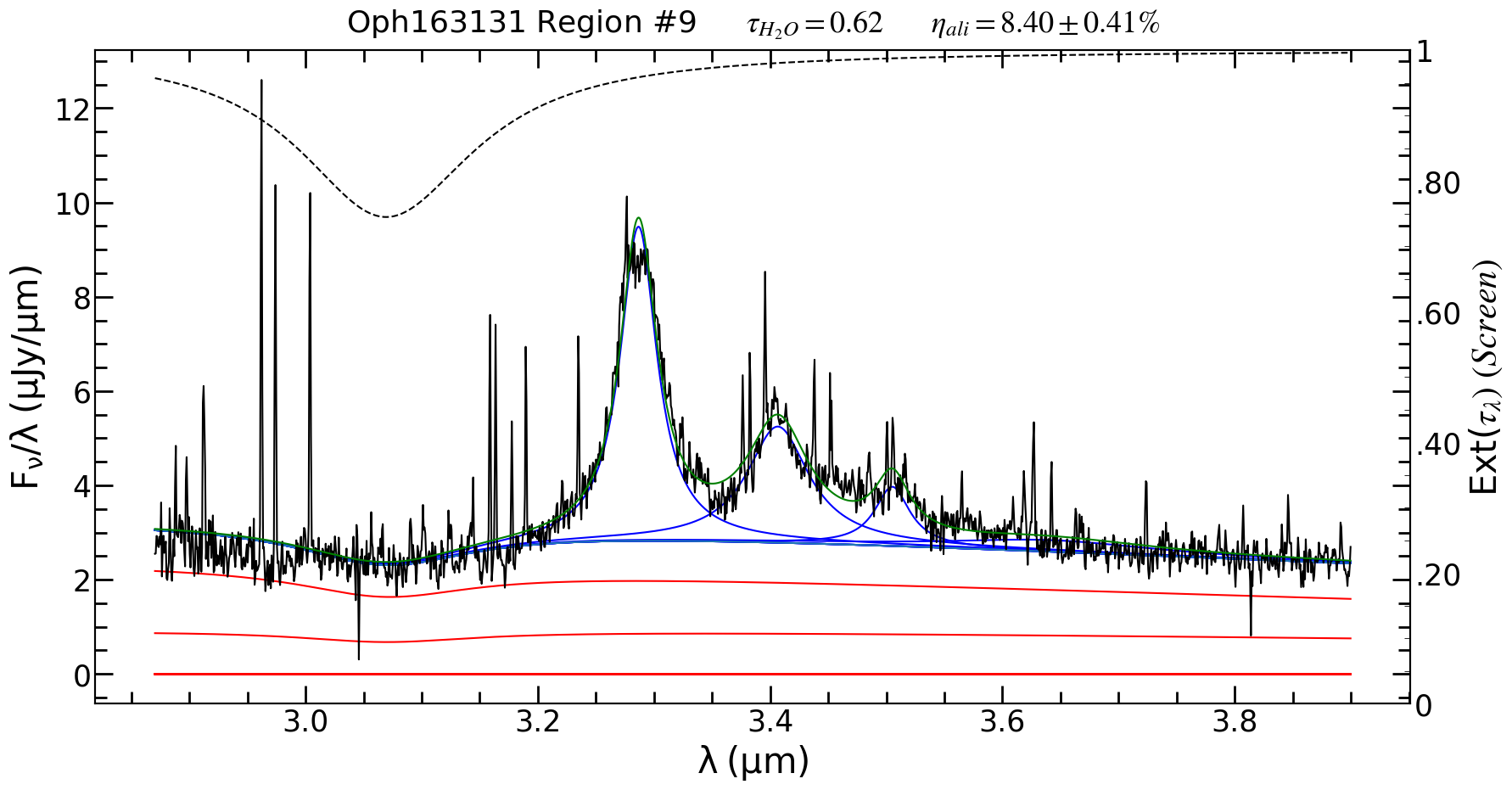}
\includegraphics[width=0.19\textwidth]{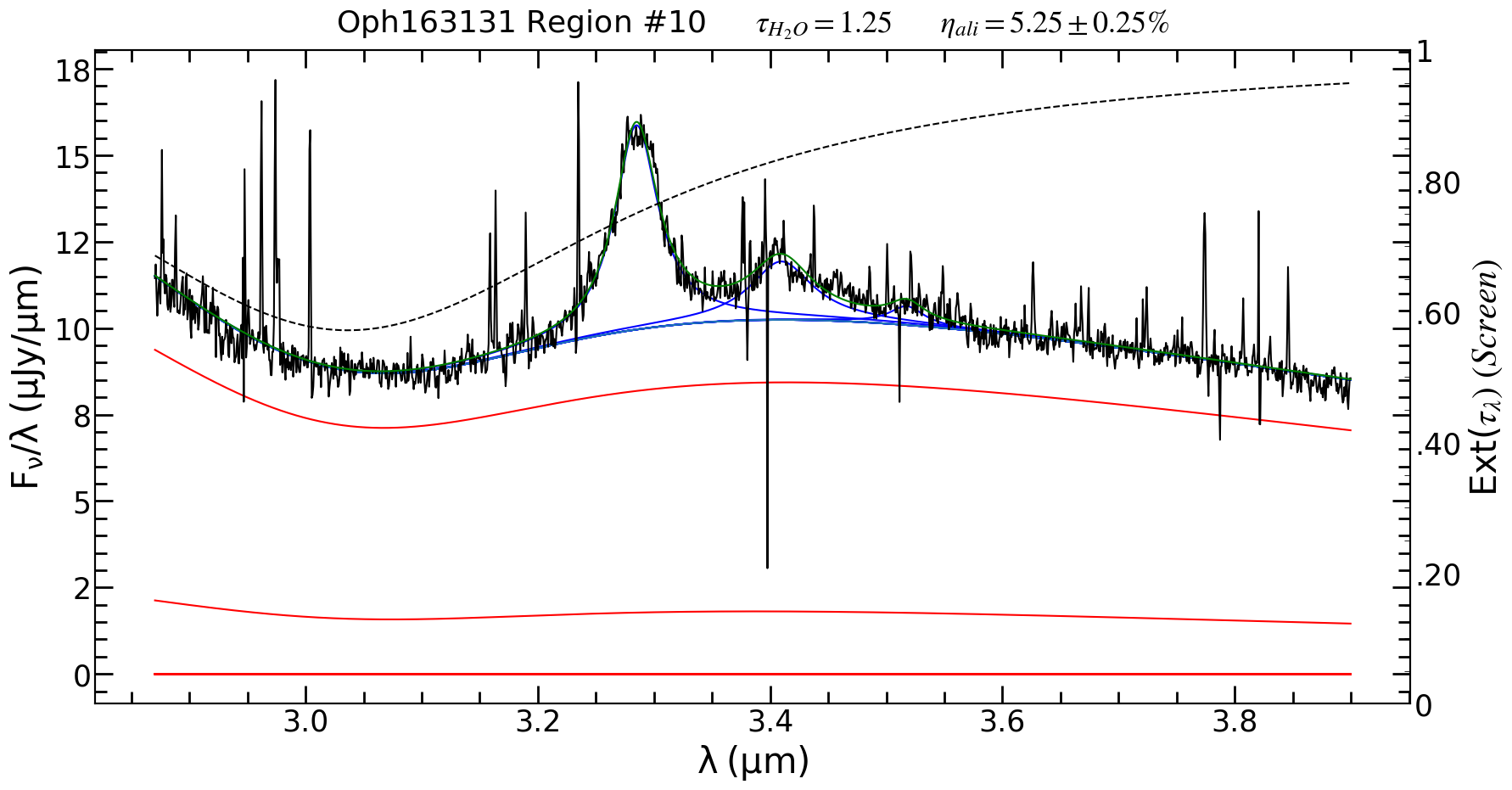}
\includegraphics[width=0.19\textwidth]{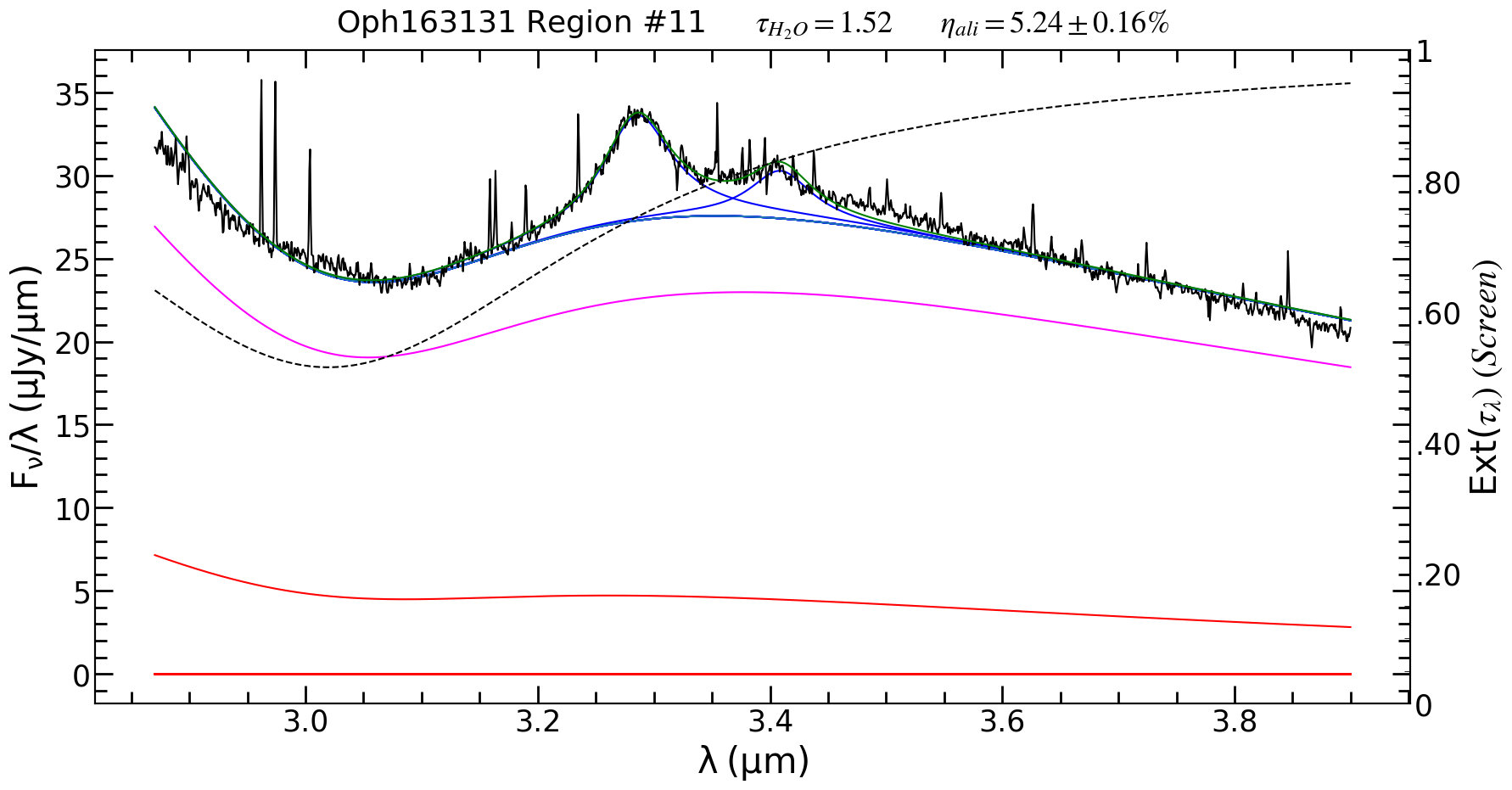}
\includegraphics[width=0.19\textwidth]{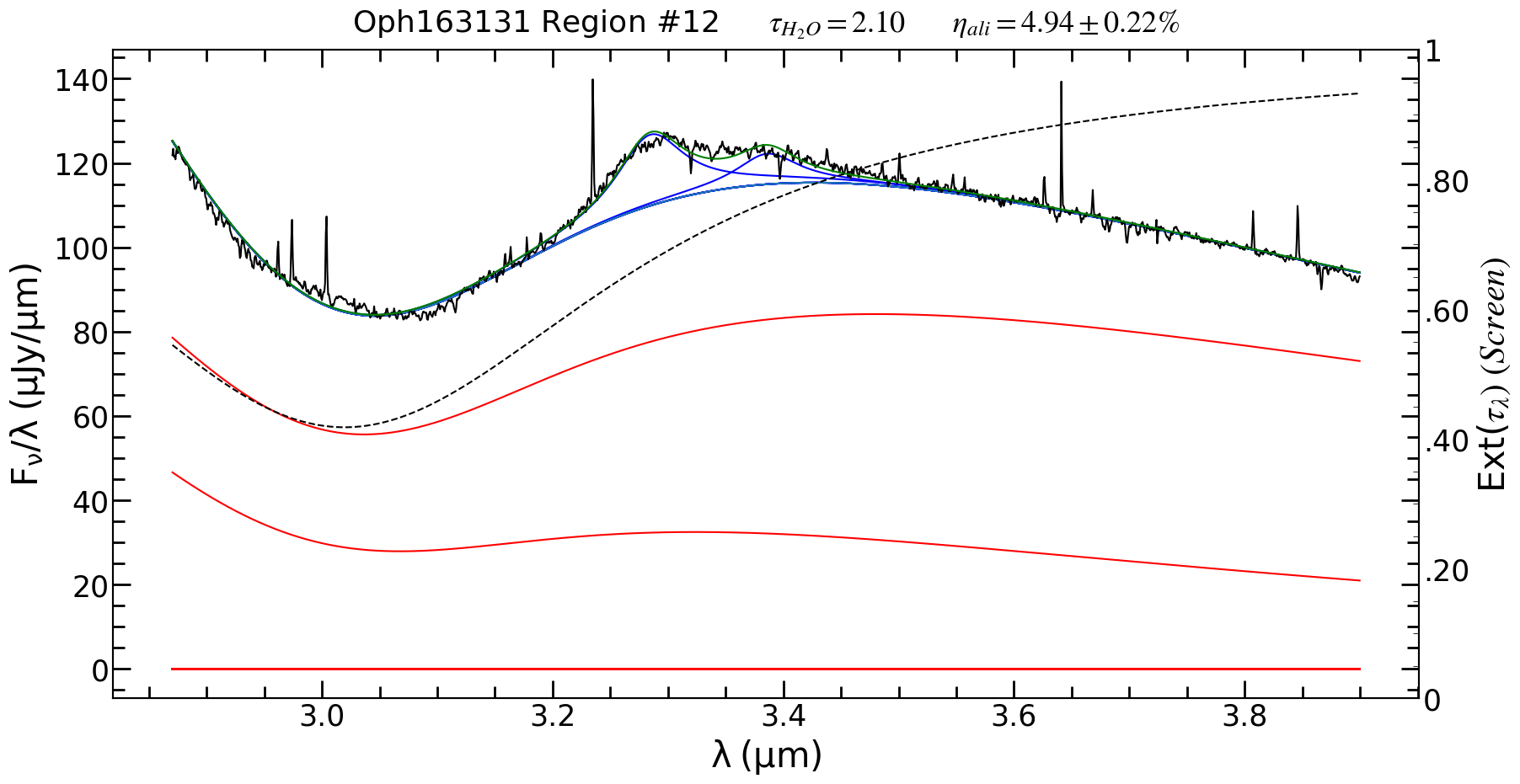}
\includegraphics[width=0.19\textwidth]{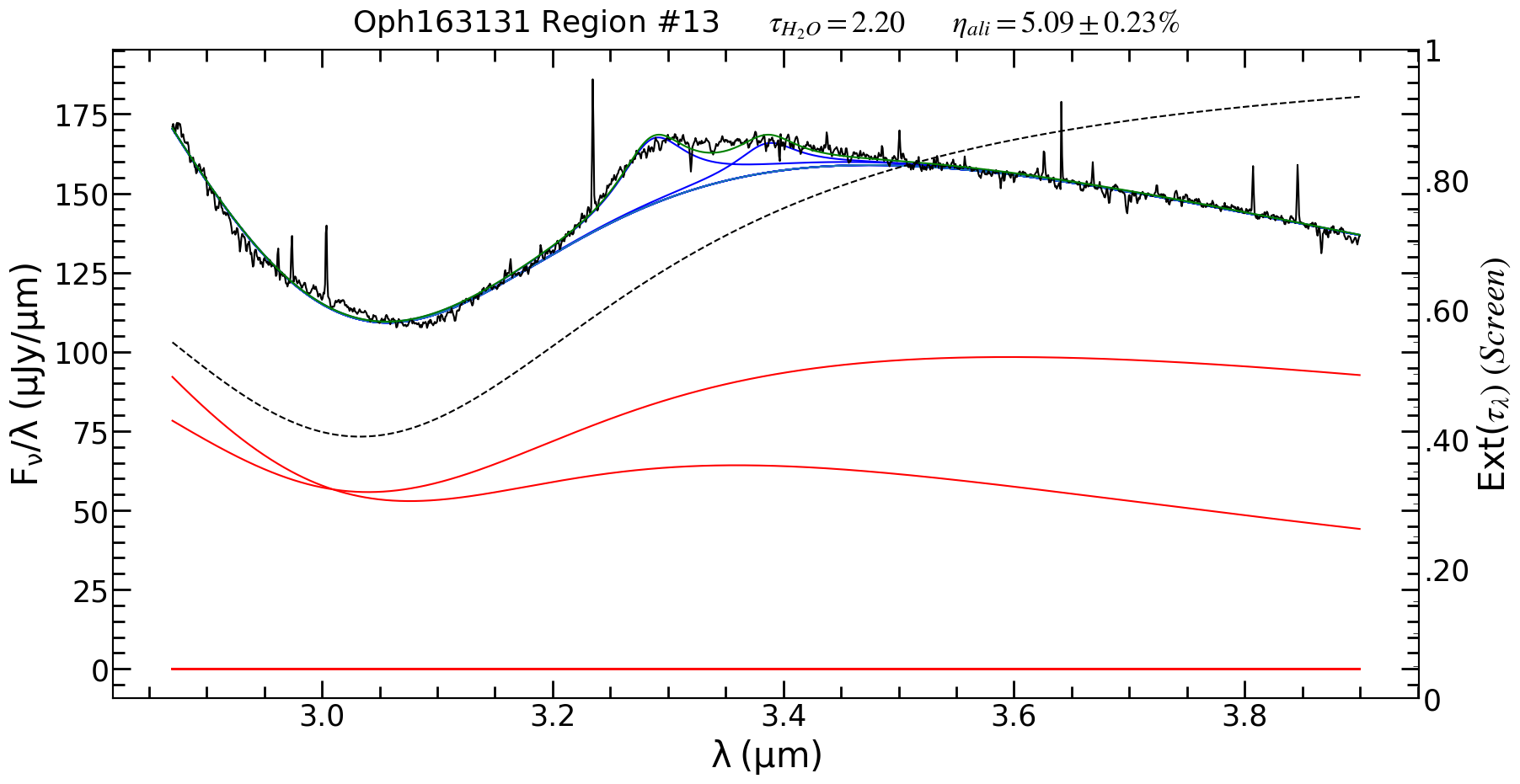}
\includegraphics[width=0.19\textwidth]{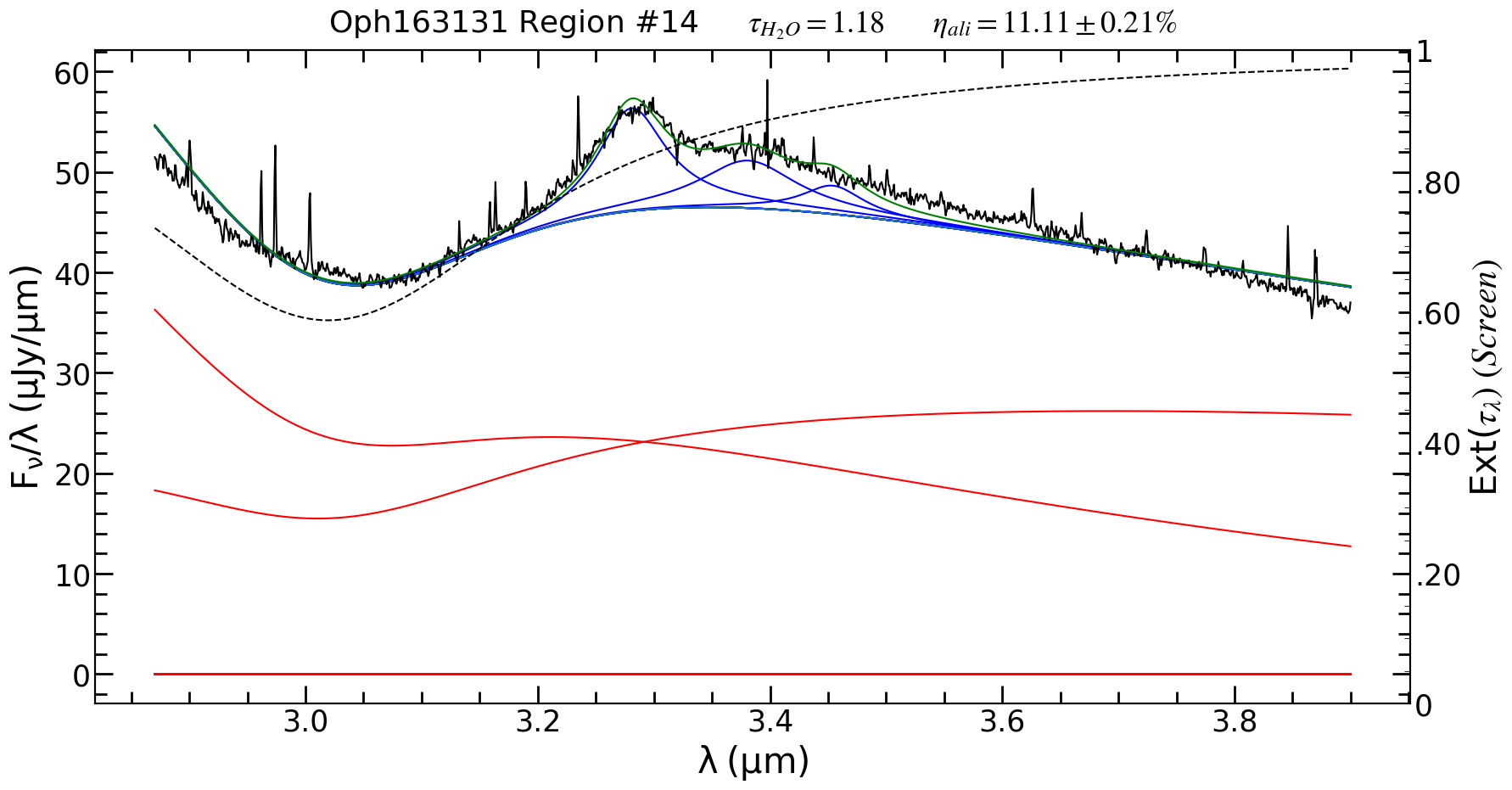}
\includegraphics[width=0.19\textwidth]{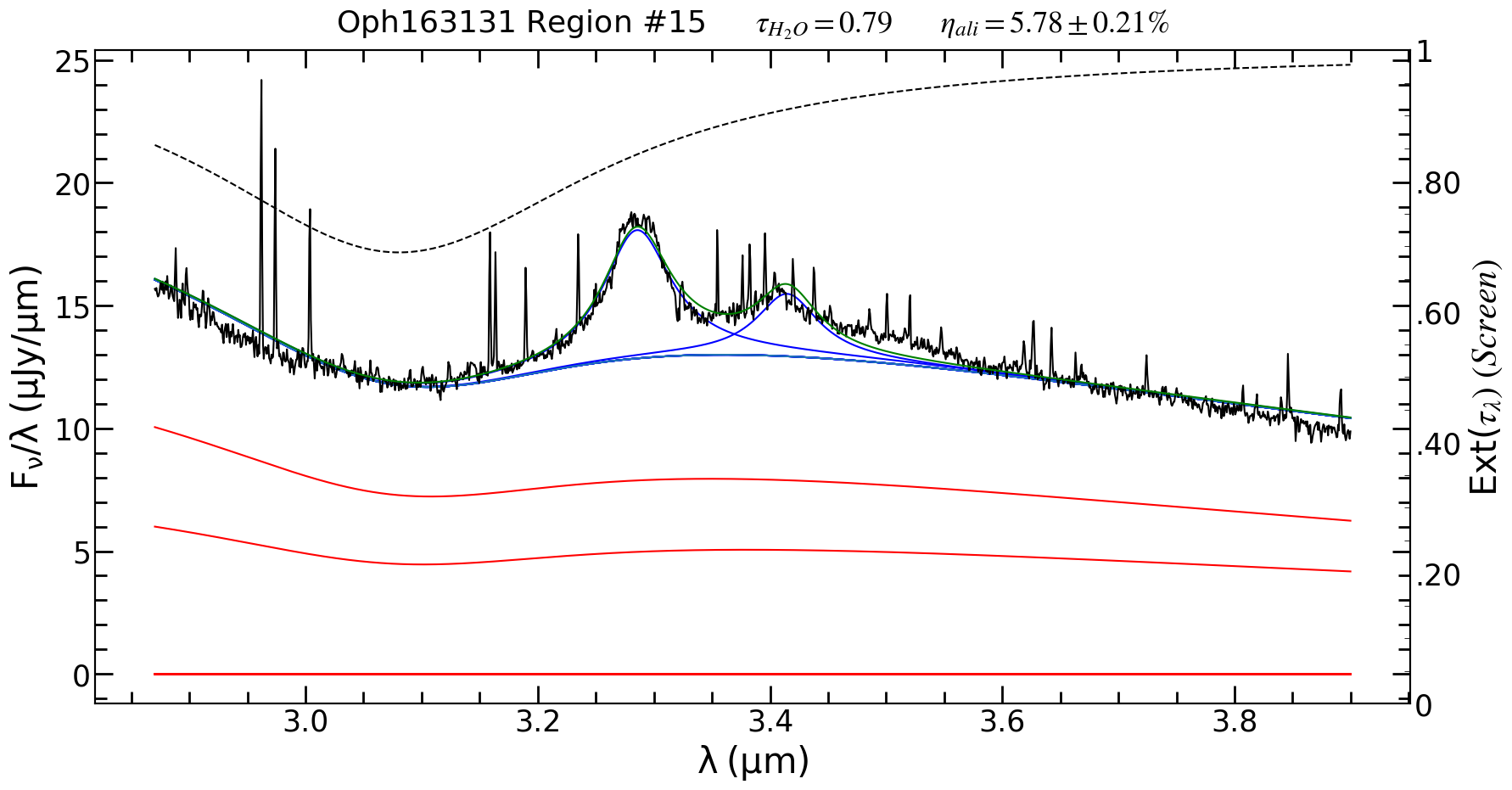}
\includegraphics[width=0.19\textwidth]{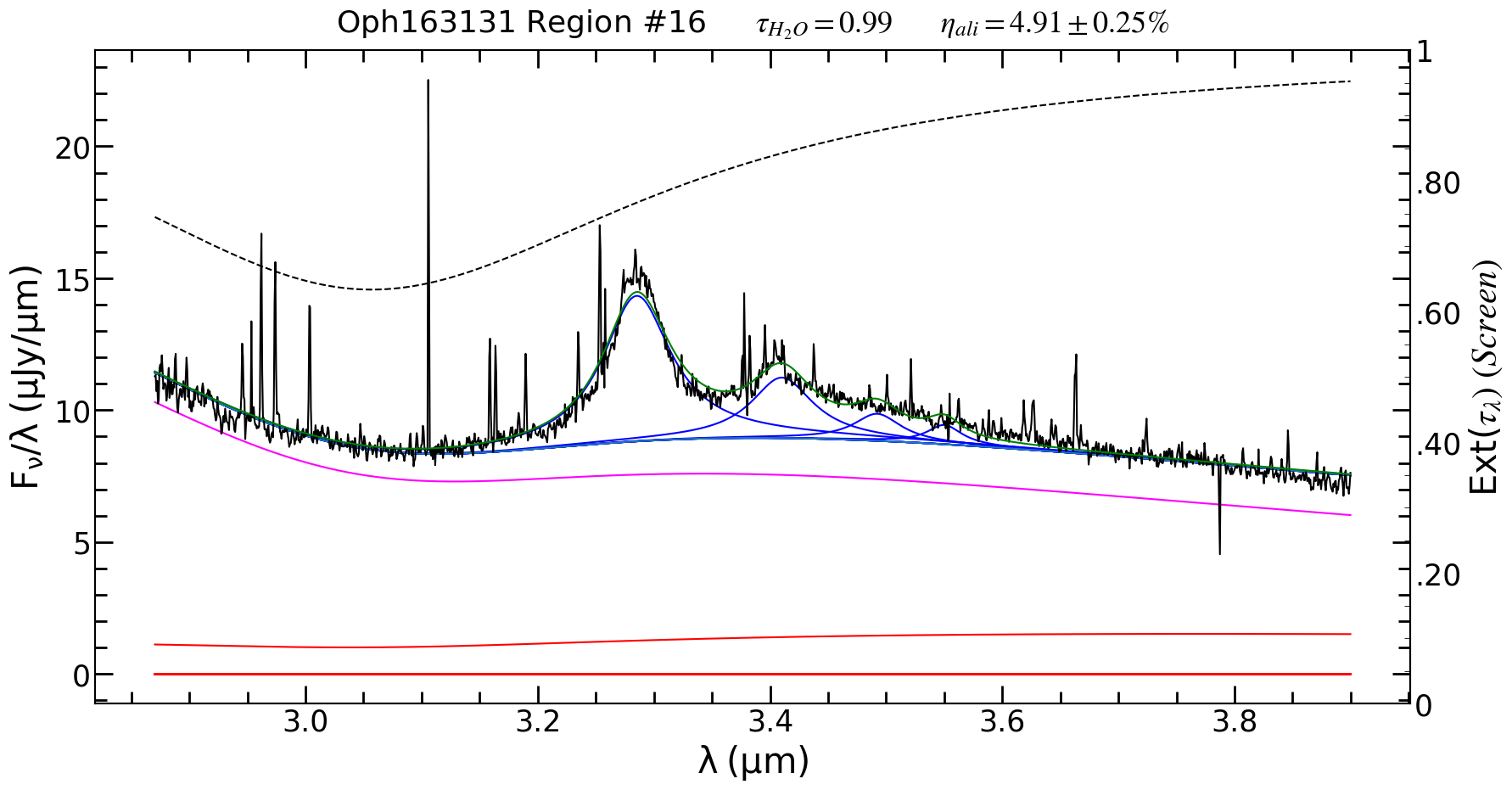}
\includegraphics[width=0.19\textwidth]{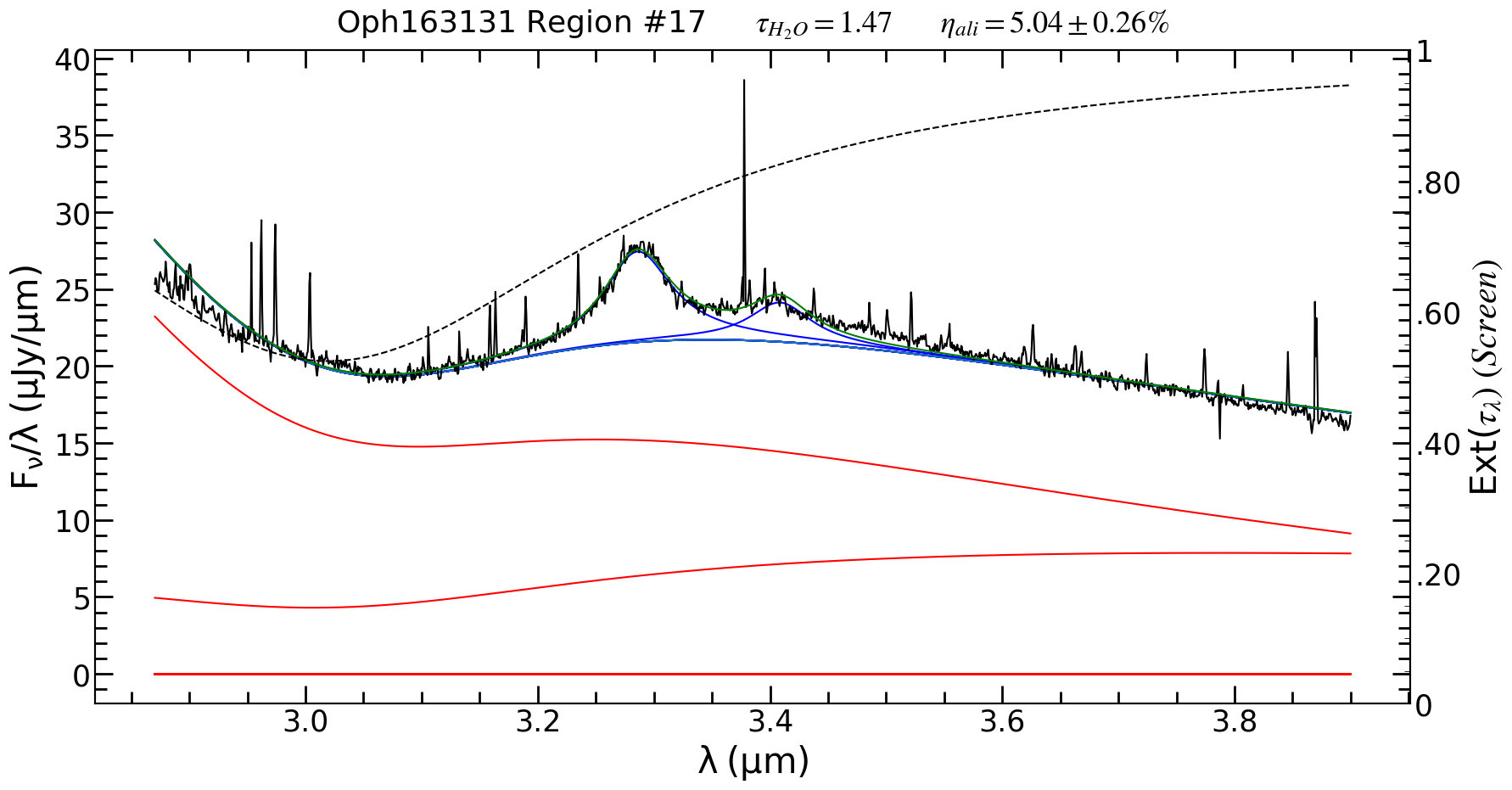}
\includegraphics[width=0.19\textwidth]{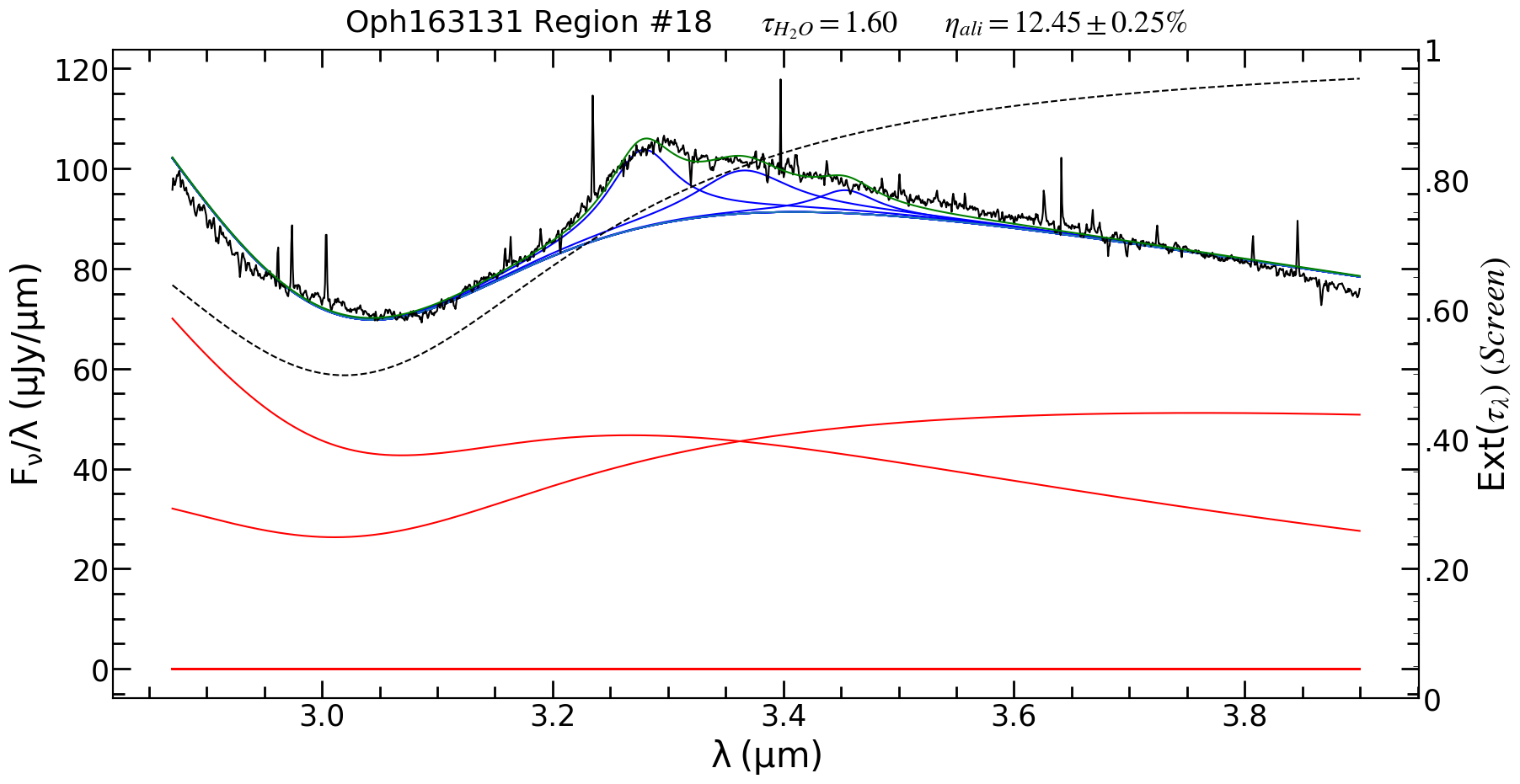}
\includegraphics[width=0.19\textwidth]{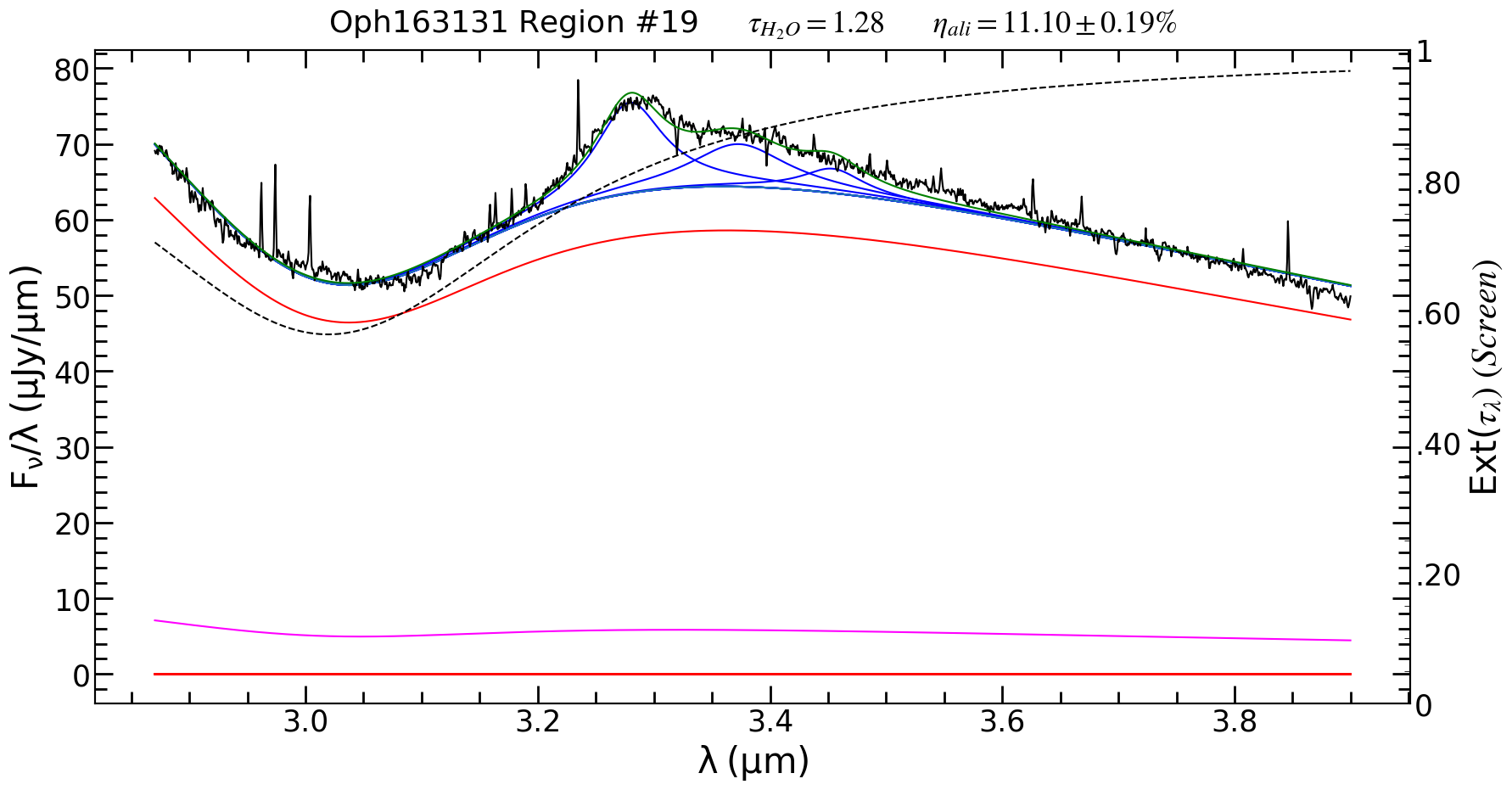}
\includegraphics[width=0.19\textwidth]{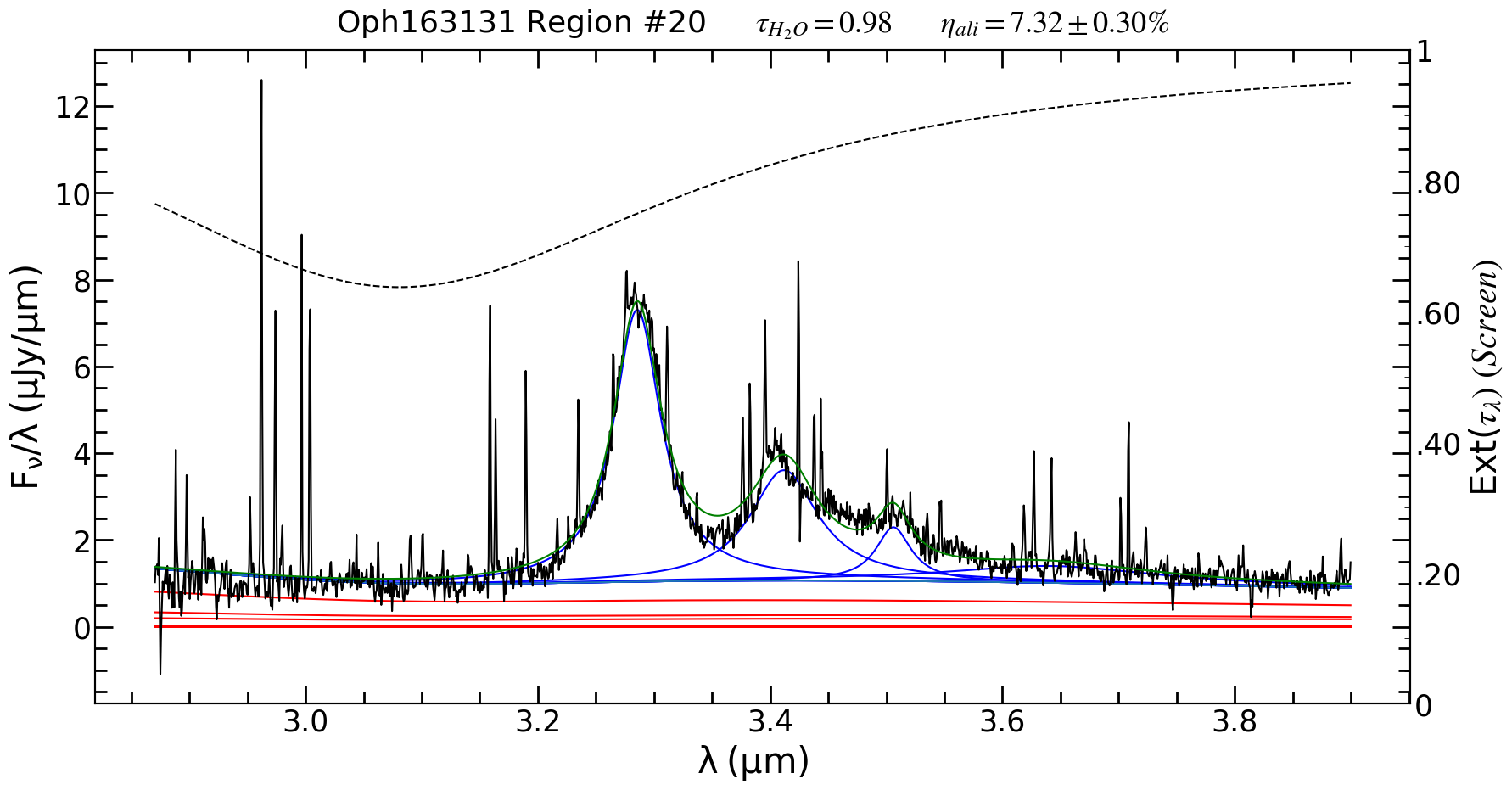}
\includegraphics[width=0.19\textwidth]{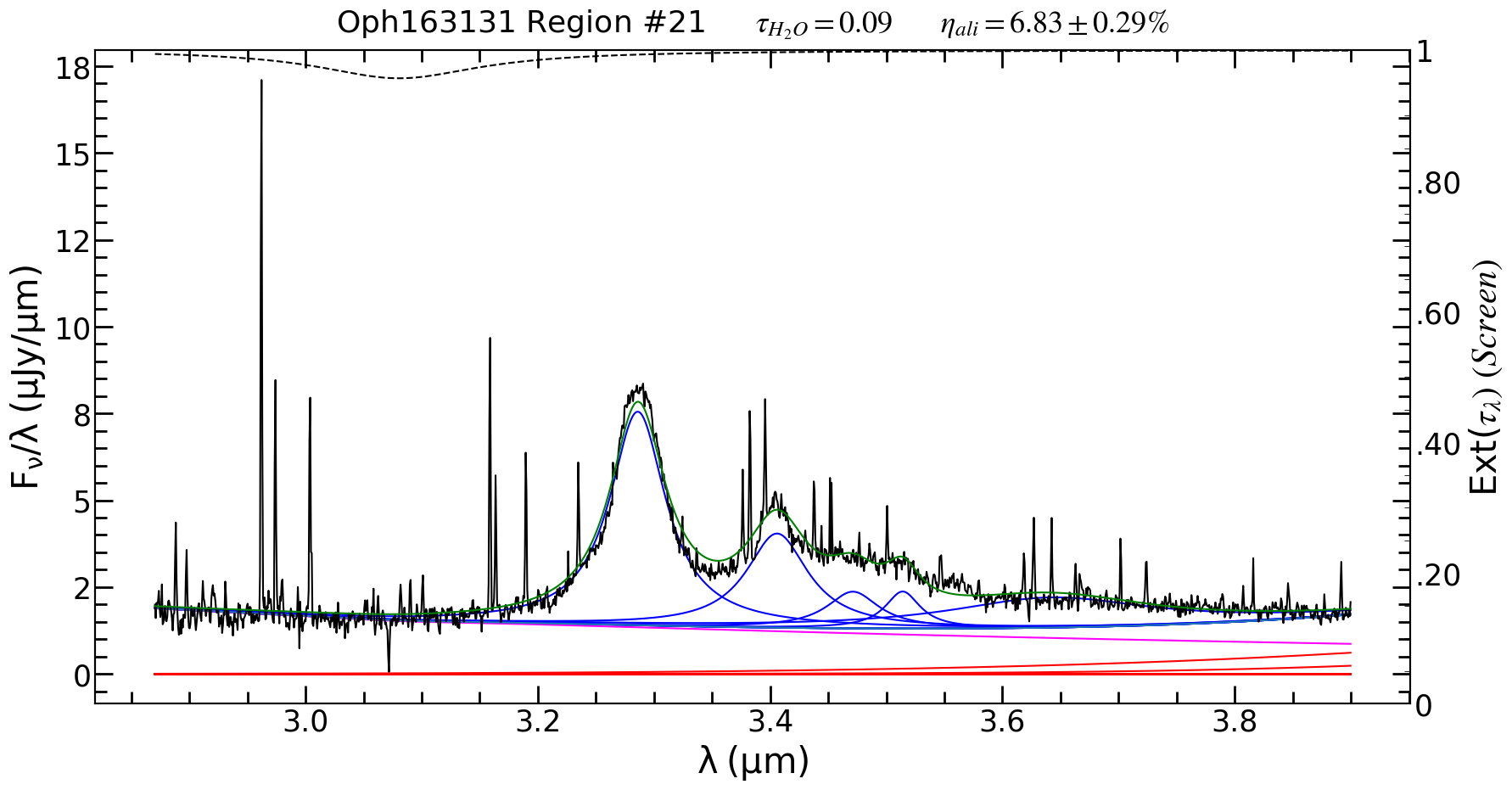}
\includegraphics[width=0.19\textwidth]{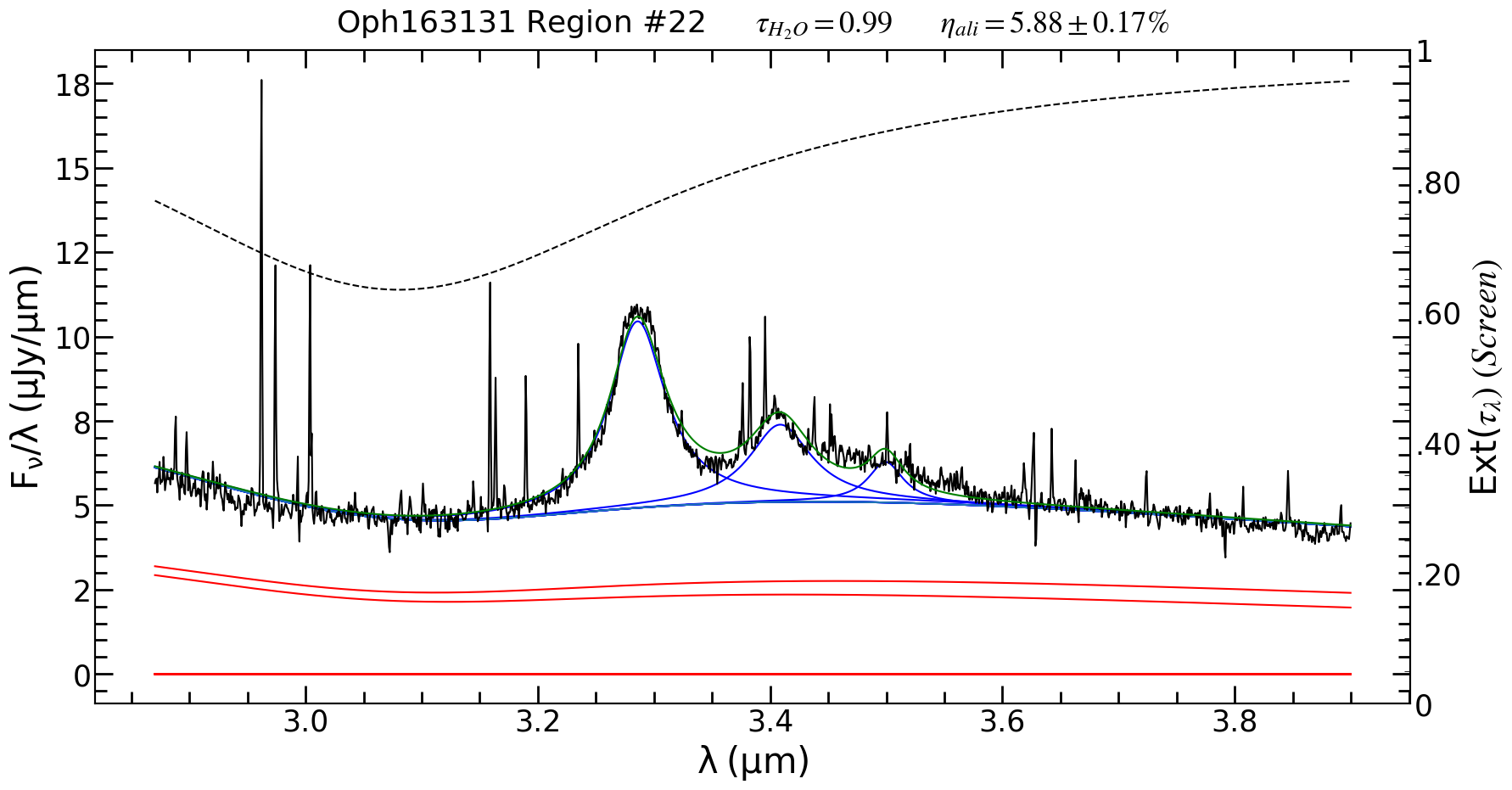}
\includegraphics[width=0.19\textwidth]{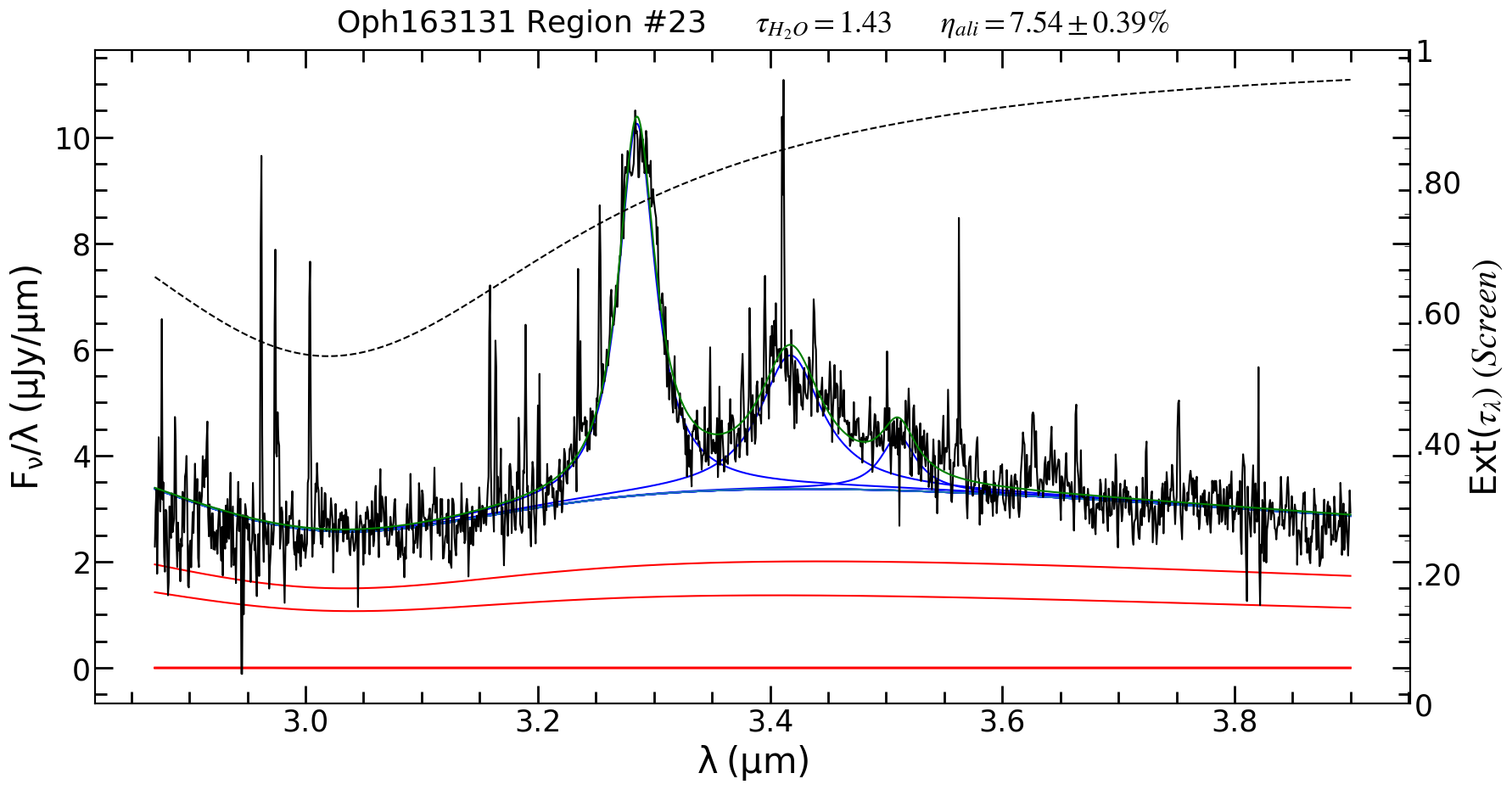}

\caption{PAHFIT results for Oph~163131 for the near-IR used to calculated the $\eta_{\rm ali}$ for all regions.}
\label{fig:OPH_Slices}
\end{figure*}

\begin{figure*}
    \centering
    
    \includegraphics[width=0.19\textwidth]{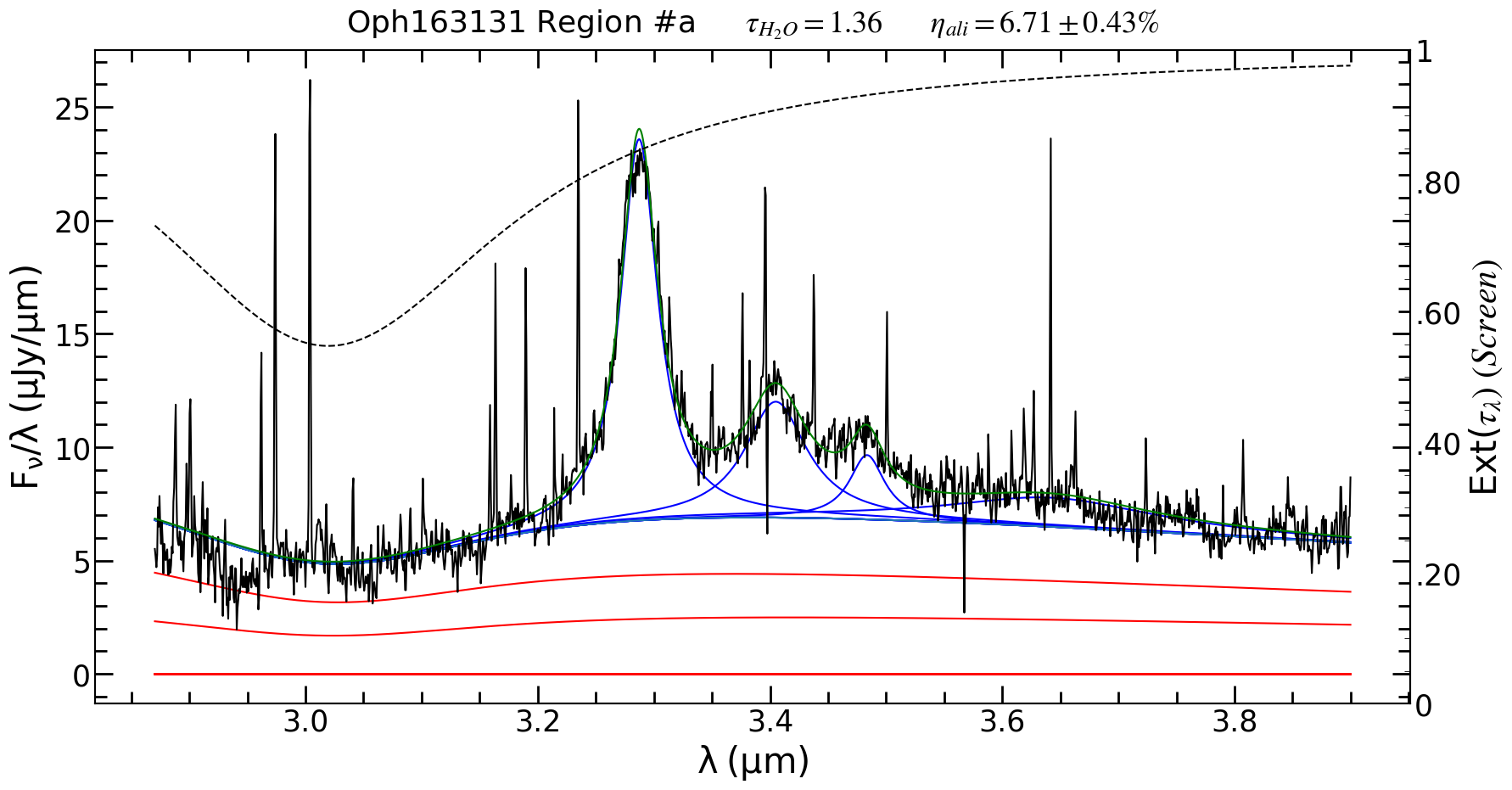}
    \includegraphics[width=0.19\textwidth]{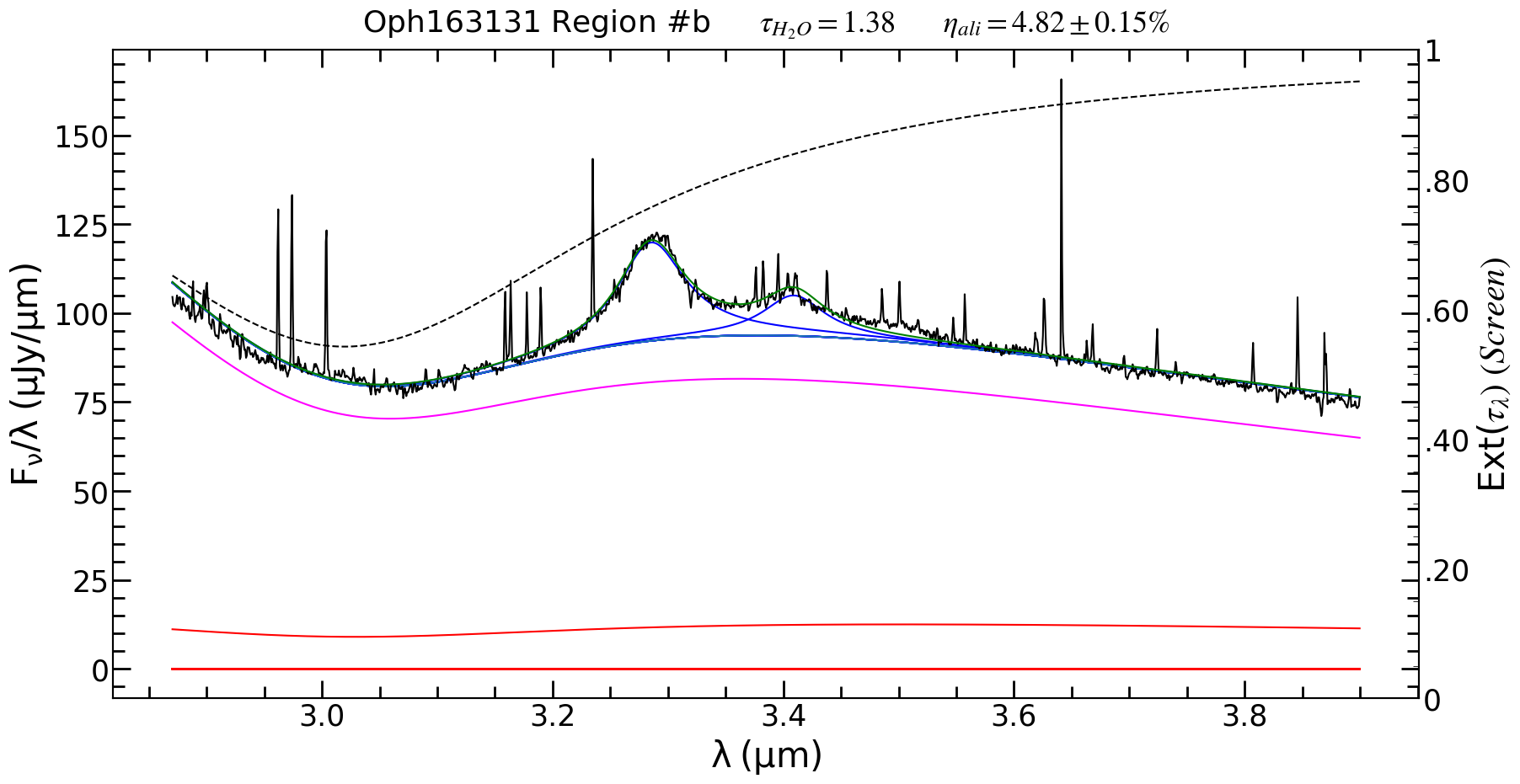}
    \includegraphics[width=0.19\textwidth]{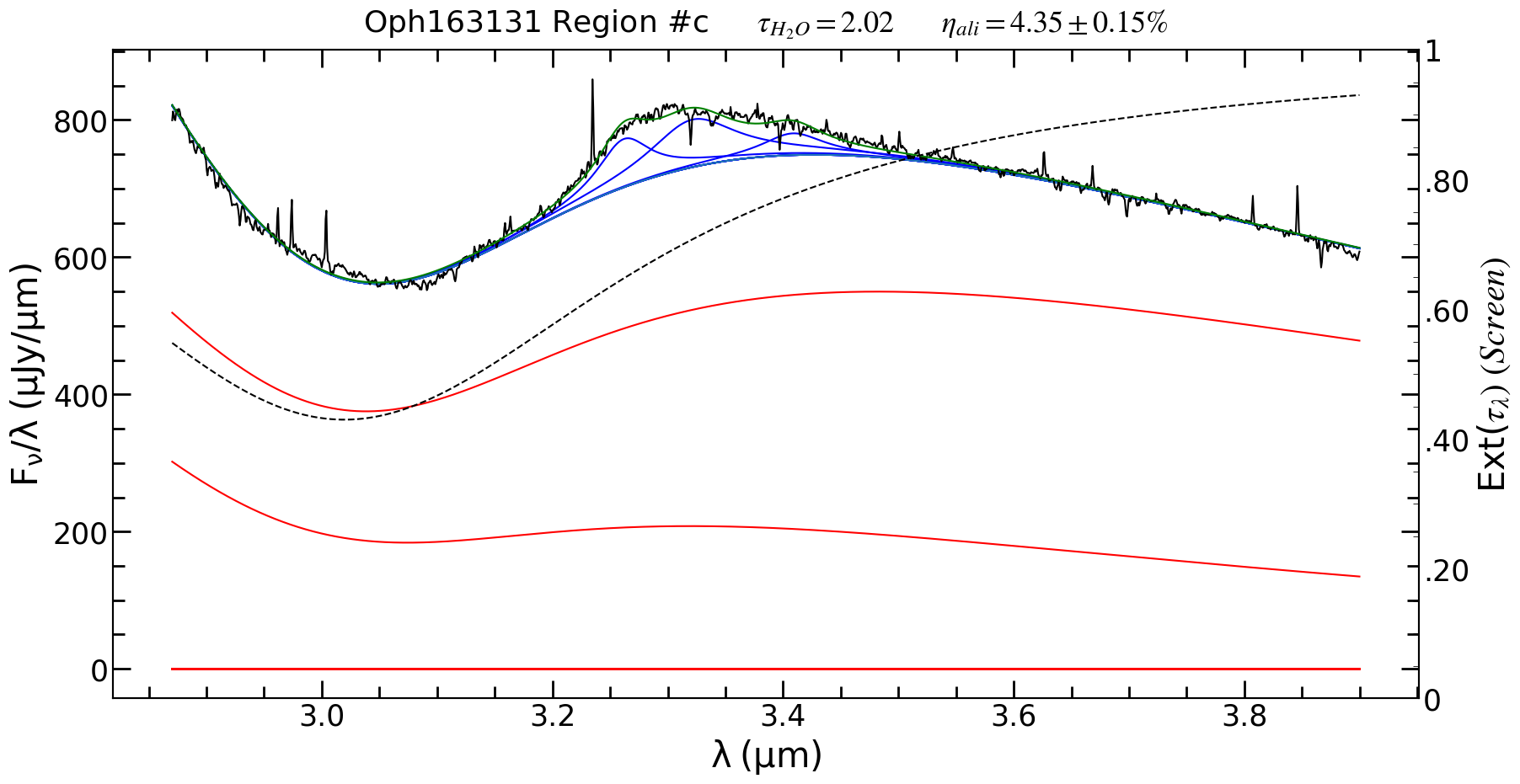}
    \includegraphics[width=0.19\textwidth]{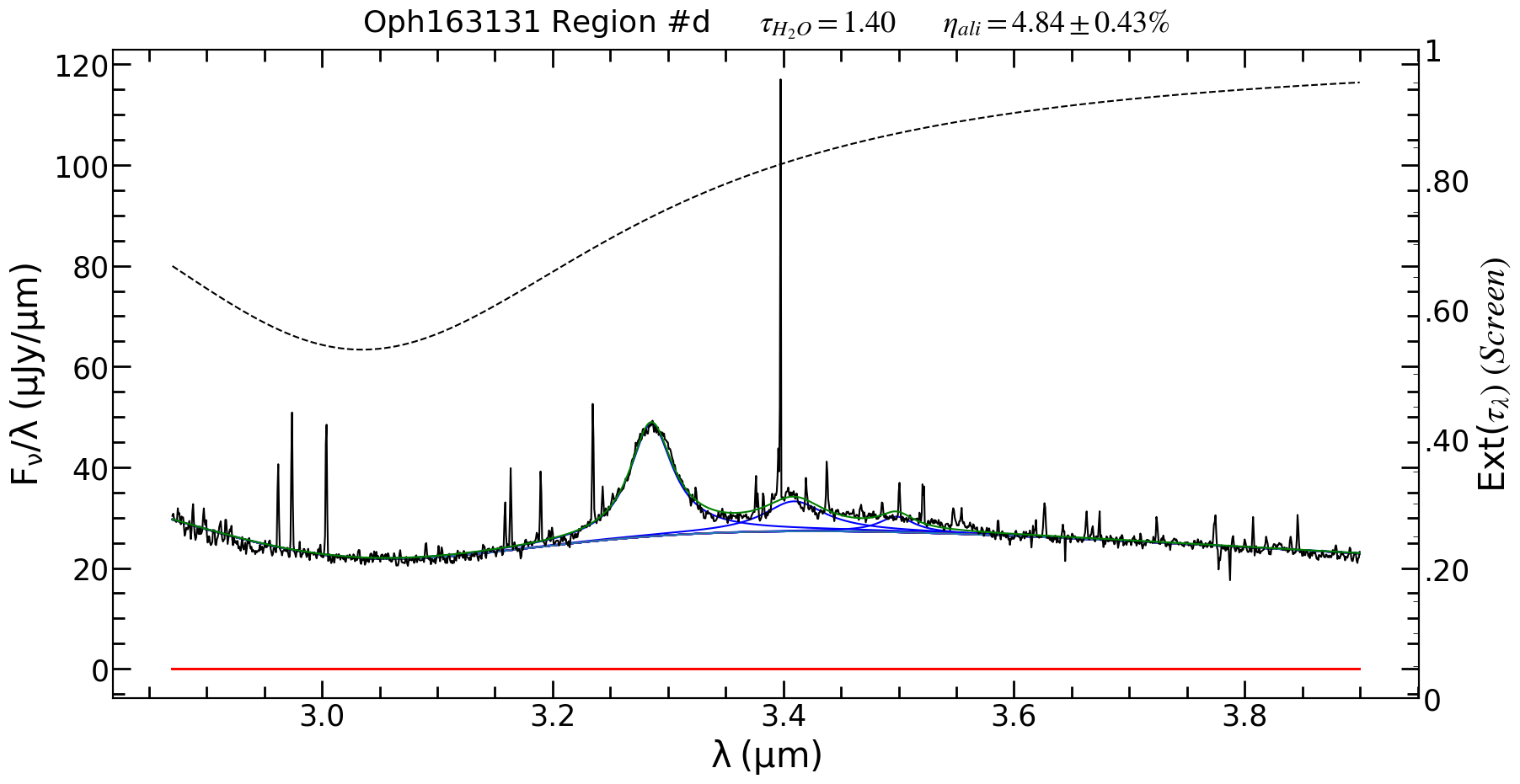}
    \includegraphics[width=0.19\textwidth]{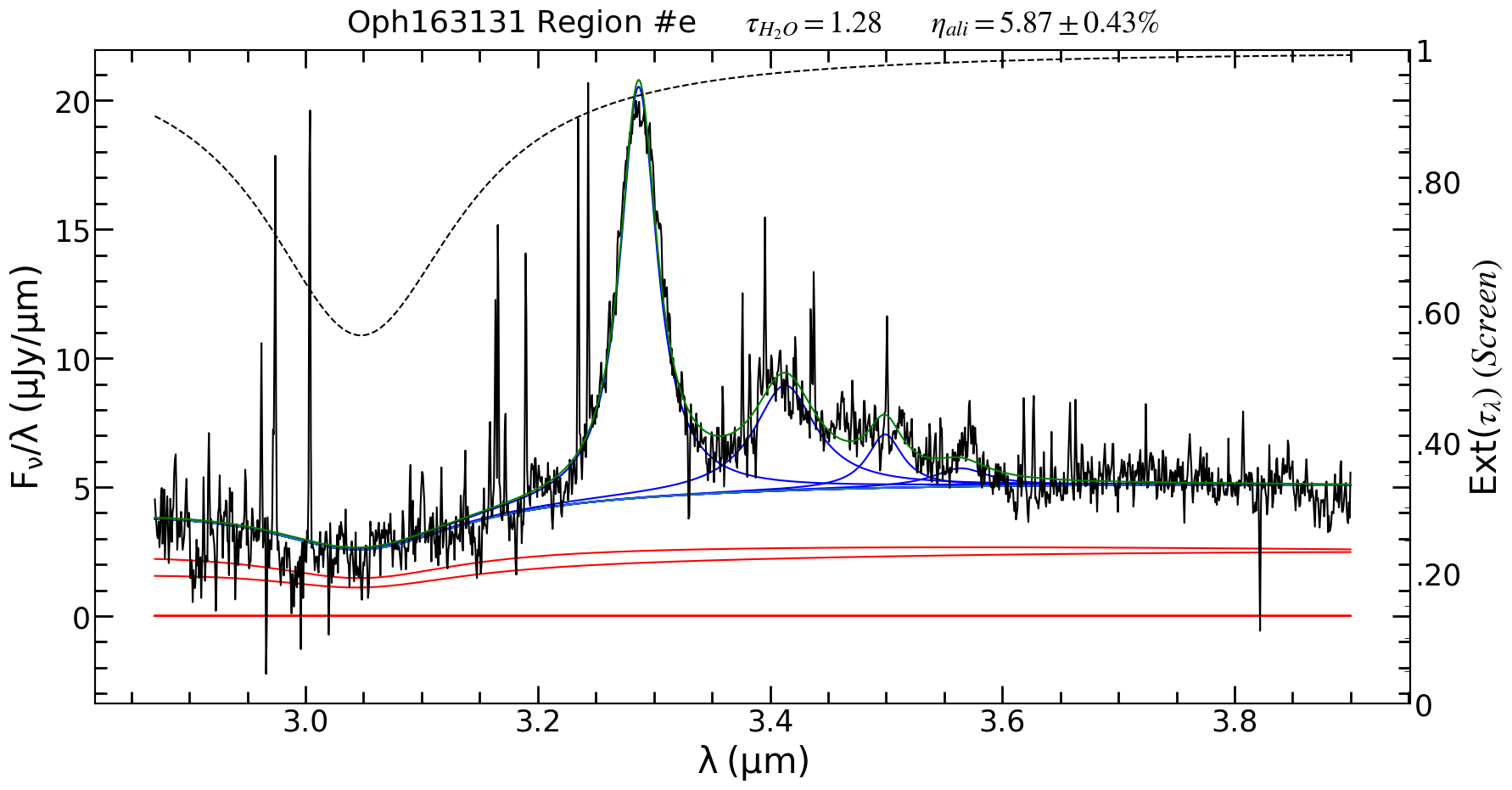}
    \includegraphics[width=0.19\textwidth]{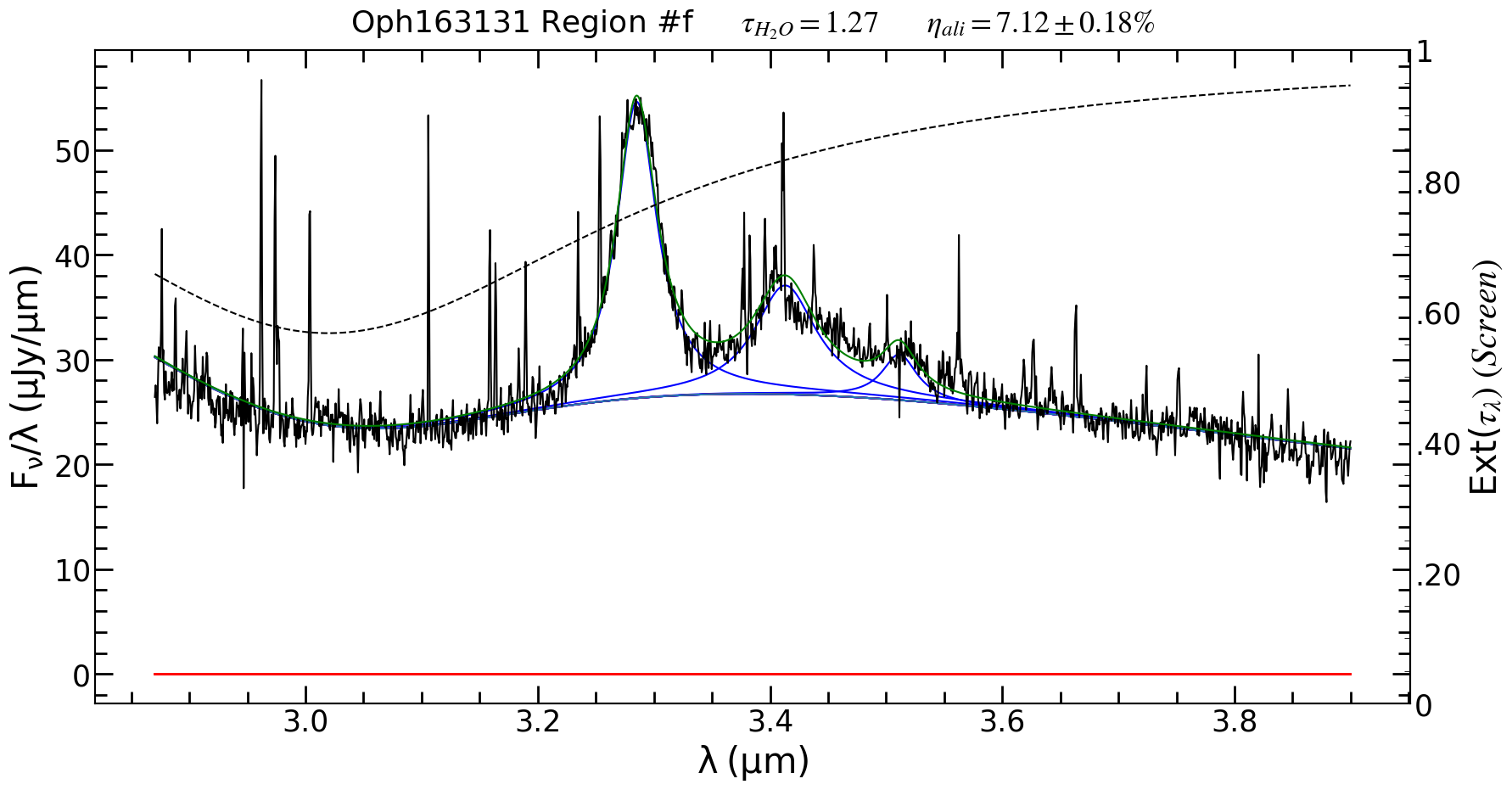}
    \includegraphics[width=0.19\textwidth]{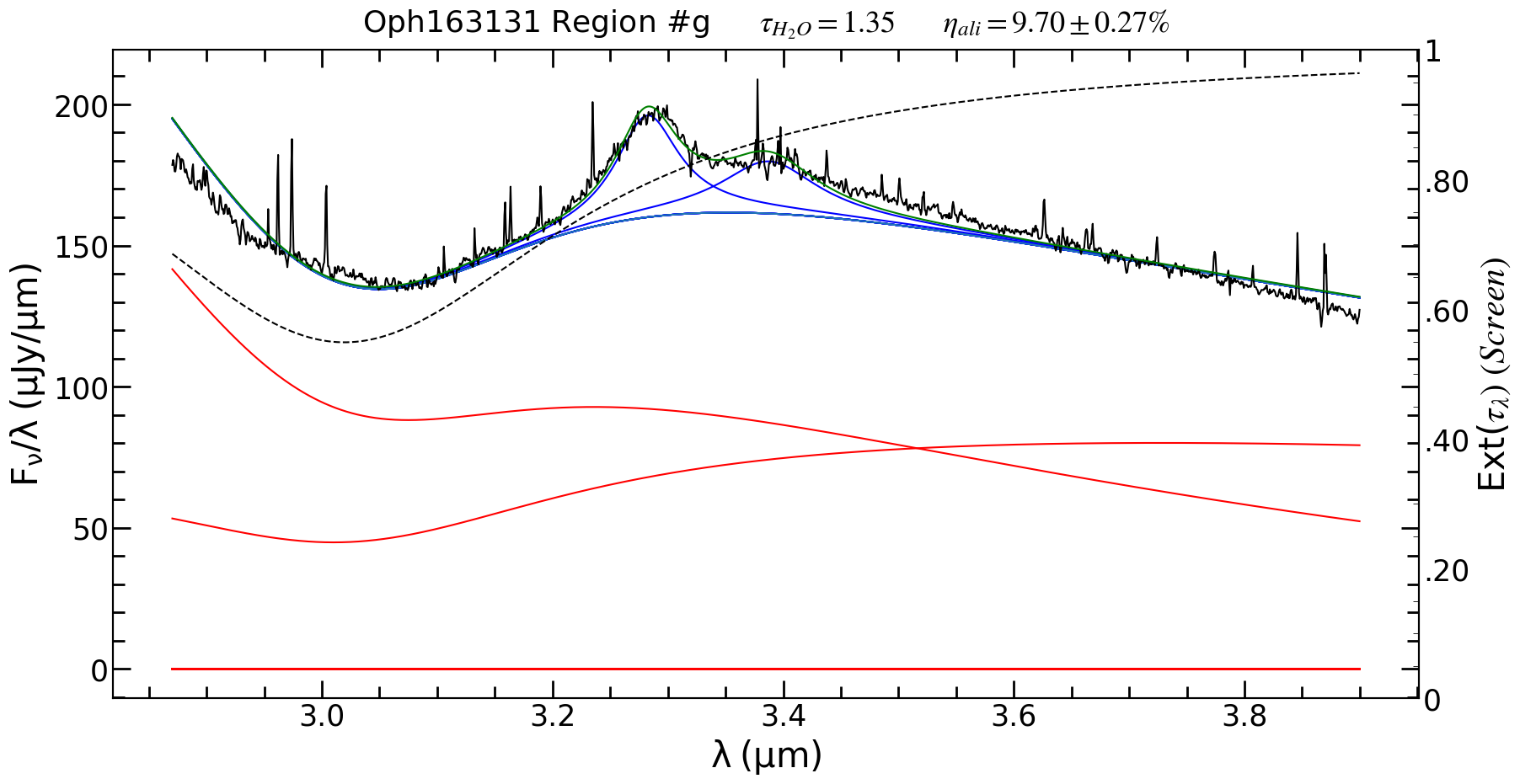}
    \includegraphics[width=0.19\textwidth]{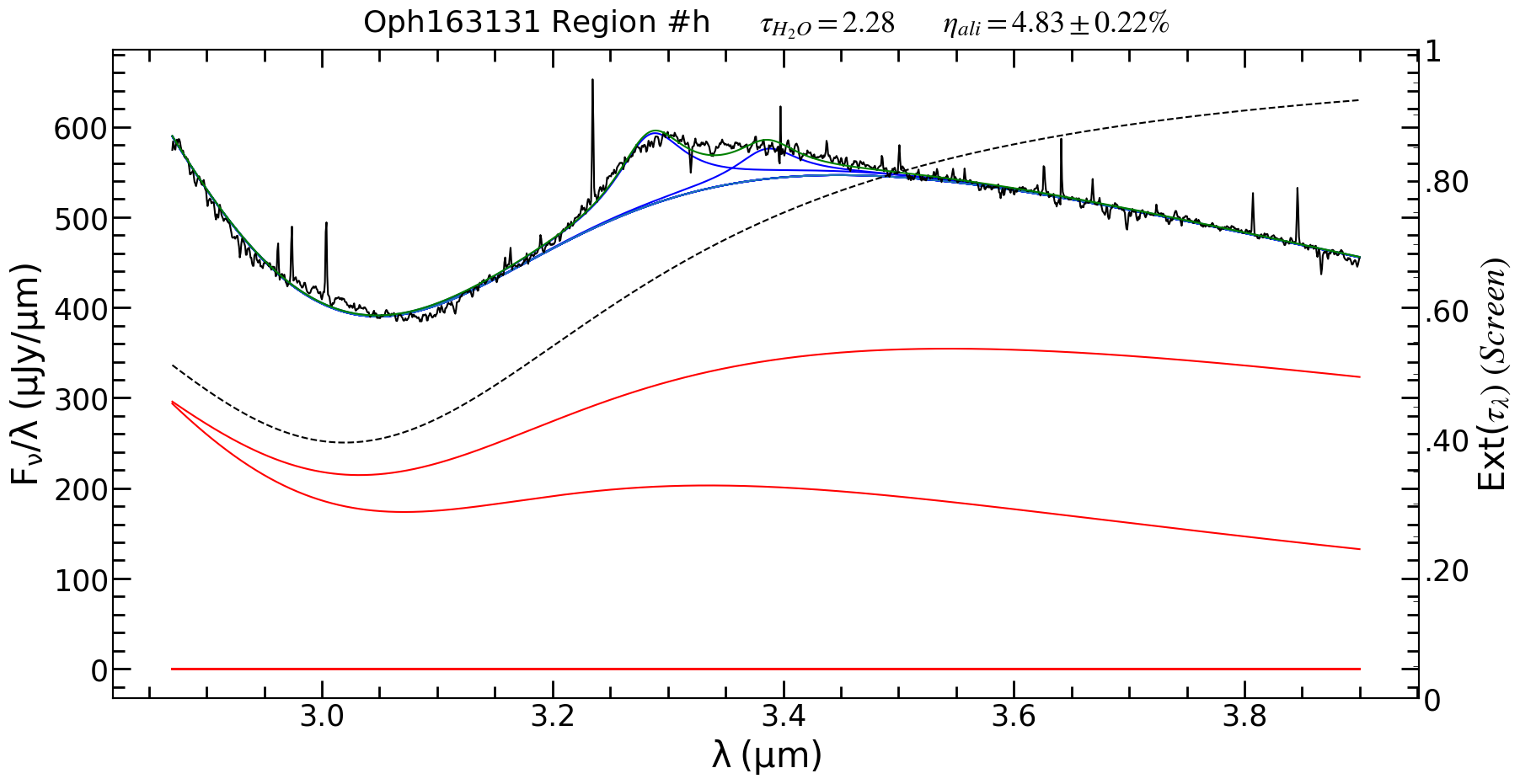}
    \includegraphics[width=0.19\textwidth]{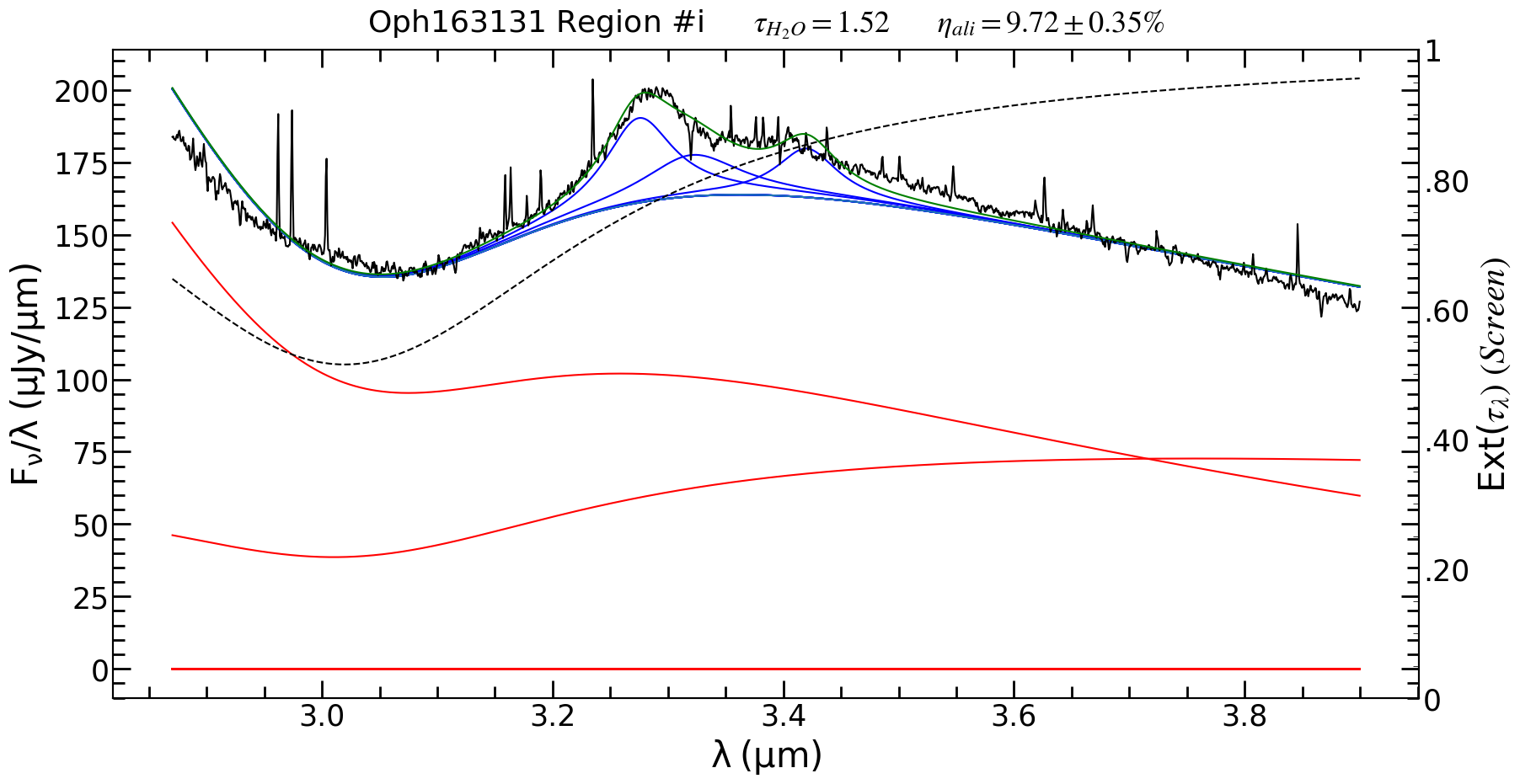}
    \includegraphics[width=0.19\textwidth]{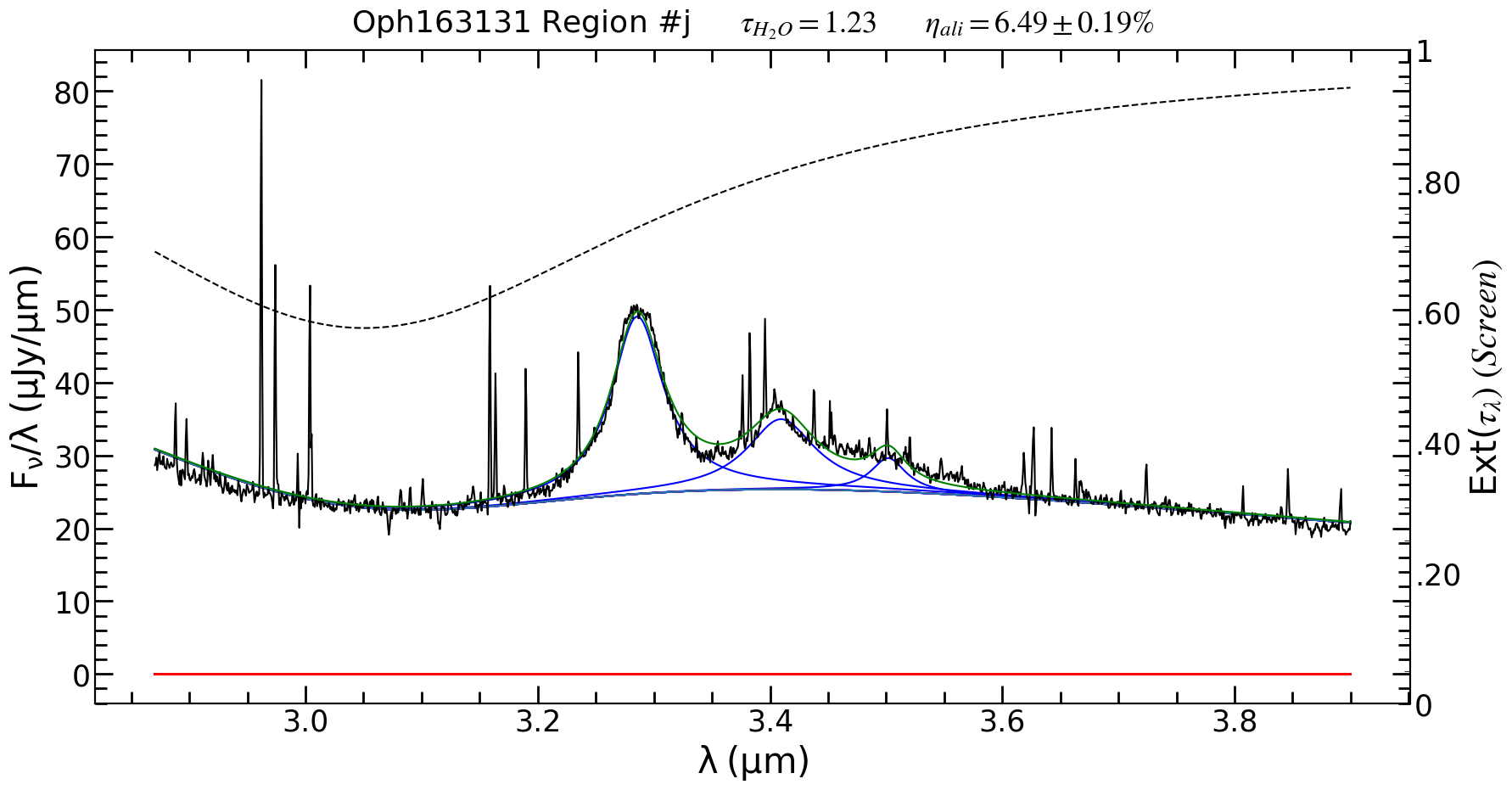}
    
    \includegraphics[width=0.19\textwidth]{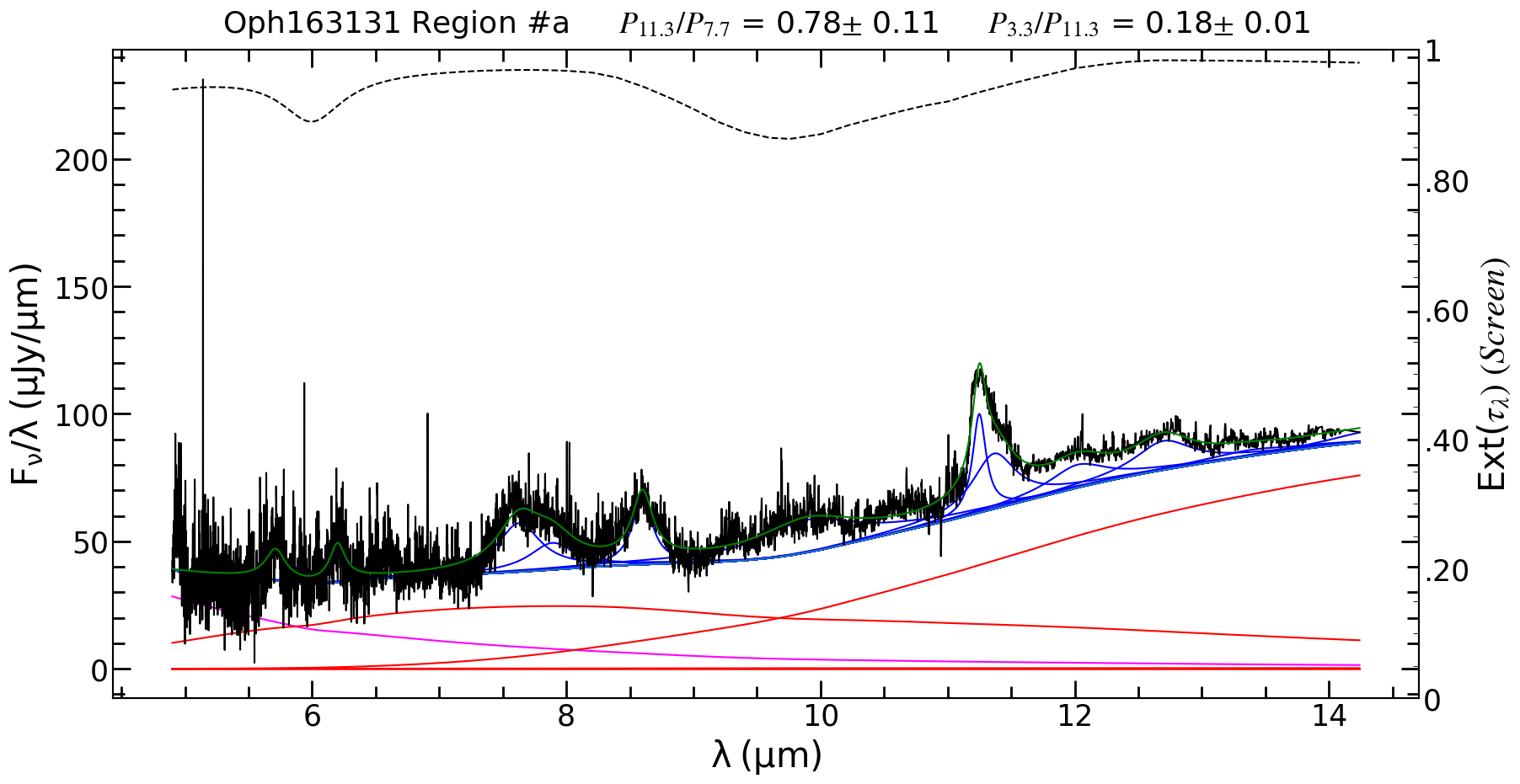}
    \includegraphics[width=0.19\textwidth]{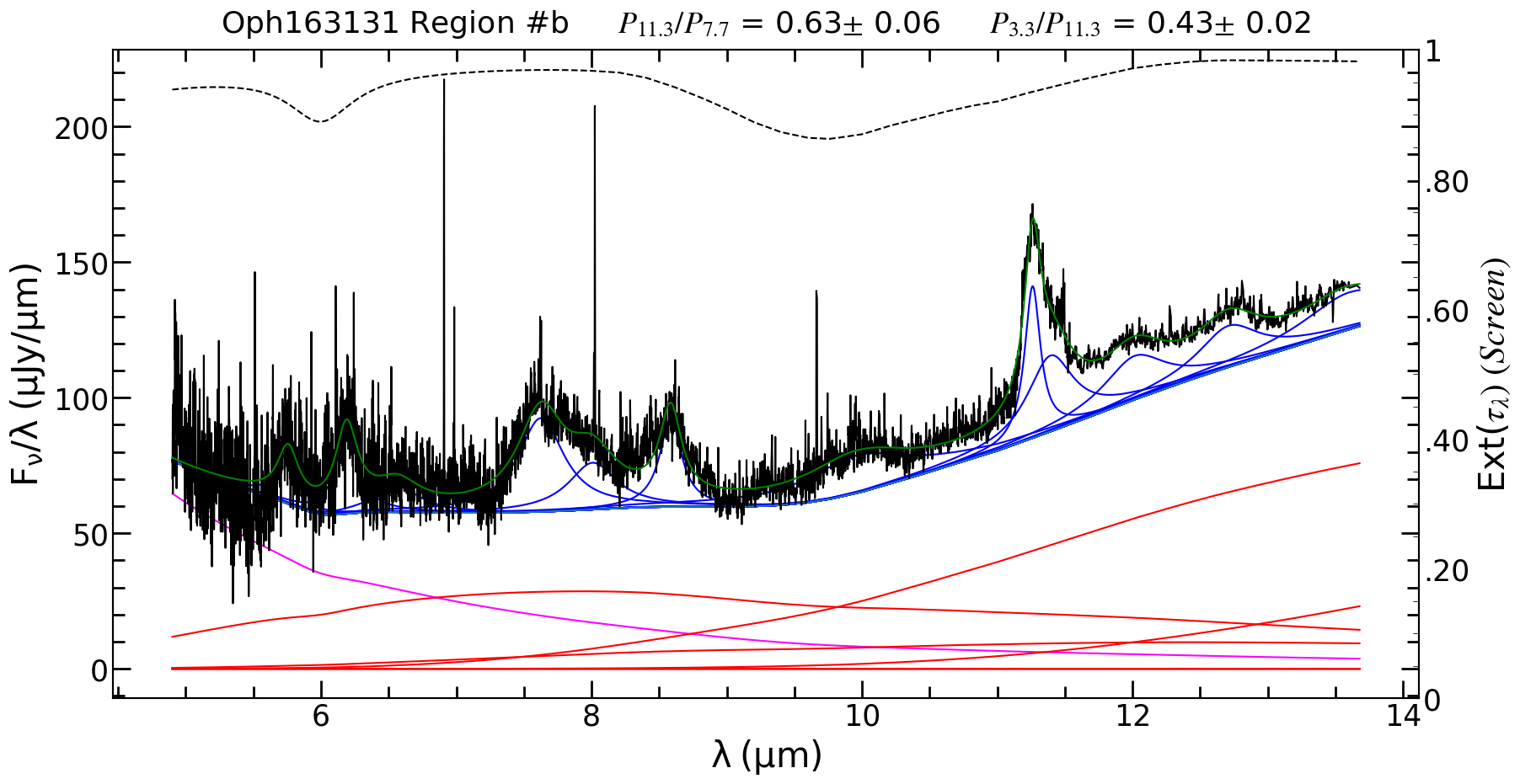}
    \includegraphics[width=0.19\textwidth]{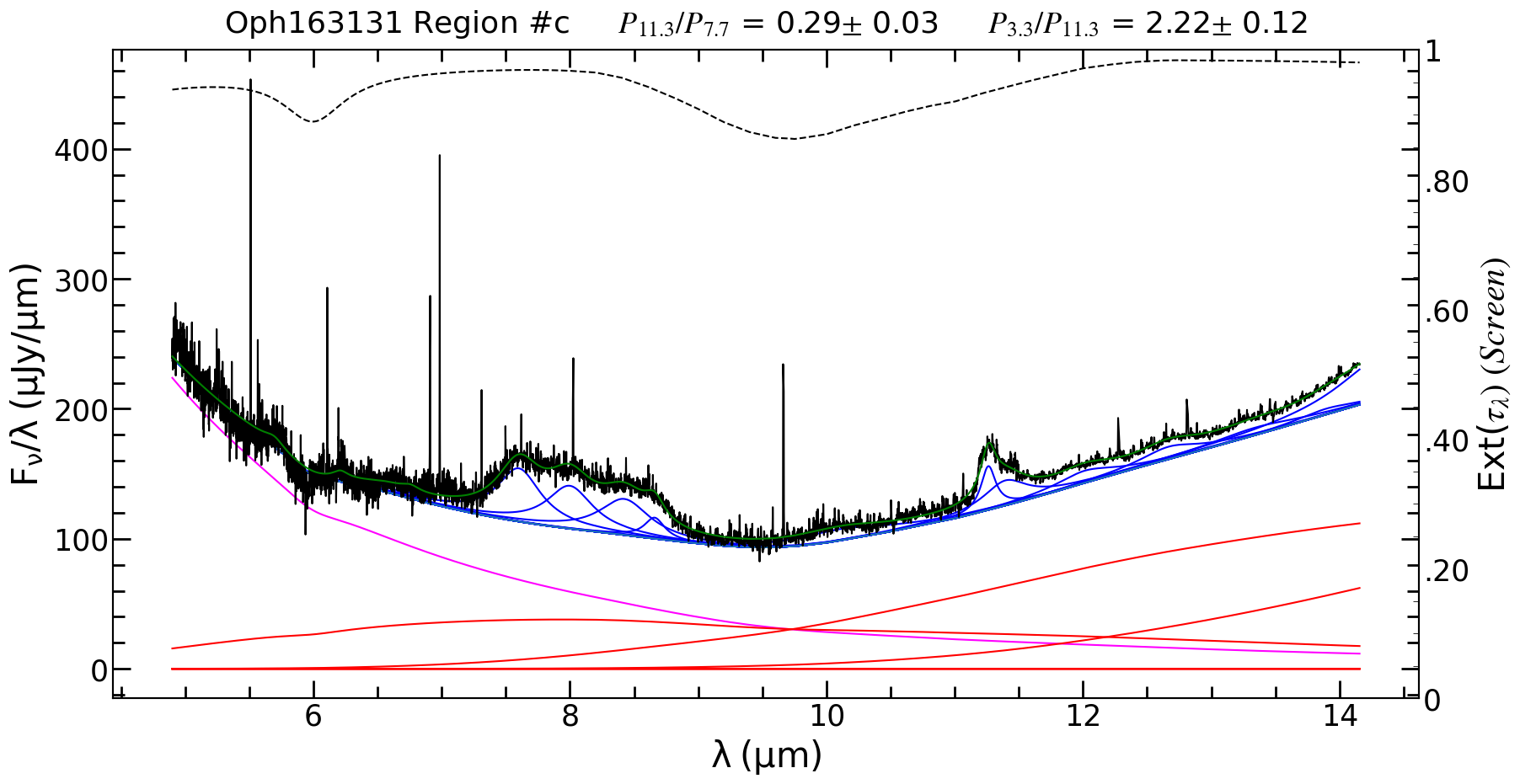}
    \includegraphics[width=0.19\textwidth]{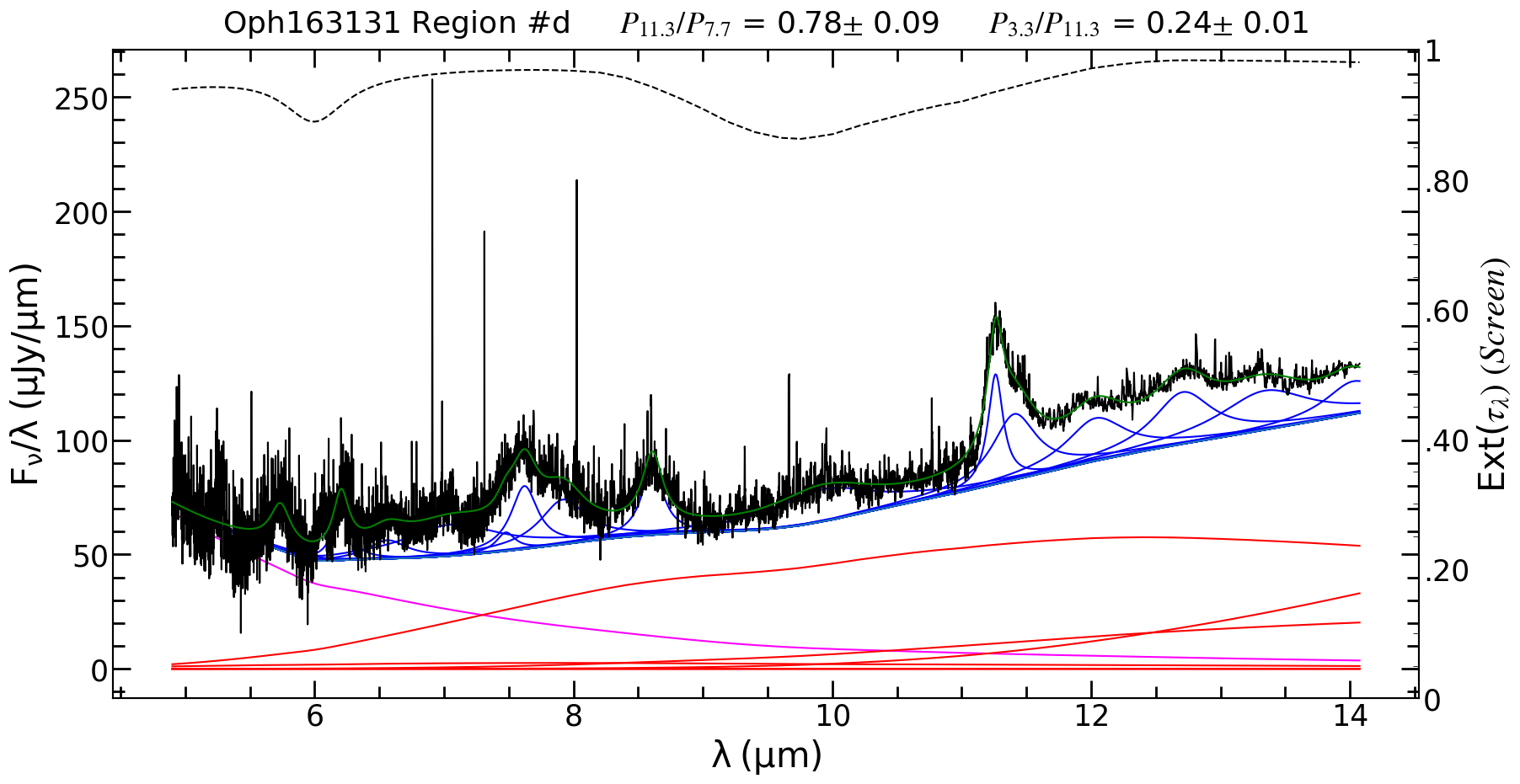}
    \includegraphics[width=0.19\textwidth]{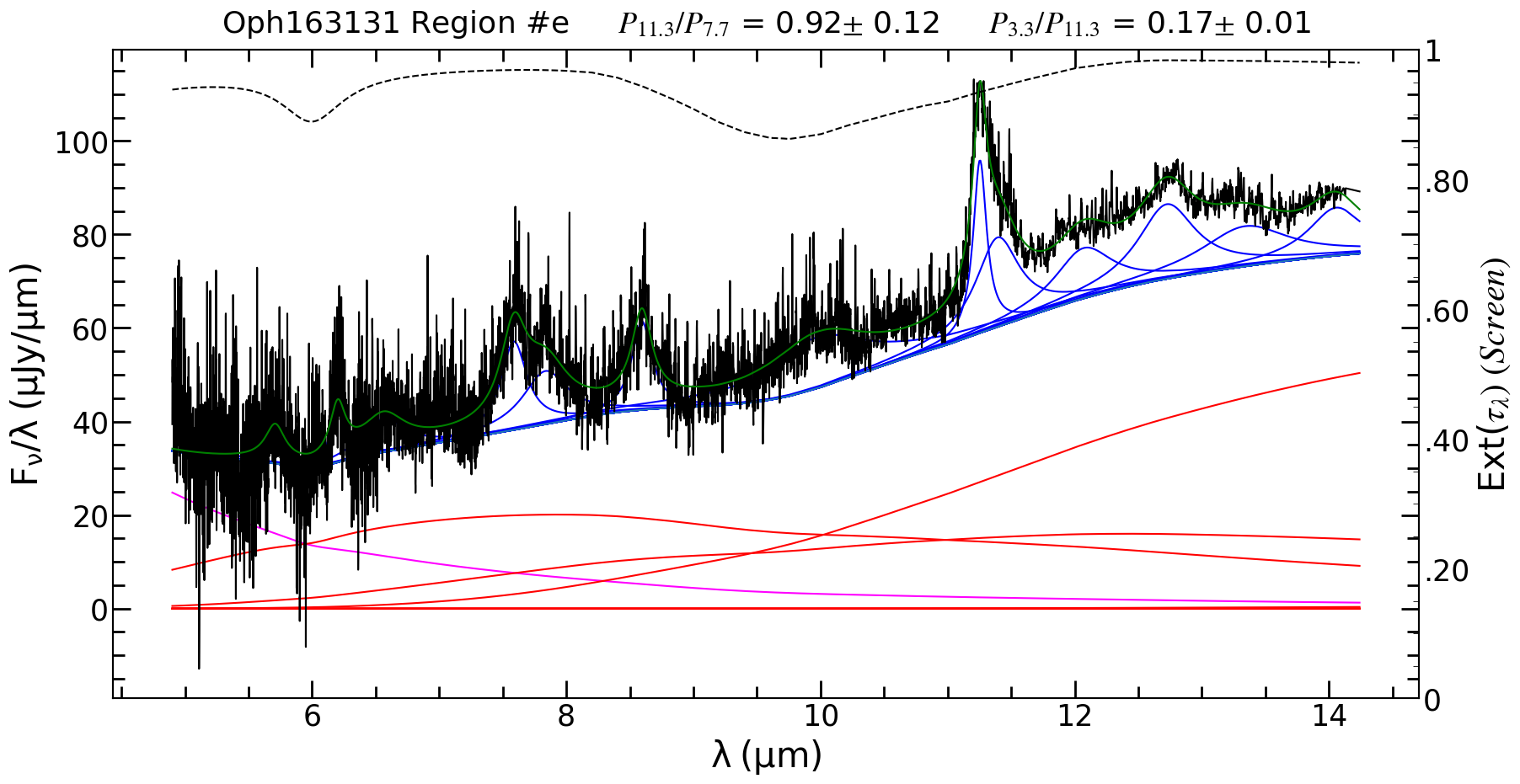}
    \includegraphics[width=0.19\textwidth]{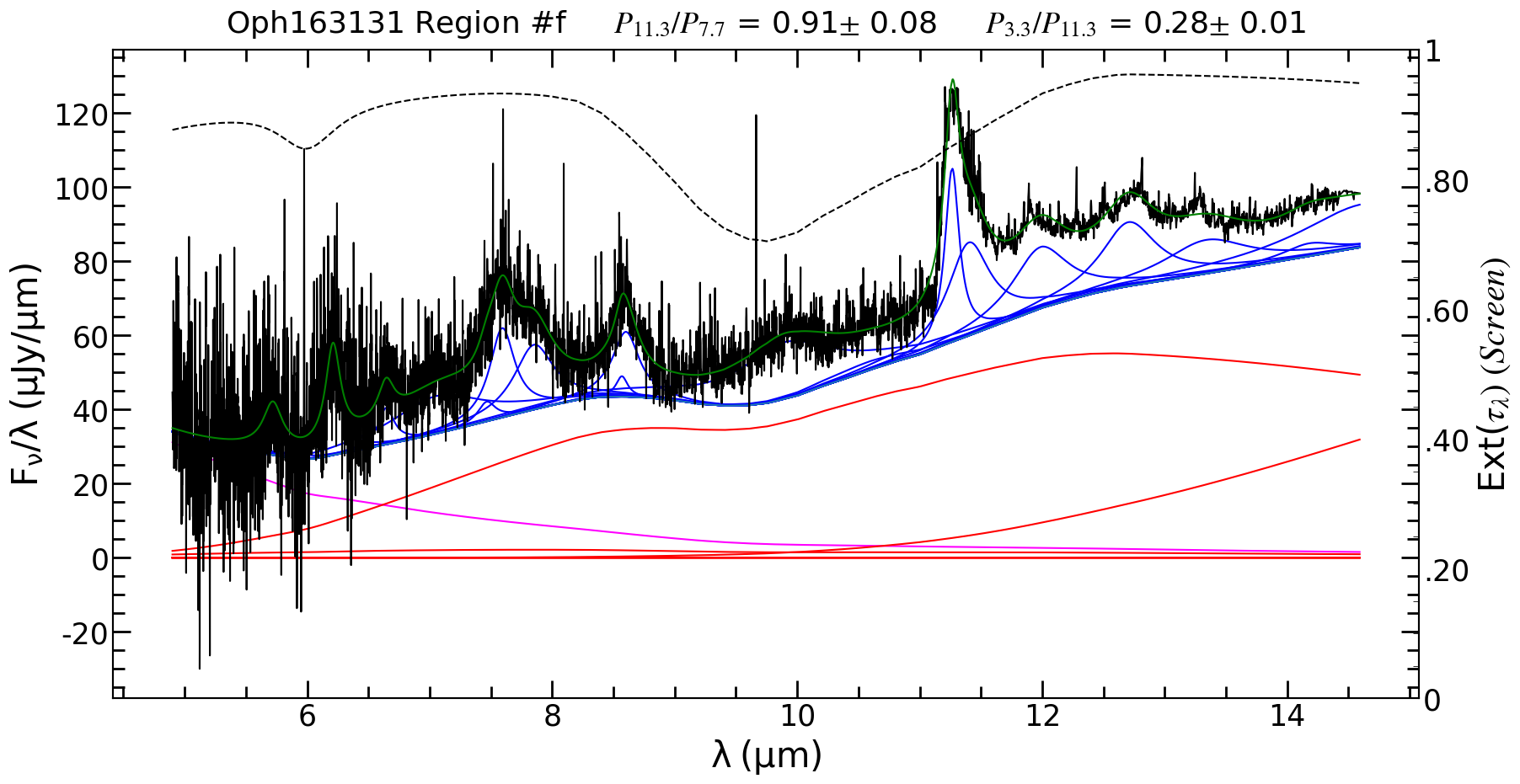}
    \includegraphics[width=0.19\textwidth]{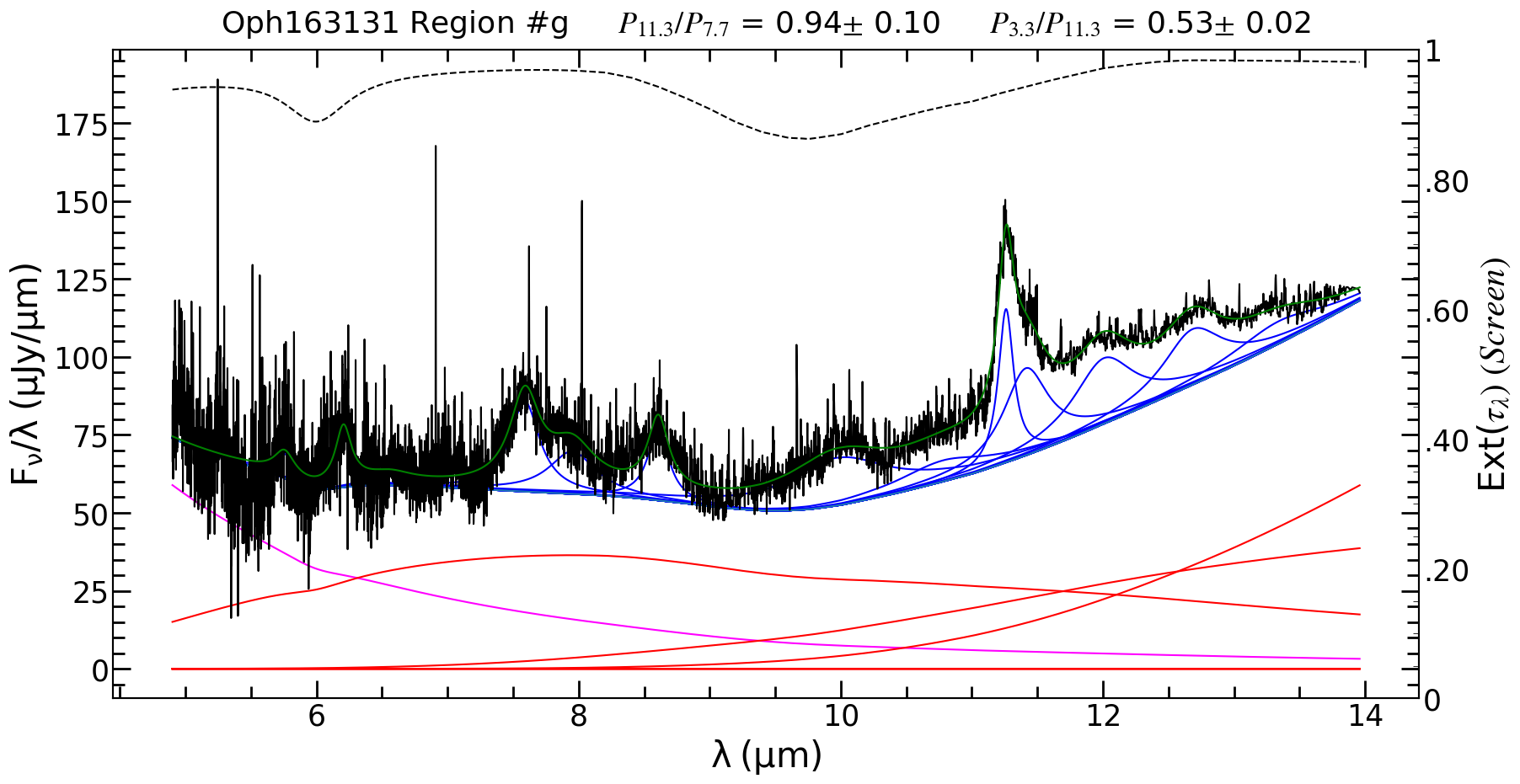}
    \includegraphics[width=0.19\textwidth]{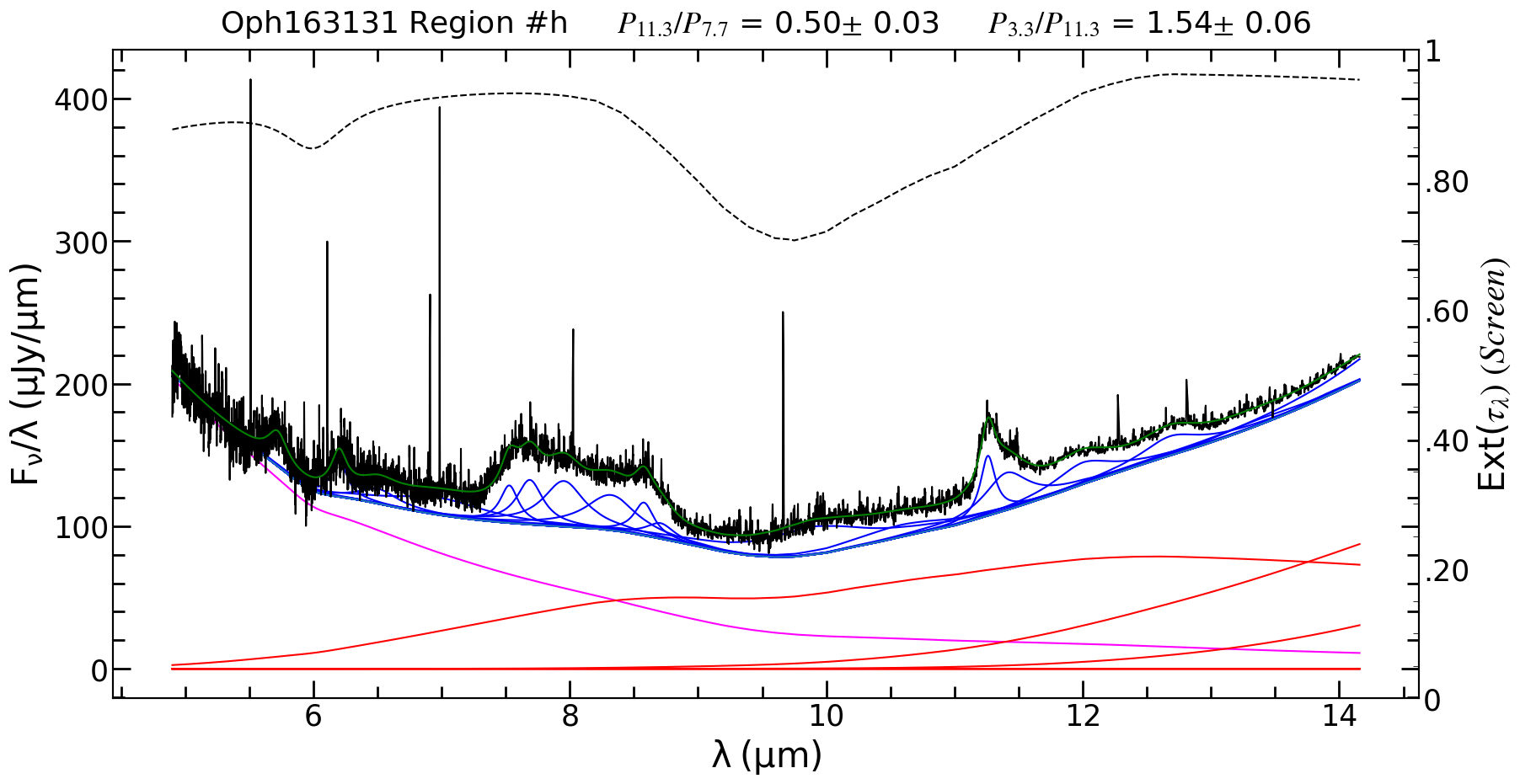}
    \includegraphics[width=0.19\textwidth]{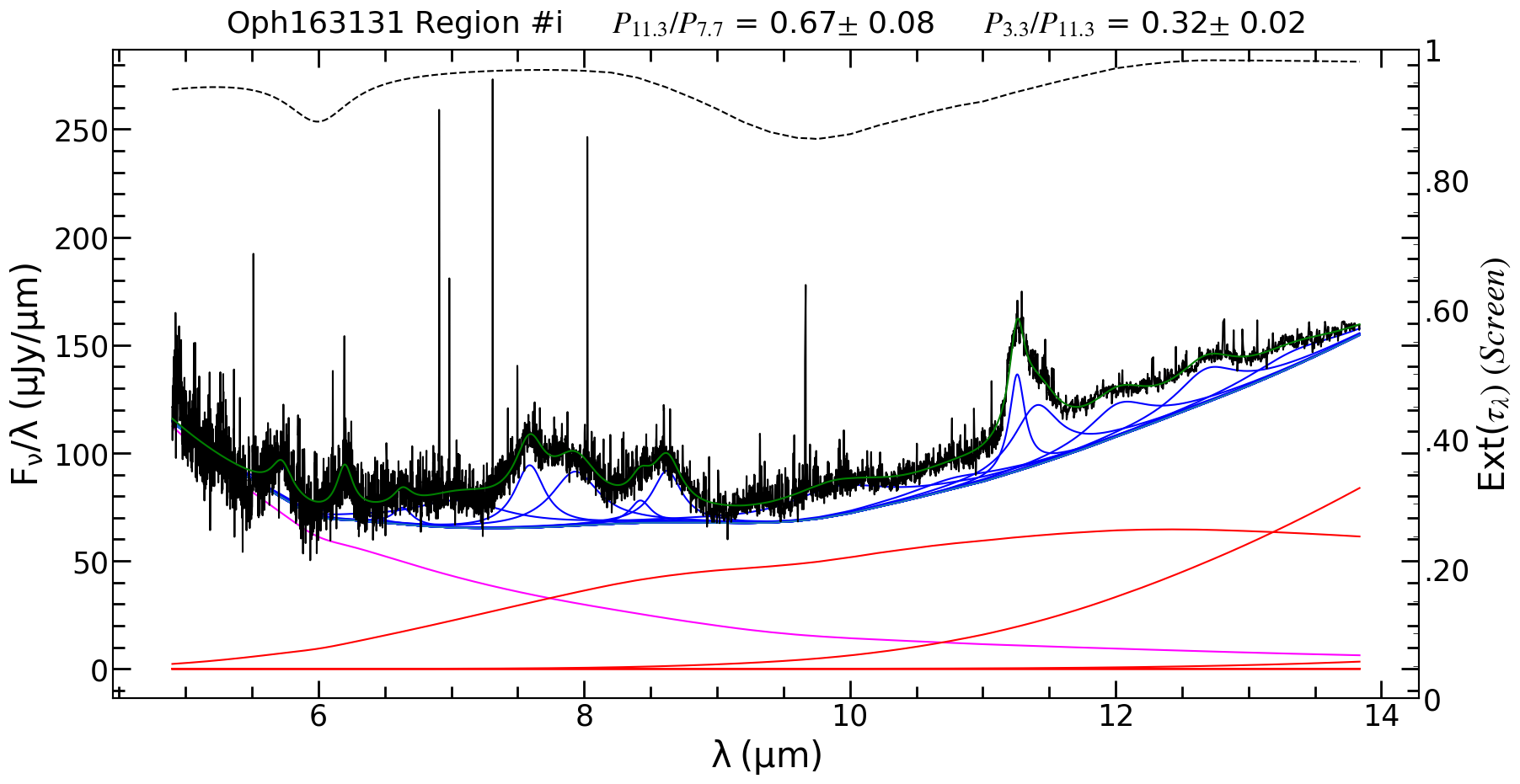}
    \includegraphics[width=0.19\textwidth]{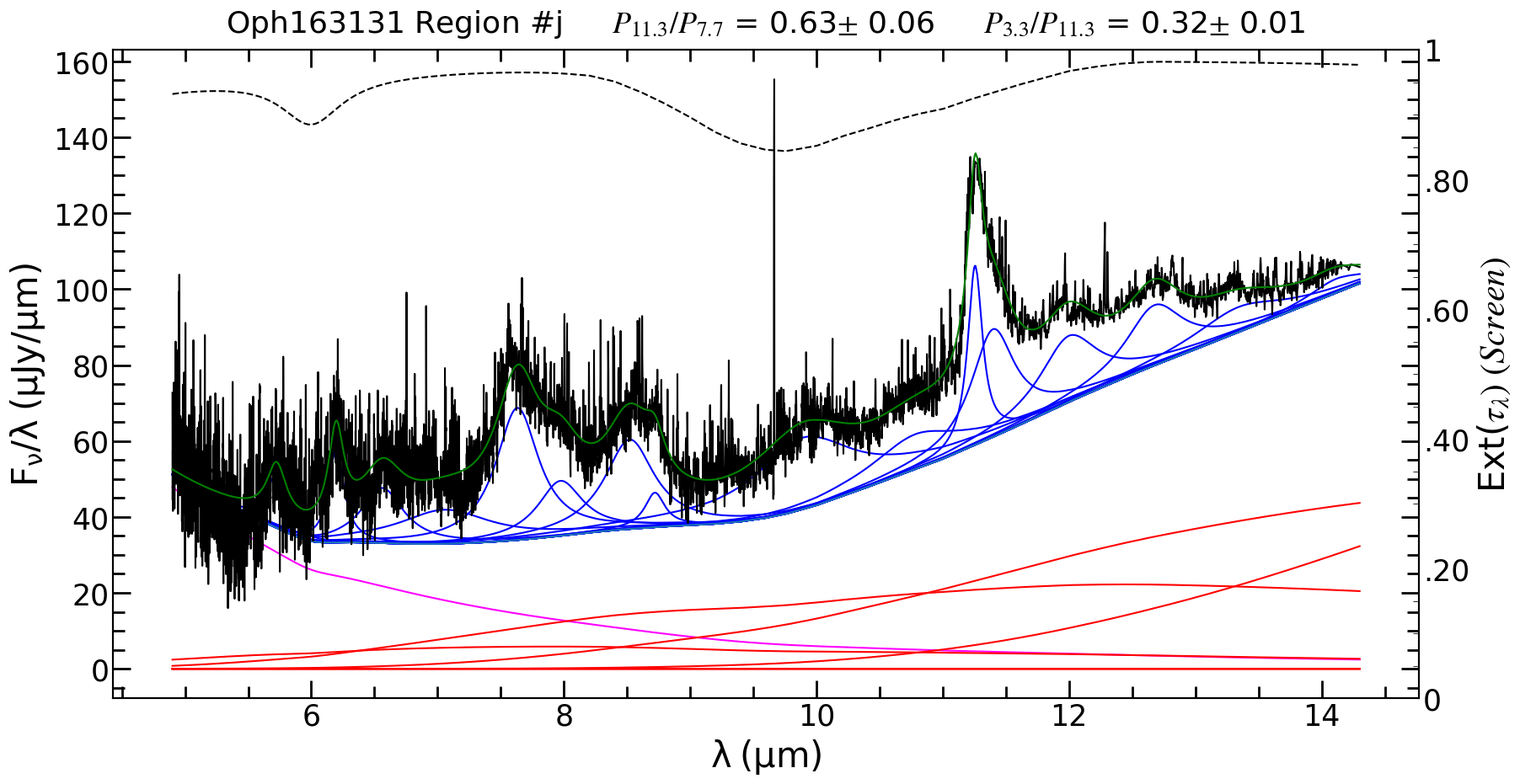}

    \caption{PAHFIT results for Oph~163131 for both the near- and mid- IR used to calculated the ratios involving 11.3~$\mu$m (3.3/11.3~$\mu$m \& 11.3/7.7~$\mu$m)}
    \label{fig:OPH163131_slices_larger}
\end{figure*}

Individual PAHFIT results \citep{PAHFIT} for the apertures slices from Appendix~\ref{apps} used to construct the spatial variation plots found in Figures~\ref{fig:spatial} \& \ref{fig:psf_just} can be seen below in Figures~\ref{fig:ESO_Slices},\ref{fig:HVTAU_Slices}, and \ref{fig:OPH_Slices}. The larger slices of Oph~163131 to sample the 11.3~$\mu$m while maintaining PSF justification \citep{STPSF} can be found in Figure~\ref{fig:OPH163131_slices_larger} for both the near- and mid- IR.

\begin{figure*}[htb!]
\centering

\includegraphics[width=0.19\textwidth]{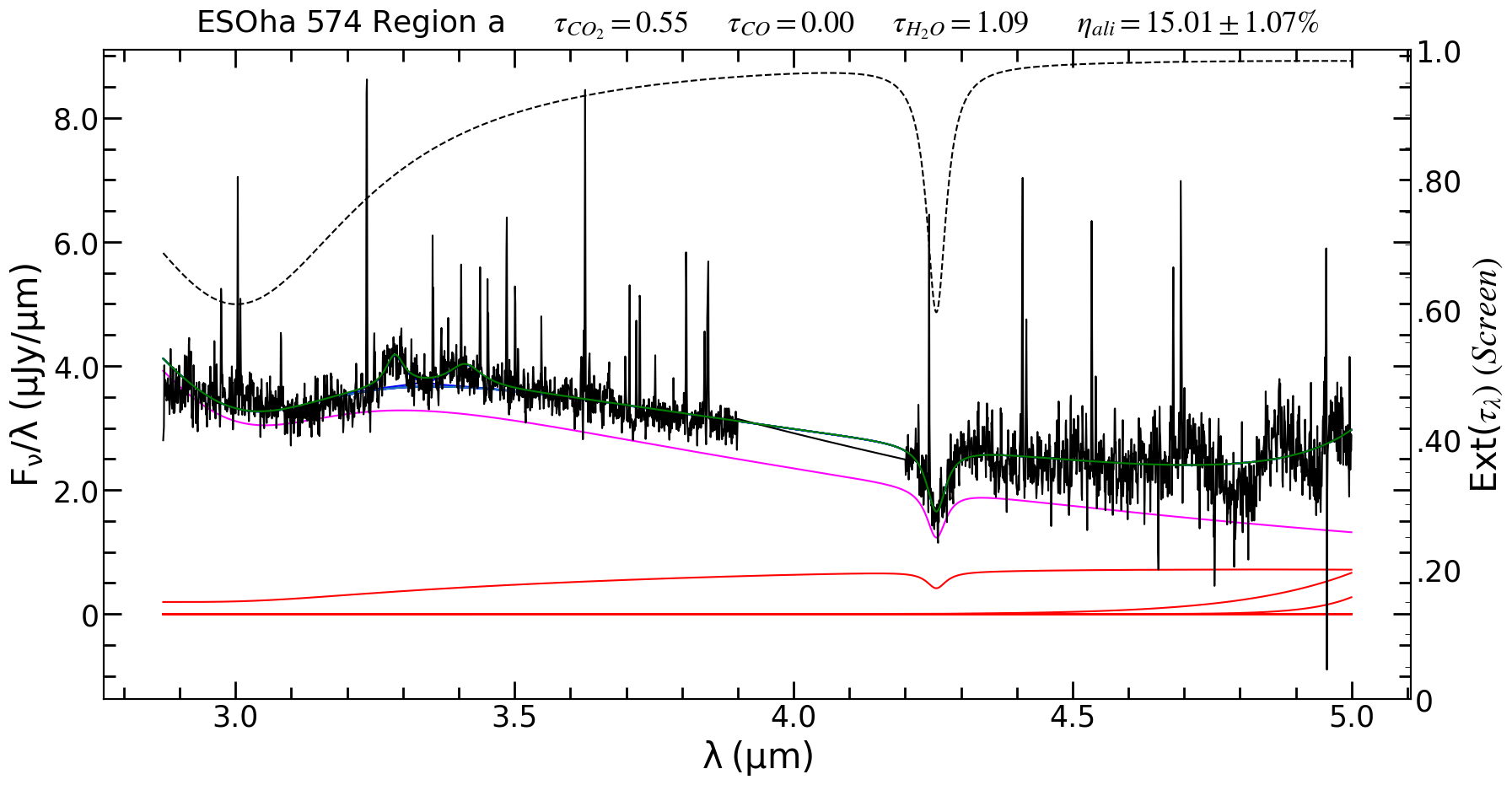}
\includegraphics[width=0.19\textwidth]{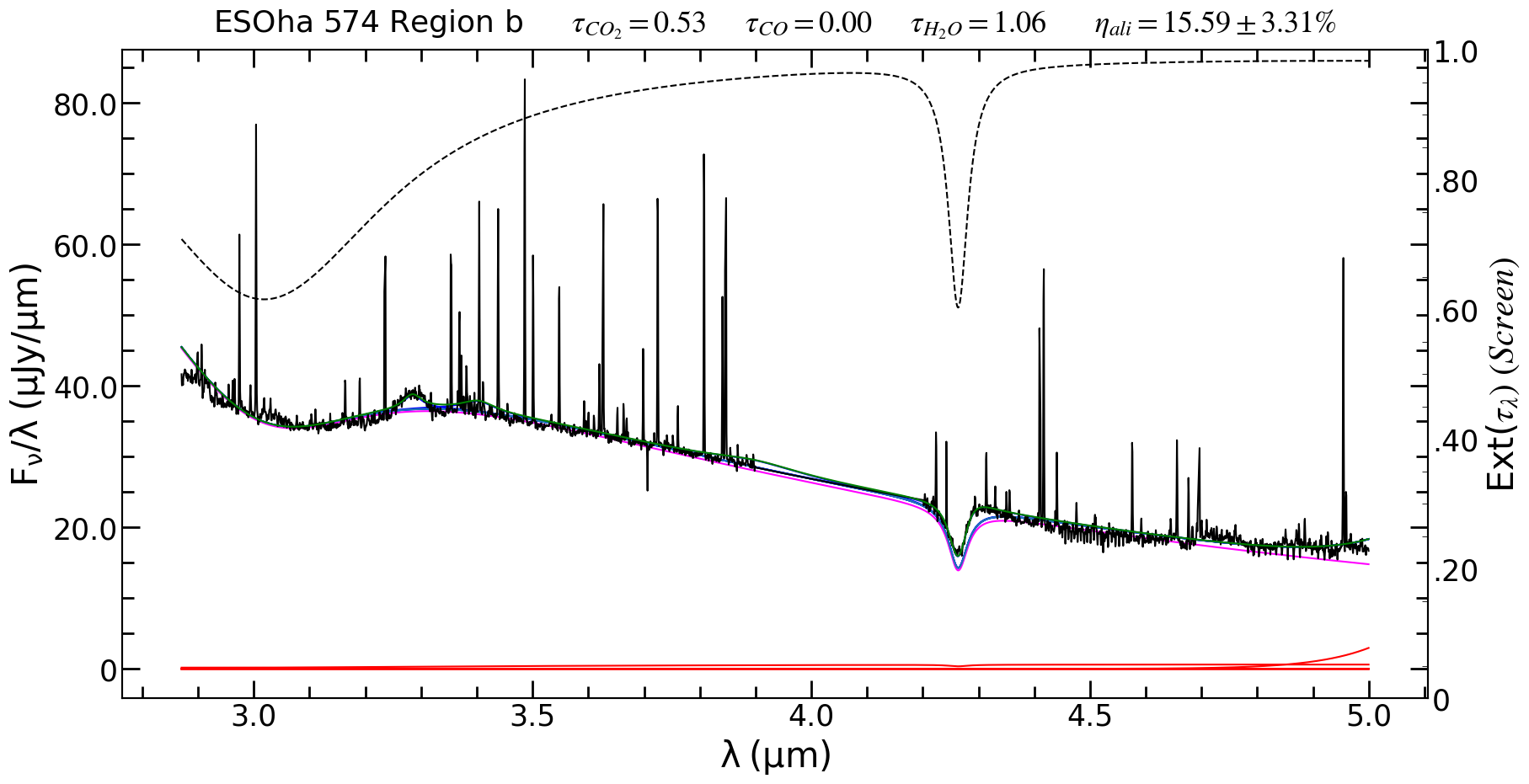}
\includegraphics[width=0.19\textwidth]{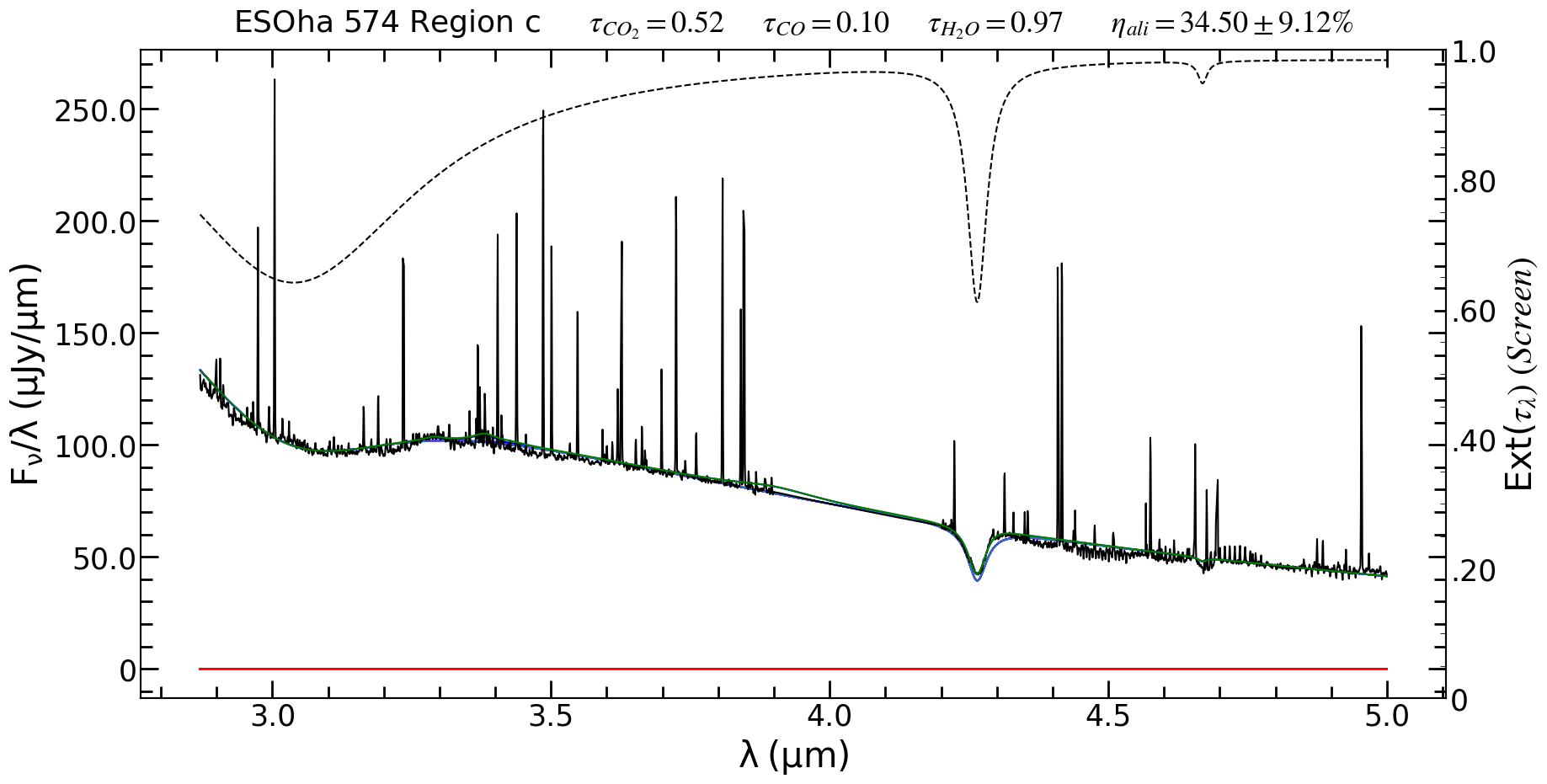}
\includegraphics[width=0.19\textwidth]{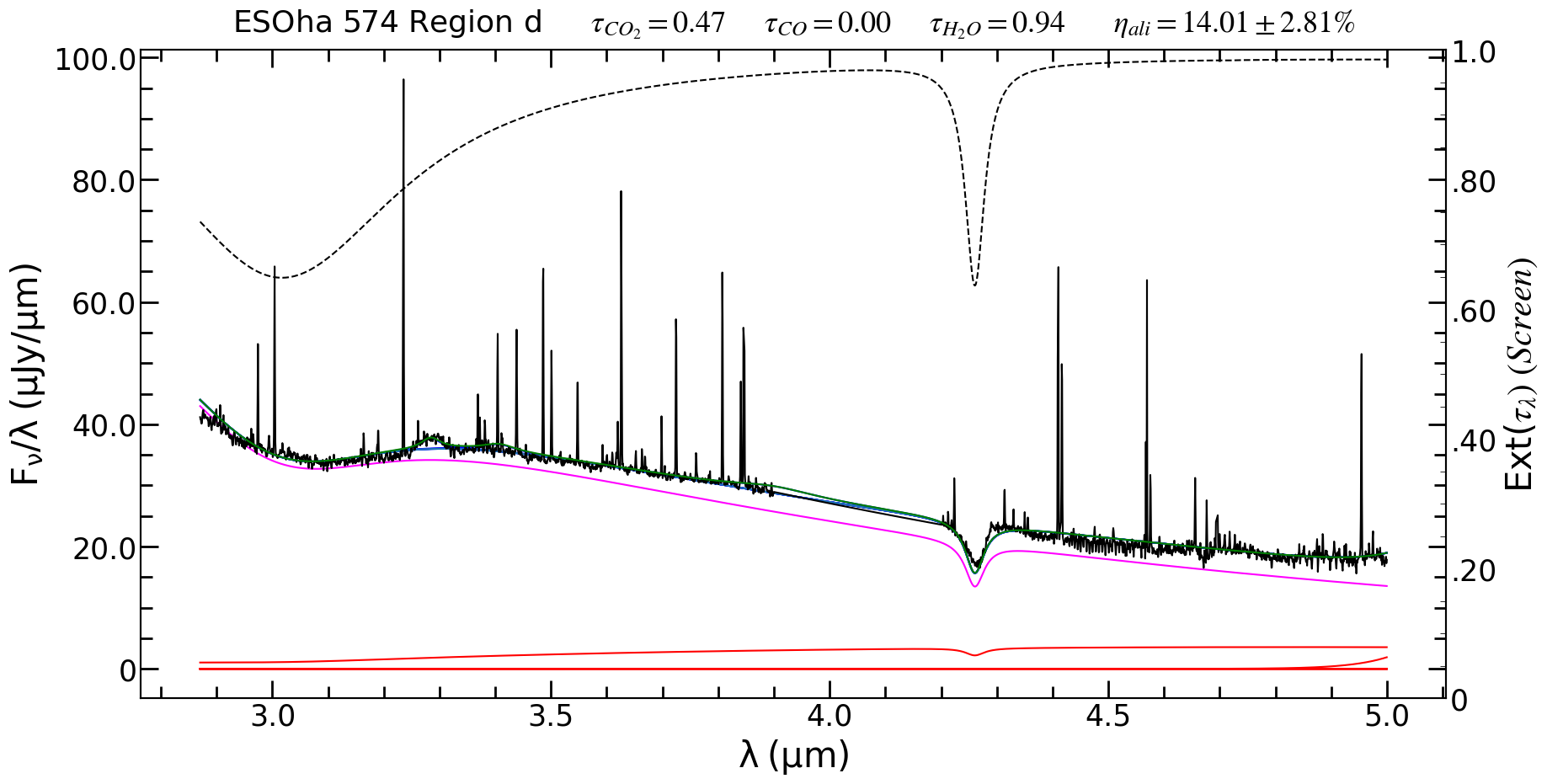}
\includegraphics[width=0.19\textwidth]{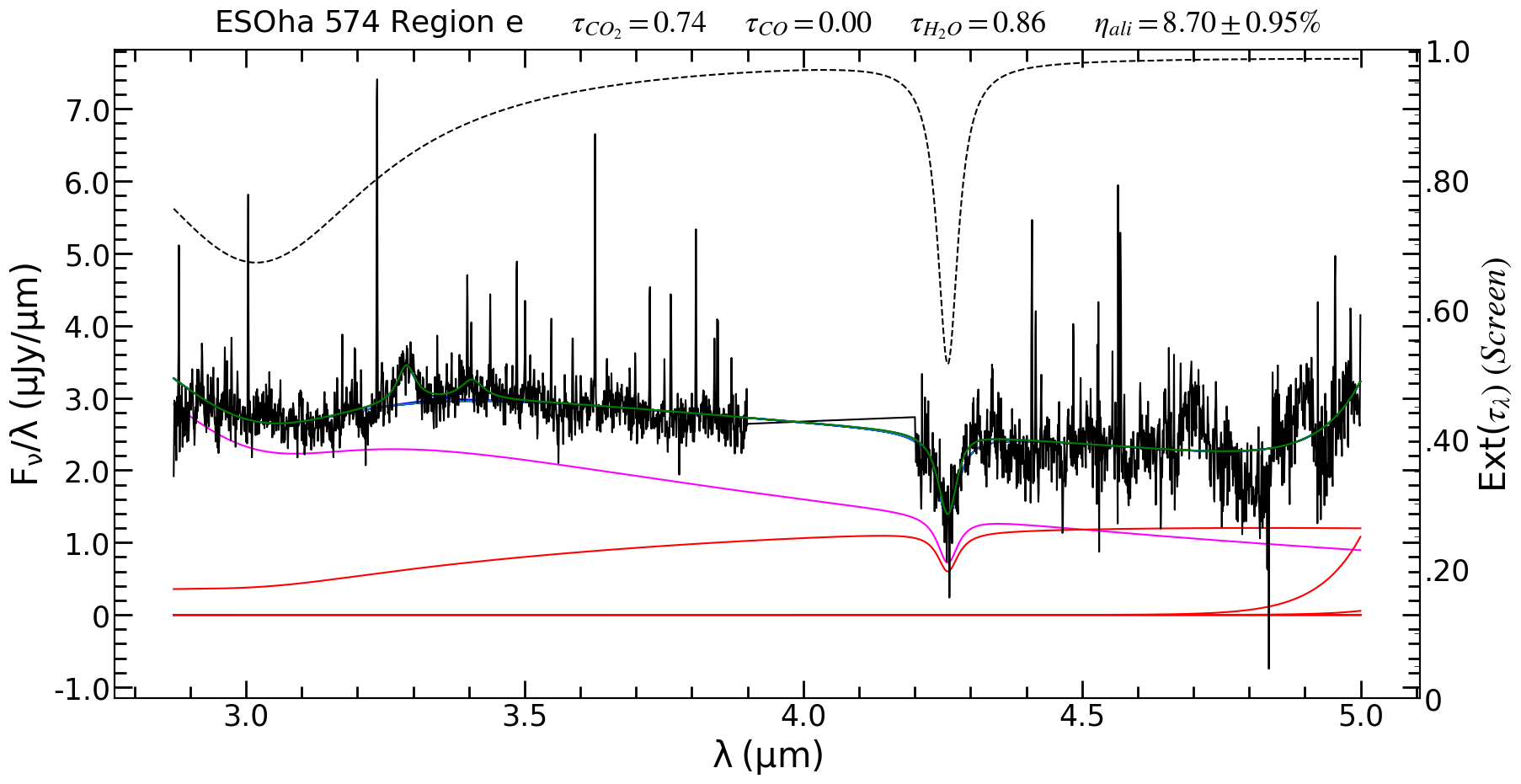}
\includegraphics[width=0.19\textwidth]{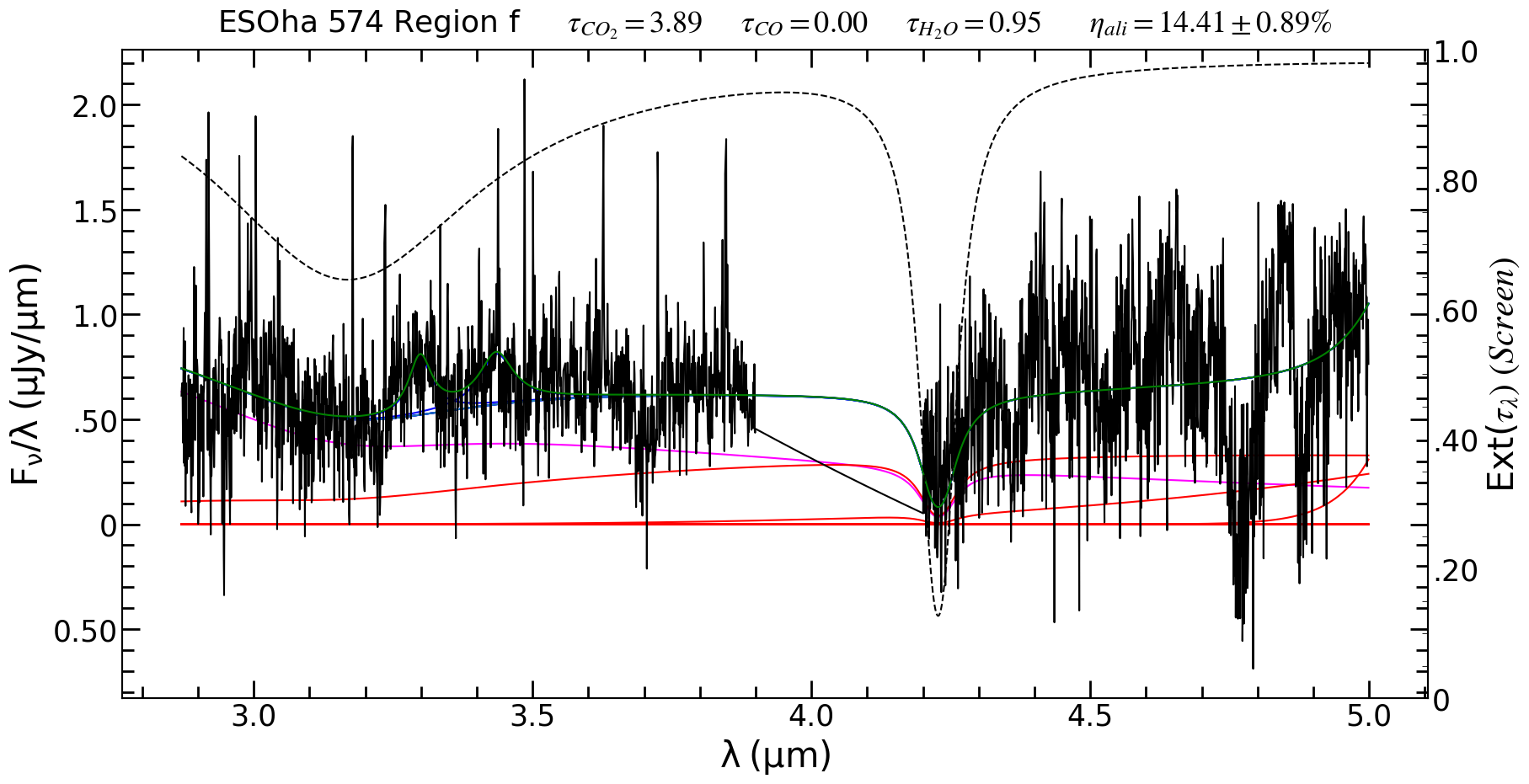}
\includegraphics[width=0.19\textwidth]{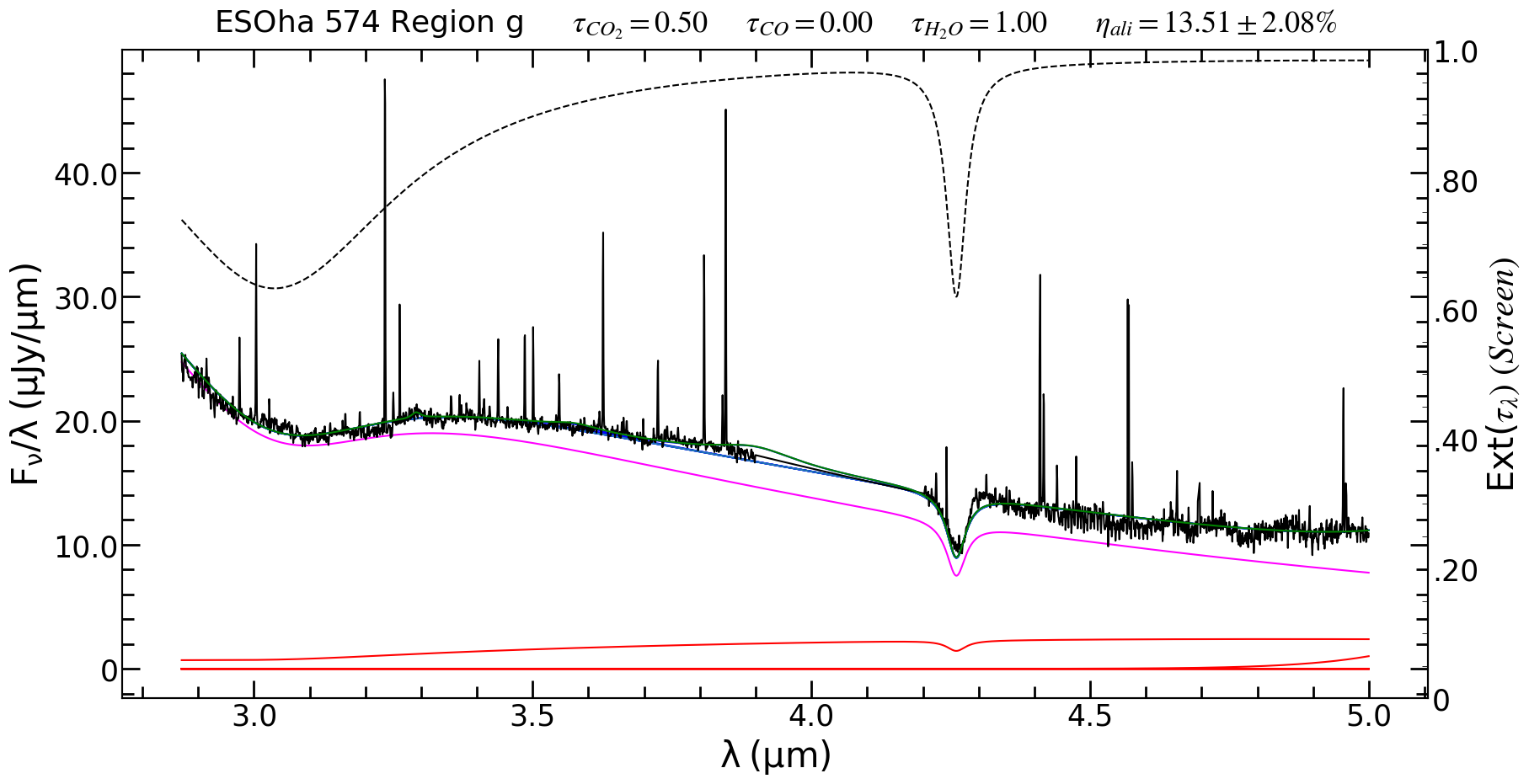}
\includegraphics[width=0.19\textwidth]{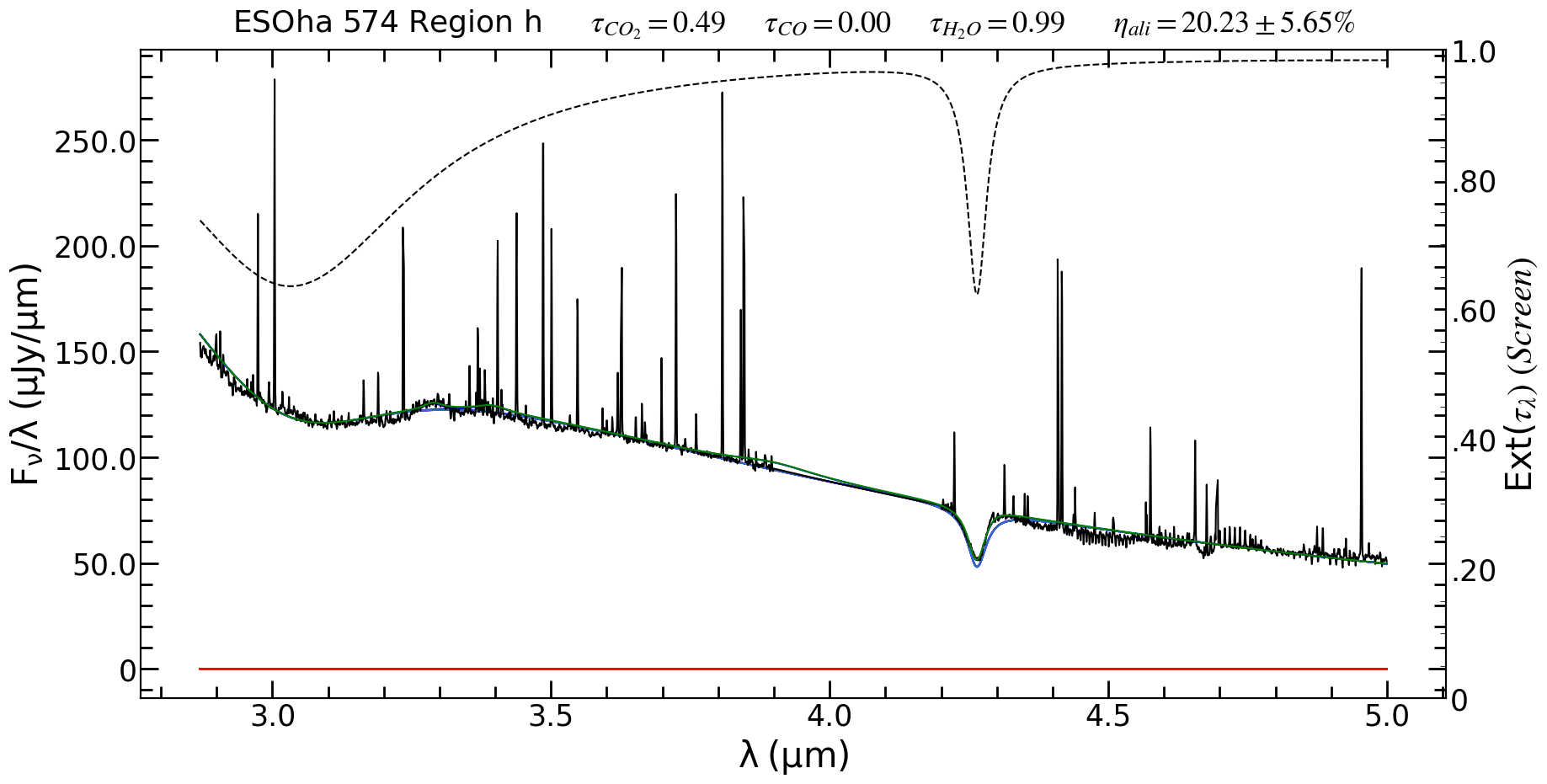}
\includegraphics[width=0.19\textwidth]{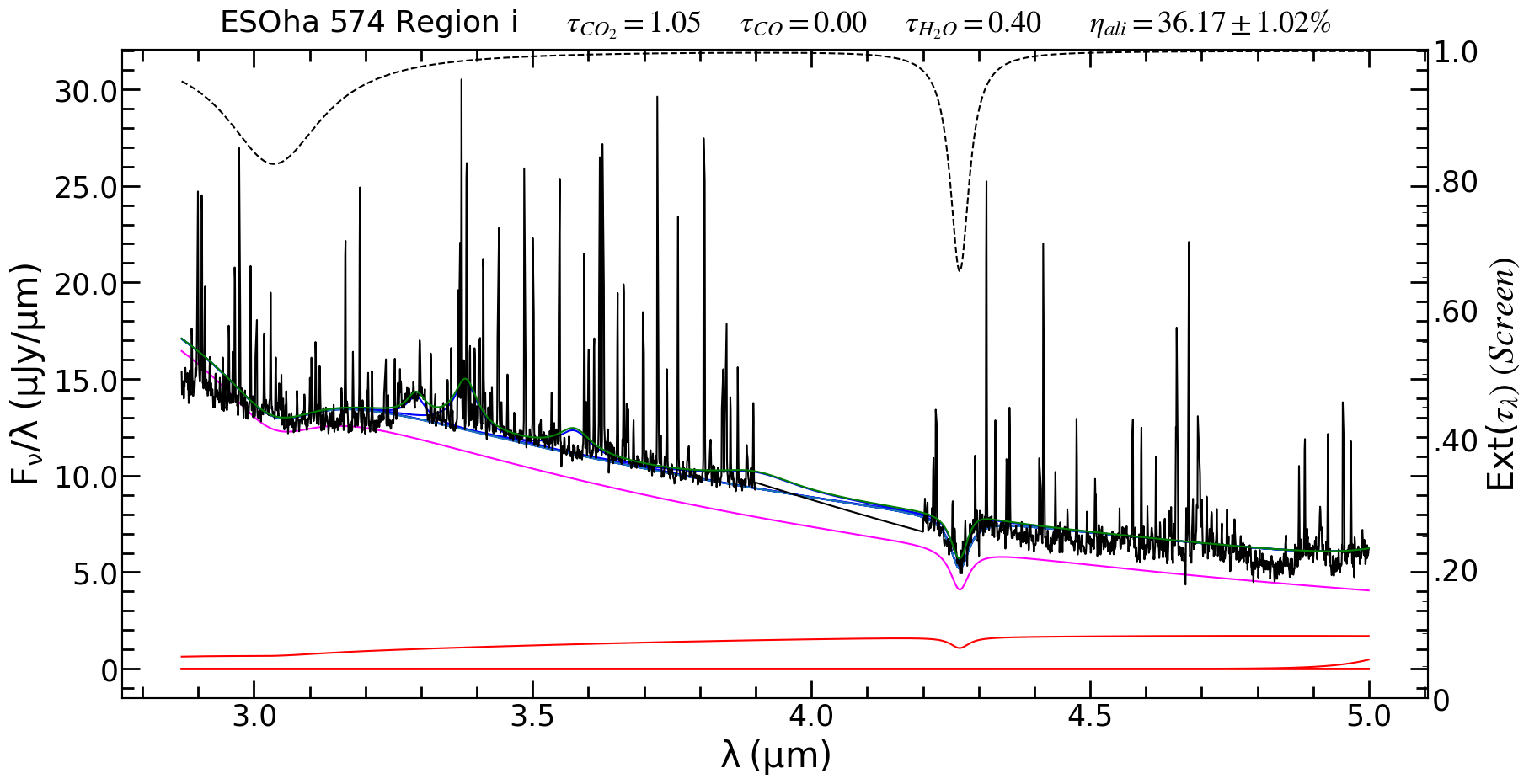}
\includegraphics[width=0.19\textwidth]{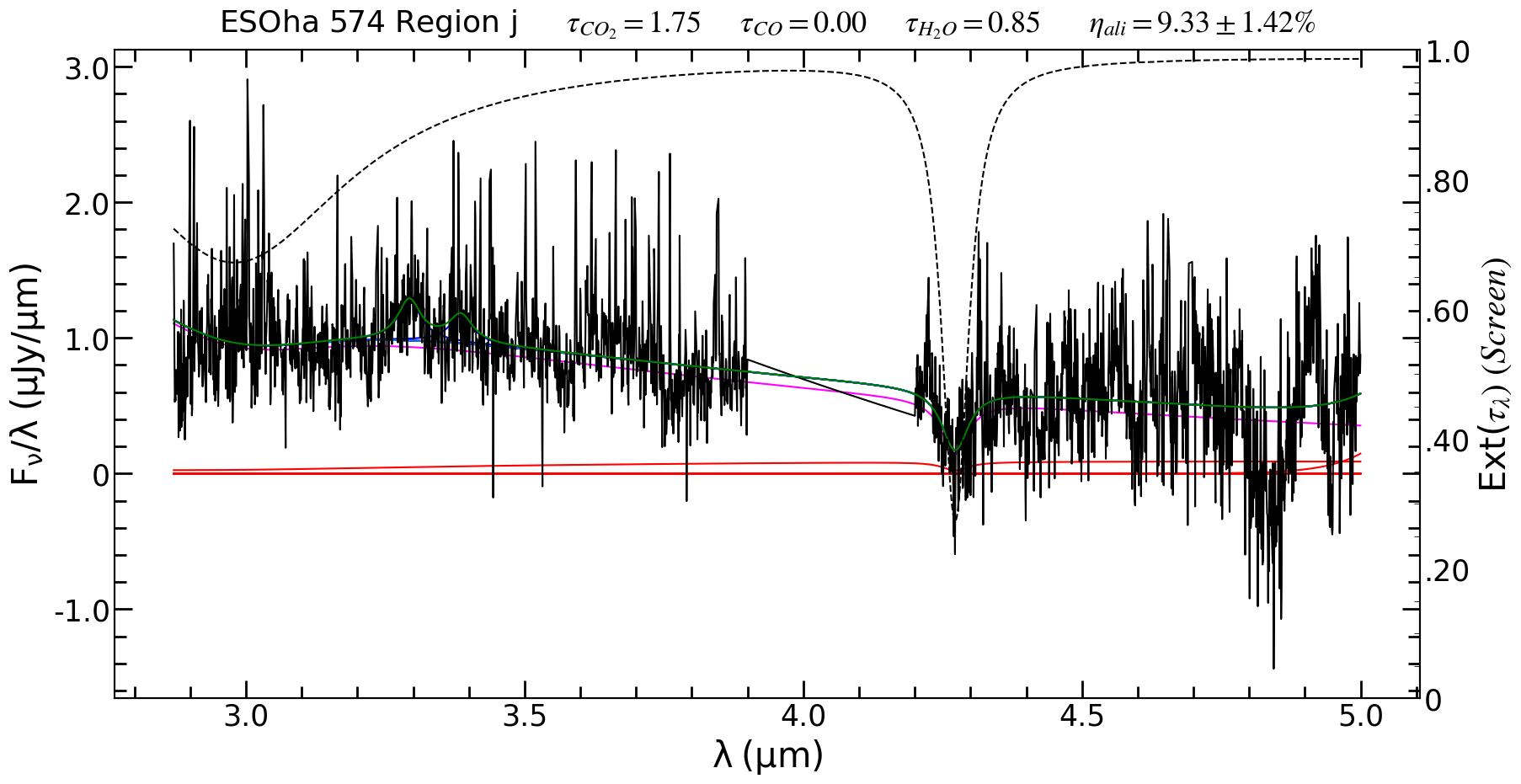}

\caption{PAHFIT results for ESO-H$\alpha$ 574 for the near-IR of all regions examined to calculate the $\eta_{\rm ali}$.}
\label{fig:ESO_Slices}
\end{figure*}

\begin{figure*}[htb!]
\centering

\includegraphics[width=0.19\textwidth]{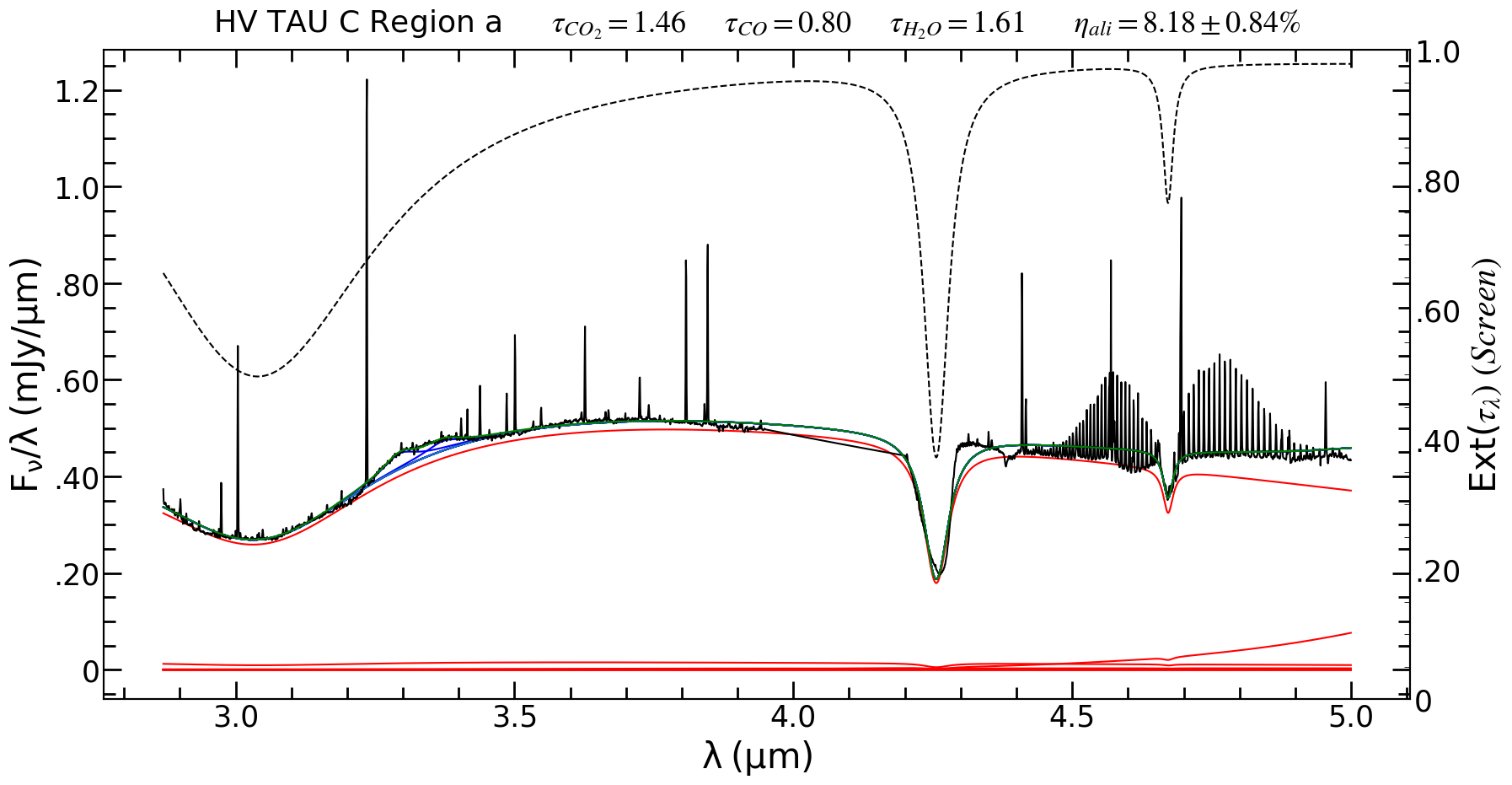}
\includegraphics[width=0.19\textwidth]{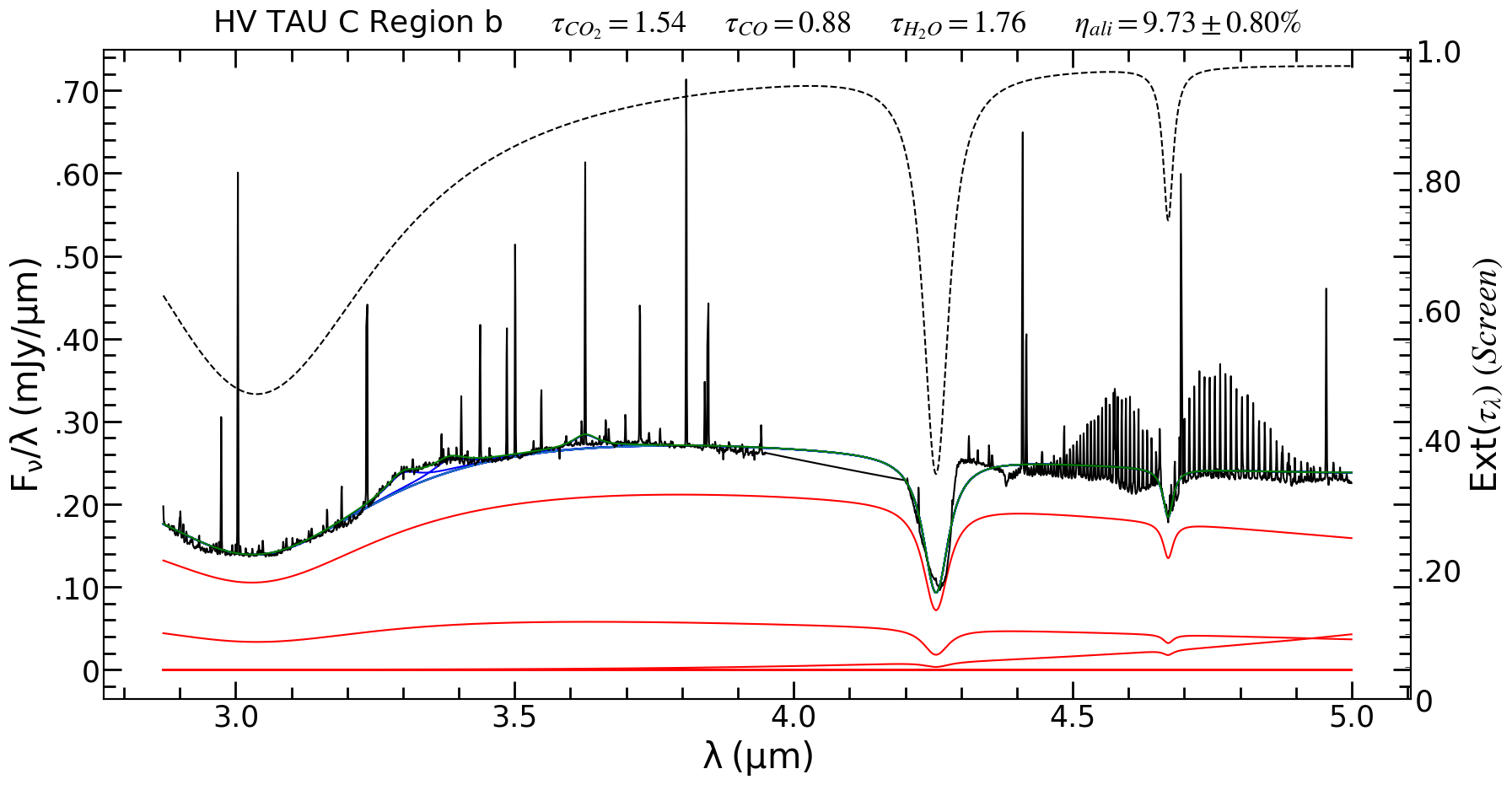}
\includegraphics[width=0.19\textwidth]{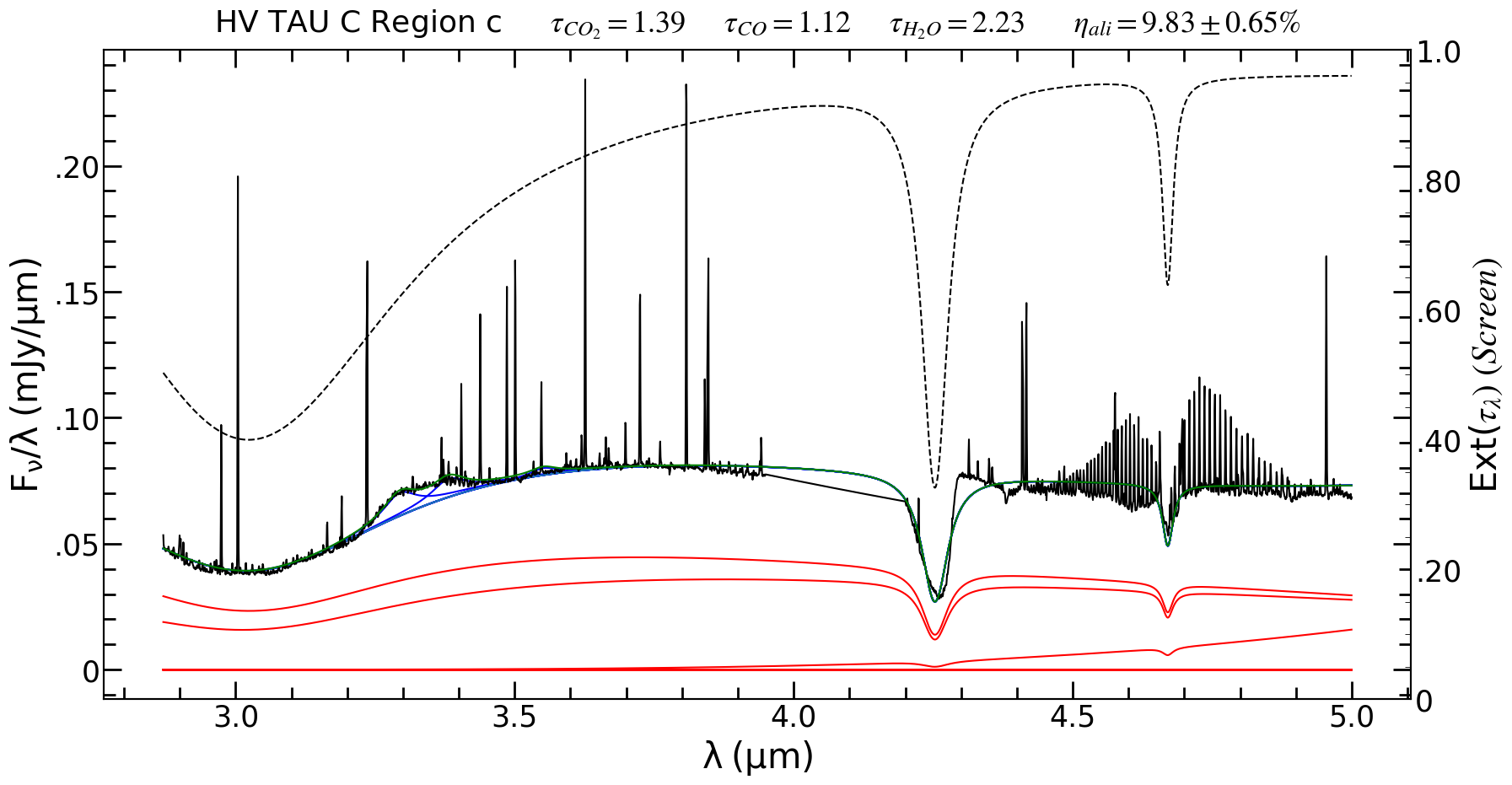}
\includegraphics[width=0.19\textwidth]{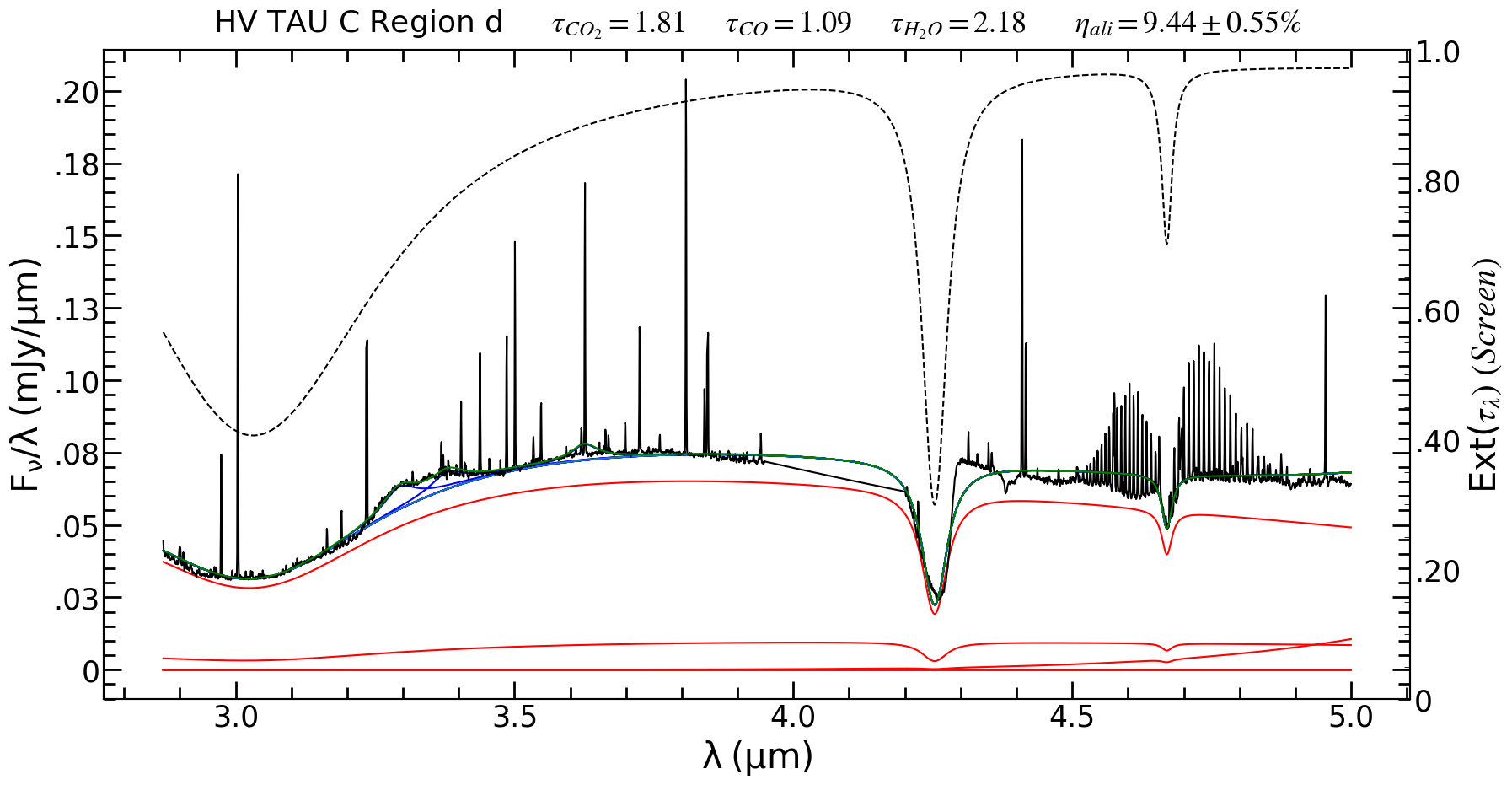}
\includegraphics[width=0.19\textwidth]{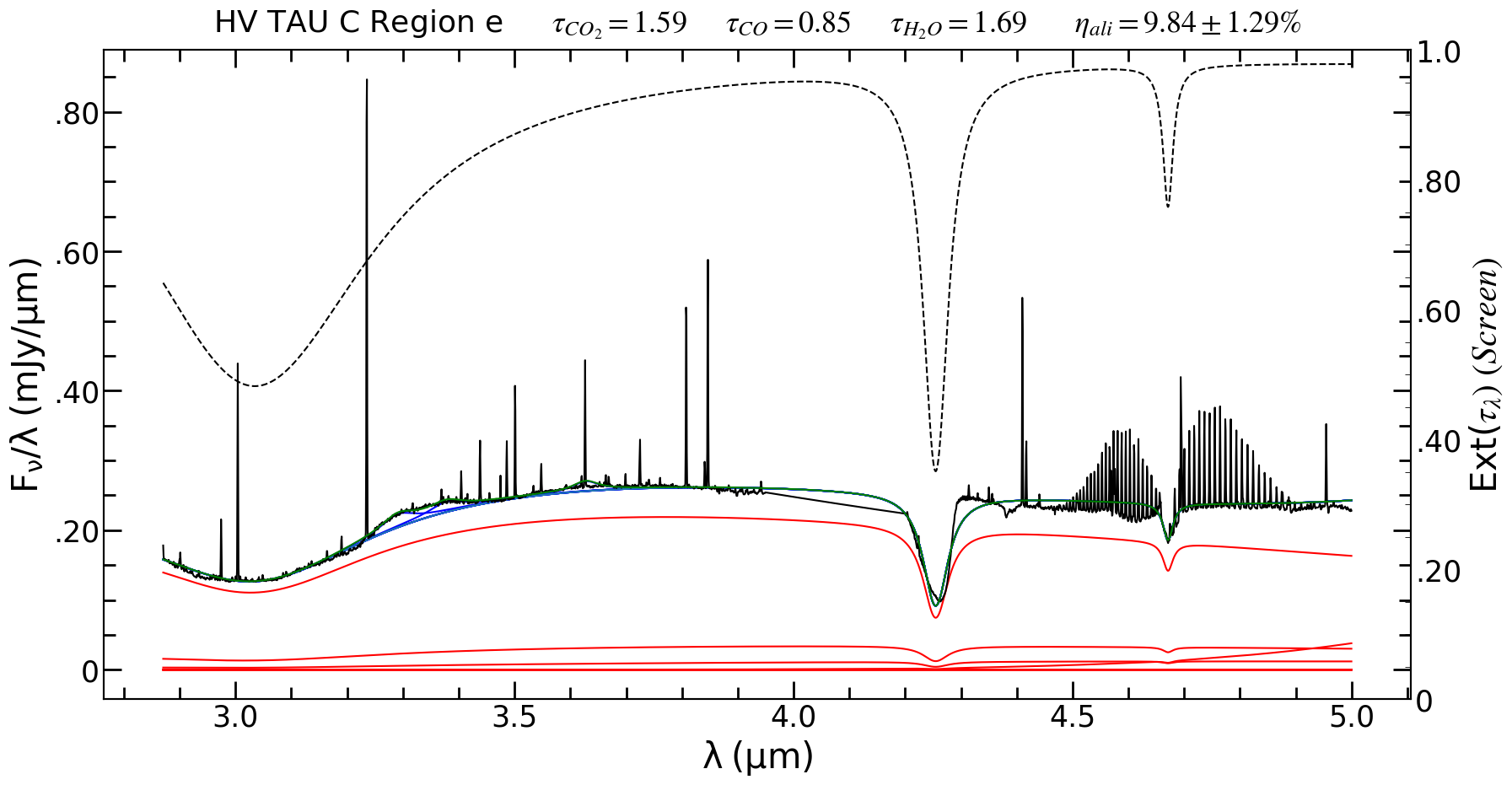}
\includegraphics[width=0.19\textwidth]{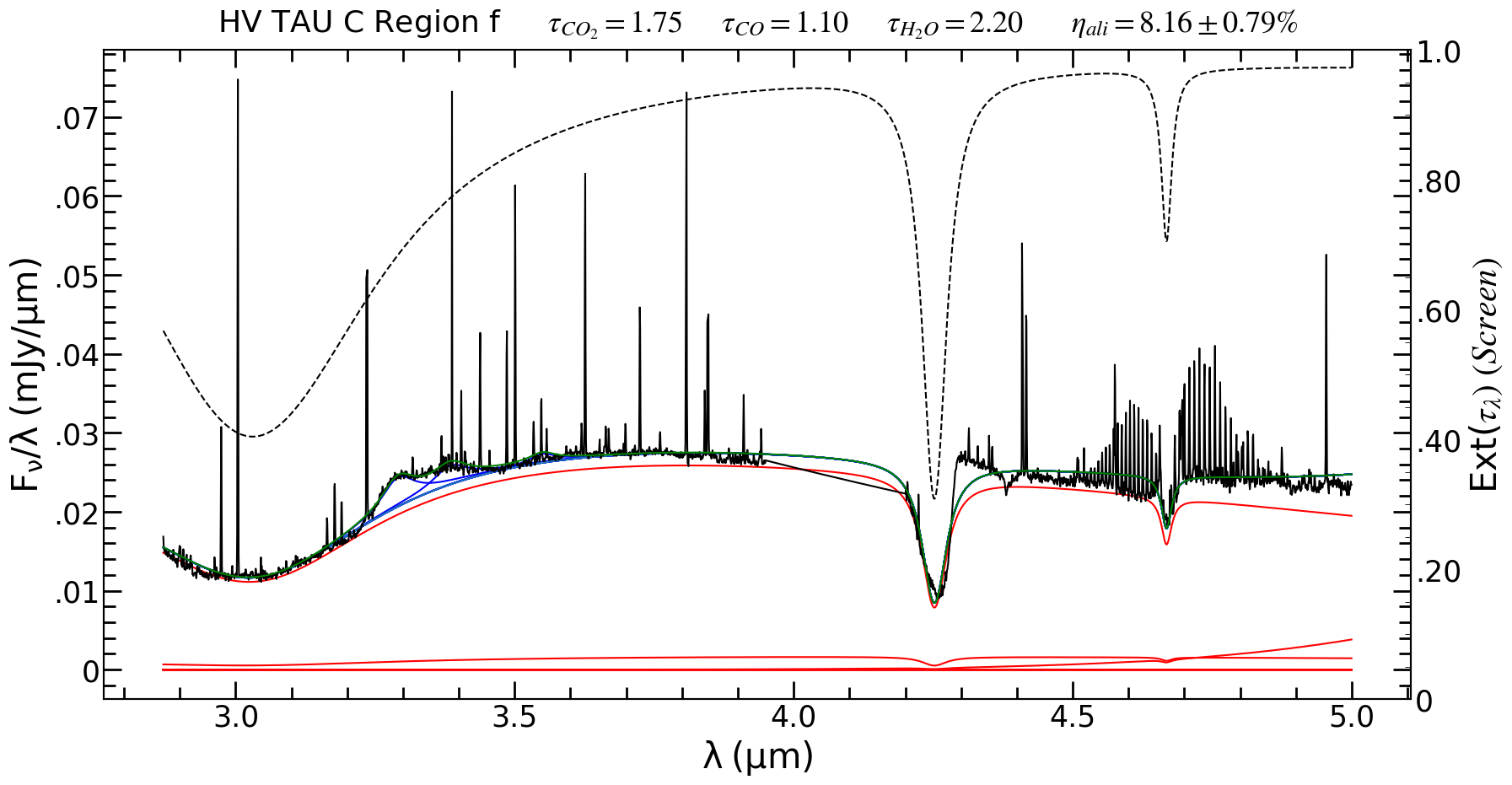}
\includegraphics[width=0.19\textwidth]{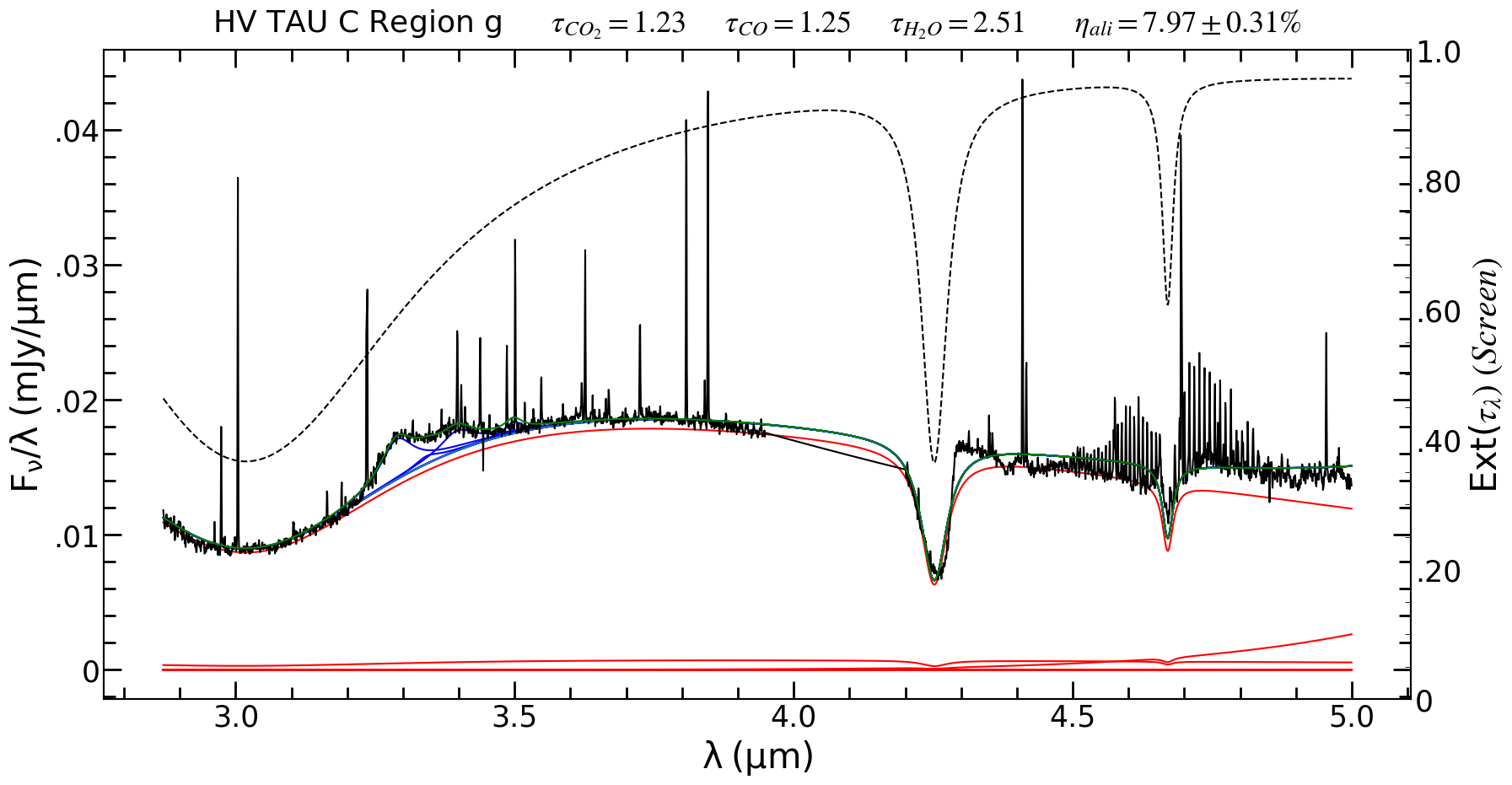}
\includegraphics[width=0.19\textwidth]{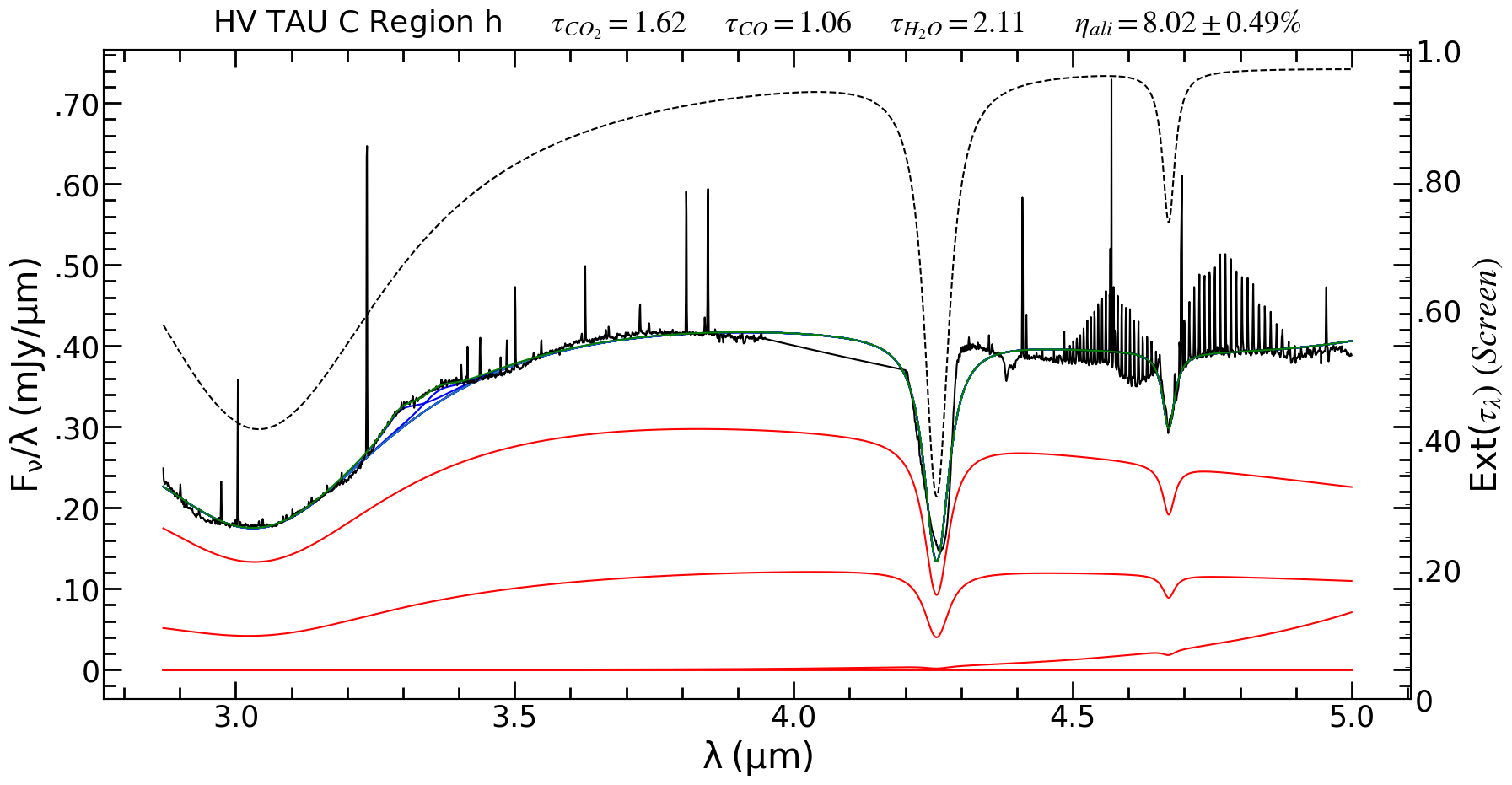}
\includegraphics[width=0.19\textwidth]{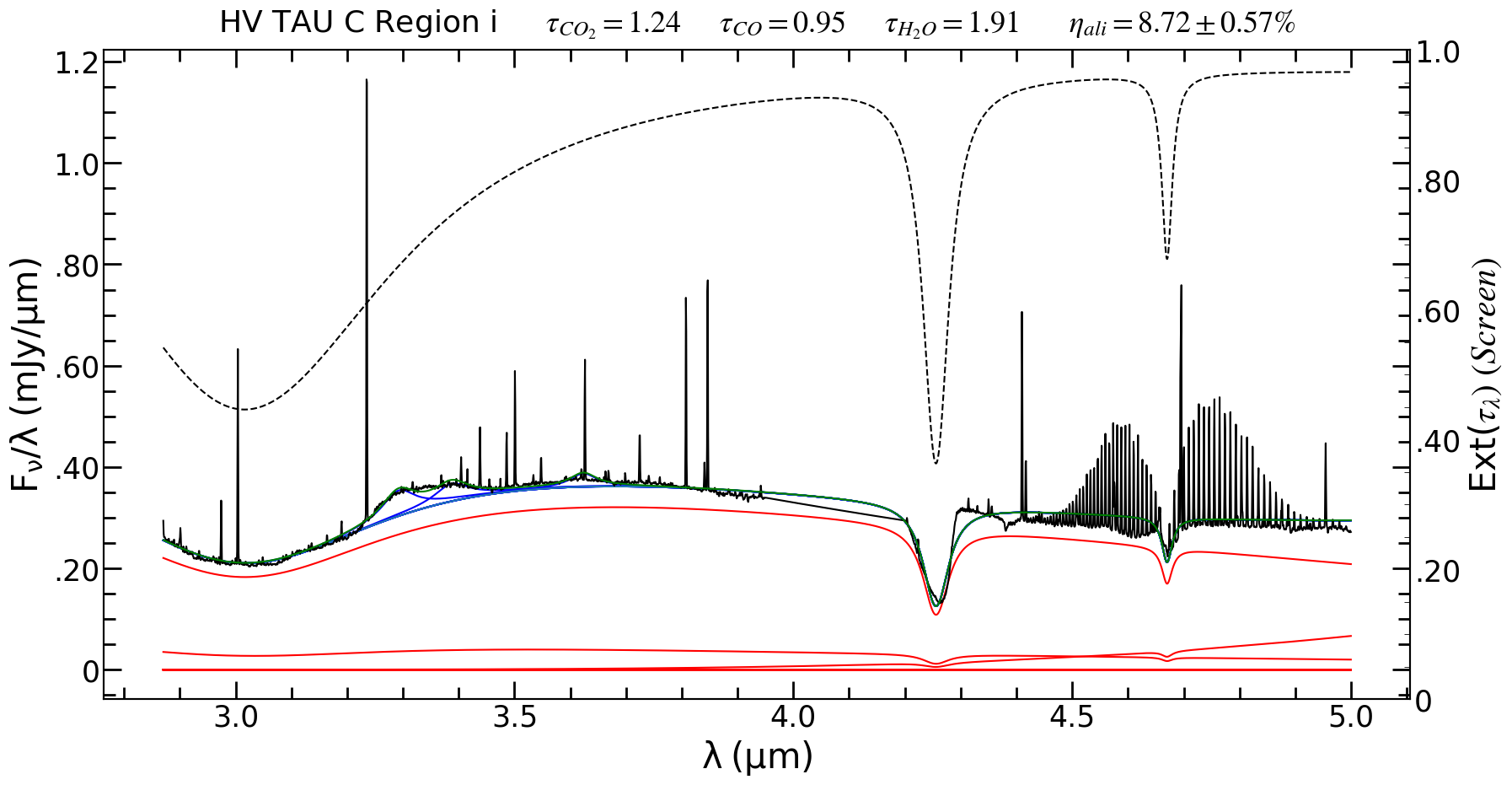}
\includegraphics[width=0.19\textwidth]{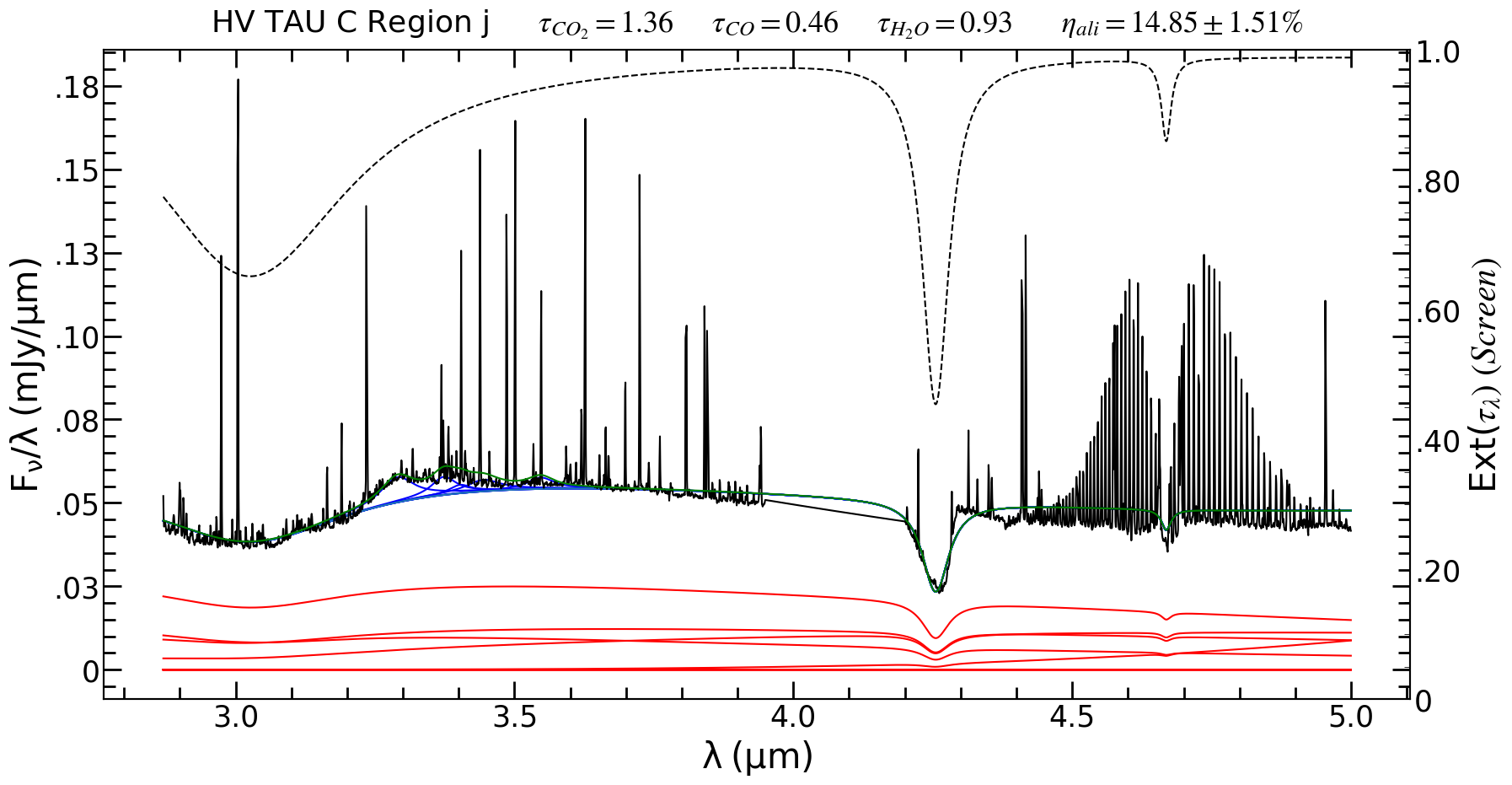}

\caption{PAHFIT results for HV~TAU~C for the near-IR all regions examined to calculate the $\eta_{\rm ali}$.}
\label{fig:HVTAU_Slices}
\end{figure*}

\bibliography{sample701}{}
\bibliographystyle{aasjournalv7}

\end{document}